 \documentclass[twoside,a4paper,11pt]{Classes/fourthterm}

\ifpdf
    \pdfcatalog { /PageMode (/UseOutlines)
                  /OpenAction (fitbh)  }
\fi

\title{Dynamical instabilities in disc-planet interactions}

\ifpdf
\author{\href{mailto:mkl23@cam.ac.uk}{Min-Kai Lin}}
\collegeordept{\href{http://www.joh.cam.ac.uk}{St. John's College}}
\university{\href{http://www.cam.ac.uk}{University of Cambridge}}
\crest{\includegraphics[width=30mm]{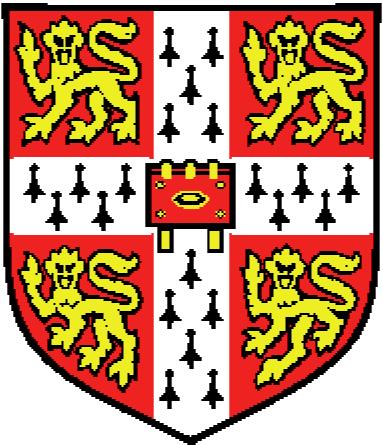}}
\else
\author{\href{mailto:mkl23@cam.ac.uk}{Min-Kai Lin}}
\collegeordept{\href{http://www.joh.cam.ac.uk}{St. John's College}}
\university{\href{http://www.cam.ac.uk}{University of Cambridge}}
\crest{\includegraphics[bb = 0 0 292 336, width=30mm]{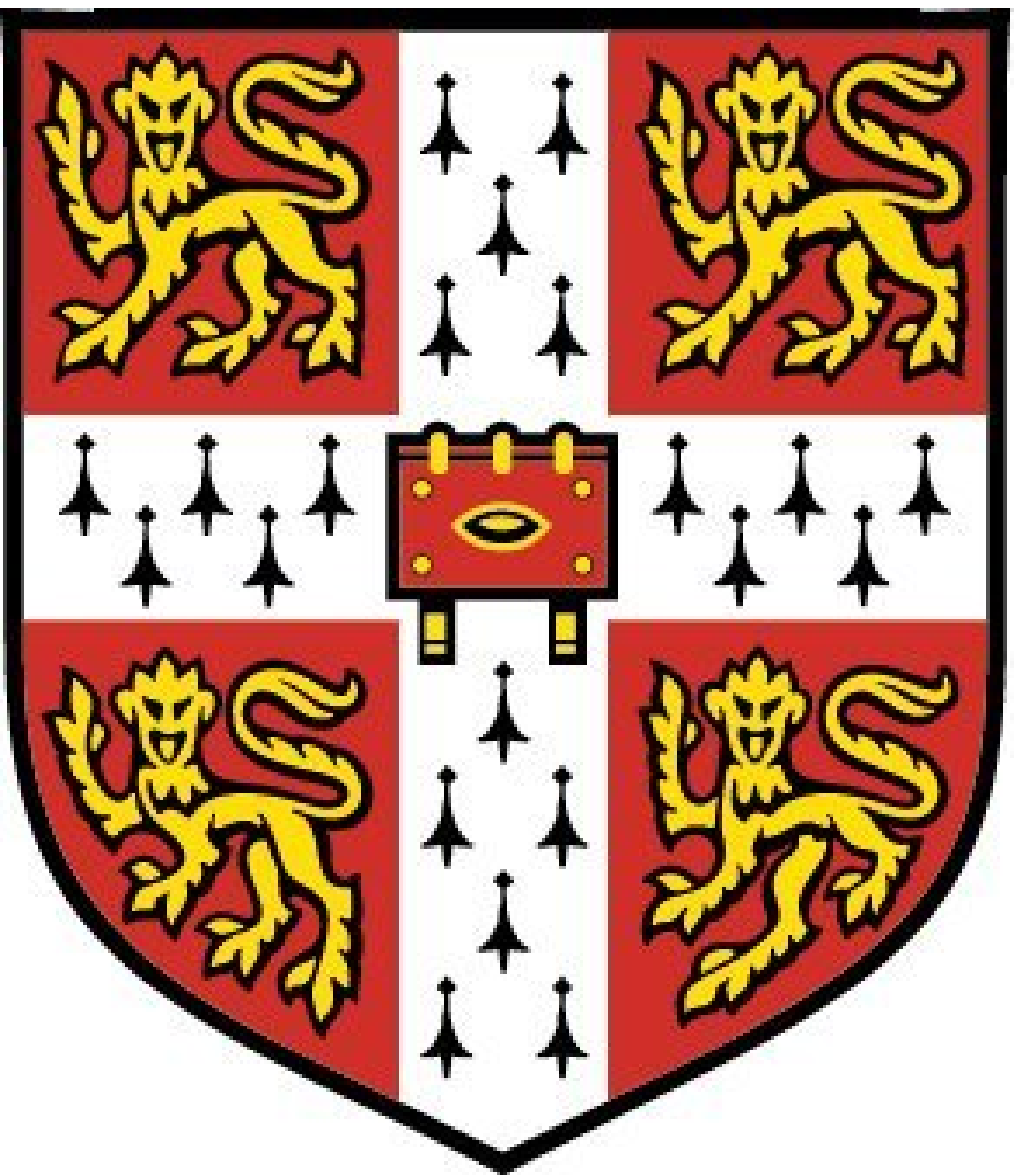}}
\fi

\renewcommand{\submittedtext}{This dissertation is submitted for the degree of}
\degree{Doctor of Philosophy}
\degreedate{September 2011}

\usepackage{StyleFiles/watermark}
\usepackage{amsmath}
\usepackage{aastex}
\usepackage{graphicx}
\usepackage{subfigure}
\usepackage{lscape}
\usepackage[font=normal,format=plain,labelfont=bf,up,textfont=it,up]{caption}
\usepackage{bm}
\usepackage{appendix}
\usepackage{setspace}

\newcommand{\p}{\partial}
\newcommand{\sbar}{\bar{\sigma}}
\newcommand{\adot}{\dot{a}}
\newcommand{\fargo}{{\tt FARGO }}
\newcommand{\zeus}{{\tt ZEUS-MP }}
\newcommand{\zeusold}{{\tt ZEUS }}

\newcommand{\bmu}{\bm{u}}
\newcommand{\lmax}{l_\mathrm{max}}
\newcommand{\mmax}{m_\mathrm{max}}

\newcommand{\he}{\mathcal{H}}

\newcommand{\dd}{\delta}

\begin{document}




\maketitle
\cleardoublepage

\setcounter{secnumdepth}{3}
\setcounter{tocdepth}{1}

\begin{romanpages}

\begin{dedication} 
Dedicated to my parents, Yen-Chu Lu and Ing-Young Lin. 
\end{dedication}




\begin{acknowledgements}      

This dissertation is the result of my own work and includes nothing which is the outcome of work done in collaboration 
except where specifically indicated. \\

The contents of Chapter \ref{paper1}, \ref{paper2} and \ref{paper3} have respectively been 
published as \cite{lin10,lin11a,lin11b}: 

\begin{itemize}

\item \emph{Type III migration in a low viscosity disc}, Lin, M.-K., Papaloizou, J.C.B, 2010, MNRAS, 405, 1473
\item \emph{The effect of self-gravity on vortex instabilities in disc-planet interactions}, Lin, M.-K., Papaloizou, J.C.B, 2011, MNRAS, 415, 1426
\item \emph{Edge modes in self-gravitating disc-planet interactions}, Lin, M.-K., Papaloizou, J.C.B, 2011, MNRAS, 415, 1445

\end{itemize}

Specifically, the stabilisation of vortex modes (\S\ref{S3.4}) was proved by J. Papaloizou. The analytical 
interpretation of edge modes (\S\ref{Fluxbal}, \S\ref{Neutedge}---\ref{spirals}) was given by J. Papaloizou.  
All semi-analytical modelling, linear calculations, hydrodynamic simulations and their results analysis 
have been performed by myself. 

I thank my supervisor Professor John Papaloizou for support, guidance and inspiration. I thank C. Baruteau
for his version of the \fargo code and the ASIAA CFD-MHD group in Taiwan 
for providing a copy of their \texttt{ANTARES} code for reference.  
 I thank members of the AFD group in DAMTP for three fruitful years.  
I thank the Isaac Newton Trust for a studentship and St John's College, Cambridge for a 
Benefactor's Scholarship. This work was also supported by an Overseas Research award. 
\clearpage
\end{acknowledgements}



\cleardoublepage


\begin{abstracts}        

Protoplanetary discs can be dynamically unstable due to
structure induced by an embedded giant planet. In this thesis, I
discuss the stability of such systems and explore the consequence of
instability on planetary migration. \\


I present semi-analytical models to understand the formation of the
unstable structure induced by a Saturn mass planet, which leads to
vortex formation. I then investigate the effect of such vortices on
the migration of a Saturn-mass planet using hydrodynamic
simulations. I explain the resulting non-monotonic behaviour in the
framework of type III planetary migration. 


I then examine the role of disc 
self-gravity on the vortex instabilities. It can be shown that  
self-gravity has a stabilising effect. Linear
numerical calculations confirms this. When applied to disc-planet
systems, modes with small azimuthal wavelengths
are preferred with increasing disc self-gravity. 
This is in agreement the observation that more vortices develop in
simulations with increasing disc mass. 
Vortices in more massive discs also resist
merging. I show that this is because inclusion of self-gravity sets a
minimal vortex separation preventing their coalescence, which would
readily occur without self-gravity. 

I show that in sufficiently massive discs vortex modes are
suppressed. Instead, global spiral instabilities develop. 
They are interpreted as disturbances associated with the
planet-induced structure, which interacts with the wider disc leading to
instability. I carry out linear calculations to confirm this physical 
picture. Results from nonlinear hydrodynamic simulations are also in
agreement with linear theory. I give examples of the effect of these
global modes on planetary migration, which can be outwards,
contrasting to standard inwards migration in more typical disc
models. 


I also present the first three-dimensional computer simulations
examining planetary gap stability.  I confirm that the results
discussed above, obtained from two-dimensional disc approximations,
persist in  three-dimensional discs. 


\end{abstracts}



\tableofcontents

\end{romanpages}




\chapter{Introduction}\label{introduction}\label{intro}
\graphicspath{{Introduction/figures/}}

The interaction between planets and their environments
---protoplanetary discs---can potentially account for 
the structure of the Solar system as well as extra-Solar planetary systems.  
The number of detected exoplanets currently 
stand at 603 (September 2011). They are found with a variety 
of orbital properties. Among these 
are the `hot Jupiters' --- Jovian mass planets orbiting close to their
host star \citep{marcy05,udry07}. The first exoplanet discovered around a Solar-type 
star, 51 Pegasi B, has a period of only 4 days \citep{mayor95}. 
It is difficult to explain their formation in situ with either the core
accretion or gravitational instability theories of giant planet formation 
\citep[see, e.g.][for recent reviews]{lissauer07,durisen07,dangelo10}.

One possibility is that hot Jupiters formed further out, where
conditions are more favourable, then through gravitational
interaction with the gaseous protoplanetary disc migrated inwards \citep{lin86}. 
Understanding disc-planet interactions for a range of 
disc and planet properties is desirable, because its application is not 
restricted to planet formation. Analogous interactions occur
for stars in black hole accretion discs, 
which has recently received attention and results from 
planetary migration applied \citep{baruteau11b,kocsis11,mckernan11}.

Embedded objects can affect stability properties of the disc, and this
can back-react on the migration of said object. Thus, a complete
theory of planetary migration should also account for possible
planet-induced instabilities. The studies presented in this thesis
consider instabilities associated with radially structured
protoplanetary discs due to a planet and their effects on
migration. 




\section{Disc-planet interactions}
Tidal interaction between the planet and disc fluid leads to angular
momentum exchange between them, and a change in the planet's orbital
radius from the star. This is planetary migration. It is classically
distinguished between type I, type II and more recently, type III
\cite[see, e.g.][for recent
reviews]{papaloizou07,masset08,paardekooper09c}.    

\subsection{Type I migration}
The response of a protoplanetary disc to an embedded low-mass planet
can be treated with linear perturbation  
analysis \citep[e.g.][]{goldreich79, goldreich80,lin93,artymowicz93,
ward97}. Disc-planet torques are exerted at
Lindblad and co-rotation resonances. In standard disc models, e.g. 
the minimum Solar mass nebula \citep{weidenschilling77,hayashi81}, the
corotation torque is negligible and the total Lindblad torque is negative, and yield 
 inward migration timescales too small compared to disc
lifetimes for the survival of protoplanetary cores. Thus, type I
migration remains an active area of study.   

Recent developments in type I migration include three-dimensional 
linear calculations \citep{tanaka02}, non-isothermal effects
\citep{paardekooper06,paardekooper08, paardekooper10b,
  paardekooper11}, 
disc self-gravity \citep{pierens05,baruteau08} and interaction with
magnetic turbulence \citep{nelson04,baruteau11c,uribe11}. However, low
mass planets do not perturb the disc significantly to render it
unstable, so type I  migration will not be the most relevant mechanism 
for disc-planet properties considered later.  


\subsection{Type II migration}
Massive planets can clear an annular gap in the disc. This requires
the disc to be perturbed nonlinearly, leading to shocks, which
amounts to requiring the planet's Hill radius to exceed the disc thickness. The
planet also needs to exert a torque on the disc, which tends to push
material away from its orbital radius, that is sufficiently large to
overcome viscous torques trying to close the gap. Gap-opening is
discussed in \cite{lin93},\cite{bryden99}, and the two criteria above
have been combined by \cite{crida06}. 

In type II migration, the planet remains in the gap 
evolving inwards with disc accretion. Migration therefore proceeds on disc
viscous timescales \citep{lin86}. The planetary masses to be
considered are gap-opening, but standard type II migration is still
not relevant because for the conditions presented later, 
migration is strongly influenced by gap instabilities
operating on much shorter, dynamical timescales.  


\subsection{Type III migration}
Type III migration is relatively new. It was first
described by \cite{masset03} and further discussed in
\cite{artymowicz04} and \cite{papaloizou05}. Recently, a series of
detailed numerical studies were performed by
\cite{peplinski08a,peplinski08b,peplinski08c}. 
This is a rapid form of
migration with timescales less than 100's of orbits,
compared to at least a few thousand for type I and type
II. Type III migration is most applicable to Saturn-mass planets
opening partial gaps in massive discs.

Type III migration is driven by flow of disc material across the
planetary gap, due to its migration in the first place. 
It is self-sustaining and can be directed inwards or
outwards. However, it requires an initial kick. A simplified
description of this mechanism, following \cite{papaloizou07}, is
outlined below.

\begin{figure}[!ht]
 \center
 \includegraphics[clip,trim=0cm 0cm 0cm 1.3cm,scale=.50]{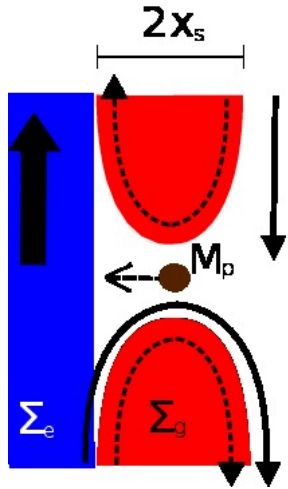}
 \caption{Illustration of inwards type III migration. Fluid crossing from the inner disc to outer disc 
executes one horse-shoe turn behind the planet. The fluid gains angular momentum which exerts a negative 
torque on the planet, encouraging further orbital decay.  \label{coorb}}
\end{figure}

The flow topology near a (partial) gap is shown in Fig. \ref{coorb}. 
If the planet circulates at fixed orbital radius $r=r_p$ then 
the flow is divided into the horseshoe region (red) where material trapped in libration, 
and the circulating region where fluid completes circular orbits (blue). 
If the planet migrates with rate $\dot{r}_p <0$ then   
 material can cross from the inner disc (blue) 
to the outer disc by executing one horseshoe turn behind the planet 
Fluid elements change orbital radius $r_p-x_s\to r_p + x_s$, gaining angular momentum and exerts
a negative torque on the planet:
\begin{align}
\Gamma_3=2\pi r_p^2\dot{r}_p\Sigma_e\Omega(r_p) x_s,
\end{align}
where $\Sigma_e$ is the density at the upstream separatrix, $\Omega(r)$ is the disc angular speed and 
$x_s$ is half the horseshoe width (see Fig. \ref{coorb}).  
The migrating mass includes the planet $M_p$, fluid gravitationally bound to the planet
$M_r$ and fluid trapped on horseshoe orbits $M_h=4\pi\Sigma_g r_p x_s$.
Assuming circular Keplerian orbits, the migration rate is then
\begin{align}\label{type3_rate}
\dot{r}_p=\frac{2\Gamma_\mathrm{L}}{\Omega(r_p) r_p (\underbrace{M_p+M_\mathrm{r}}_{M_p^\prime}-\delta m)}
\end{align}
where $\Gamma_L$ is the total Lindblad torque and
\begin{align}\label{delta_m_dens}
\delta m =& 4\pi\Sigma_e r_p x_s-M_{h}
= 4\pi r_p x_s (\Sigma_e-\Sigma_g)
\end{align}
is called the co-orbital mass deficit (a more precise definition is used in
Chapter \ref{paper1}). $\delta m$ essentially compares the mass 
supplying the torque to the mass being moved, so migration speed increases with
$\delta m$. 

Instabilities associated with the flow around the coorbital region 
are expected to modify the standard type III picture above. For the disc-planet systems to be  
considered, the planetary gap becomes unstable. Hence, type III migration
is the most relevant form of disc-planet interaction.     
Instabilities contributing to $\Sigma_e$ ($\Sigma_g$) will favour (disfavour)
the mechanism. 
Fundamentally, the mechanism is angular momentum exchange due to direct gravitational 
scattering of disc material by the planet. Hence, instabilities that supply material for
the planet to scatter may be interpreted as a modified type III migration. 
Explicit examples of these are described in subsequent Chapters.

\section{Instabilities in structured astrophysical discs}
Instabilities in astrophysical discs have important applications. 
The magneto-rotational instability \citep[MRI,][]{balbus91} 
and gravitational instability \citep[GI,][]{gammie01} are robust paths
to turbulent transport of angular momentum in accretion discs. The
baroclinic instability \citep[BI,][]{klahr03,peterson07a,peterson07b} 
has also been invoked, but BI is probably more relevant to planetesimal formation. 
See \cite{armitage10} for a recent review of these processes in protoplanetary discs.

Instability may also be associated with steep surface density or
vortensity gradients\footnote{The term vortensity is used for the ratio of
  vorticity to surface density.}. This is commonly called the disc Rossby wave or vortex   
instability \citep{lovelace99,li00}, as it results  
from unstable interaction between Rossby waves across a vortensity
extremum. The mechanism is similar to instability in tori
\citep{papaloizou84,papaloizou85,papaloizou87}.  
They lead to vortex formation in the nonlinear regime
\citep{hawley87,li01}. Recently, linear calculations have been extended
to magnetic discs \citep{yu09,fu11} and 3D nonlinear simulations have
begun \citep{meheut10}. 

The original analysis was applied to non-self-gravitating
discs. Analogous instabilities exist in structured self-gravitating
particle discs \citep{lovelace78, sellwood91} 
and gaseous  discs \citep{papaloizou89, papaloizou91}. These discs
support global modes in addition to the localised vortices.

\section{Previous studies and relation to present work}
The stability of planetary gaps has
been studied with fix-orbit simulations \citep{koller03, li05, valborro07}. 
The consequence of instability on migration was noted in simulations 
with very low disc viscosity, when 
migration becomes non-monotonic \citep{ou07, li09,yu10}. 
In Chapter \ref{paper1}, the effect of vortex instabilities on migration is considered in 
detail for Saturn mass planets (previous studies consider smaller masses), 
and it is shown that results can be understood in the framework of type III migration.

Most works on disc-planet interactions ignore disc self-gravity. Simulations by \cite{li09} 
and \cite{yu10} did include self-gravity, but its effect was not explored. \cite{lyra09} also simulated 
the vortex instability in a self-gravitating disc, but for artificial internal edges. 
Chapter \ref{paper2} 
extends the linear theory for the vortex instability to account for disc self-gravity, and extensive simulations 
performed to examine the effect of varying the strength of self-gravity. \citeauthor{lyra09} noticed
that more vortices form and persist longer in calculations with self-gravity switched on. The results in 
Chapter \ref{paper2} explain this observation. 

Global instabilities have been shown to exist in self-gravitating discs 
with prescribed gap profiles \citep{meschiari08}. 
 Chapter \ref{paper3} considers the stability of gaps self-consistently opened 
by a planet in self-gravitating discs. Analytic discussion, originally
given by \citet{sellwood91} for particle discs, is extended to gaseous  discs. 
Chapter \ref{paper3}---\ref{edge_migration} also presents the first simulations 
of the effect of such global modes on planetary migration. 

Planetary gap stability has been studied in two-dimensional discs only (
although \cite{meheut10} have simulated the vortex instability for artificial density
bumps in three-dimensional discs). 
Chapter \ref{3d} presents the first three-dimensional simulations of planetary
gap stability in non-self-gravitating and self-gravitating discs, as well as their
effects on migration. Chapter \ref{3d} confirms 2D results 
in Chapters \ref{paper2}---\ref{edge_migration}.

\section{Modelling disc-planet systems}\label{3d_models}
The studies presented in this thesis are based on direct 
simulations. The general numerical setup and governing equations are 
presented here. Specific disc models and initial conditions are
described in each subsequent Chapter. 

\begin{figure}
  \centering
  \begin{tabular}{l}
    \includegraphics[width=0.5\linewidth]{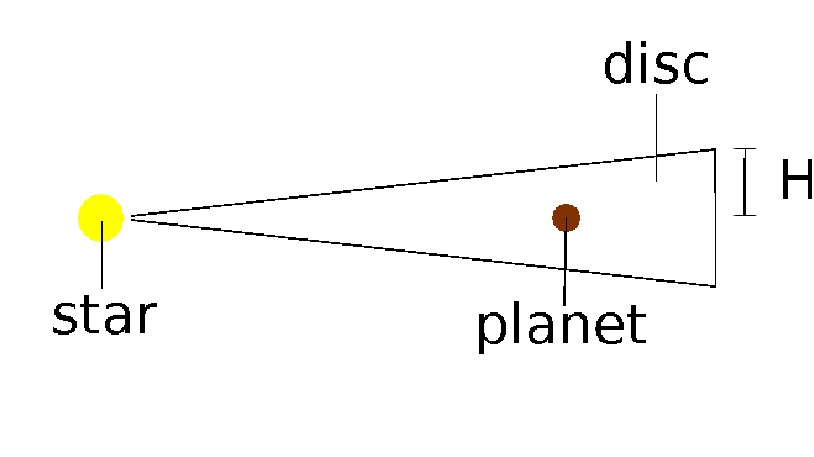}
  \end{tabular}
  \begin{tabular}{r}
    \includegraphics[width=0.4\linewidth]{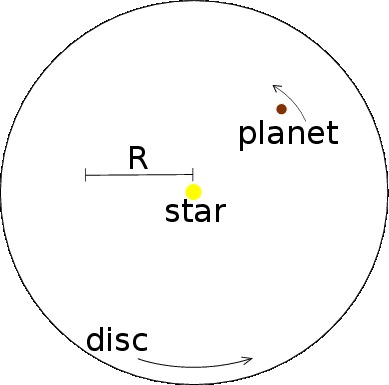}    
  \end{tabular}
  \caption{Schematic illustration of a protoplanetary disc with an embedded
    planet. Left: edge-on view. Right: face-on
    view.\label{dp_system_schematic}} 
\end{figure}

The system is a fluid disc of mass $M_d$ with an
embedded planet of mass $M_p$, both rotating about a central 
star of mass $M_*$. A schematic representation is shown in
Fig. \ref{dp_system_schematic}, in which $H$ denotes the pressure
scale-height and $R$ is the cylindrical distance from the star.
The disc is governed by the Navier-Stokes equations in a non-rotating
frame in spherical polar co-ordinates $\bm{r}=(r,\theta,\phi)$ centred
on the central star: 
\begin{align}\label{3d_gov_eq}
  &\frac{\p\rho}{\p t}+\nabla\cdot(\rho \bmu)=0,\\
  & \frac{\p\bmu}{\p t}+\bmu\cdot\nabla\bmu= -\frac{1}{\rho}\nabla p +
  \frac{1}{\rho}\nabla\cdot\bm{T} - \nabla{\Phi_\mathrm{eff}} \label{3d_mom_eq}  
\end{align}
where $\rho$ is the mass density, $\bmu=(u_r, u_\theta, u_\phi)$
is the velocity field, $p$ is the pressure, $\bm{T}$ is the viscous
stress tensor and $\Phi_\mathrm{eff}$ is an effective potential. 
The azimuthal velocity is also written 
$u_\phi = R\Omega$ where $\Omega$ is the angular velocity. For thin
and low mass discs, the angular speed is well approximated by the
Keplerian profile: 
\begin{align}\label{keplerian}
  \Omega\simeq \Omega_k(R) \equiv \sqrt{\frac{GM_*}{R^3}},
\end{align} 
where $G$ is the gravitational constant. 


\nomenclature[a]{$M_d$}{Mass} 
\nomenclature[b]{$M_p$}{Mass}
\nomenclature[c]{$M_*$}{Mass}
\nomenclature[a]{$H$}{Semi-thickness or pressure scale-height } 
\nomenclature[a]{$\rho$}{Density} 
\nomenclature[a]{$\bmu$}{Velocity}
\nomenclature[a]{$\Omega$}{Angular velocity}
\nomenclature[a]{$\Omega_k$}{Keplerian angular velocity}
\nomenclature[a]{$p$}{Pressure}
\nomenclature[a]{$\bm{T}$}{Viscous stress tensor}

\subsection{Units}
Units such that $G=M_*=1$ are adopted in numerical
calculations, and time will be quoted in initial orbital periods of the
planet, $P_0=2\pi/\Omega_k(r_p(0))$, where $r_p(t)$ is the planet's
orbital radius.  

\subsection{Equations of state}\label{EOS}
Two equations of state (EOS) are adopted: \emph{barotropic} and
\emph{locally isothermal}. These close the system of equations above
without the need for an energy equation to describe the
thermodynamics. Neglecting the energy equation prohibits more 
realistic modelling but does allow tractable analysis and faster
computational simulations. 

For a barotropic disc, the pressure is a function of density only:
\begin{align}\label{eos_barotropic}
  p =p(\rho),\quad c_s \equiv \sqrt{\frac{dP}{d\rho}},
\end{align}
where $c_s$ is the sound-speed. Eq. \ref{eos_barotropic} is 
used in analytical discussions, where general results may be derived without
explicitly specifying the functional form of $p(\rho)$. 

In a locally isothermal disc the sound-speed is a specified function of
space: 
\begin{align}\label{eos_localiso}
  p =c_s^2\rho,\quad c_s = h R\Omega_k(R),
\end{align}
where $h=H/R$ is the disc aspect-ratio. 
Eq. \ref{eos_localiso} results from
vertical hydrostatic balance between pressure
forces and stellar gravitational force, assuming
$c_s=c_s(R)$ and that the vertical extent
of the disc is small compared to the local
radius. 

Eq. \ref{eos_localiso} with constant $h$ is commonly used
in disc-planet studies, and is adopted in the numerical
studies later. The temperature profile of the
disc is fixed in space by Eq. \ref{eos_localiso}, even when a planet
is present. It is possible to modify Eq. \ref{eos_localiso} to account
for expected heating as disc material falls into the planet
potential (see \S\ref{planet_motion}). 

\nomenclature[a]{$c_s$}{Sound-speed}
\nomenclature[a]{$h$}{Aspect-ratio}
\nomenclature[z]{$\Phi_\mathrm{eff}$}{Effective gravitational potential} 

\subsection{Gravitational potentials}
The effective potential is: 
\begin{align}
  \Phi_\mathrm{eff} = \Phi_* + \Phi_p + \Phi + \Phi_i,
\end{align}
where
\begin{align}
  \Phi_* = -\frac{GM_*}{r} 
\end{align}
is the stellar potential and
\begin{align}
  \Phi_p = -\frac{GM_p}{\sqrt{|\bm{r} - \bm{r}_p|^2 + \epsilon_p^2}}
\end{align}
is the \emph{softened} planet potential, where $\bm{r}_p$ is the
vector position of the planet and $\epsilon_p$ is the softening
length. 

The gravitational potential
$\Phi$ arising from the disc material is given via the
Poisson equation 
\begin{align}\label{3d_poisson}
  \nabla^2 \Phi = 4\pi G \rho.
\end{align}
It is uncommon to include $\Phi$ in $\Phi_\mathrm{eff}$ 
for protoplanetary discs because typically  $M_d\ll M_*$. 
Models neglecting $\Phi$ in the hydrodynamic equations are called 
\emph{non-self-gravitating} whereas models including $\Phi$ are called
\emph{self-gravitating}. Solving  
Eq. \ref{3d_poisson} usually adds substantial
cost to numerical simulations. 

$\Phi_i$ is the indirect potential due to the disc and the
planet, which arises from the non-inertial reference frame:
\begin{align}\label{indirect_potential}
  \Phi_i(\bm{r}) = \int \frac{G\rho(\bm{r}')}{r'^3}\bm{r}\cdot\bm{r}'
  d^3\bm{r}' + \frac{GM_p}{|\bm{r}_p|^3}\bm{r}\cdot\bm{r}_p. 
\end{align}
The indirect potential accounts for the acceleration of the
co-ordinate origin relative to the inertial frame. This is an
unimportant term for the dynamics of concern. 

\nomenclature[c]{$\Phi_*$}{Star potential}
\nomenclature[b]{$\Phi_p$}{Planet potential}
\nomenclature[b]{$\epsilon_p$}{Softening length for planet potential}
\nomenclature[b]{$\bm{r}_p$}{Planet position}
\nomenclature[a]{$\Phi$}{Gravitational potential}
\nomenclature[z]{$\Phi_i$}{Indirect potential}

\subsection{Viscosity}
The viscous stress tensor is given by
\begin{align}\label{stress_tensor}
  \bm{T} = \mu\left[\nabla\bmu + \left(\nabla\bmu\right)^{T} -
  \frac{2}{3}\left(\nabla\cdot\bmu\right)\bm{1}\right],
\end{align}
where superscript $^T$ denotes transpose. In Eq. \ref{stress_tensor},
$\mu = \rho\nu$ is the \emph{shear viscosity} and $\nu$ is the
\emph{kinematic viscosity}. Note that zero bulk viscosity has been
assumed.  Explicit expressions of $\bm{T}$ in different co-ordinate
systems may be found in \cite{tassoul78}. 

Eq. \ref{stress_tensor} represents the effects of turbulent angular momentum
transport in accretion discs. For a rotationally supported disc,
angular momentum must be transported outwards if disc material is to be
transported inwards and accreted onto the star.  It is often easier to
use Eq. \ref{stress_tensor} and prescribe $\nu$ to 
circumvent detailed modelling of turbulence. For example, if turbulence
results from the MRI then one would need to consider the
magnetohydrodynamic equations.  


The disc models presented in subsequent Chapters adopt constant
kinematic viscosity, so that $\nu$ is a parameter independent 
of space or time. This includes the \emph{inviscid disc}, for
which $\nu=0$. The present work is concerned with shear instabilities
that lead to large-scale structure, rather than fully fledged
turbulence. A constant $\nu$ model is an adequate approach to describe
the level of viscosity (or turbulence) and its effect on the
instabilities of interest. 


Although not used in the present work, it is worth mentioning the more
frequently used 
\emph{$\alpha$-disc models}, in which one assumes  
\begin{align}\label{alpha_viscosity} 
  \nu = \alpha c_s H, 
\end{align} 
where $\alpha\lesssim1$ is a dimensionless parameter
\citep{shakura73} characterising turbulence of
unspecified origin. Due to its popularity, it has become common
practice to quote $\alpha$ as a non-dimensional measure of  
turbulence strength in an accretion disc. If a particular physical
process is responsible for the turbulence, 
one may accordingly define an associated $\alpha$ and
compare to $\alpha$'s due to other turbulence sources.    


\nomenclature[a]{$\mu$}{Shear viscosity}
\nomenclature[a]{$\nu$}{Kinematic viscosity}
\nomenclature[a]{$\alpha$}{Dimensionless viscosity parameter}


\subsection{Motion of the planet}\label{planet_motion}
The embedded planet feels the gravitational force from the
protoplanetary disc, in addition to that from the star. 
The motion of the planet through the disc is therefore given by
Newton's second law: 
\begin{align}
  \frac{d^2\bm{r}_p}{dt^2} = -\nabla(\Phi + \Phi_i + \Phi_*). 
\end{align}
The contribution due to the disc is given by
\begin{align}\label{disc_planet_force}
  \left(\frac{d^2\bm{r}_p}{dt^2}\right)_\mathrm{disc} =
  -\int\frac{G\rho(\bm{r})(\bm{r}_p-\bm{r})}{\left(|\bm{r}_p-\bm{r}|^2 +
      \epsilon_p^2 \right)^{3/2}}d^3\bm{r},  
\end{align}
where the distance between the planet and a fluid element has been
softened to prevent a singularity in numerical computations. 
\nomenclature[b]{$r_h$}{Hill radius}
\nomenclature[b]{$x_s$}{Half-width of co-orbital region}
\nomenclature[b]{$q$}{Mass relative to central star}
\nomenclature[b]{$\Omega_p$}{Angular speed}
\nomenclature[b]{$P_0$}{Initial orbital period}
 Disc-planet forces arising near the planet are subject to numerical
 errors because of the diverging potential and finite resolution. To
 overcome this issue one can apply a Gaussian envelope to  
 disc-planet forces within the Hill radius $r_h =
 (M_p/3M_*)^{1/3}r_p$ of the planet.  This procedure is called
 tapering. 

In some of the simulations a 
modified locally isothermal equation of state is adopted, with sound-speed given by
\begin{align}\label{pepeos}
  c_s=\frac{hRh_pd_p}{[(hR)^{7/2} + (h_pd_p)^{7/2}]^{2/7}}\sqrt{\Omega_k^2 +
  \Omega_\mathrm{kp}^2}, 
\end{align}
where $\Omega_\mathrm{kp}^2 = GM_p/d_p^3$ and $d_p = \sqrt{
|\bm{r}-\bm{r}_p|^2 + \epsilon_p^2}$ is the softened distance to the
planet. Increasing the parameter $h_p$ increases the temperature near 
$\bm{r}_p$, thereby decreasing the mass built up near the planet and
hence numerical errors from this region. This EOS is taken from
\cite{peplinski08a}, in which $h_p$ has the physical meaning of the
aspect-ratio of the circumplanetary disc. However, the use
of Eq. \ref{pepeos} here is entirely numerical. 


\subsection{Two-dimensional approximation}\label{2d_approx}
Protoplanetary discs are thin because $h\ll 1$. A good approximation
is the \emph{razor-thin disc}. The fluid equations 
(Eq. \ref{3d_gov_eq}) are integrated over the vertical extent of the
disc, assuming no vertical motions and the fluid radial and azimuthal
velocities are independent of the cylindrical $z$. The planet is
restricted to move in the midplane ($z=0$).   

The co-ordinate system becomes $(r,\phi)$ cylindrical
polars by setting $\theta=\pi/2$.  The 
three-dimensional density $\rho$ is replaced by 
surface  density $\Sigma$, $\bmu$ becomes the two-dimensional 
velocity  $(u_r,u_\phi)$ and $p$ is re-interpreted as the vertically 
integrated pressure. In the integral expressions above
, the expression for the mass element $\rho d^3\bm{r}$ is replaced
by $\Sigma d^2\bm{r}$.    

The 2D hydrodynamic equations are
\begin{align}
  &\frac{\p\Sigma}{\p t}+\nabla\cdot(\bmu\Sigma)=0,\notag \\
  &\Sigma \left(\frac{\p\bmu}{\p t}+\bm{u}\cdot\nabla\bm{u}\right) =
  -\nabla p -\Sigma\nabla\Phi_\mathrm{eff}  + \bm{f} \label{2d_governing}, 
\end{align} 
where $\bm{f}$ is the 2D viscous force. All quantities in
Eq. \ref{2d_governing}  are evaluated at $z=0$. The kinematic
viscosity is now related to the shear viscosity by $\mu=\nu\Sigma$. 
Explicit expressions for $\bm{f}$ are given in \cite{masset02}. 

The Poisson equation for a 2D fluid is
\begin{align}\label{2d_poisson_diff}
  \nabla^2 \Phi = 4\pi G \Sigma\delta(z)
\end{align}
where $\delta(z)$ is the Dirac delta function. Note that 
a 2D disc still has a 3D potential. 
Eq. \ref{2d_poisson_diff} can be used for analytical discussions, but it
cannot be used in numerical computations. For 2D numerical work, the 
midplane disc potential is given by 
\begin{align}\label{2d_poisson_int}
  \Phi = -\int
  \frac{G\Sigma(r^\prime,\phi^\prime)r^\prime
  dr^\prime d\phi^\prime}{\sqrt{r^2+r^{\prime 2} -
      2rr^\prime\cos{(\phi-\phi^\prime)}+ \epsilon_g^2}}, 
\end{align}
where $\epsilon_g$ is a softening length. 
In a 3D disc, material is spread over the vertical extent, which
dilutes the disc and planet gravitational forces compared to the case
where all the disc material is confined in the plane. Thus in 2D,
softening accounts for the vertical extent of the disc, in addition to 
preventing numerical divergence. 

\nomenclature[a]{$\Sigma$}{Surface density}
\nomenclature[a]{$\epsilon_g$}{Softening length for gravitational
  potential in two-dimensions}

\subsection{Stability parameters}

The  \emph{vortensity profile} and \emph{Toomre $Q$ parameter} 
are of fundamental importance to disc
stability. They arise naturally in linear stability analysis of 2D
discs. Vortensity concerns shear instabilities and Toomre
$Q$ concerns gravitational instability.  

\subsection{Vortensity}
The vortensity $\bm{\eta}$ is 
\begin{align}
  \bm{\eta} \equiv \frac{\nabla\times\bm{u}}{\rho} =
  \frac{\bm{\omega}}{\rho},
\end{align}
where $\bm{\omega}$ is the absolute vorticity, defined from the
velocity in a non-rotating frame. In 2D models, $\rho$ is
replaced by $\Sigma$ and there is only one component of vortensity,
$\bm{\eta}_\mathrm{2D}=\eta\hat{\bm{z}}$ where 
\begin{align}
  \eta = \frac{1}{r\Sigma}\left[\frac{\p}{\p r}\left(ru_\phi\right) -
    \frac{\p u_r}{\p\phi}\right].   
\end{align}
Furthermore, if the 2D disc is axisymmetric ($\p/\p\phi=0$) or has zero
radial velocity, then $\eta$ can be written as
\begin{align}
  \eta = \frac{\kappa^2}{2\Omega\Sigma},
\end{align}
where
\begin{align}
  \kappa^2 = \frac{1}{r^3}\frac{\p}{\p r}\left(r^4\Omega^2\right) 
\end{align}
is the square of the epicycle frequency. $\kappa$ is the
oscillation frequency of a fluid element perturbed about a circular
orbit. The existence of vortensity extrema allow shear 
instabilities \citep[e.g.][]{lovelace99}.

\nomenclature[a]{$\bm{\omega}$}{Absolute vorticity}
\nomenclature[a]{$\eta$}{Vortensity in two-dimensional discs}
\nomenclature[a]{$\kappa$}{Epicycle frequency}

\subsection{Toomre parameter}
The Toomre parameter is 
\begin{align}\label{toomre_q}
  Q = \frac{c_s\kappa}{\pi G \Sigma},
\end{align}
and was originally derived in a linear analysis of the gravitational
stability of razor-thin discs \citep{toomre64}. Hence, the surface 
density appears in Eq. \ref{toomre_q} rather than $\rho$. 

The Toomre $Q$ is often used to describe 
 the strength of disc self-gravity. The razor-thin disc is
gravitationally unstable to axisymmetric, local perturbations when 
$Q<1$, although general gravitational instabilities may appear at
somewhat larger values \citep[e.g.][]{papaloizou89,papaloizou91}. 




\chapter{Vortex-induced type III migration}\label{paper1}
\graphicspath{{Chapter1/figures/}}


A sufficiently massive planet in a protoplanetary disc will induce
spiral shocks extending close to the planet's orbital radius.
Vortensity generation across shock tips results in thin high
vortensity rings outlining the gap. 
However, vortensity rings are unstable if the disc viscosity is
sufficiently small \citep{valborro07} and lead to vortex formation.  



%

This Chapter considers the effect of such vortices on 
planetary migration and is organised as follows. 
\S\ref{modeleq}  describes
the disc-planet models and parameters. The
formation of vortensity rings is studied with numerical simulations
and a semi-analytical model in \S\ref{rings}---\S\ref{model}. 
The linear stability problem is discussed in
\S\ref{Dynstab}. Nonlinear hydrodynamic 
simulations of giant planet migration as a function of viscosity are
presented in \S\ref{mig_visc}, high-lighting the effect of vortices at
low viscosity. The inviscid case is examined in detail in
\S\ref{inviscid}, including simulations exploring the effect of
disc mass, temperature and planetary mass. Additional simulations
concerning numerical issues are described in \S\ref{addsim}. 
\S\ref{conc} concludes this Chapter.

\section{Disc-planet setup}\label{modeleq}
The protoplanetary disc is modelled as 
razor-thin and non-self-gravitating. The appropriate hydrodynamic
equations are thus Eq. \ref{2d_governing}
 with the disc potential $\Phi$ neglected. 
The general setup is described here and problem-dependent parameters are
described in the following sections. 
The initial orbital radius of the planet is $r_p(t=0)=2$. 

The disc has initially uniform surface density $\Sigma=\Sigma_0\times10^{-4}$
. Throughout simulations, a  uniform kinematic viscosity
$\nu=\nu_0\times 10^{-5}$ 
is imposed, with $\nu_0$ being a
dimensionless constant. The locally isothermal equation of state is adopted
(\S\ref{EOS})  with constant aspect-ratio $h$. 
Discs with $\Sigma_0=1$---9, $\nu_0=0$---1 and $h=0.03$---0.06 are
considered. Planetary masses of $q\equiv M_p/M_*=10^{-4}$---$10^{-3}$ are used. If
$M_*=1M_\odot$, then $q=3\times10^{-4},\,10^{-3}$ corresponds  to
Saturn and Jupiter, respectively.  The planet position in
the disc plane is denoted $(r_p,\phi_p)$ and its potential is softened with
a softening length $\epsilon_p = 0.6hr_p$.  



\subsection {Vortensity conservation}\label{secvort}
The vortensity $\eta$ is of fundamental importance to the issue to gap
stability. For a 2D disc the vortensity is simply
\begin{equation*}\label{vortensity}
  \eta =  \frac{\omega}{\Sigma} \equiv \frac{\omega_r+2\Omega_p}{\Sigma},
\end{equation*}
where $\omega=\hat{\bm{z}}\cdot\nabla\times\bm{u}$ is the $z$
component of the absolute vorticity and $\omega_r$ is the vertical
component of the relative vorticity seen in a rotating frame with
angular speed $\Omega_p$. 

It is well known that in inviscid barotropic flows without shocks, the
vortensity $\eta = \omega/\Sigma$  is conserved for a  fluid particle
($D\eta/Dt=0$). When a locally isothermal equation of state with
spatially varying sound-speed as is used here is adopted, vortensity
is no longer strictly conserved. However, the adopted sound-speed
profile varies on a global scale so that  when phenomena are
considered on a local scale, vortensity is  conserved to a good
approximation in the absence of shocks and viscosity. Thus, analysing
the vortensity distribution for inviscid discs allows one to track
fluid elements as their orbits are perturbed by the planet.

\subsection{Numerical method for nonlinear simulations}
The \fargo code is used to numerically evolve the hydrodynamics equations
describing the disc in response to the planet potential
\citep{masset00a,masset00b}. The code and its extended versions are
publicly available \footnote{\url{http://fargo.in2p3.fr/}}.  
\fargo uses a finite-difference scheme
with van Leer upwind advection, similar to the popular \zeusold code
\citep{stone92}, but \fargo employs a modified algorithm for 
azimuthal transport that allows for large time steps. 
The 2D disc is divided into $N_r\times N_\phi$ zones. The grid is
uniformly spaced in radius and azimuth. Wave damping boundary
conditions are imposed at disc boundaries 
\citep{valborro06} and $2\pi$-periodic boundary condition imposed in 
azimuth. Since vortex formation occur near the planet, the phenomenon is
unaffected by distant radial boundary conditions.

A fifth order Runge-Kutta method was used to integrate the planet's
equation of motion when migration is considered.  
Simulations ran with halved time-step show 
very similar results to the corresponding standard set-up.  
 The RK5 
integrator is sufficiently accurate to capture the vortex-planet
interaction, which can be regarded as a two-body problem. 

\nomenclature[e]{$N_r,\,N_\theta,\,N_\phi$}{Number of zones in radial,
vertical and azimuthal directions, respectively}

\section{Ring structures in disc-planet interactions}\label{rings} 

\subsection{ Vortensity generation by shocks}\label{fiducial} 
Shocks are induced by a sufficiently massive planet and 
vortensity conservation is broken in the presence of shocks. 
A high-resolution numerical disc-planet simulation demonstrates this below.  
The disc is thin and inviscid ($h=0.05,\,\nu=0$), occupying $r\in[1,3]$ 
with $\Sigma_0=1$. The planet has mass
$M_p=2.8\times10^{-4}M_*$, its potential is introduced at $t=0$ and
switched on to its full value over $5P_0$. The planet is held on a
fixed circular Keplerian orbit. 
 The computational domain is divided into
$N_r\times N_\phi=1024\times3072$ zones, corresponding to  radial and
azimuthal resolution of $\Delta r\simeq 0.02r_h$ and 
$r\Delta\phi\simeq0.05r_h$. 

\begin{figure}[!ht]
  \center
  \includegraphics[width=0.67\linewidth]{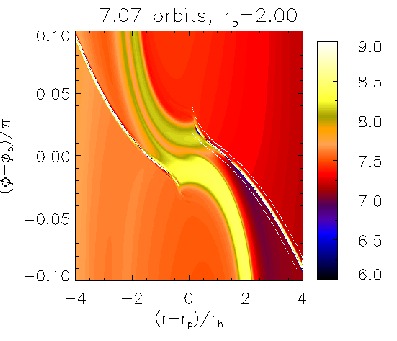}\caption{Vortensity
    generation and destruction across shocks induced by a Saturn-mass
    planet in an inviscid disc. The plot shows $\ln{\eta} =\ln{(\omega/\Sigma)}$.
    Regions of increased vortensity are clearly visible as half horseshoes
    above and below the planet. The increase occurs as material passes through
    parts of the shock fronts that extend  into  the co-orbital
    region. Thick white lines indicate the shock front, whereas thin
    white lines above (below) the outer shock indicate where pre-shock
    (post-shock) quantities are measured.} 
  \label{vortxy_020} 
\end{figure}

Fig.\ref{vortxy_020} shows the vortensity field at
$t=7.07P_0$ close to the planet. Vortensity is
generated/destroyed as material passes through the two spiral shocks.
For the outer shock, vortensity
generation occurs for fluid elements executing a
horseshoe turn ($|r-r_p|\lesssim r_h$) while  vortensity is reduced for
fluid that passes by the planet, but the change is smaller in magnitude
in the latter case. The 
situation is similar for the inner shock, but some post-shock material
with increased vortensity continues to pass by the planet. 
Note that a pre-shock fluid element that would be  on a horseshoe trajectory,
may in fact pass by the planet after crossing the shock.
 It is
clear that vortensity rings originate from passage through shock fronts interior
to the  co-orbital  region that would correspond to the horseshoe region
for free particle motion.

The streams of high vortensity eventually move around the whole orbit outlining the
entire co-orbital region. Fig. \ref{vortxy_020} shows  that they are generated
along a small  part of the shock front  of length  $\sim r_h$. This  results in
thin rings with  a similar  radial width.
The fact that they originate from horseshoe
material can  enhance the contrast  as post-shock
inner disc horseshoe material is mapped  from $r-x$ to  $r+x$ 
to become adjacent to post-shock outer disc material passing by the planet, and
the two trajectories experience opposite vortensity jumps. 


\subsection{Long term evolution}

\begin{figure}[!ht]
  \center
  \includegraphics[width=1.0\linewidth]{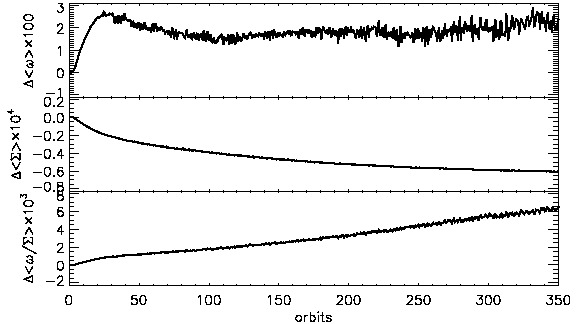}\caption{Long term
    evolution of co-orbital properties for a Saturn-mass planet held on
    fixed orbit. Vorticity, density and vortensity 
    perturbations are expressed relative to their  values at
    $t=0$. Angle brackets denote an average over the co-orbital region
    defined by the annulus $r=[r_p-2.5r_h,r_p+2.5x_s]$. 
    \label{coorb_plot_0}}
\end{figure}

Fig. \ref{coorb_plot_0} shows the long-term evolution of
average co-orbital properties from a corresponding lower resolution run
($N_r\times N_\phi = 256\times768$). The co-orbital region is taken to be
the annulus $[r_p- x_s, r_p+x_s]$, with $x_s=2.5r_h$ as is typically
measured from hydrodynamic simulations for intermediate or massive
planets \citep{artymowicz04b,paardekooper09} . In Appendix
\ref{xslimit}, it is shown that in the pressureless limit, $x_s
\lesssim 2.3r_h$, comparable to the value adopted above.

Vorticity generation occurs within $t\lesssim25P_0$,
after which it remains approximately steady. 
It is important to note that subsequent vortensity increases is in
narrow rings and fluctuations are due to instabilities associated with
the rings. The average surface density falls as the planet
opens a gap. In fact, gap formation is a requirement for 
consistency with vortensity rings. 
Fig. \ref{coorb_plot_0} reflects modification of
co-orbital properties on dynamical timescales due to shocks, so
migration mechanisms that depend on co-orbital structure will be
affected. 

\subsection{ Location of vortensity generation 
for different planet masses}
 
\begin{figure}[!ht]
  \center
  \includegraphics[width=0.75\linewidth]{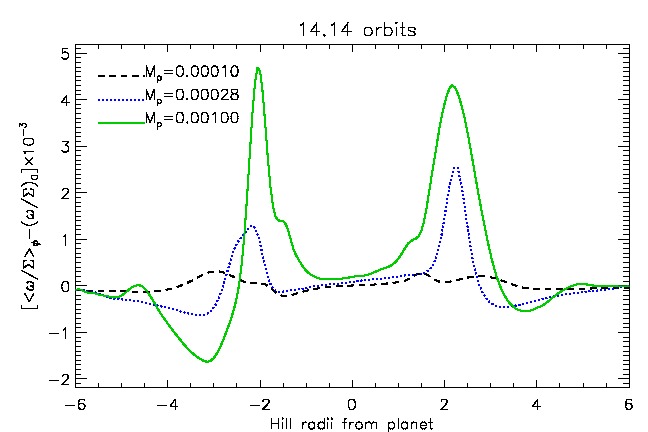}\caption{Azimuthally
    averaged vortensity perturbation for different planet masses $M_p$
    (in units of the central mass). The resolution for these runs were  
    $N_r\times N_\phi =
    256\times768$.} \label{compare2_020}  
\end{figure}

Vortensity ring formation has been observed 
in disc-planet simulations by \cite{koller03} and \cite{li05}, but there 
the rings are generated by shocks \emph{outside} the co-orbital
region whereas here rings are inside it. This  
is because the authors used a smaller
planet mass. To illustrate this, 
simulations with $M_p/M_*=10^{-4},\,2.8\times10^{-4}$ and
$10^{-3}$ are compared in Fig. \ref{compare2_020}, which show azimuthally
averaged vortensity perturbations induced after $t=14.14P_0$.  

The intermediate and high mass cases are
qualitatively similar, having $\Delta(\omega/\Sigma)$ maxima at
$r-r_p\sim\pm 2r_h$ and minima at $r-r_p\sim3 r_h$. The magnitude of
$\Delta(\omega/\Sigma)$ increases with $M_p$ because higher masses
induce stronger shocks with higher Mach number. As the half-width of
the horseshoe region is $x_s=2.5 r_h$ for such masses, vortensity 
rings are co-orbital features. 

The low mass $M_p=10^{-4}M_*$ case has much smaller
$|\Delta(\omega/\Sigma)|$. There is a vortensity maximum at $r-r_p=-3r_h$
but nothing of similar magnitude at $r-r_p>0$. 
\cite{paardekooper09} 
found the co-orbital half-width, in the limit of zero softening
for low mass planets, to be:
\begin{align*}
  x_s=1.68(q/h)^{1/2}r_p.
\end{align*}
For $q\equiv M_p/M_* = 10^{-4},\,h=0.05$, $x_s=2.33r_h$.
Non-zero softening gives a smaller $x_s$. Hence, vortensity
rings for low mass planets occur outside their  co-orbital region as is
confirmed and they are very much weaker. Note that
in \cite{li09} significantly 
longer timescales are simulated. 


In the standard type III migration described by \cite{masset03}
, the migrating system is the
planet plus the co-orbital material . If vortensity
rings are co-orbital features, they should migrate with the
planet. On the other hand, if they reside just outside the horseshoe
region, the migrating system must work against these rings to
migrate. Because type III migration is driven by the contrast between
co-orbital and circulating  material, whether vortensity rings count
as the former or the latter can affect type III migration. 
Furthermore, the fact that vortensity rings lie closer to the
planet (in units of Hill radius) as $M_p$ increases imply that
instabilities associated with the rings will interact more readily
with more massive planets.


\section{A semi-analytic model for co-orbital vortensity generation by
  shocks}\label{model} 
A semi-analytic model describing 
shock-generation of vortensity is presented in order to provide an
understanding of the physical processes involved. 
 More specifically, the outer spiral shock in Fig. \ref{vortxy_020} is 
modelled. 
Three pieces of information are required: the  pre-shock flow field,  the
shock front location  and the vortensity change
undergone by material as it passes through the shock.

\subsection{Flow field}\label{pre_shock_flow}
The pre-shock flow field is calculated in the
shearing sheet approximation \cite[e.g.][]{paardekooper09}.
A local Cartesian co-ordinate system $(x,y)$ is adopted, which
co-rotates with the planet at angular speed $\Omega_p$. The co-ordinates
$x,\,y$ correspond to radial and azimuthal directions,
respectively. Velocities in the 
shearing sheet are thus relative to the planet. The unperturbed 
velocity field in the shearing sheet is assumed Keplerian  
$\bm{u}=(0,-3\Omega_p x/2)$. 

In the pre-shock flow, pressure forces are assumed negligible 
 compared to the planet potential. This ballistic
approximation is appropriate for a slowly varying  supersonic flow as
is expected to be appropriate for the pre-shock fluid. Pressure should
also be insignificant because giant planet masses are being considered. 
Fluid elements then behave like test particles moving under the potential: 
\begin{align}
  \Phi_{\mathrm{tot}} =  -
  \frac{GM_p}{\sqrt{x^2+y^2+\epsilon_p^2}}-\frac{3}{2}\Omega_p^2x^2. 
\end{align}
The particle dynamic equations are those listed in Appendix
\ref{xslimit}, except here the planet potential is softened. 
Indirect potentials are neglected in this treatment for simplicity,
and viscosity ignored for consistency (Fig. \ref{vortxy_020} is an
inviscid simulation). In this section, the equations of motion are
integrated numerically. In Appendix \ref{alt_linear}, a linearised
version of this calculation is presented.


A fluid particle's trajectory in a steady state flow  is written 
as  $ x=x(y)$ and the corresponding velocity field as 
$\bm{u}=\bm{u}(y).$ Noting that on a particle trajectory
$\frac{D}{Dt}=u_y\frac{d}{dy}$, a  particle trajectory can be described by
the following system of three simultaneous first order differential
equations:  
\begin{align}
  \label{semi-an}
  &\frac{du_y^2}{dy}=-4\Omega_pu_x-\frac{2GM_py}{(x^2+y^2+\epsilon_p^2)^{3/2}}\equiv \mathcal{Q},\\ 
  &\frac{d}{dy}\left(u_xu_y^2\right)=
  u_y\left[3\Omega_p^2x-\frac{GM_px}{(x^2+y^2+\epsilon_p^2)^{3/2}}\right] 
  +u_x\mathcal{Q}+2\Omega_pu_y^2,\\
  &\frac{d}{dy}\left(xu_y^2\right)=x\mathcal{Q}+u_xu_y.
  \label {semi-an1}\end{align}
Eq.\ref{semi-an}---\ref{semi-an1} are equivalent to
Eq. \ref{horseshoe}, but 
this particular form is chosen so that the RHS of 
Eq. \ref{semi-an}---\ref{semi-an1} remain finite at a horseshoe
U-turn, where $u_y=0$. The explicit expressions for $du_y/dy$ involve
divisions by $u_y$, which would be numerically inconvenient close to
the U-turn. The state vector 
$\mathbf{U}(y)=[u_y^2,\,u_xu_y^2,\,xu_y^2]$ is solved for a
particular particle in $x>0$, with the boundary condition 
\begin{align}
  (u_y,\, u_x,\, x)\to(-3\Omega_p x_0/2, \, 0, \, x_0) \,\,\mathrm{as}\,\,
  y\to\infty , 
\end{align}
where $x=x_0$ is the particle's unperturbed path (or impact
parameter). A simple fourth 
order Runge-Kutta integrator was used. 
The totality of paths obtained by considering
a continuous range of $x_0$ then constitutes the flow field.  Having
obtained the velocity field, vortensity conservation is used to obtain
the surface density, 
\begin{align}
  \frac{\omega}{\Sigma} = \frac{\omega_0}{\Sigma_i},  
\end{align}
where the unperturbed absolute vorticity in the shearing sheet is
$\omega_0 =\Omega_p/2$ and $\Sigma_i$ is the surface density at the
initial particle position. Since the unperturbed surface density is
uniform, all particles have the same vortensity  at the 
beginning of their trajectory. When a particle reaches position
$(x,y)$, the surface density there is given by
\begin{equation}
  \Sigma(x,y) = \frac{2\Sigma_i}{\Omega_p}[\omega_r(x,y)+2\Omega_p].
\end{equation}
Numerically, the relative vorticity  is calculated by relating it to
the circulation through  
\begin{equation}
  \omega_r\simeq\frac{1}{\Delta A}\oint\bm{u}\cdot\mathbf{dl}
\end{equation}
where the integration is taken over a closed loop about the point of
interest and $\Delta A$ is the enclosed area. This avoids errors due
to  numerical differentiation on the uneven grid in $(x,y)$  generated
by solving the above system.


\subsection{Location of the shock front}
A shock forms because the 
planet presents an obstacle to the flow. Where the relative flow is
subsonic,  the presence of this obstacle can be communicated to the
fluid via sound waves. In 
 supersonic regions,  the fluid is unaware of the planet via
sound waves (but the planet's gravity is felt).  An estimate the
location of the boundary separating these two regions can be made by
specifying an appropriate characteristic curve or ray defining a sound
wave.  This is a natural location for shocks. Applying this 
idea to Keplerian flow, \cite{papaloizou04} obtained a good match
between the predicted theoretical shock front and the wakes associated with a low mass
planet (which perturbs the Keplerian flow by a negligible amount). For general
velocity field $\bm{u}$ 
 the characteristic curves satisfy the equation 
\begin{align}\label{shock_front}
  \frac{dy_s}{dx} =
  \frac{\hat{u}_y^2-1}{\hat{u}_x\hat{u}_y-\sqrt{\hat{u}_x^2+\hat{u}_y^2-1}},
\end{align} 
where $\hat{u}_i= u_i/c_s$. A brief derivation of
Eq. \ref{shock_front} is given in Appendix \ref{shockfront}. The sign
of the square root has been chosen so that the curves have negative
slope in the domain $x>0$   with $u_y<0$ (as required by the outer
shock in Fig. \ref{vortxy_020}). The fluid flows from 
$y>y_s$ (super-sonic) to $y<y_s$ (sub-sonic). Fluid located at $y=y_s$
begins to know about the planet through pressure waves
\citep{papaloizou04}.  

The solution to Eq. \ref{shock_front} for Keplerian flow is 
\begin{align}\label{keplerian_shock}
  \hat{y} = -\frac{\hat{x}}{2}\sqrt{\hat{x}^2 - 1} +
  \frac{1}{2}\ln{\left(\sqrt{\hat{x}^2 - 1} + \hat{x}\right)},  
\end{align}
where $\hat{x}=x/x_i,\,\hat{y}=y/x_i$ with $x_i$ being the starting
point of the curve \citep{papaloizou04}. 
In Keplerian flow, the rays defining the shock fronts originate from
the  sonic points at $x=x_i=\pm2H/3,\,y=0$. 

In a  general flow,  sonic points  ($|\mathbf{u}|=c_s$) at
which the rays may start, lie on curves and can occur for
$x<2H/3$. This is possible, for example, if the flow is perturbed
by a large planet mass and fluid particles move supersonically across
a U-turn as they are accelerated by the planet. In this case, shocks
can extend close to the planet's orbital radius.  
 The starting sonic  point for solving  Eq. \ref{shock_front}
that is eventually adopted has $x=0$ 
 (the lowest dotted curve in  Fig. \ref{shock_curve2_}, see
 \S\ref{theoretical_shock}).   


\subsection{Vorticity and vortensity changes across a
  shock}\label{sec_vortensity_jump} 
The jump in absolute vorticity $[\omega]$ is readily obtained by
resolving the fluid motion parallel and perpendicular to the shock
front  \citep[eg.][]{kevlahan97}. As the energy equation is neglected,
shocks are locally  isothermal and the expression for $[\omega]$ here
differs from those of  \citet{kevlahan97}, accordingly a brief
derivation of $[\omega]$ is presented in  Appendix
\ref{vortjump}. The result for a steady 
shock is 
\begin{align}\label{vorticity_jump}
  [\omega]=-\frac{(M^2-1)^2}{M^2}\frac{\p u_\perp}{\p S}+(M^2-1)\omega
  -\left(\frac{M^2-1}{u_\perp}\right)\frac{\p c_s^2}{\p S},
\end{align}
where $u_\perp$ is the pre-shock  velocity component perpendicular to 
the shock front,  $M=u_\perp/c_s$ is perpendicular Mach number and
$\omega$ is the pre-shock absolute vorticity. $\p/\p S$ is the derivative
along the shock (increasing $S$ is taken to be moving away from the planet, see Fig. \ref{vortgen_shock}). 
It is important to note that for Eq. \ref{vorticity_jump} to hold, 
the direction of increasing $S,$ the direction of positive 
$u_{\perp},$ and the vertical direction should form a right handed
triad. 


The vortensity jump $[\omega/\Sigma]$ follows immediately from
Eq. \ref{vorticity_jump} as
\begin{align}\label{vortensity_jump}
  \left[\frac{\omega}{\Sigma}\right]=-\frac{(M^2-1)^2}{\Sigma M^4}\frac{\p
    u_\perp}{\p S}-\left(\frac{M^2-1}{\Sigma M^2u_\perp}\right)\frac{\p c_s^2}{\p S},
\end{align}
which reduces to the expression derived by \cite{li05} if
$c_s=\mathrm{constant}$ (and a sign change due to different
convention). The sign of  $[\omega/\Sigma]$ depends mainly on the
gradient of $u_\perp$ (or $M$) along the shock. For the outer shock
induced by the planet $u_\perp<0,$ so the width of the increased vortensity 
rings produced is determined by  the length along the shock where $|M|$ is
increasing. Note that $[\omega/\Sigma]$ does not depend explicitly on
the  pre-shock vortensity, unlike the absolute vorticity jump.

The last term on the RHS of
Eq. \ref{vorticity_jump}---\ref{vortensity_jump} is produced by   the
local isothermal equation of state. For the $c_s\propto 
r^{-1/2}$ profile adopted in this study, the sound-speed 
varies slowly in the region of vortensity
generation, which occurs locally. Hence, the contribution from
the $\p c_s^2/\p S$ term to the vorticity/vortensity jump is not
important here (as checked numerically). However,
Eq. \ref{vorticity_jump}---\ref{vortensity_jump} are valid for any
sound-speed profile. Should a $c_s$ profile varying on a local
scale be adopted, then $\p c_s^2/\p S$ could be important.
  

\subsection{Comparing the semi-analytic model to
  numerical simulations}\label{theoretical_shock}
\begin{figure}[!ht]
  \centering
  \includegraphics[scale=0.75]{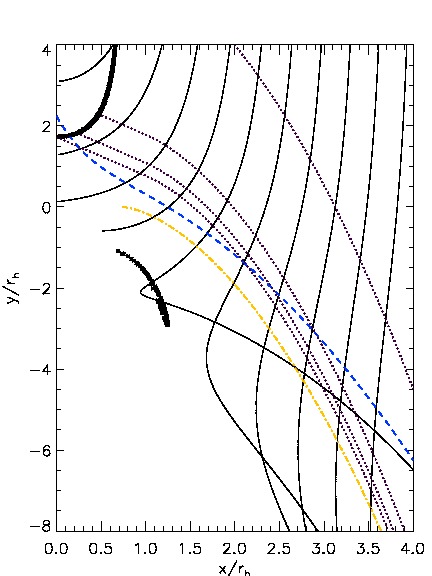}\caption{Solid
    lines: particle paths from the reduced zero-pressure 
    momentum equations (Eq. \ref{semi-an} - \ref{semi-an1});
    thick lines: curves composed of sonic points 
    on which $|\mathbf{u}|=c_s$;  
    dotted lines: theoretical shock fronts; dash-dot line: solution to
    Eq. \ref{shock_front} for Keplerian flow; dashed line:  
    polynomial fit to simulation shock front. The actual shock front
    begins  at a sonic point  
    around $x=0.2r_h$.}
  \label{shock_curve2_}
\end{figure}

The three subsections above can be combined and compared
to results of hydrodynamic simulations.
First, Fig. \ref{shock_curve2_} illustrates the particle paths that constitute
the flow field together with the  shock
fronts obtained by assuming coincidence with the characteristic curves
that are obtained from the semi-analytic model. A polynomial fit to
the simulation shock front is also shown. 
Particle paths cross for $x > r_h,\,y < -2r_h$
so  that the neglect of pressure becomes invalid.  Accordingly  the
solution to Eq. \ref{shock_curve2_}  should not be trusted in this
region.  Fortunately, vortensity generation occurs within $r_h$ from the planet
, where pre-shock particle paths do not cross.

In Fig. \ref{shock_curve2_}, the estimated shock location is
qualitatively good and tends to the Keplerian solution further away. 
The important feature is that the shock can
extend close to $x=0$, across horseshoe orbits. If the flow were
purely Keplerian there could be no significant vortensity generation
close to $x=0$ because the flow becomes sub-sonic and no shock
occurs there. Only circulating fluid can be shocked in that 
case. This implies that for low mass planets where the flow is
nearly Keplerian,  shock-generation of vortensity cannot occur close 
to the planet's orbital radius. Shock-generation of vortensity inside
the co-orbital region is only possible for sufficiently massive
planets that induce large enough non-Keplerian velocities. 

\begin{figure}[!ht]
  \center
  \includegraphics[width=\linewidth]{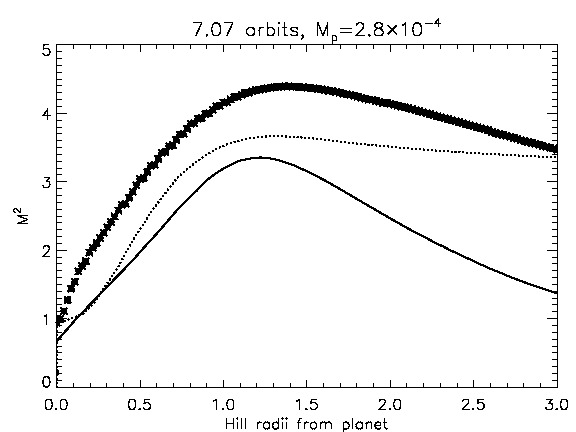}\caption{Square of the
    perpendicular Mach number along the outer spiral shock  illustrated in
    Fig. \ref{vortxy_020}.
    Asterisks: simulation data; dotted line:  $M^2$ obtained
    from the simulation shock front
    with the semi-analytic flow; solid line: $M^2$
    obtained from the semi-analytic shock front and flow. The planet
    mass is in units of $M_*$.}
  \label{sc3_machsq_020}
\end{figure}

The key quantity determining vortensity generation is the
perpendicular Mach number. Fig. \ref{sc3_machsq_020} compares $M$ from
simulation and model.  Although the semi-analytic model gives a shock  Mach number
that is somewhat smaller than found from the simulation, all curves have $|M|$
increasing from $x=0\to 1.3r_h$ which is the important domain for vortensity
generation.  Thus Eq. \ref{vortensity_jump}
implies  vortensity generation  in this region for all cases. 

\begin{landscape}
  \begin{figure}
    \center
    \subfigure[\label{mp2.8}]{\includegraphics[width=.5\linewidth]{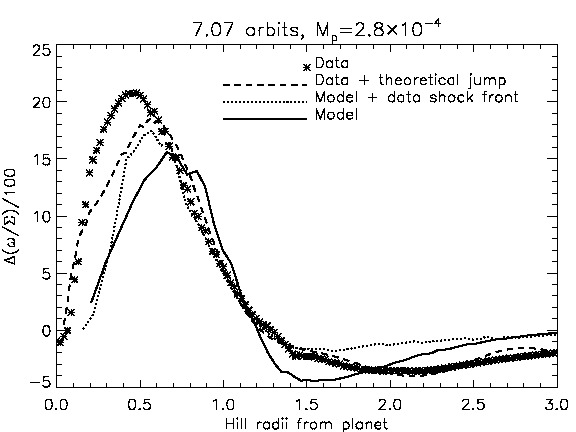}}\subfigure[\label{mp2}]{\includegraphics[width=.5\linewidth]{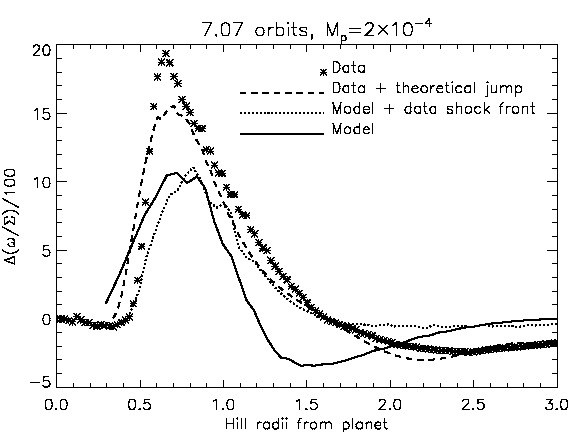}}   
    \caption{Semi-analytic  and actual vortensity jumps across a shock in the
      co-orbital region. Asterisks are measured from the simulations. The
      dashed lines were obtained by using pre-shock simulation data coupled
      with the jump condition  specified by Eq.\ref{vortensity_jump}. The  dotted lines 
      were obtained using the semi-analytic flow field together with
      the location of the  shock front obtained from the simulation.  The solid
      curves  correspond to the semi-analytic model for both the flow
      field and shock front. The planet mass is in units of $M_*$. \label{vort_jump}}
  \end{figure}
\end{landscape}

Fig. \ref{mp2.8} illustrates various  combinations of
semi-analytic modelling  and simulation data that have been used to
estimate the  vortensity jump across the shock.
The qualitative similarities  
between the various curves confirms that
vortensity generation  occurs  within co-orbital
material about one Hill radius away from the planet. 
It is shocked as it undergoes horseshoe turns.

Assuming material is mapped from $x\to -x$ as it switches to
the inner leg of its horseshoe orbit, the outer spiral
shock will produce a vortensity ring peaked at $r-r_p\sim -0.5r_h$ of
width $O(r_h)$ ($=O(H)$). Similarly
 the inner shock is expected to produce
a vortensity ring peaked at $r-r_p\sim 0.5r_h$ . As a
fluid element moves away from the U-turn region, $|r-r_p|$ increases,
but it remains on a horseshoe orbit. Thus,
thin vortensity rings are natural features of the co-orbital
region for such planet masses.

The model has also been tested for $M_p=2\times10^{-4}M_*$ in
Fig. \ref{mp2}.  There is still  a good qualitative 
match between the simulation and  the model;
even though  lowering $M_p$  makes  the zero-pressure
approximation, adopted to determine the semi-analytic flow field, less good. 
Decreasing $M_p$ shifts vortensity generation
away from the planet in the semi-analytic model, as is also observed in 
the hydrodynamic simulation
(Fig. \ref{compare2_020}). In this case, there is no vortensity
generation in $r-r_p<0.5r_h$ but vortensity rings are still
co-orbital (with  peaks at $\sim0.7r_h$).

\subsection{A recipe for vortensity rings}
The modelling above can be used to construct the
azimuthally averaged vortensity profile of the shock-modified
co-orbital region of a giant planet.  
However, this calculation is only illustrative due to
 the highly nonlinear nature of shocks and flows near a giant planet, plus the
various assumptions already made in estimating the pre-shock flow and
shock-front. 
It is usually more practical to run nonlinear 
disc-planet simulations to obtain accurate vortensity profiles. 

To make progress, further assumptions are made: 
\begin{enumerate}
\item Vortensity of a fluid element is only changed across the shock
  and conserved elsewhere ( see \S\ref{secvort}).  
\item Symmetric spiral shocks. The jump in vortensity across the
  shock, $[\omega/\Sigma]$, for $x<0$ (inner spiral shock) is the  
  same as for the outer spiral shock.  This is consistent with
  the shearing sheet.  
\item The shearing sheet model above is used to obtain
  $[\omega/\Sigma]$. The vortensity jump is then applied to the 
  unperturbed vortensity profile of the global model. 
  Assuming Keplerian rotation, this is
  $\eta_0=r^{-3/2}/(2\Sigma)$ with $\Sigma$ being uniform.  
\end{enumerate}

In Fig. \ref{mp2.8}---\ref{mp2}, the horizontal axis 
corresponds to the $x$-coordinate of the shock front. A fluid element 
which shocks at a particular $x$ originated from a location 
$(x_0,\,y>0)$. Fluid particles with $x_0 < x_s\equiv2.5r_h$ are
assumed to be on horseshoe orbits and the shock does not change the
nature of such trajectories. Horseshoe particles are then mapped to
$-x_0$  well after executing the horseshoe turn. The inner vortensity
ring is then attributed to the outer spiral shock, and similarly the
outer ring results from the inner spiral shock (see Fig. \ref{vortxy_020}). Particles with $x_0>x_s$ are
assumed to circulate after the shock. 

Both librating and circulating particles can be re-shocked so    
the vortensity peaks and troughs are expected to increase in amplitude
to some extent. Eventually after a few returns the system approaches
a steady state (Fig. \ref{coorb_plot_0}), implying  
${\partial u_{\perp}}/{\partial S}\rightarrow 0$ (recall from
Eq. \ref{vortensity_jump} that the vortensity jump is mainly
attributed to this term). However, the shock front would have changed
since what was post-shock becomes pre-shock, as material
returns to the shock. This detail shall be neglected and only the
initial ring formation as a result of the first shocking will be
considered below. 

The post-shock vortensity of a particle originating from $x_0>0,y>0$
and which shocks at $x$ is
\begin{align}
  \left(\frac{\omega}{\Sigma}\right)_{x} =
  \left(\eta_0\right)_{x_0}
  +\left[\frac{\omega}{\Sigma}\right]_{x}. 
\end{align}
This new vortensity value is conserved following the particle after it
passes through the shock. If the origin of the particle satisfied
$x_0 < x_s$ it will be located at $(-x_0,\,|y/r_h|\gg1)$ well after it
executes the horseshoe turn. If $x_0 >x_s$, it continues to circulate
and  will be located at $(x_0,\,y<0)$. Thus,  
\begin{align}
\left(\frac{\omega}{\Sigma}\right)_{x} = \left\{ 
\begin{array}{l l}
  \left(\omega/\Sigma\right)_{-x_0} & \quad \mbox{$x_0<x_s$}\\
  \left(\omega/\Sigma\right)_{x_0} & \quad \mbox{$x_0>x_s$.}\\
\end{array} \right.
\end{align}
A similar argument is applied to the inner shock. The particle
position is then transformed to the global cylindrical co-ordinates
using $r_0= x_0 + r_p$.

\begin{figure}[!ht]
  \center
  \includegraphics[width=.75\linewidth]{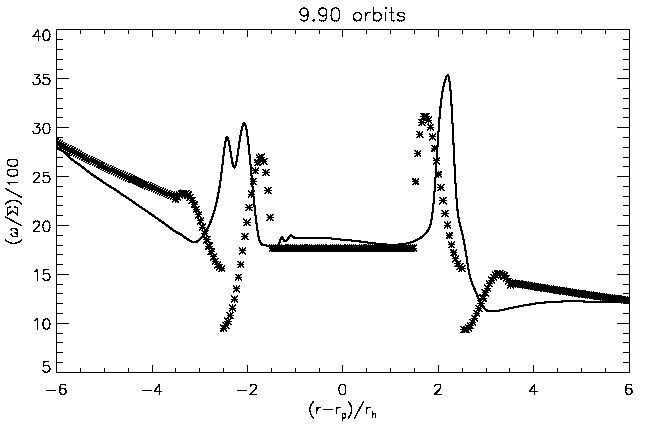}
  \caption{Vortensity rings constructed using the modelling from
    \S\ref{pre_shock_flow}---\ref{sec_vortensity_jump}
    (asterisks). The solid line is the azimuthally averaged vortensity
    profile taken from a hydrodynamic
    simulation. The double peak in the solid line, near the inner
    vortensity ring, shows that fluid elements do not return to
    exactly the same radius after being shocked. This is because the
    snapshot is taken early on in the simulation, during gap-opening
    by the planet, where material is pushed away from the
    planet.
\label{vort1d_temporary}}  
\end{figure}

Fig. \ref{vort1d_temporary} compares this procedure to
simulation results. The snapshot is chosen at the point when  the
rings  first occupy the complete azimuth. The central region
correspond to particles not crossing the shock so their vortensity
is unmodified. 
However, strictly speaking the vortensity distribution still changes
because particles switch between  $r_p\pm x_0$ conserving  their
vortensity and the unperturbed vortensity distribution is
non-uniform in the global disc.  
This is a minor effect that is expected to lead to 
a uniform distribution due to phase mixing. To deal with  this, the  
vortensity everywhere in this region is simply replaced by the
unperturbed  vortensity at the planet's location 
$r_p^{-3/2}/(2\Sigma)$. 

The predicted profile in Fig. \ref{vort1d_temporary} leads to the
expectation of sharp vortensity rings adjacent to the planet, with
troughs exterior to them. In comparison to simulation data, although
the semi-analytic  model reproduce the essential features (e.g.
vortensity peaks and troughs with the correct relative positions), 
the predicted distribution is slightly shifted towards the planet (by
$0.2r_h$). The model does not reproduce  the troughs very well  due to
inadequate modelling of the pre-shock flow and the shock position away
from the co-orbital region.   


The modelling shows that it is possible to 
predict the vortensity ring profile induced by a giant planet, from first 
principles. When combined with centrifugal balance, the corresponding 
equilibrium surface density profile is a gap (see
\S\ref{basic_state}). This is one method to calculate gap
profiles induced by giant planets in low viscosity discs.

\section{Dynamical stability of vortensity rings}\label{Dynstab} 
Now that the origin of vortensity rings 
is established and understood, one can proceed to
a linear stability analysis on the shock-modified protoplanetary disc
models described above. This is an important issue as instability can lead to
their breaking up into mobile non-axisymmetric structures, or
vortices, which can affect the migration of the planet significantly. 

The analysis applies to a two-dimensional
disc and locally isothermal perturbations, for consistency with
hydrodynamic simulations presented above and later on.
The governing equations are Eq. \ref{2d_governing} without
viscous and potential terms (Euler equations). 
The planet potential is assumed to contribute to the background gap profile only.


\subsection{Basic background  state}\label{basic_state}
In order to perform a linear analysis, one needs to define 
an appropriate background equilibrium structure to
perturb. 
The basic state should be axisymmetric and time independent with no
radial velocity ($\p/\p\phi=\p/\p t=u_r=0$). Suppose the vortensity
profile $\eta(r)$ is known (e.g. via shock modelling as above).
Recall the angular velocity is related to the vortensity by
\begin{align*}
\eta(r) = \frac{1}{r\Sigma}\frac{d}{dr}\left(r^2\Omega\right), 
\end{align*}
and the radial momentum equation gives
\begin{align}\label{hydrostatic}
  r\Omega^2 = \frac{1}{\Sigma}\frac{dp}{dr} + \frac{GM_*}{r^2}.
\end{align}
Note that the planet potential is absent in Eq. \ref{hydrostatic}. The
only role of the planet is to generate vortensity rings, after which the fluid is
assumed to be in centrifugal balance with stellar gravity and pressure
gradients. This is precisely the assumption to be validated. 
Other source terms in the momentum equation, e.g. viscosity, are neglected
for simplicity.     

\begin{figure}[t]
  \center
  \includegraphics[width=.98\linewidth]{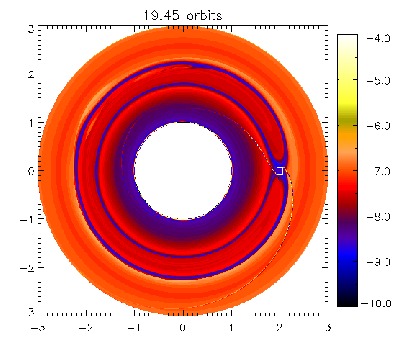}
  \caption{Vortensity rings in the co-orbital region. 
  $\ln{(\Sigma/\omega)}$ is plotted, so the rings 
   correspond to small values (blue). 
      The centre of the small square  marks the planet
    position. This snapshot is taken from a high-resolution
    ($N_r\times N_\phi=1024\times3072$) inviscid simulation.
  \label{vort_055}}
\end{figure}

Using these together with the locally isothermal equation of state,
$p=h^2GM_*\Sigma/r,$ 
a single equation for $\Sigma$ can be derived:
\begin{align}\label{basic_density}
  h^2\frac{d^2\ln{\Sigma}}{dr^2} = \frac{h^2-1}{r^2}  +
  \frac{2\Sigma\eta}{\sqrt{GM_*}}
  \left(\frac{1-h^2}{r}+h^2\frac{d\ln{\Sigma}}{dr}\right)^{1/2} 
  -\frac{2h^2}{r}\frac{d\ln{\Sigma}}{dr}.
\end{align}
Eq. \ref{basic_density} is solved to obtain $\Sigma$, with  $\eta(r)$ 
taken as an azimuthal average from the fiducial simulation described
in \S\ref{fiducial}, at a time at which vortensity rings have 
developed. These structures are essentially axisymmetric  apart from
in the close neighbourhood of the planet. They are illustrated 
in Fig. \ref{vort_055}. The boundary condition is $\Sigma=10^{-4}$
(the unperturbed uniform density value) at $r=1.1$ and $r=3$.  
These conditions are consistent with the fact that shock-modification
of the surface density profile only occurs near the planet ($r_p=2$). 

\begin{figure}
  \center
  \subfigure[$\Sigma$\label{custom_dens}]{\includegraphics[width=0.49\linewidth]{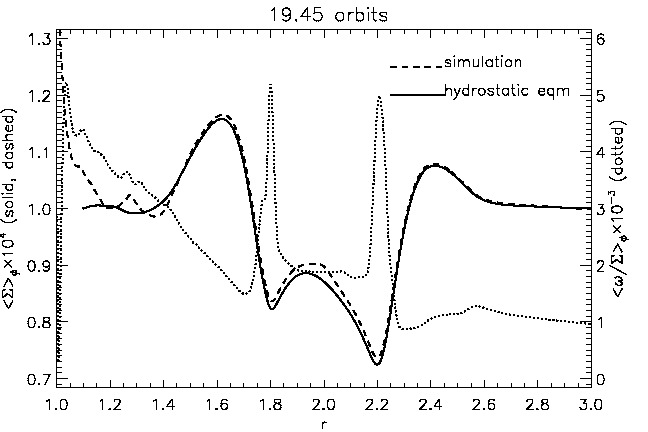}} 
  \subfigure[$\Omega/\Omega_k$\label{custom_urad}]{\includegraphics[width=0.49\linewidth]{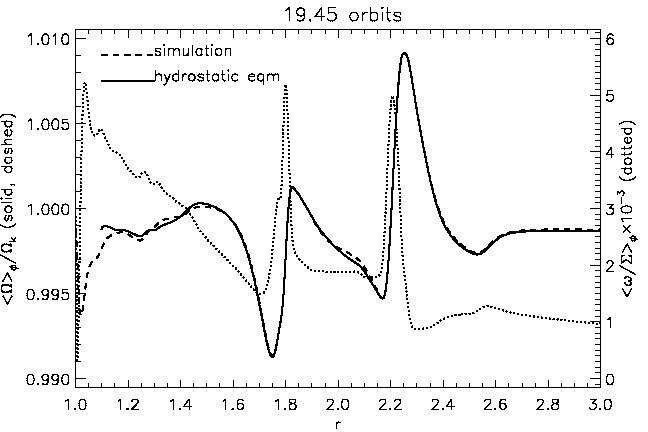}} 
  \caption{A comparison of the axisymmetric surface density and  
    angular velocity (scaled by Keplerian speed) profiles  
    obtained by solving Eq. \ref{basic_density} (solid lines), with the same quantities
    obtained by azimuthally averaging simulation data 
    (dashed lines). The azimuthally averaged vortensity profile from
    simulation data, used as input in solving Eq. \ref{basic_density}, is
    also shown (dotted).  The co-orbital region of the planet is $r=[1.77,2.22]$.} 
\end{figure}

A comparison between surface density  
and angular velocity  profiles
obtained by solving Eq. \ref{basic_density} with those obtained  as
azimuthal averages from the corresponding simulation is made in
Fig. \ref{custom_dens} and Fig. \ref{custom_urad}.  The agreement is
generally very good indicating that the adoption  of the basic
axisymmetric state  defined by the simulation vortensity profile,
together with centrifugal balance (Eq. \ref{hydrostatic}),  for
stability analysis should be a valid procedure. This validates
neglecting the planet potential, even though it was responsible for
ring generation. 

Fig. \ref{custom_dens} shows that  vortensity rings reside  in the
horseshoe region just inside the gap. At  a surface density extrema
Eq. \ref{basic_density} implies that
\begin{align*}
  h^2\frac{d^2\ln{\Sigma}}{dr^2} = \frac{h^2-1}{r^2} +
  2\Sigma\eta\sqrt{\frac{1-h^2}{GM_*r}}.
\end{align*}  
Since $h\ll1$, if $\eta$ is sufficiently
large then $d^2\Sigma/dr^2 > 0$. Hence vortensity maxima nearly coincides 
with surface density minima. In Fig. \ref{custom_dens}, local
vortensity minima
and maxima are separated by $O(H)$ ($\simeq 0.1$ in dimensionless units).
As vortensity rings originate from 
spiral shocks, Eq. \ref{basic_density} demonstrates the  link 
between shocks and gap formation. Given the double-ringed vortensity
profile $\eta(r)$, which may be estimated by modelling shocks
, one can solve Eq. \ref{basic_density} for the
axisymmetric  surface density profile $\Sigma(r)$. One then finds 
that in order for the rings to be in hydrostatic equilibrium, a gap in
the surface  density must be  present around the planet's orbital
radius, which is in between the vortensity peaks. Sufficiently massive
planets induce strong shocks, implying larger vortensity maxima, and
hence deeper surface density minima or gaps.  


\subsection{Linearised equations}
Having established the ring-basic state, a linear analysis can now be
performed to determine its stability. 
The hydrodynamic variables are written as
\[
\begin{pmatrix}
\begin{matrix}
\Sigma \\
u_r\\
u_\phi
\end{matrix}
\end{pmatrix}
=
\begin{pmatrix}
\begin{matrix}
\Sigma \\
0\\
u_\phi
\end{matrix}
\end{pmatrix}
+
\begin{pmatrix}
\begin{matrix}
\delta\Sigma(r) \\
\delta u_r(r)\\
\delta u_\phi (r)
\end{matrix}
\end{pmatrix}
\times \exp{i(\sigma t + m\phi)},
\]
where the first term on the RHS corresponds to
the basic background state, $\dd$ denote small Eulerian perturbations, 
 $\sigma$ is a complex frequency and $m,$ the azimuthal mode number,  is a positive
integer. The linearised  equation  of motion gives
\begin{align}
  \delta u_r &= -\frac{c_s^2}{\kappa^2 - \bar{\sigma}^2}
  \left(i\bar{\sigma}\frac{dW}{dr} + \frac{2im\Omega W}{r}\right)\label{velocities1}\\
  \delta u_\phi &= \frac{c_s^2}{\kappa^2 - \bar{\sigma}^2}
  \left(\Sigma\eta\frac{dW}{dr} + \frac{m\bar{\sigma}}{r}W\right)\label{velocities2}
\end{align}
where
$W \equiv \delta\Sigma/\Sigma$
is the relative surface density perturbation, recall 
$\kappa^2 = 2\Omega\Sigma\eta$
is the epicycle frequency  expressed in terms of the  vortensity $\eta,$ and
$\bar{\sigma} \equiv \sigma +m\Omega(r) = \sigma_R + m\Omega(r) + i\gamma$ 
is the Doppler-shifted frequency with  $\sigma_R$ and $\gamma$ being
real. In this notation, $\gamma <0$ corresponds to an exponential
growth of perturbations and therefore instability. 
Substituting  Eq. \ref{velocities1}---\ref{velocities2} into the
linearised continuity equation 
\begin{align}\label{lin_cont}
  i\bar{\sigma} W = -\frac{1}{r\Sigma}\frac{d}{dr}\left(r\Sigma\delta
    u_r\right) - \frac{im}{r}\delta u_\phi,
\end{align} 
yields  a  governing equation for $W$  of the form
\begin{align}\label{goveq_gen}
  \frac{d}{dr}\left(\frac{rc_s^2\Sigma}{\kappa^2 -\bar{\sigma}^2}\frac{dW}{dr}\right)
  + \left\{\frac{m}{\bar{\sigma}}\frac{d}{dr}
    \left[\frac{c_s^2\kappa^2}{\eta(\kappa^2-\bar{\sigma}^2)}\right]
    -r\Sigma
    -\frac{c_s^2m^2\Sigma}{r(\kappa^2-\bar{\sigma}^2)}\right\}W = 0.
\end{align}
Specifically, for the locally isothermal equation of state, this is
\begin{align}\label{goveq}
  \frac{d}{dr}\left(\frac{\Sigma}{\kappa^2 -\bar{\sigma}^2}\frac{dW}{dr}\right)
  + \left\{\frac{m}{\bar{\sigma}}\frac{d}{dr}
    \left[\frac{\kappa^2}{r\eta(\kappa^2-\bar{\sigma}^2)}\right]
    -\frac{r\Sigma}{GM_*h^2}
    -\frac{m^2\Sigma}{r^2(\kappa^2-\bar{\sigma}^2)}\right\}W = 0.
\end{align}

Eq. \ref{goveq} is  an eigenvalue problem  
for the complex eigenvalue $\sigma.$ The co-rotation radius 
$r_c$ is where $\sigma_R + m\Omega(r_c)=0$. Co-rotation  
resonance occurs at $\bar{\sigma} = 0$ which requires 
$\gamma =0$ for it to be on the real axis.   
Then for the equation  to remain regular,  the gradient of the  terms
in square brackets must vanish at co-rotation. This results in the 
requirement that 
the gradient of $\eta r$ vanish there. Because the sound speed varies
with radius, this is slightly different from
the condition that the gradient of $\eta$ should vanish which applies
in the barotropic or strictly isothermal case
\citep{papaloizou84,papaloizou85,papaloizou89}. However, because
$\eta$ varies rapidly in the region of interest and  
the modes of interest are locally confined in radius, this difference is of no
essential consequence. Lindblad resonances occur when
$\kappa^2-\bar{\sigma}^2 = 0$, but as is well known, and can be seen
from formulating a governing equation for $\delta u_r$ rather than
$W$, these do not result in a singularity. 

\nomenclature[d]{$W$}{Fractional surface density perturbation}
\nomenclature[d]{$S$}{=$c_s^2W + \dd\Phi$, where $\dd\Phi$ is the
  potential perturbation}
\nomenclature[d]{$\sigma$}{Complex frequency ($=\sigma_R + i\gamma$)}
\nomenclature[d]{$m$}{Azimuthal wavenumber}
\nomenclature[d]{$\sbar$}{Doppler shifted frequency ($=\sigma +
  m\Omega$)}
\nomenclature[d]{$D$}{$=\kappa^2  - \sbar^2$}


\subsection{Simplification of the governing ODE}\label{corot_modes}
It is useful to simplify Eq. \ref{goveq}  to gain further
insight. To do this, consider modes localised around the co-rotation circle
such that the condition 
$\kappa^2\gg|\bar{\sigma}^2|$ is satisfied. Beyond this region
the mode amplitude is presumed to be exponentially small.
Now, the 
ratio of the second to last to last  term in Eq. \ref{goveq_gen} is
\begin{align*}
  \frac{r^2\kappa^2}{m^2c_s^2}\sim\frac{1}{m^2h^2}.
\end{align*}
For a thin disc $h\ll1$, so for $m=O(1)$ this ratio is
large and the last term in Eq. \ref{goveq}  can be neglected. This is 
also motivated by the fact that only low $m$ modes  are
observed in simulations. Doing this and replacing
$\kappa^2-\bar{\sigma}^2$ by $\kappa^2$, 
Eq. \ref{goveq_gen} reduces to the simplified form
\begin{align}\label{goveq2}
  \frac{d}{dr}\left(\frac{rc_s^2\Sigma}{\kappa^2}\frac{dW}{dr}\right)
  +\left\{\frac{m}{\bar{\sigma}}\frac{d}{dr}
    \left[\frac{c_s^2}{\eta}\right]
    -r\Sigma\right\}W = 0,
\end{align}
which is valid for any fixed $c_s$  profile. 

Localised modes described by the approximations above have been
called `co-rotation modes' \citep{lovelace99}. Suppose the
basic state has a localised vortensity extremum at $r_c$. Then far away
from $r_c$, vortensity gradients are small and the $-r\Sigma W$ term
in Eq. \ref{goveq2} dominates, implying exponentially decaying 
solutions. Conversely, near $r_c$ vortensity gradients are large and
the term proportional to $1/\sbar$ dominates. Evaluating
Eq. \ref{goveq2} for a neutral mode ($\gamma=0$) at $r_c$ implies the
balance 

\begin{align*}
 \left[\left(\frac{rc_s^2\Sigma}{\kappa^2}\right)^\prime\frac{dW}{dr}\right]_{r_c}   
  +\left[\left(\frac{rc_s^2\Sigma}{\kappa^2}\right)\frac{d^2W}{dr^2}\right]_{r_c} 
+\left[\frac{1}{\Omega^\prime}
  \left(\frac{c_s^2}{\eta}\right)^{\prime\prime}\right]_{r_c}W(r_c)
\sim 0,
\end{align*}
where $^\prime$ denotes $d/dr$, and $\bar{\sigma}^\prime =
m\Omega^\prime$ has been used. Assuming the mode is symmetric about 
$r_c$ so that $W^\prime(r_c)=0$, one deduces

\begin{align}\label{localisation_condition}
 \left.\frac{W^{\prime\prime}}{W}\right|_{r_c} \sim
 \left[\frac{-2\Omega\eta}{rc_s^2\Omega^\prime}
 \left(\frac{c_s^2}{\eta}\right)^{\prime\prime}\right]_{r_c},  
\end{align}
where $\kappa^2 = 2\Omega\Sigma\eta$ has been
used. Eq. \ref{localisation_condition} says that the degree of
localisation, measured by $W^{\prime\prime}(r_c)$, is proportional
to $(c_s^2/\eta)^{\prime\prime}|_{r_c}$, so the more
sharply peaked the vortensity extremum is, the more localised the mode
will be. 

\subsection{Necessity of extrema}
Vortensity extrema must exist for unstable modes. 
 Multiplying Eq. \ref{goveq2}
by $W^*$ and integrating between $[\,r_1,r_2]$ assuming, consistent
with a sharp exponential decay, that  $W=0$ or
$dW/dr=0$ at these  boundaries, yields
\begin{align}\label{neccessary}
\int^{r_2}_{r_1}\frac{m}{\bar{\sigma}}\left(\frac{c_s^2}{\eta}\right)^\prime|W|^2dr
=\int^{r_2}_{r_1}r\Sigma|W|^2 dr + \int^{r_2}_{r_1}\frac{r\Sigma
  c_s^2}{\kappa^2}|W^\prime|^2 dr
\end{align}
Since the RHS is real, 
the imaginary part
of the LHS must vanish. For general complex $\sigma$ this implies that
\begin{align}\label{vortmin}
  \gamma\int^{r_2}_{r_1}\frac{m}{(\sigma_R+m\Omega)^2+\gamma^2}
  \left(\frac{c_s^2}{\eta}\right)^\prime|W|^2dr
  =0.
\end{align}
Thus for a growing mode ($\gamma\neq0$) to exist, one requires
$(c_s^2/\eta)^\prime =0$ at some $r$ in $[\,r_1,r_2]$. For
the locally isothermal equation of state, $c_s^2$ varies on a scale
$O(r)$, but $\eta$ varies on  a scale  $H\ll r$, thus given  
that the range of relative variation of the vortensity  $\eta$ is of
order unity, one infers that $\eta$ needs to  have stationary
points in order  for there to be unstable modes. 
 Referring back to Fig. \ref{custom_dens} it is clear
that the basic state being considered satisfies the necessary
criterion for instability.


\subsection{Limit on the growth rate}
The approximation for co-rotational modes,
$\kappa^2\gg|\bar{\sigma}^2|$, implicitly assumes
the growth rate $|\gamma|/\Omega\ll1$, since
$\kappa \sim\Omega$. A more quantitative estimate
of the upper limit to the growth rate can be used to provide bounds on 
initial guess solutions when solving the eigenvalue problem
numerically. Let 
\begin{align}
  W = g\bar{\sigma},
\end{align}
Eq. \ref{goveq2} then gives an alternative governing equation:
\begin{align}\label{goveq3}
  \left(\frac{r\Sigma c_s^2\bar{\sigma}^2}{\kappa^2}g^\prime\right)^\prime
  &=
  \bar{\sigma}\left[r\Sigma\bar{\sigma} -m\left(\frac{c_s^2}{\eta}
      + \frac{r\Sigma c_s^2\Omega^\prime}{\kappa^2}\right)^\prime\right]g\\
  &=\bar{\sigma}\left[r\Sigma\bar{\sigma} -m\left(\frac{c_s^2\Sigma}{2\Omega}\right)^\prime\right]g,\notag
\end{align}
where the second line is simplified from the first line by using
$\kappa^2 = (1/r^3)d(r^4\Omega^2)/dr=2\Omega\Sigma\eta$. Multiplying
Eq. \ref{goveq3} by $g^*$, integrating and neglecting boundary terms
gives
\begin{align}\label{integral_relation}
-\int_{r_1}^{r_2}\frac{r\Sigma c_s^2 \bar{\sigma}^2}{\kappa^2}|g^\prime|^2 dr = 
\int_{r_1}^{r_2}\bar{\sigma}\left[r\Sigma\bar{\sigma}
  -m\left(\frac{c_s^2\Sigma}{2\Omega}\right)^\prime\right]|g|^2 dr.
\end{align}
The real part of Eq. \ref{integral_relation} is
\begin{align}\label{realpart}
  \gamma^2\int_{r_1}^{r_2} r\Sigma\left(\frac{c_s^2}{\kappa^2}|g^\prime|^2 +
    |g|^2\right)dr =
  \int_{r_1}^{r_2}\left[r\Sigma\bar{\sigma}_R^2\left(\frac{c_s^2}{\kappa^2}|g^\prime|^2 + 
      |g|^2\right) -
    m\bar{\sigma}_R\left(\frac{c_s^2\Sigma}{2\Omega}\right)^\prime|g|^2\right]dr,  
\end{align}
where $\sbar_R= \sigma_R + m\Omega$. The imaginary part of
Eq. \ref{integral_relation} implies (assuming $\gamma\neq0$),
\begin{align}\label{impart}
\int_{r_1}^{r_2}\left[2r\Sigma\bar{\sigma}_R\left(\frac{c_s^2}{\kappa^2}|g^\prime|^2 +
  |g|^2\right) -
m\left(\frac{c_s^2\Sigma}{2\Omega}\right)^\prime|g|^2\right]dr = 0.
\end{align} 
If the second term of the integrand in Eq. \ref{impart} is negligible (justified
below),  Eq. \ref{impart} reduces to
\begin{align}\label{impart2}
\int_{r_1}^{r_2}\left[2r\Sigma(\sigma_R +
  m\Omega)\left(\frac{c_s^2}{\kappa^2}|g^\prime|^2 + 
  |g|^2\right)\right]dr = 0,
\end{align} 
which implies that $\sigma_R + m\Omega = 0$ somewhere in $[r_1,r_2]$,
since the other terms are non-negative. This confirms the co-rotation
point $r_c$, such that $\sigma_R=-m\Omega(r_c)$, is inside the
domain.

Eq. \ref{realpart} and Eq. \ref{impart} can be combined to give:
\begin{align}\label{semi1}
\int_{r_1}^{r_2} r\Sigma(\gamma^2+\sigma_R^2)\left(\frac{c_s^2}{\kappa^2}|g^\prime|^2 +
  |g|^2\right)dr =& \int_{r_1}^{r_2} m^2\Omega^2r\Sigma\left(\frac{c_s^2}{\kappa^2}|g^\prime|^2 +
  |g|^2\right)dr \\\notag
& - \int_{r_1}^{r_2} m^2\Omega\left(\frac{c_s^2\Sigma}{2\Omega}\right)^\prime |g|^2 dr.
\end{align} 
Note that in Eq. \ref{semi1}, the LHS and the first term
on RHS are non-negative. The second integral on RHS, involving
$(c_s^2\Sigma/2\Omega)^\prime$, may be positive or negative. Far away
from the gap edge the disc is smoothly varying and gradients are
small. Near the gap edges, the disc varies rapidly over local
scale-heights. One can expect the integral
involving $(c_s^2\Sigma/2\Omega)^\prime$ makes a small contribution
in Eq. \ref{semi1}, due to small gradients and internal cancellations 
when the integral is performed. 
One can also
compare coefficients of $|g|^2$ in the integrands on RHS of
Eq. \ref{semi1}:
\begin{align}\label{approximate}
\tau \equiv  \left|\frac{m^2\Omega^2r\Sigma}{m^2\Omega\left(\frac{c_s^2\Sigma}{2\Omega}\right)^\prime}\right|.
\end{align}
If the disc varies on a  length-scale $L\sim O(r)$ then
$\tau=O(h^{-2})$, whereas if $L\sim O(H)$ then $\tau=O(h^{-1})$. In either
case, the ratio is large for thin discs ($h\ll1)$. Inserting the
actual disc profiles obtained from numerical simulation gives $\tau\sim
90$. The last term on RHS of Eq. \ref{semi1} can be safely
neglected. This also justifies Eq. \ref{impart2}.  

Next,
consider the inequality
\begin{align}
  0\geqslant m^2\int_{r_1}^{r_2} (\Omega-\Omega_+)(\Omega -
  \Omega_-) r\Sigma \left(\frac{c_s^2}{\kappa^2}|g^\prime|^2 +
    |g|^2\right)dr,
\end{align}
where $\Omega_+,\,\Omega_-$ are the maximum and
minimum angular speeds in the range of integration. 
Using Eq. \ref{semi1} together with Eq. \ref{impart}, 
and neglecting the integrals involving $(c_s^2\Sigma/2\Omega)^\prime$
in both equations, one can show that
\begin{align}\label{semicircle}
  \gamma^2 + \left[\sigma_R + \frac{1}{2}m(\Omega_+ +
    \Omega_-)\right]^2 \leqslant m^2\left(\frac{\Omega_+-
      \Omega_-}{2}\right)^2.
\end{align}
The complex eigenfrequency $\sigma$ is contained inside a
circle centred on the real axis at $-m(\Omega_+ +
\Omega_-)/2$ with radius $m(\Omega_+ -
\Omega_-)/2$. Furthermore, if the mode is localised about
$r_c$ with characteristic width $\Delta>0$, so the range of
integration may be taken as $r_1=r_c - \Delta,\,r_2=r_c +
\Delta$,  and the angular speed
is monotonically decreasing then
\begin{align*}
  \Omega_\pm \simeq \Omega(r_c) \mp \Omega^\prime(r_c)\Delta.
\end{align*}
The term in square brackets in Eq. \ref{semicircle} is then $\sigma_R
+ m\Omega(r_c)$, which is zero by the definition of $r_c$. This means
the growth rate is limited to   
\begin{align}\label{ratelimit}
  \gamma^2
  \leqslant
  \left(m \Omega^\prime(r_c)\Delta \right)^2.
\end{align}
The growth of co-rotational modes is
limited by local shear. Strong shear may enable higher growth rates.  
Inserting Keplerian shear and $\Delta = H(r_c)=hr_c$ in
Eq. \ref{ratelimit} gives
\begin{align}\label{growth_time}
  \frac{|\gamma|}{\Omega(r_c)} \leq  \frac{3}{2}hm. 
\end{align}
For $h\ll 1$ and $m=O(1)$, growth rates are at most a fraction of 
the local angular speed. The growth timescale is
$|\gamma|^{-1}$. When measured in Keplerian orbital periods at
 $r_p$ , the growth time is
\begin{align*}
  \frac{|\gamma|^{-1}}{P_0} \geq  \frac{1}{3\pi h
    m}\left(\frac{r_p}{r_c}\right)^{-3/2}, 
\end{align*}
where Keplerian rotation has been used. 
Vortex formation will be shown to be associated with vortensity minimum at gap edges
.  For the reference 
case later (Fig. \ref{custom_dens}), the vortensity minimum close to  
the inner gap edge is located at $r_c=1.7$. The planet is at $r_p=2$
and the aspect-ratio is $h=0.05$. Inserting these for the $m=3$ mode
in Eq. \ref{growth_time} yields a growth time $|\gamma|^{-1}\gtrsim
0.6P_0$, so the instability can grow on dynamical timescales. 


\subsection{Numerical solution of the eigenvalue problem}
The eigenvalue problem Eq. \ref{goveq} is 
solved using a shooting method  that employs an adaptive Runge-Kutta
integrator and a  multi-dimensional Newton method \citep{press92}.

The shooting method works as follows, beginning with a trial value for
$\sigma$. The previous analysis (and nonlinear simulations
described later) hints that a good guess would
be $\sigma_R = -m\Omega(r_0)$, where $r_0$  is a vortensity
minimum. The imaginary part of $\sigma$ should be a small
fraction of $\Omega(r_0)$ and negative for unstable modes. The
governing equation is then evolved as an initial value problem, from
the inner boundary (with imposed boundary conditions) 
to the outer boundary. At the outer boundary, one
compares the evolved solution to the required outer boundary condition. The
eigenvalue is accordingly adjusted, using the Newton method, and the initial value
problem solved again to obtain a new solution with outer boundary
values closer to the desired conditions. The process is repeated until
convergence.    

For low $m$ ($\leq 3$) , unstable modes mainly comprise an evanescent disturbance
near co-rotation (vortensity minimum) and the simple
boundary condition $W = 0$ applied at the inner boundary $r/r_p=0.55$
and the outer boundary $r/r_p=1.5$ produces good results.  
As  $m$ increases,  the  Lindblad resonances (where
$\kappa^2=\sbar^2$) approach  $r=r_0$ 
and a  significant portion of the mode is wave-like 
requiring  the application of outgoing  radiation boundary conditions.   
These  are determined using the  WKBJ  approximation (see eg.
\cite{korycansky95}). 
Recognising the governing equation (Eq. \ref{goveq}) as
\begin{align*}
  \frac{d}{dr}\left(A\frac{dW}{dr}\right) = -BW,
\end{align*}
the WKBJ solution is given by
\begin{align*}
  W&\propto\frac{1}{(AB)^{1/4}}\exp{\pm i\int\sqrt{\frac{B}{A}}dr}. 
\end{align*}
However, high $m$, wave-like modes are in fact irrelevant because
these are not the fastest growing modes. 



\subsection{Fiducial case}
Some example solutions are presented to illustrate the 
instability of gap edges. Hydrodynamic simulations indicate the   
ultimate  dominance of small $m$ values. 
One class of mode is associated with the inner vortensity ring while 
another is associated with the outer ring. 

\begin{landscape}
\begin{figure}
  \center
  \subfigure[Inner edge]{\includegraphics[width=0.5\linewidth]{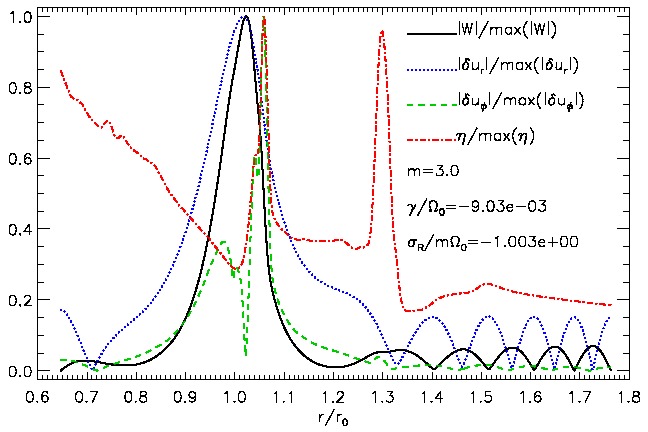}}\subfigure[Outer edge]{\includegraphics[width=0.5\linewidth]{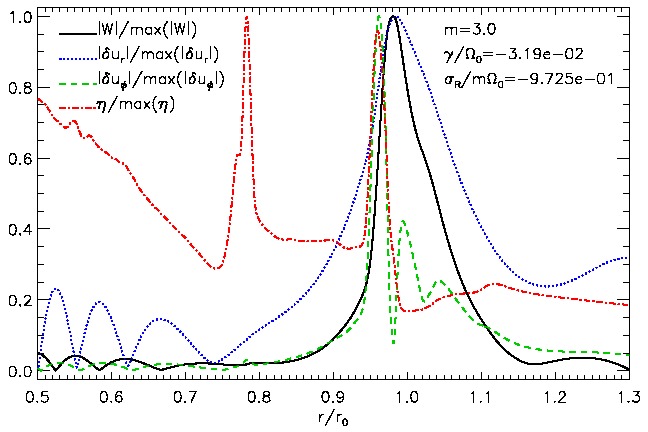}}
  \caption{$m=3$ eigenmodes obtained  with the  boundary condition
    $W=0.$ The 
    perturbed quantities are plotted as $|W|$ (solid, black),
    $|\delta u_r|$ (dotted,blue),  
    $|\delta u_\phi|$ (dashed, green) and scaled by their maximum
    values in $r=[1.1,3.0]$. The background vortensity profile is
    also shown (dash-dot, red). $r_0=1.7,\,2.3$ 
    correspond to the inner and outer vortensity minima,
    respectively.\label{s1h005}} 
\end{figure} 
\end{landscape}

As a typical example of the behaviour that is found for low $m,$ each
type of eigenmode  for  $h=0.05$ and  $m=3$  
is shown in Fig. \ref{s1h005}. The background vortensity profile is also shown.
The instabilities ($\gamma<0$) are associated with the  vortensity minima at the inner
and outer  gap edge, as has also been 
observed by \cite{li05} in simulations. 
The modes are evanescent around co-rotation and the  vortensity peaks 
behave like walls through which the instability scarcely penetrates.
The mode decays away from $r=r_0$. For $m=3$,
Lindblad resonances occur at $r_L/r_0=0.76,\,1.21$  from which 
waves travelling away from co-rotation are emitted. 
However,
the oscillatory amplitude is at most $\simeq 20\%$ of that at $r=r_0$. Hence for  low-$m$ 
the dominant effect of the instability will be due to perturbations near co-rotation. 
Increasing $m$ brings $r_L$ even closer to $r_0$, 
waves  then propagate through the planetary gap. 
The growth timescale of the inner mode with $m=3$ is  $\sim14P_0$.  
The outer mode has a growth rate that is about three times faster. 
Since the instability
grows on dynamical timescales, non-linear interaction of vortices are expected
to occur within few tens of orbits, and to affect planet migration 
if migration  were also on similar or longer timescales.

After the onset of linear instability and the formation of several vortices,
it has been observed that non-linear effects 
cause them to eventually 
merge into a single vortex \citep{valborro07} which interacts with the planet. 
In the fiducial simulation to be presented,
vortex-induced rapid migration begins within $55P_0$ 
which is compatible with the characteristic  growth 
times found  from linear theory. 


\begin{figure}
  \center
  \includegraphics[width=0.99\linewidth]{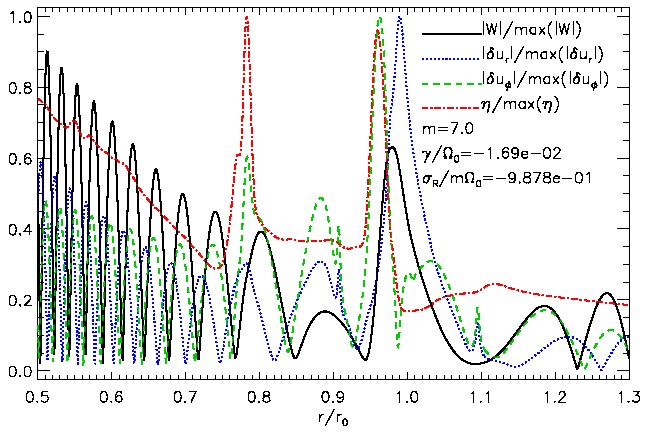}
  \caption{$m=7$ eigenmodes associated with the outer edge, obtained
    with WKBJ boundary condition. The
    perturbed quantities are plotted as $|W|$ (solid, black), $|\delta
    u_r|$ (dotted, blue), $|\delta u_\phi|$ (dashed, green) and scaled by
    their maximum 
    values in $r=[1.1,3.0]$. The background vortensity profile is also
    shown (dash-dot, red). $r_0=2.3$ is the radius of the outer vortensity
    minima.   
    \label{s1h005m7}}
\end{figure}

Fig. \ref{s1h005m7} shows the eigenfunctions
 for the $m=7$ mode of the outer ring. The equivalent mode was not found for
the inner edge because high-$m$ are quenched. Radiative boundary conditions were 
adopted in this case. Although the WKBJ condition is the appropriate
physical boundary condition 
, its application here is uncertain because the boundaries cannot be
considered `far' from the gap edge.
However, solutions are actually not sensitive to 
boundary conditions as reported by \cite{valborro07}.  Note the two spikes in 
$\delta u_r$ and $\delta u_\phi$ at $r/r_0\simeq0.90,\,1.09$ which
correspond to Lindblad resonances. These are not singularities as can
be seen from $W(r)$ which is smooth there; other eigenfunctions were
calculated from  
the numerical solution for $W$ and thus may be subject to numerical
errors. Increasing $m$ increases the amplitude in the 
wave-like regions of the mode, but the growth rate is smaller than for
$m=3$. As the instability operates on dynamical timescales, low-$m$
modes will dominate over high-$m$ modes, particularly through
non-linear evolution and interaction of the former.  

\subsection{Dependence on  $h$}\label{rates_h}
The aspect ratio $h$, equivalent to  disc temperature, can affect
stability properties. Lowering $h$ increases shock strength and
therefore the
magnitude of vortensity jump, but does not affect ring locations and
widths. The basic state formed with lower $h$ should be more
unstable because gradients are larger in magnitude. $h$ is
also explicit in Eq. \ref{goveq}, appearing as
$-r\Sigma W/(GM_*h^2)$. Recalling from \S\ref{corot_modes} that this
term is responsible for exponential decay, then decreasing $h$ should
increase localisation. So lowering $h$ should favour
development of co-rotational modes. 

In the discussion below, $h=0.05$ will be abbreviated as $h_5$ and
similarly for other values. The linear problem was solved for
$h_3$---$h_6$. In each case, a disc-planet simulation was
performed to generate the basic state. Table \ref{rates_table} compare
growth rates of $m=1$---3 modes with $W=0$ boundary
condition. Localised modes as those in Fig. \ref{s1h005} were
found, except for the inner edge for $h_6$ where no unstable modes
where found.  
This means that only sufficiently extreme profiles support
co-rotational modes.   As $h$ is lowered, 
the inner edge becomes more unstable with
$|\gamma_\mathrm{in}|(h_3)\simeq(1.5$ 
~---2$)\times|\gamma_\mathrm{in}|(h_5)$. 

\begin{table}
  \caption{Growth rates of linear modes scaled by local angular speeds for $m=1$---3 for
    the inner (subscript `i') outer (subscript `o') gap edges. 
    \label{rates_table}}
\begin{center}
  \begin{tabular}{|c|ll|ll|ll|}
    \hline
    & $m=1$  & & $m=2$ & & $m=3$ &  \\ 
    \hline
    $h$ & $-10^3\gamma_\mathrm{i}/\Omega_\mathrm{i}$ 
    & $-10^3\gamma_\mathrm{o}/\Omega_\mathrm{o}$
    & $-10^3\gamma_\mathrm{i}/\Omega_\mathrm{i}$ 
    & $-10^3\gamma_\mathrm{o}/\Omega_\mathrm{o}$ &
    $-10^3\gamma_\mathrm{i}/\Omega_\mathrm{i}$ 
    & $-10^3\gamma_\mathrm{o}/\Omega_\mathrm{o}$ \\
    \hline
    0.03 & 8.405 &  8.466  & 15.72 &  16.14  & 20.82 &  22.33 \\
    0.04 & 7.804 &  13.51  & 14.21 &  25.62  & 18.20 &  35.01 \\
    0.05 & 5.586 &   13.42  & 9.298 &   24.79 & 9.090 &  31.95 \\
    0.06 & $0.00287$ & 4.304  & $0.00437$ & 6.526 
    & $0.00233$ & 5.741 \\
    \hline
  \end{tabular}
\end{center}
\end{table}

Generally, lowering $h$ is destabilising. The exception to this trend
is the outer edge of $h_3$, being more 
stable than $h_4$ and $h_5$. This is an artifact from azimuthally
averaging the simulation data to generate the background profile. 
Hydrodynamic simulations for $h_3$ indicate, that vortensity rings 
 actually become unstable \emph{before} reaching the full azimuth.   
By the time the rings reach the full $2\pi$, vortices already appear
adjacent to vortensity rings. Radial variations are smoothed during
an azimuthal average because vortex/ring are separated by a curve
$r=r(\phi)$. 

\begin{figure}
  \center
  \includegraphics[width=0.99\linewidth]{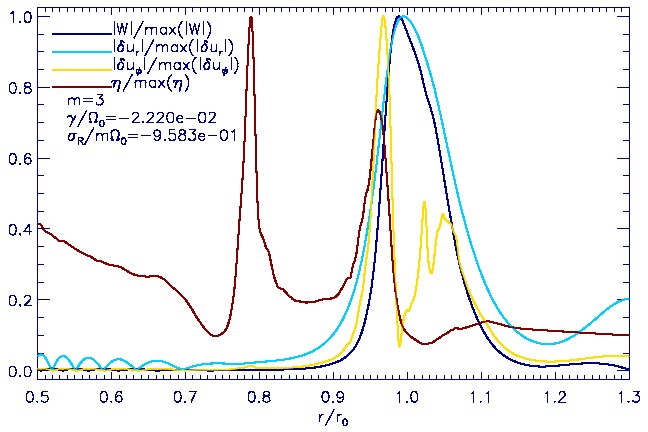}
  \caption{Eigenmodes for $m=3,\,h=0.03$ associated with the outer
    vortensity ring. The horizontal axis is scaled by the radius
    of local vortensity minimum. The boundary condition $W=0$ was
    applied in this case. \label{s1h003}} 
\end{figure}

Fig. \ref{s1h003} show solutions and the  background
vortensity profile for $h_3$. As explained above, the   
outer vortensity ring in $h_3$ is apparently less sharp compared to
$h_5$ (Fig. \ref{s1h005}). The outer edge is \emph{apparently} more
stable. In this case, the azimuthally averaged
vortensity profile is not an appropriate representation of the 
outer edge, having already become non-axisymmetric.  
Nevertheless, the solution in Fig. \ref{s1h003} indicate more
localisation compared to $h_5$ (Fig. \ref{s1h005}, where the amplitude
of the wave region is relatively larger than $h_3$). 
This is expected from the increased importance of the exponential
decay term governing the linear response (Eq. \ref{goveq}). 


Table \ref{rates_table} also show the outer edge is typically more
unstable, consistent with simulations  that vortices first form near
the outer ring. There is increasing difference in growth rates
between  inner and outer minima as $h$ increases. This may be because
as $h$ increase, there is less vortensity jump (in magnitude) across
weaker shocks. Hence, the background vortensity variation ( 
before the planet is introduced) is
more pronounced, so the inner/outer edge asymmetry is
enhanced.  Conversely, for small $h$ strong shocks are produced, the
final vortensity profile is dominated by vortensity jumps 
across shocks, and less affected by the background profile. 



\subsection{Growth rates and $m$}

Hydrodynamic simulations will show that low $m$
co-rotational (vortex) modes are most relevant to co-orbital 
disc-planet interactions. For completeness though, 
stability as a function of $m$ is briefly discussed. 
Table \ref{rates_table} show $|\gamma|$ typically increase with
$m$.  However, rates cannot increase indefinitely. Waves can be stabilising
by carrying energy away from  co-rotation. The co-rotational
disturbance itself has positive energy provided co-rotation is at a 
vortensity minimum (see Chapter \ref{paper2}). Consequently
$|\gamma|$ eventually decrease as $m$  increases. One expects
$|\gamma|$ to peak as a function of $m$ \citep{li00,valborro07}, and
a cut-off at some maximum $m$. 

\begin{figure}
  \center
  \subfigure[Inner
  edge]{\includegraphics[width=0.49\linewidth]{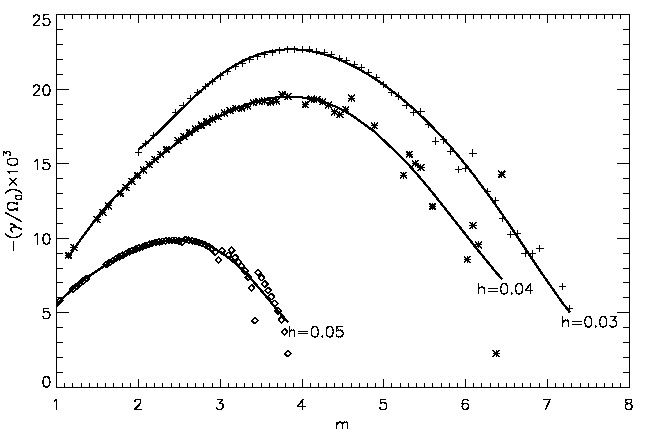}} 
  \subfigure[Outer
  edge]{\includegraphics[width=0.49\linewidth]{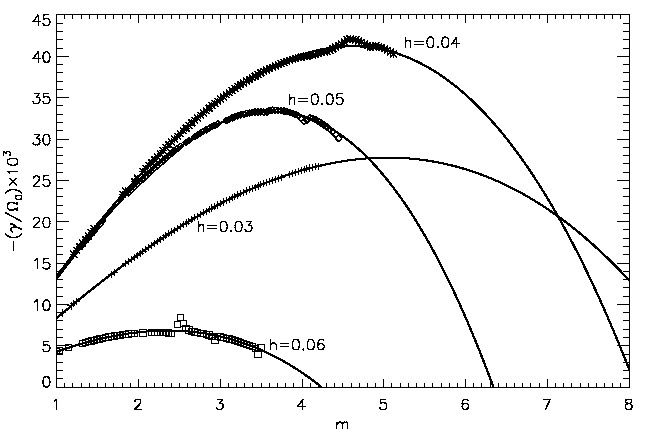}} 
  \caption{ Growth rates as a function of azimuthal wave number $m$
    and aspect-ratio $h$. 
    $\Omega_0$ is background angular speed at corresponding vortensity
    minima. The $h=0.03$ 
    case for the outer edge is spurious because the disc has already become unstable
     when the profile is taken. For the inner
    edge, generally being more stable, modes were not found for $h=0.06$.
    \label{rates}}
\end{figure}

The linear problem, Eq. \ref{goveq}, is solved subject to WKBJ boundary
conditions and non-integer $m$, to map out dependence of $\gamma$ on
azimuthal wave number. 
Only integer values of $m$ make physical sense, but this is not 
required to solve Eq. \ref{goveq}. Results are  shown in
Fig. \ref{rates} with polynomial fits and extrapolation. 

Inner edge growth rates in the standard disc $h_5$ peaks at $m=2$---3
and extrapolation suggest a cut-off for $m\geq5$. Decreasing to $h_4$,
$|\gamma|$ peaks around $m=4$ with growth rate roughly twice the peak
of $h_5$. Higher-$m$ modes are enabled as temperature is lowered. 
 Lowering $h$ generally destabilise the system
while increasing localisation. 
A similar behaviour was found for the outer edge (except for
the spurious $h_3$). 

Boundary effects became
apparent as higher $m$ were considered. This is reflected in the
sporadic growth rates beyond the most unstable mode (for the inner
edge) and the inability to find satisfactory solutions for the outer
edge. This is because the boundaries were not very far
from gap edges.  
In these cases, extrapolation to higher $m$ was performed, assuming
there is some cut-off at high $m$. Fortunately, these very high $m$
modes are not the most unstable, nor do they show up in nonlinear
simulations.

\cite{valborro07} performed a similar analysis for gaps opened
by a Jovian-mass planet in a $h_5$ disc, and obtained growth rates  
almost an order of magnitude larger than here. 
\citeauthor{valborro07} also found peak growth rates
around $m=4$---6. For disc-planet interactions, 
 increasing planetary mass has the same effect as lowering
$h$. Results here are then consistent with that of
\citeauthor{valborro07}, that lowering $h$ destabilises the system
and shifts the modes to higher $m$. 


\section{Simulations of fast migration driven by vortex-planet
  interaction}\label{mig_visc} 
Disc-planet simulations are now examined, where the planet is
free to migrate,  
so that $r_p=r_p(t)$. Vortices that form near 
the gap edge move through the co-orbital region
and cause torques to be exerted on the planet.
The interaction between these vortices and the planet
can be interpreted as non-monotonic type III migration. 
In the original description of type III
migration by \cite{masset03}, the type III torque increases with the
co-orbital mass deficit $\delta m$ defined as: 
\begin{align}\label{delta_m}
  \delta m = 8\pi r_p B_p \left[x_sw(-x_s)-\int^0_{-x_s}w(x)dx\right]
\end{align}
where $w=\Sigma/\omega=\eta^{-1}$, $B_p=\frac{1}{2r}\partial_r
(r^2\Omega)$ is the Oort constant evaluated at the planet radius
$r_p$, and the co-ordinate $x=r-r_p$. Hence the inverse vortensity
$\Sigma/\omega$ will be central to the discussion.

Simulations described  below have computational domain $r=[\,0.4, 
4.0\,]$ with resolution $N_r\times N_\phi=768\times 2304$ uniformly 
spaced in both directions. The initial surface density scale is chosen to be 
$\Sigma_0=7.0$, corresponding to a few times the  
value appropriate to the minimum mass Solar nebula in order to achieve
smooth rapid migration when a typical viscosity $\nu_0=1$ is used 
\citep{masset03}. The initial radial velocity of the disc is set to be  
$u_r=-3\nu/2r$ as expected for a steady accretion disc 
\citep{lyndenbell74}, while the initial azimuthal velocity is slightly
sub-Keplerian to achieve centrifugal balance with pressure and stellar gravity. 
 The planet is introduced in Keplerian circular
orbit. For most of these simulations the full planet potential is
applied from  $t=0$, but similar results were obtained when the potential is
switched on over $5P_0$. 
 In those cases, vortices
were also observed to form.  Switching on the planet potential over several
orbits does not weaken the instability, and vortex-planet interactions
still occur.
 
Type III migration is numerically challenging due to its potential 
dependence on flow near the planet, one issue being the numerical  
resolution. \cite{dangelo05} reported the suppression of type III 
migration in high resolution simulations. The main migration feature
discussed below is brief phases of rapid migration due to
vortex-planet interaction, which does not depend on conditions very
close to the planet. Lower resolution simulations with 
$N_r\times N_\phi= 192\times576,\,256\times768$ both show such
behaviour, thus the higher resolution described below is sufficient to
study this interaction.

\subsection{Dependence of the migration rate on viscosity}
The previous sections showed that ring structures formed by a 
Saturn-mass planet can be linearly unstable. Other fixed-orbit
simulations of disc-planet interactions show that a dimensionless
viscosity of order $10^{-5}$ suppresses instability and subsequent
vortex  formation \citep[e.g.][who used a Jupiter-mass
planet]{valborro07}. 
As a consequence  studies using viscous discs
typically yield smooth migration curves. This motivates a study of
type III migration as a function of viscosity. In this section the planet mass
is fixed to be $M_p=2.8\times10^{-4}M_*$. 

\begin{figure}[!ht]
  \centering
  \includegraphics[trim=0cm 0cm 6cm 4cm, clip,
  width=1.0\linewidth]{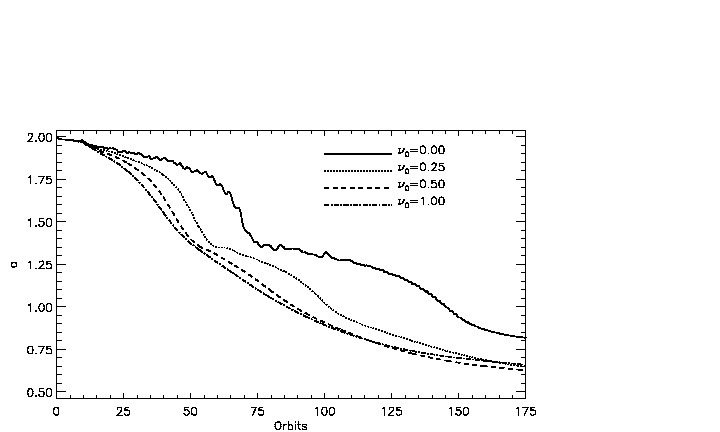} 
  \caption{Migration as a function of viscosity.
    Migration for $\nu_0=1$ indicates classic type III migration. As
    viscosity is lowered, migration becomes non-smooth. The 
    inviscid case has two distinct migration timescales: the
    initial slower migration followed by a sudden drop in $a$. 
    \label{orbit2_visc}} 
\end{figure}

Fig. \ref{orbit2_visc} shows the orbital semi-major axis  $a(t)$
for viscosities $\nu_0=0$ to $\nu_0=1$.  As the orbit is very
nearly circular, $a(t)$ is always close to the instantaneous orbital
radius $r_p(t)$. A case with $\nu_0=10^{-3}$ was considered, which
showed almost identical $a(t)$ curve to $\nu_0=0$. The numerical 
viscosity is thus between $10^{-8}\lesssim \nu 
\lesssim 10^{-6}$ (from test simulations, see also Fig. \ref{orbit2_visc}). 
It is expected to be much smaller than
the typically adopted physical viscosity of $\nu=10^{-5}$ in
disc-planet simulations.

\nomenclature[b]{$a$}{Semi-major axis}
\nomenclature[b]{$e$}{Eccentricity}

With the standard viscosity $\nu_0=1$, $a$ is halved in less than
$100P_0$, implying classic type III migration \citep{papaloizou06}.
Comparing different $\nu_0$, $a(t)$ is indistinguishable for 
$0<t\lesssim15P_0$, since viscous timescales are much longer than
the  orbital timescale.  At $t=20P_0$, $|\adot|$ increases with
$\nu_0$.  It has been shown that in the limit $\nu\to 0$, the
horse-shoe drag on a planet on fixed orbit  is  $\propto \nu$
\citep{balmforth01,masset02}. However, in this case $\adot\neq 0$ so
there  is a much  larger  rate-dependent  torque responsible for type
III migration, for which explicit dependence on viscosity has not been
demonstrated analytically.  

Migration initially accelerates inwards
($\adot,\,\ddot{a}<0$) and subsequently  slows  down at $r \sim 1.4$
(independent of $\nu_0$). For $\nu_0=1.0,\,0.5$ migration proceeds
smoothly, decelerating towards the end of the simulation
at which point $a$ has decreased by a factor of $\sim 2.7$.  
Migration curves  for  $\nu_0=1.0$ and $\nu_0=0.5$ are quantitatively similar. 
Lowering  $\nu_0$ further  enhances the deceleration at $r \sim 1.4$
until in the inviscid limit  the  migration stalls before eventually
restarting. 

Despite  
differences in detail, the overall extent of the
orbital decay in all of these cases  is similar. This is
expected in the model of  type III migration where the torque is 
due to circulating fluid material switching from $r_p-x_s\to
r_p+x_s$. 
In this model the extent of the  orbital decay should not depend on the nature of the
flow across $r_p$, but only on the amount of disc material
participating in the interaction, or equivalently the disc mass
and this does not depend on $\nu$.
On the other hand, the flow may not be a smooth function
of time with migration proceeding  through  a series  of  fast and slow episodes
as observed in Fig. \ref{orbit2_visc}. 

The interpretation above is only valid if migration proceeds via the type
III mechanism. As indicated by Eq. \ref{delta_m}, the torque depends
on the contrast between the co-orbital region and the flow just
outside. If this contrast is small, e.g. due to viscous diffusion when
large $\nu$ is employed, then the type III mechanism cannot operate at
all. 


\subsection{Stalling of type III migration }
\begin{figure}[!ht]
 \centering
 \includegraphics[width=1.0\linewidth]{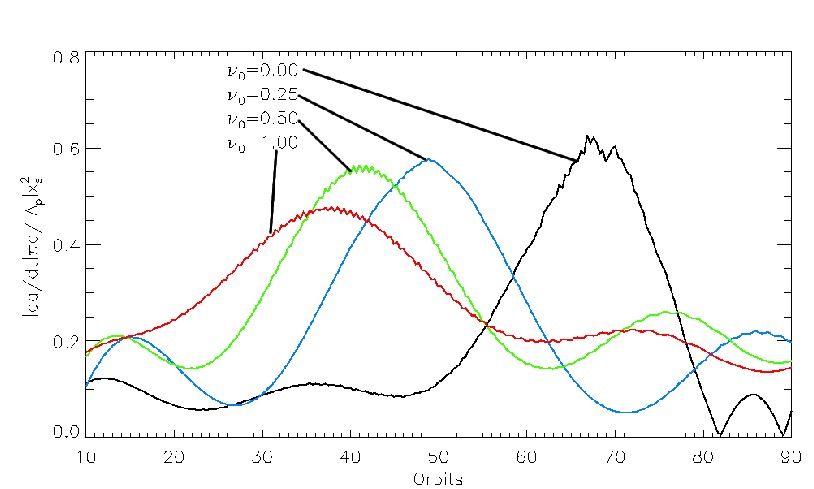}
 \caption{Evolution of $\chi$, the ratio of libration-to-migration
   timescale as a function of viscosity. Classic, smooth type III
   migration can be analytically described in the $\chi\ll 1$ regime.    
   Libration time is measured at
   $r_p-r=x_s$ and migration timescale is that across $x_s$, with
   $x_s = 2.5r_h$.    
   \label{timescales_plot}}
\end{figure}

Classic type III migration is fast because it is
self-sustaining \citep{masset03}. The issue discussed here  is what
inhibits the growth of $|\adot|$ seen in Fig. \ref{orbit2_visc}? 
Descriptions of type III migration usually assume that  the
libration time at $r_p-x_s$ is much less than the time to migrate
across the co-orbital region \citep{masset03}. This implies that 
\begin{align}\label{slow_migration}  
  \chi\equiv\frac{|\adot|\pi a}{|A_p|x_s^2}\ll 1,
\end{align}
where $A_p={1/2}({\p\Omega}/{\p r})$ at $r_p(t)$ ($\simeq a(t)$).    
\cite{papaloizou06}
present a similar critical rate with the same dependence on
$\Omega,\,a,\,x_s$. If Eq. \ref{slow_migration} holds, co-orbital
material is trapped in libration on horseshoe orbits and migrates 
with the planet. When $\chi\gtrsim 1$ the horseshoe  region shrinks to
a tadpole, and material is trapped in libration about the L4 and L5
Lagrange points \citep[as observed by][]{peplinski08b}. This can tend
to remove the co-orbital mass deficit $\delta m$ 
\footnote{The process of shrinking from a partial gap that extends
  nearly the whole azimuth to one with a smaller azimuthal extent, can
  be regarded as gap filling.} which reduces the migration  torque.
Comparing $\chi$ for cases shown in Fig. \ref{timescales_plot}, it is
clear that  migration with $\chi \ll 1$ 
does not always hold, with
$\mathrm{max}(\chi)\sim0.6$ being 
comparable for different $\nu_0.$ 

\begin{figure}[!ht]
  \center
  \includegraphics[scale=.55,clip=true,trim=0cm 0.8cm 3.3cm
  0cm]{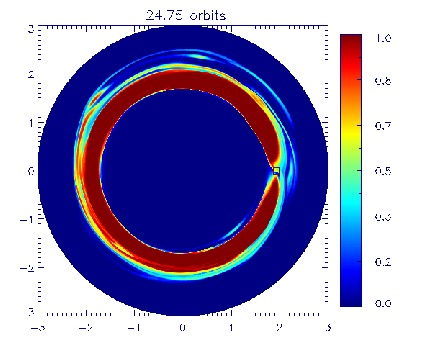}
  \includegraphics[scale=.55,clip=true,trim=1.22cm 0.8cm 0cm 0cm ]{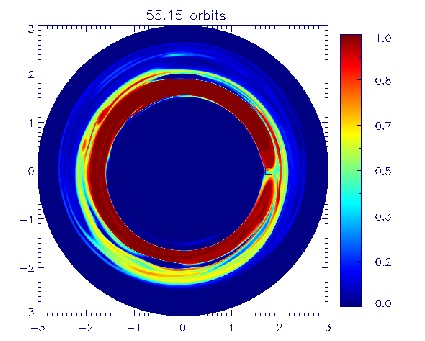}\\
  \includegraphics[scale=.55,clip=true,trim=0cm 0cm 3.3cm
  0cm]{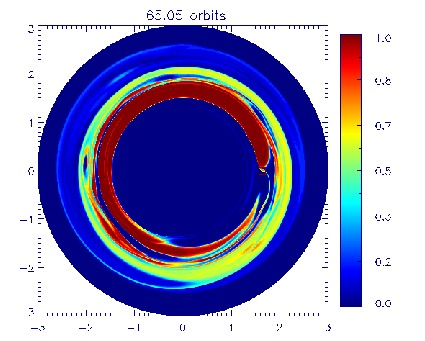}
  \includegraphics[scale=.55,clip=true,trim=1.22cm 0.0cm 0cm 0cm]{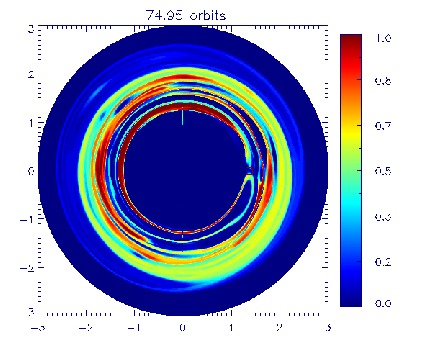}
  \caption{Loss of horseshoe (co-orbital) material during fast
    migration. Snapshots are taken from the inviscid case
    ($\nu_0=0)$. The colour bar shows the surface density of a tracer
    fluid (on a linear
    scale), initially  placed in the co-orbital region 
    of the planet.\label{lostHS}}  
\end{figure}

By following the evolution of a 
passive scalar in Fig. \ref{lostHS}, it can be seen that horseshoe
material indeed no longer migrates with the planet when $|\adot|$ is
large.  This occurs for all $\nu$ but only the low viscosity cases
exhibit stalling. Hence, while horseshoe material is lost due to 
fast migration, it is not responsible for stopping it. 
By examining the inviscid case later,  
the stopping  of migration will be shown to be due to the flow of a
vortex across the  co-orbital region, where some of it becomes
trapped in libration.  

\subsection{The connection between vortensity and fast migration} 
The difference between the value
of the inverse vortensity  $\Sigma/\omega$ evaluated
in the co-orbital region and the value  associated with material that
passes 
from one side of the co-orbital region to the other, defines the co-orbital mass
deficit $\delta m$ in \cite{masset03}. 
It is often assumed that the vorticity is slowly varying so that
the difference in the values of inverse vortensity  reduces, to within a scaling
factor, simply to the difference in the values of the surface density (Eq. \ref{delta_m_dens}). 

\begin{figure}[!ht]
  \center
  \subfigure[$t=10.25P_0$\label{visc1d_evolA}]{\includegraphics[width=0.5\linewidth,clip=true,trim=0cm 0cm 0cm .99cm]{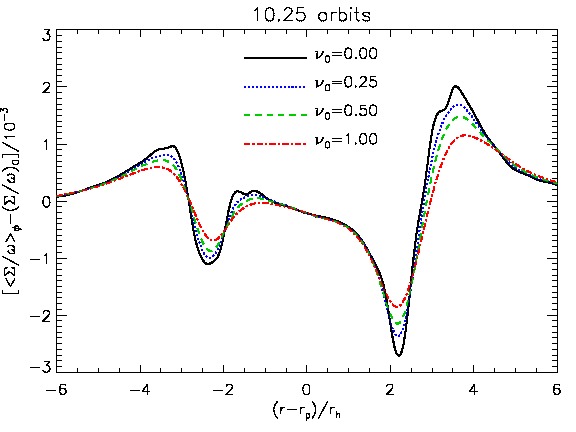}}\subfigure[$t=50.20P_0$\label{visc1d_evolB}]{\includegraphics[width=0.5\linewidth,clip=true,trim=0cm 0cm 0cm .99cm]{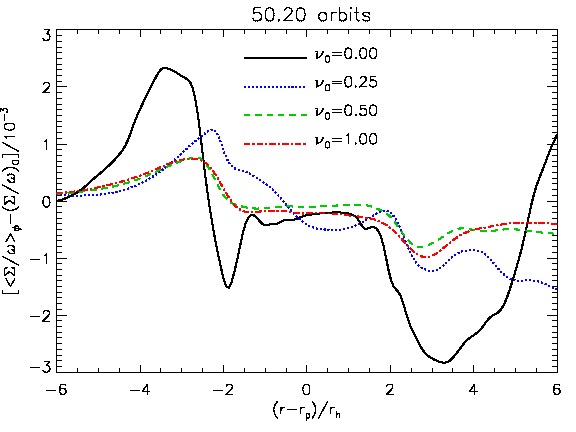}}\\  
  \subfigure[$t=65.41P_0$\label{visc1d_evolC}]{\includegraphics[width=0.5\linewidth,clip=true,trim=0cm 0cm 0cm .99cm]{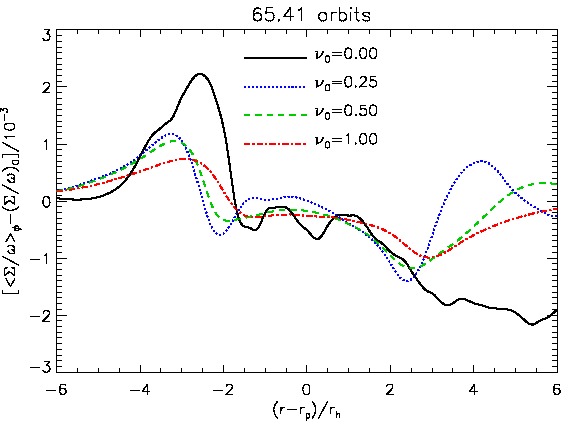}}\subfigure[$t=170.77P_0$\label{visc1d_evolD}]{\includegraphics[width=0.5\linewidth,clip=true,trim=0cm 0cm 0cm .99cm]{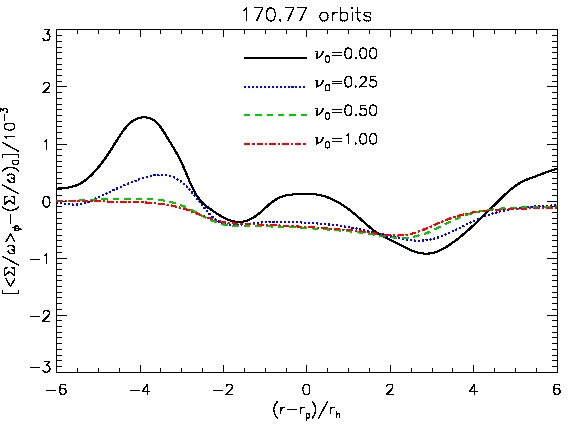}}   
\caption{Azimuthally averaged $\Sigma/\omega$-perturbation for
  different viscosities for the case $\Sigma_0 = 7$ and
  $M_p=2.8\times10^{-4}M_*$. The co-orbital disc structure in terms of
  $\Sigma/\omega$ strongly correlates to migration behaviour.
  \label{visc1d_evol}} 
\end{figure}

Although  \citet {masset03} assumed
steady, slow migration in the low viscosity limit,  
 which clearly does
not hold when the disc is unstable, it is nevertheless
useful to examine the evolution of $\Sigma/\omega$ in relation to the
migration of the planet. 
Fig. \ref{visc1d_evol} shows the azimuthally averaged
$\Sigma/\omega$-perturbation following planet migration. Introducing
the planet modifies the co-orbital structure on orbital
timescales. Vortensity rings develop
at $r-r_p\simeq\pm 2r_h$ within $t\lesssim10P_0$
(Fig. \ref{visc1d_evolA}), as modelled previously. Note that
 vortensity rings were not accounted in \citeauthor{masset03}'s model. 

Increasing $\nu$ reduces the rings' amplitude, but their  locations
are unaffected. Taking the length-scale of interest as $l=r_h\simeq
0.1$, for $\nu_0=1$ the viscous timescale is $t_d=l^2/\nu\simeq
56P_0$. Hence at $t\lesssim10$ viscous diffusion is not significant
even locally. Thus ring-formation is not sensitive to the value of
$\nu$ within the range concerned.  Note the
correspondence between the similarity of  the $\Sigma/\omega$ profiles 
and similarity in $a(t)$ for different $\nu$ in the initial
phase. That  is, the co-orbital disc structure strongly affects
migration \citep{masset03}.   
Dependence on the value of the viscosity  is seen  beyond  $t=50P_0$
(Fig. \ref{visc1d_evolB}), producing much smoother (and similar)
profiles for $\nu_0=0.5$ and  $\nu =1.0.$ 



In Fig. \ref{visc1d_evolB} only the inviscid case
is still in slow migration, and only this disc retains the
inner ring (low $\Sigma/\omega$). This  suggests that the
inner vortensity ring inhibits inward migration.  In terms of $\delta
m$, for $\nu\neq0$ the planet resides in a gap (co-orbital
$\Sigma/\omega$ is less than that at the inner separatrix, or $\delta
m>0$) whereas in the inviscid case $\delta m\sim 0$.

Consider the
$\nu=0$ case. The outer vortensity ring has widened to $\sim 2r_h$ (c.f.
Fig. \ref{visc1d_evolA}). It is centred at 
$3r_h$ so that  co-orbital dynamics may not account for it. However,
inward migration implies a flow of material across $r_p$ from the interior region.
 The increased region of
low $\Sigma/\omega$ exterior to the planet may be due to this flow.
 Notice the high-$\Sigma/\omega$ ring at $+4r_h$ in
Fig. \ref{visc1d_evolA} is no longer present in
Fig. \ref{visc1d_evolB} because this ring is not co-orbital and
therefore does not migrate with the planet.

At $t=65P_0$ (Fig. \ref{visc1d_evolC}) the $\nu_0=0.5$ and $\nu=1.0$
cases continue smooth migration (Fig. \ref{orbit2_visc}) with
qualitatively unchanged profiles. For $\nu_0=0.25$, characteristic
vortensity double-rings re-develop after a stalling event at
$t\sim60P_0$ (Fig. \ref{orbit2_visc}). The peaks and troughs of
$\Sigma/\omega$ recover forms that are close to those 
in the initial phase (Fig. \ref{visc1d_evolA}). At this time the
inviscid case is in rapid migration, and there is a large
$\Sigma/\omega$ just interior to the planet, i.e. a positive co-orbital mass deficit.

Fig. \ref{visc1d_evolD} shows  the final
$\Sigma/\omega$-perturbation profiles . Viscous cases are again in slow
migration, and have  much  smoother profiles. 
This can be due to
diffusion (since this is the late stage of evolution) and/or migration 
across the background,  in both situations there is little disc
material carried by the planet. 
The inviscid case is also in slow 
migration but retains the double-ring structure. This indicates two
possible co-orbital configurations which slow down type III
migration.

Fig. \ref{visc1d_evol} shows a connection between
vortensity evolution and the state of migration.
A correlation between migration (fast, slow) and
co-orbital structure is apparent. The effective action of viscosity
appears to be through its  modification of  the disc structure, rather
than  associated viscous torques  acting on the co-orbital region.


\subsection{Evolution of the co-orbital region}
\begin{landscape}
\begin{figure}
  \center
  \subfigure[$\nu_0=0.5$\label{nu0d5}]{\includegraphics[width=0.49\linewidth]{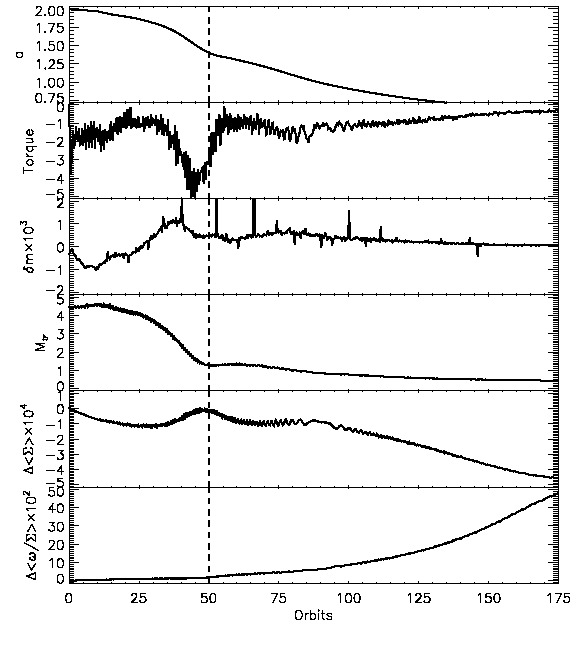}} 
  \subfigure[$\nu_0=0$\label{nu0}]{\includegraphics[width=.49\linewidth]{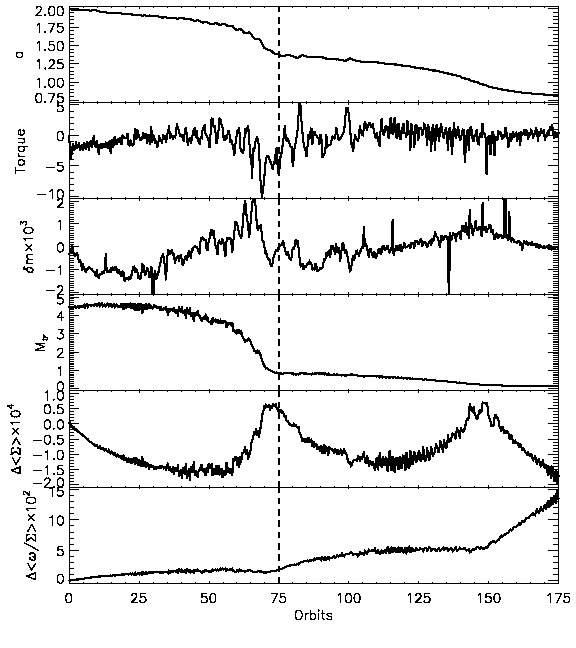}}
  \caption{Effect of viscosity  on the evolution of co-orbital disc
    properties of a migrating planet. 
    The time evolution is illustrated for the model with $\nu_0=0.5$ (left panel)
    and for the inviscid model (right panel).  From top to bottom of
    each panel the evolution of the semi-major axis, $a(t),$ the
    co-orbital mass deficit $\delta m$, the disc-planet torque, the tracer mass 
    $M_\mathrm{tr}$,  the surface density $\langle\Sigma\rangle$ and the vortensity
    $\langle \omega/\Sigma \rangle$ are plotted. Angle brackets denote space
    averaging over the annulus $[r_p- 2.5r_h,r_p+2.5r_h]$ and $\Delta$
    denotes perturbation relative to $t=0$. The vertical line
    indicates stalling of migration. 
    \label{coorb_plot}}  
\end{figure}
\end{landscape}

Fig. \ref{coorb_plot} illustrates the  evolution of the following
quantities associated with the co-orbital region of the migrating planet 
\begin{enumerate}
\item The  gravitational torque acting on the planet due to fluid with  $|r-r_p| \le 2.5r_h$,
  excluding fluid within $r_h$ of the planet. This includes co-orbital material and
  orbit-crossing fluid, the latter being responsible for the type III 
  torque.  
\item The  co-orbital mass deficit $\delta m$ is computed from
  azimuthally averaged, one-dimensional disc profiles.  The
  separatrices are taken to be at $|r-r_p| = 2.5r_h.$    
\item $M_\mathrm{tr}$, the mass of a passive scalar initially placed
  such that  $|r - r_p|= 2 r_h$ \footnote{Due to the uncertainty in the horseshoe half-width
  $x_s$, the tracer was initially placed well within the region
  defined by $x_s=2.5r_h$ to ensure it  is
  co-orbital.}. Note that  $M_\mathrm{tr}=\mathrm{const.}$ if it migrates with
  the planet.
\item The average density $\langle\Sigma\rangle$ and vortensity
  $\langle\omega/\Sigma\rangle$ of the region $|r-r_p| < 2.5r_h.$
\end{enumerate}

The evolution of viscous and inviscid cases  is  qualitatively
similar up to the  stalling indicated by vertical lines in
Fig. \ref{coorb_plot}. Just prior to this 
there is rapid migration associated with a large negative torque,  
which was found to originate from
material crossing the planet's orbit. Co-orbital torques
are oscillatory and the period (amplitude) is longer (larger) for  $\nu_0=0$
than for $\nu_0=0.5.$  During rapid migration phases, the inviscid torque is twice
as negative than in the viscous case. While the torque in a viscous disc
remains negative, torques in an inviscid disc can be positive
due to the formation of  large-scale vortices.  Note that the torque
does not originate from within the Hill sphere since it is excluded
from the summation.

Migration is slow until
sufficient difference builds up between co-orbital and circulating
flow at which point  
there is  a sudden flow-through the co-orbital region. At the same time
there is significant loss of the original horse-shoe material
($M_\mathrm{tr}$ decreases by $\sim 80\%$). The flow-through is
reflected in $\nu_0=0.5$
($\nu_0=0$) by a $\sim 17\%$ (36\%) increase in
$\langle\Sigma\rangle$ from $t=25P_0\to50P_0$ ($t=50P_0\to75P_0$). In
the case of zero viscosity, 
this material is in the form of a high surface density vortex.
Notice in
Fig. \ref{nu0}  that there is a repeated episode where there is a fall
followed by a rise  in
$\langle\Sigma\rangle.$ However,  in the late stages of 
 $\nu_0 =0.5,$
$\langle\Sigma\rangle$ decreases monotonically and the
migration is slower.


The co-orbital mass deficit is initially negative\footnote{
Ring structure in the vicinity of the separatrix means that the sign of $\delta 
m$ is sensitive to the adopted value of $x_s.$  Here, 
$\delta m$ is regarded as a representation of gap depth, so
$x_s=2.5r_h$ is fixed.}   
as the rings develop. It subsequently  increases resulting in
the onset of  type III migration  and is most positive during  the
following rapid migration phase, with  peak values $\delta m\times10^3
\sim1,\,2$ for 
$\nu_0=0.5,\,0$ respectively. This is a few times larger than the
planet mass. $\delta m $ then falls
to $+0.5\times 10^{-3}$ in the viscous case but to $\lesssim 0$ 
without viscosity. In the inviscid case, material 
flowing into the co-orbital region  removes the co-orbital mass
deficit and  
type III is  suppressed; migration experiences a more abrupt
stall. $\delta m$  increases 
again for $\nu_0=0$ while it remains approximately constant for
$\nu_0=0.5$ and decreases towards the end. Type
III migration can re-start in the inviscid disc but  such  behaviour
is  not observed  for large viscosity. 
Type III is  not operating in the late stages
of  the viscous case, in contrast to the inviscid evolution where
fast type III 
migration  is recurrent, faster
than in the cases with applied viscosity,
and is associated with large values of $\delta m$. 

The above discussion shows that the magnitude of the
applied  viscosity  is  significant in
determining the character of the migration.
This is because the form of the flow through the co-orbital
region is sensitive to the choice of viscosity.  In particular, for
sufficiently high viscosity,  this flow is smooth and there is less
disruption of the co-orbital region.   

\section{The inviscid disc and parameter studies}\label{inviscid}
In the inviscid disc, the development of 
vortices significantly affect migration. 
 Three phases  of migration are apparent from  Fig. \ref{orbit2_visc}:
a) $t\lesssim 45P_0$ (slow migration); b) $60P_0 \lesssim t\lesssim
70P_0$ (rapid migration, or a jump in orbital radius); c)  $t\sim
75P_0$ 
(stalling) and d) $75P_0\lesssim t\lesssim 
110P_0$ (second phase of slow migration).  
Typical migration rates at various times are
$\adot(25)\sim-3\times10^{-3}$; $\adot(65)\sim -2\times 10^{-2}$
and $\adot(85)\sim -2\times 10^{-3}$.
The vortex-induced rapid migration (phase b) 
is almost an order of magnitude faster than
the phases a) and d). Hence the latter is referred to as as slow
migration, but more accurately they are migration during gap
formation. They are not necessarily slow in comparison to type I or
type II migration. 

\begin{figure}
  \center
  \includegraphics[scale=.53,clip=true,trim=0cm 0.84cm 1.04cm
  0cm]{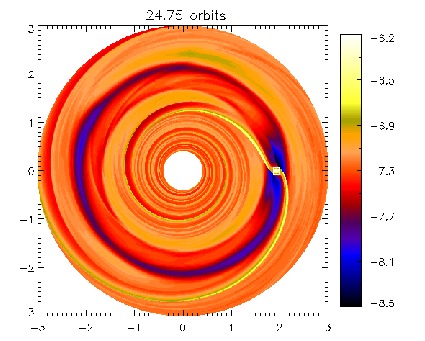}\includegraphics[scale=.53, clip=true, trim=1.22cm 0.84cm 1.04cm 0cm]{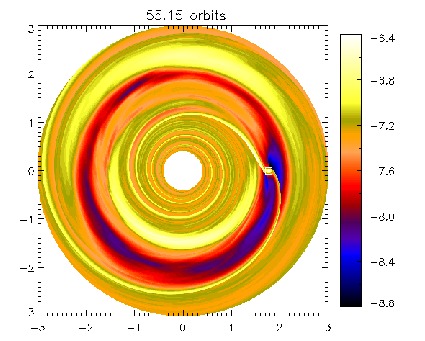}\\  
  \includegraphics[scale=.53,clip=true,trim=0.0cm 0.0cm 1.04cm
  0cm]{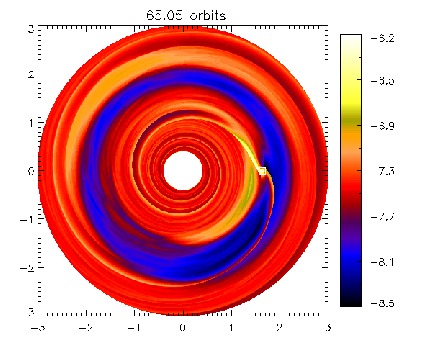}\includegraphics[scale=.53, clip=true, trim=1.22cm 0.0cm 1.04cm 0cm]{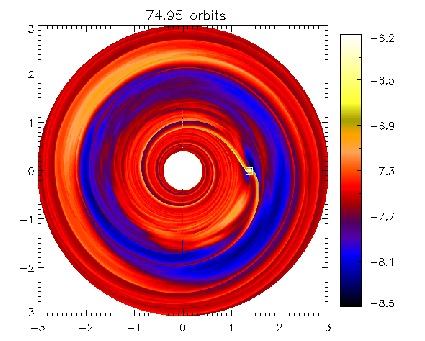}
  \caption{ The evolution of $\ln{\Sigma}$ from early slow
    migration to stalling  ($\sim 75$ orbits). The disc is inviscid. 
    \label{overview_density}}
\end{figure}

Different  migration phases correspond to different disc
structures. Fig. \ref{overview_density} shows the global
surface density evolution. 
At $t=24.75P_0,$ during the  slow migration (phase a), the planet resides in a
partial gap ($|r-r_p|\lesssim 2.3r_h$) with surface density $\sim 20\%$
lower than the uniform background, $\Sigma_0.$ 
A partial gap is necessary for the  type III  migration 
mode \citep{papaloizou06b} but not sufficient. 
A surface density asymmetry ahead/behind the 
planet is needed to provide the net co-orbital torque \citep{artymowicz04}.
At this stage, non-axisymmetry in the gap is too weak. 
Further azimuthal density asymmetry  has developed  in the gap by
$t=55P_0$ and there  is a factor of $\sim 3$   variation in the  gap
surface density, but the asymmetry is still limited.

Strong asymmetry can be provided  by large-scale vortices near the gap edges 
of azimuthal extent $\Delta \phi\sim\pi$ (Fig. \ref{overview_density}). 
The outer (inner) vortex rotates clockwise (counter-clockwise)
relative to the planet so they are not co-orbital.  Their origin was explained
as a natural consequence of the instability of the 
vortensity rings (see \S\ref{Dynstab} above). Their occurrence near gap
edges has a strong influence on the type III  migration mode.

At $t=65.05P_0$  Fig. \ref{overview_density} shows that 
the planet is just inside the inner gap edge. 
The surface  density contrast in the neighbourhood of  the planet 
is largest as the  inner vortex enters the  co-orbital region from behind
the planet. 
It exerts a net negative torque  as the material crosses  the 
planet orbit and enters (is scattered into)  the exterior disc, and
this snapshot corresponds to  fast migration (phase b). At $t=75P_0$ the
migration stalls and the planet  
no longer resides inside a gap. 
The planet has effectively left its gap by scattering vortex material outwards. 
This completes a single vortex-planet episode. During this time, the
outer vortex simply circulates around the original outer gap edge 
and does not influence the  co-orbital dynamics, though
it contributes  an oscillatory torque on the planet.

\begin{figure}
  \centering
  \includegraphics[scale=.435,clip=true,trim=0cm 0cm 3.1cm
  0.0cm]{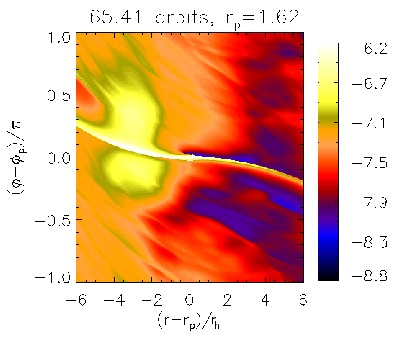}\includegraphics[scale=.43,clip=true,trim=1.4cm 0cm 3.1cm
  0cm]{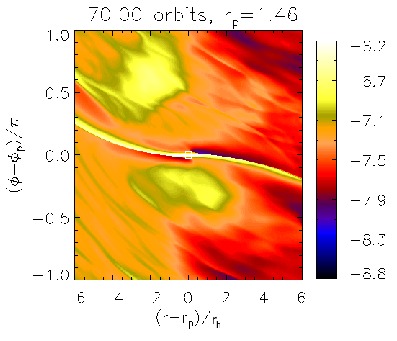}\includegraphics[scale=.43,clip=true,trim=1.4cm 0cm 0.0cm
  0cm]{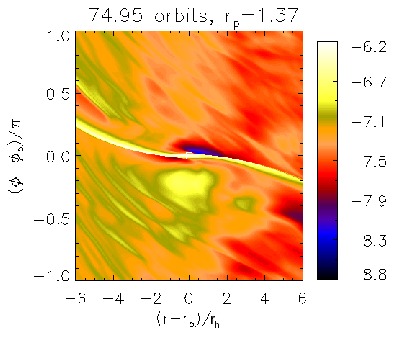}
  \caption{ Illustration of the surface density evolution
    from the start of  rapid migration to
    stalling at $t=75P_0.$ Maps of $\ln{\Sigma}$ are plotted.
    \label{stop_2d}}
\end{figure}

The vortex-planet interaction is  magnified 
in Fig. \ref{stop_2d}. At $t=65P_0$ the vortex
circulates at $\sim r_p - 3r_h$ and has radial extent $\sim 3r_h$. 
The gap depth is largest and  the migration is fast. 
As the planet migrates  inwards, the vortex splits with some 
material entering  the co-orbital
region while the rest continues  to circulate ($t=70P_0$).
Vortex material becomes trapped just behind the planet at
$t=75P_0$. (For large viscosity cases, no trapping is seen.) 
The flow of vortex material into the co-orbital region is gap filling,
thereby lowering the co-orbital mass deficit, discouraging type III
migration. Part of the original horseshoe material is replaced with  
vortex material. Another view is that rapid migration in the
first place was provided by the vortex just interior to the co-orbital
region. When this vortex is no longer there, rapid migration cannot be
sustained. 


\subsection{Evolution of inverse vortensity}

\begin{figure}
  \centering
  \includegraphics[scale=.55,clip=true,trim=0cm 0.84cm 3.25cm
  0cm]{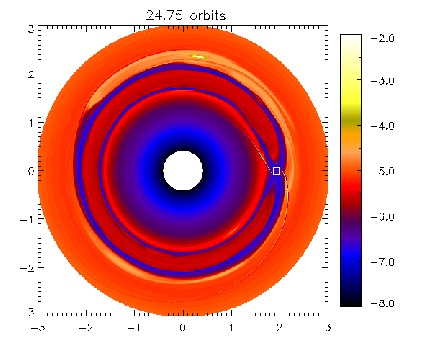}\includegraphics[scale=.55,clip=true,trim=1.22cm 0.84cm 0.99cm 0.cm]{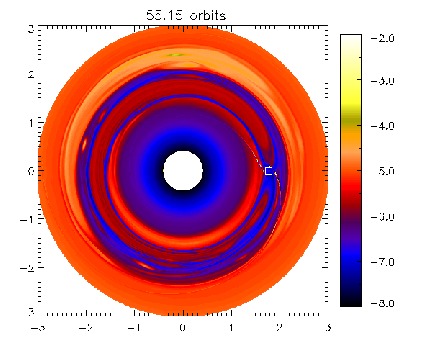}\\
  \includegraphics[scale=.55,clip=true,trim=0cm 0.84cm 3.25cm
  0cm]{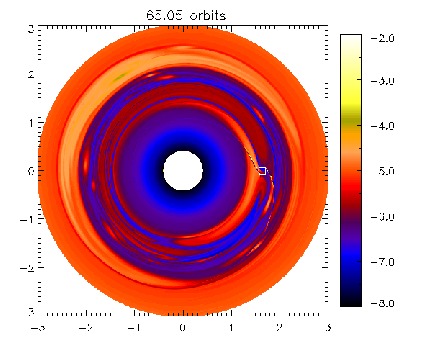}\includegraphics[scale=.55,clip=true,trim=1.22cm 0.84cm 0.99cm 0.cm]{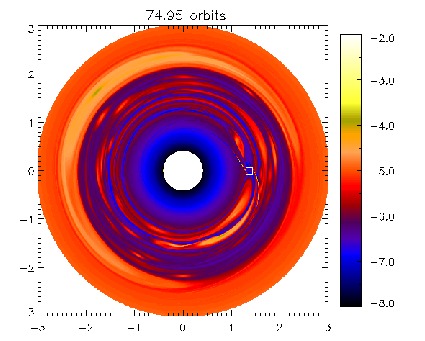}\\
  \includegraphics[scale=.55,clip=true,trim=0cm 0cm 3.25cm
  0cm]{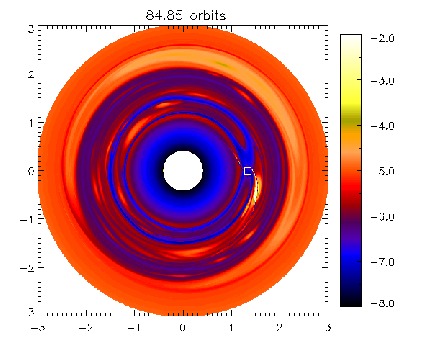}\includegraphics[scale=.55,clip=true,trim=1.22cm 0.cm 0.99cm 0.cm]{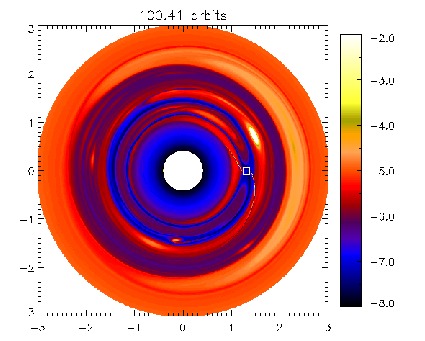}  
  \caption{Maps of the
    inverse vortensity  $\ln{(\Sigma/\omega)}$ spanning the
    time interval   from early slow
    migration to  stalling ($\sim 75$ orbits), then to the  second phase  of slow
    migration. The passage of the inner vortex material to the outer
    disc leads to formation of smaller vortices around the outer gap
    edge (e.g. at $t=75P_0$). \label{invort_evol}}
\end{figure}

It is useful to examine the evolution of $\Sigma/\omega$, or inverse vortensity, since it defines 
co-orbital mass deficit $\delta m$ that drives the  type III torque. 
This is illustrated in Fig. \ref{invort_evol}.
In inviscid discs $\Sigma/\omega$ is 
approximately conserved following a fluid element/streamline so it also tracks material.
 By $t=55P_0$ the inner vortex has formed via
nonlinear evolution of the gap edge instability (\S\ref{Dynstab}) 
and begins to interact with the 
co-orbital region. Vortensity conservation implies that the small
vortices (red blobs) near the outer ring were part of the inner
vortex, consistent with inward type III migration (see
Fig. \ref{invort_evol}). Each time this vortex passes by the planet,
some of its material breaks off and crosses 
the planet orbit from behind, thereby
exerting a negative torque. Rapid migration at $t=65P_0$ 
occurs when the bulk vortex body flows across  the 
co-orbital region. The inner ring is 
disrupted and no longer extends $2\pi$ in azimuth. 

This contrasts with  the usual type III scenario where material simply 
transfers from inner to outer disc leaving the co-orbital region unaffected. 
In the inviscid disc, disruption is \emph{necessary} due
to the existence of vortensity rings of much higher vortensity than the vortex. 
Having type III migration while maintaining vortensity conservation means
that vortex material must 
cross the  planet orbit without changing its vortensity. 
This would not be possible if a  ring structure is maintained. 
In this sense, the high vortensity rings oppose the type III mode. 
Hence, migration is slow until significant ring disruption occurs 
that is associated with the vortex flowing  across.

When migration stalls at $t=75P_0$, Fig. \ref{invort_evol} shows
that the vortensity rings are much less
pronounced compared to the initial phase. 
The vortex splits into several smaller patches circulating in 
the original gap. 
At the planet's new
radius, material of high $\Sigma/\omega$ fills the new co-orbital region, 
corresponding to lower co-orbital mass deficit. 
However, by $t=85P_0$ new vortensity rings are setup   near the new 
orbital radius and are qualitatively similar to those present in
the first ring basic state. 
During the build up to the  second rapid migration phase
, there is a single vortex associated with inner gap edge ($t=100P_0$). 
This  simply leads to  a repeat of the first rapid migration episode.
However, the presence of a vortex inside the co-orbital region, from 
the first rapid migration phase lowers the co-orbital mass deficit 
(Fig. \ref{nu0}) and hence reduces the migration rate for the second rapid
phase (at $t\sim 140P_0$).

\subsection{Effect of surface density scale}\label{varmass}
By changing the parameter $\Sigma_0$, and therefore the disc mass, the 
disc-on-planet torque is scaled accordingly. In the case of
vortex-induced migration, changing $\Sigma_0$ directly affects the
mass contained in a vortex. Here, the inviscid simulations above are
repeated with different surface density scales. Simulations here had a
resolution $N_r\times N_\phi = 512\times1536$.    

\begin{figure}[!ht]
  \center
  \includegraphics[width=\linewidth]{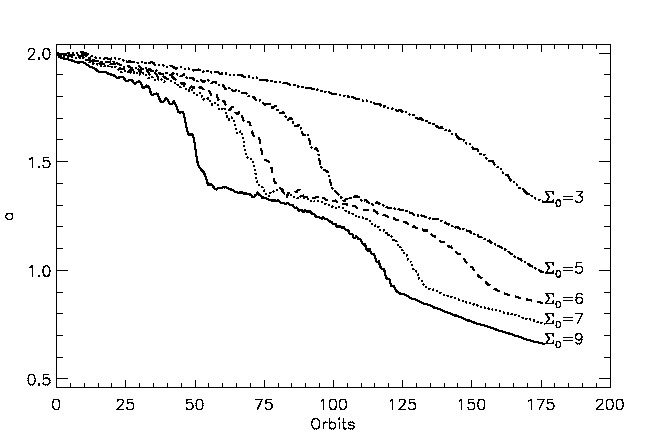}
  \caption{Vortex-induced migration in inviscid discs with 
    different surface density scales. The extent of 
    rapid migration is independent of the  initial surface density
    scaling. The fiducial case is
    $\Sigma_0=7$. \label{orbit2_varmass}} 
\end{figure}

Fig. \ref{orbit2_varmass}
shows migration as a function of $\Sigma_0$. 
For $\Sigma_0=5$---9
the planet migrates by the same amount during the first rapid migration 
phase and stalls at the same radius \footnote{This contrasts with the
  classic smooth migration scenarios. For example, in type I
  migration the disc-planet torque scales with surface density so in a
  given time interval, planets should migrate further in discs with
  higher surface density.}.  
This is also observed for a
second rapid migration phase for $\Sigma_0=6$---9. 
Although the largest $\Sigma_0$ case is unphysical due to the 
lack of inclusion  of disc self-gravity,
results are consistent with the notion that  
rapid migration is initiated by sufficient contrast between co-orbital
(gap) and circulating fluid  (vortex), which can be measured by $\delta m$.  

The jump in orbital radius, should it occur, is
independent of $\Sigma_0$. As the interaction involves the vortex
flowing from the gap edge across the co-orbital region; its change in
specific angular momentum is independent of density, because the
co-orbital region size is fixed by planet mass. The results then
suggest the vortex needs to grow to a mass $M_v$ (or surface density), 
 only dependent
on the planet mass, in order to scatter
the planet. Since the vortex forms at the gap edge, $M_v$ can be
linked to $\delta m$ because the co-orbital mass deficit depends on the
edge surface density.

As the vortex originates from instabilities with growth rate
independent of $\Sigma_0$, increasing $\Sigma_0$ means less time is needed 
for the vortex to build up to critical mass or density (to affect
migration). Hence, 
increasing surface density only shortens the `waiting time' before
rapid migration. However, if the density is too low, e.g.  
$\Sigma_0=3$ then vortex-induced rapid migration may never occur. 

Consider an inner vortex of mass 
$M_v$  formed by instability at $r_v=r_p - b r_h$ ($b>0$) with
width $a r_h$. 
$M_v$ is clearly limited by the amount of material 
that can be gathered into the vortex, so that 
$M_v < 2\pi a r_h r_v \Sigma$. 
Taking $M_v/M_p=3.5$ as critical for rapid migration \footnote{
The vortex mass can be estimated by monitoring the drop in total mass
of an annulus interior to the inner gap edge, during rapid migration.}
, this means 
\begin{align}
  \Sigma_0 > \frac{3.5M_p}{2\pi a f_0(1-b f_0)r_p^2}\times10^{4},
\end{align}
where $f_0=(M_p/3M_*)^{1/3}$.  Taking representative values from
simulations 
of $a=3,\,b=4$ and $r_p\sim2$ gives $\Sigma_0 > 3.5.$ 
For such cases rapid, vortex-induced migration was indeed
observed (Fig.\ref{orbit2_varmass}), but not for $\Sigma_0=3$. 
This is similar to the usual requirement that in order for
type III migration to  
occur for  intermediate planets, the disc should be
sufficiently massive \citep{masset03}.  In the inviscid case the
limitation is specifically due to the maximum possible vortex mass,
which is responsible for most of the migration. 


\subsubsection{Critical co-orbital mass deficit}
\begin{figure}[!ht]
  \center
  \subfigure[$\Sigma_0=9$]{\includegraphics[width=0.5\linewidth]{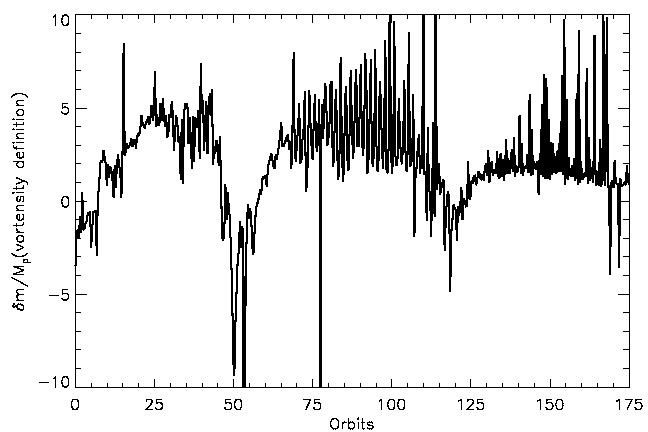}}\subfigure[$\Sigma_0=7$]{\includegraphics[width=0.5\linewidth]{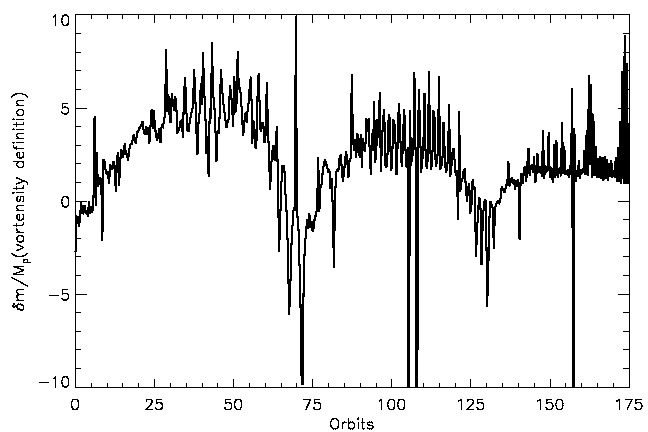}}
  \subfigure[$\Sigma_0=5$]{\includegraphics[width=0.5\linewidth]{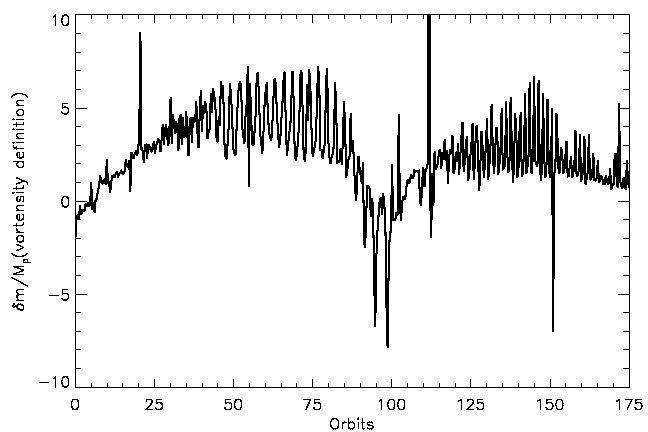}}\subfigure[$\Sigma_0=3$]{\includegraphics[width=0.5\linewidth]{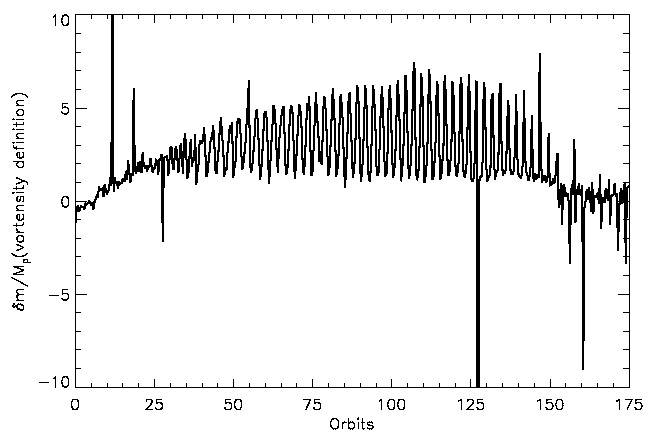}}
  \caption{Evolution of the coorbital mass deficit as defined
    by inverse vortensity. The average
    $\Sigma/\omega$ for  coorbital material  is taken over 
    $r-r_p=[-2.5r_h,0], \phi = [\phi_p,\phi_p-\pi/4]$; and that of circulating
    material is taken over
    $r-r_p=[-6,-2.5]r_h, \phi = [\phi_p,\phi_p-\pi/4]$. Very similar
    behaviour was obtained when
    comparing the co-orbital and circulating surface densities instead
    of the inverse vortensity. \label{deltam}}
\end{figure}

In order to link type III migration and vortex-planet scattering,  
one can measure the co-orbital mass deficit $\delta m$, which amounts
to comparing average inverse vortensity of co-orbital fluid just
behind the planet, to that of circulating fluid just inside the   
inner separatrix but also behind the planet:
\begin{align*}
  \delta m  = 2\pi\sqrt{GM_*} x_s r_p^{-1/2}
  (\langle\Sigma/\omega\rangle_\mathrm{circ} -
  \langle\Sigma/\omega\rangle_\mathrm{coorb}).
\end{align*}
This is a simplified version of the definition in \cite{masset03}, by
inserting Keplerian rotation for the Oort constant in
Eq. \ref{delta_m}. Results are shown in Fig. \ref{deltam}. The
oscillatory nature of $\delta m $ reflects a vortex circulating at   
 the gap edge, $\delta m$ maximising when the vortex is
within the patch of fluid where averaging is done. As it grows, the 
vortex contributes to 
$\langle\Sigma/\omega\rangle_\mathrm{circ}$ hence favouring type III migration.
Cases with rapid migration
share the same evolution of $\delta m$. $\delta m/M_p$ increases up to $\sim
4$--5 before the vortex first induces fast migration. For $\Sigma_0=
3$, which does not show such a  rapid migration phase,
typically $\delta m\lesssim 5M_p$ with smaller amplitude variation.   
During the vortex-planet scattering event, $\delta m$ rapidly decreases
and migration stalls when $\delta m \lesssim 0$. This is because as
the vortex flows across the co-orbital radius, it contributes to
$\langle\Sigma/\omega\rangle_\mathrm{coorb}$, lowering
$\delta m$. 



\subsection{Effect of background profile}

Simulations so far employed a flat initial surface density profile.
In this case the stalling and re-start 
of migration can be expected because the vortex scatters the planet
out of its original gap into a previously unperturbed region. 
If the unperturbed surface density is flat, subsequent migration
is a repeat of that occurred previously.
However, if the unperturbed surface density increases  with decreasing
radius, then the vortex-induced inwards migration moves the planet
into a region of higher surface density. The background
contribution to $\langle\Sigma/\omega\rangle_\mathrm{circ} $ may
outweigh the vortex contribution to
$\langle\Sigma/\omega\rangle_\mathrm{coorb}$ just after the vortex-planet
interaction.  In this case, migration may not stall.

\begin{figure}[!ht]
  \center
  \includegraphics[width=\linewidth]{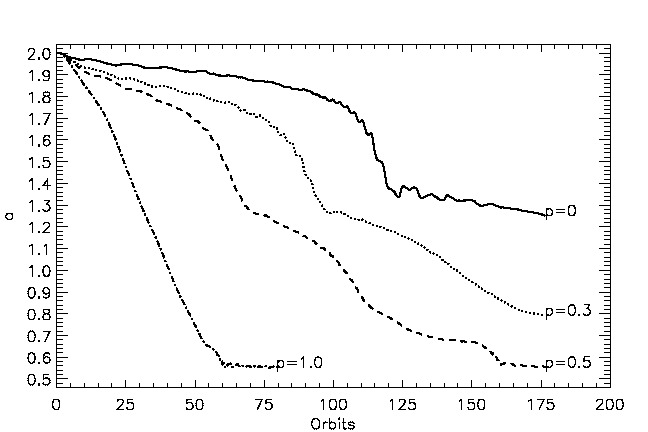}
  \caption{Migration in inviscid discs with variable initial surface
    density slopes ($\Sigma\propto r^{-p}$). 
    \label{orbit2_varp}}
\end{figure}

To investigate this, simulations with initial surface density profiles
\begin{align}
  \Sigma = 5\left(\frac{r}{2}\right)^{-p}\times10^{-4}
\end{align}
were performed with $p=0,\,0.3,\,0.5,\,1.0$. Migration is
compared in Fig. \ref{orbit2_varp}. 
 Note that for $p>0$,  the co-orbital mass
  deficit is positive at $t=0$, and there is migration from
$t=0$. With $p=1$ this initial migration is large and the planet
falls to the inner boundary, having no time to open a gap and allow 
vortex formation. 
For $p=0.5$ there is 
vortex-induced inwards migration at $t\sim 50P_0$.  
The wait until vortex-planet interaction is shorter for $p=0.5$ than
$p=0$. 
 $p=0.5$ does experience deceleration at $t\sim70P_0$ but unlike
$p=0$, does not stall for long. In this case, gap-filling by the
vortex does not effectively stall type III migration.
The shortened duration of stalling with increasing $p$ is consistent
with the discussion above regarding the variation with the surface
density scale (being higher in $r<r_p$ for $p=0.5$ than for $p=0$).   


\subsection{Effect of disc temperature}
Linear calculations showed that the growth rate of unstable
co-rotational (vortex) modes, $|\gamma|$, generally increase when the
aspect-ratio $h$ is lowered. Migrating cases with
$h=0.04$---0.06 are presented here.

\begin{figure}[!ht]
  \center
  \subfigure[h=0.04]{\includegraphics[scale=.5,clip=true,trim=0cm .5cm
    2.44cm 0cm ]{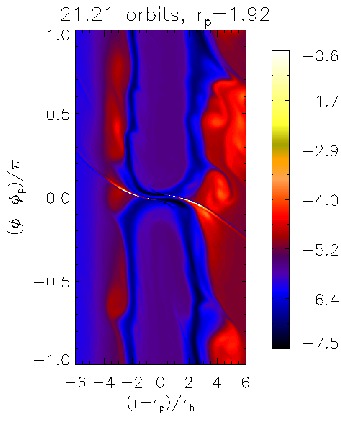}}\subfigure[h=0.05]{\includegraphics[scale=.5,clip=true,trim=1.35cm .5cm
    2.44cm
    0cm]{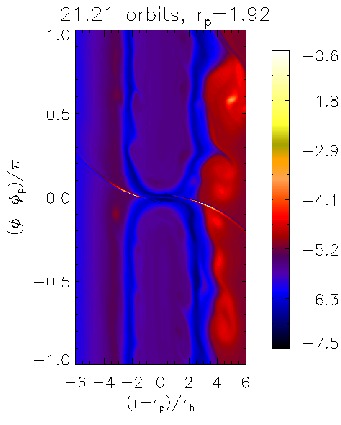}}\subfigure[h=0.06]{\includegraphics[scale=.5,clip=true,trim=1.35cm .5cm  
    0cm 0cm ]{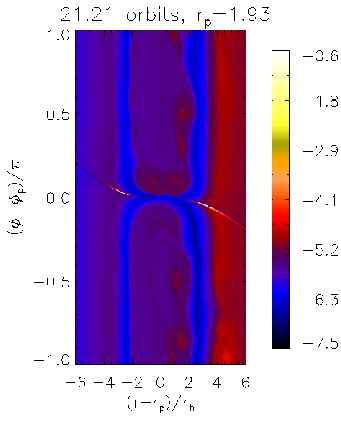}}
  \caption{Distribution of $\ln{(\Sigma/\omega)}$ near the  planet's
    co-orbital region, as a function of disc aspect-ratio $h$. \label{vortxy_varh}}  
\end{figure}

Fig. \ref{vortxy_varh} show snapshots of $\Sigma/\omega$ 
as a function of $h$ and agree with linear theory in that
co-rotational modes dominate the region. Both edges become more
unstable for lower $h$.  
A $m=4$ mode on $h=0.04$ inner edge can be seen, which has peak growth
rate from theory (Fig. \ref{rates}). A weak $m=2$ mode is observed for
$h=0.05$ inner edge, consistent with linear theory peak rates
($m=2$--3 is most unstable for $h=0.05$). As the planet tends to migrate
inwards initially, one expects interaction with inner edge vortices to occur
sooner for lower $h$. 
Disturbances at outer
edges have higher growth rates, and have already undergone merging
into large vortices. 

\begin{figure}[!ht]
  \center
  \includegraphics[width=\linewidth]{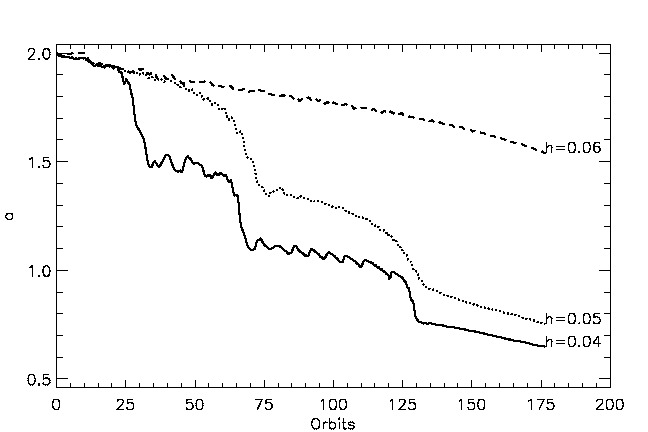}
  \caption{Vortex-induced migration in inviscid discs with different aspect-ratio $h$ but
    same initial uniform density $\Sigma=7\times10^{-4}$.\label{orbit2_varh}}
\end{figure}

Fig. \ref{orbit2_varh} show corresponding migration curves. 
Vortensity rings can act to inhibit type III migration. For example, 
in Fig. \ref{orbit2_varh} the stronger vortensity rings associated
with $h=0.04$ results in almost no migration prior to vortex-induced
migration. 
 In $h=0.04$,  
rapid migration begins at $t=25P_0$, roughly half the waiting time
than $h=0.05$. The extent of orbital decay during rapid migration is
unchanged, consistent with the notion that the vortex implies a
co-orbital mass deficit, for which a critical value reached 
enables rapid migration and determines migration extent.  The critical
value is simply attained sooner with decreasing $h$. 

More rapid migration episodes occur in $h=0.04$ than $h=0.05$, since
$h=0.04$ has higher growth rates so the wait between vortex
interaction is shorter. 
The slow phase of $h=0.04$ is also non-smooth compared to $h=0.05$ due
oscillatory torques exerted on the planet by vortices. This is not
apparent in the migration curve of $h=0.05$ because perturbations have
smaller amplitudes at a given time.  

Notice in $h=0.04$ the waiting times between rapid migration 
lengthens with time. This is probably because
the vortex forming region has width of order $H=hr$ which is decreasing, as
is the co-orbital width  ($x_s\propto r_p)$. The vortex flows across a smaller
region as it scatters the planet, which can account for smaller jumps in radius in
later interactions. 


In $h=0.06$ no rapid migration occurs within simulation
time, in agreement with linear theory where co-rotational modes
for the inner edge were not found,  and with Fig. \ref{vortxy_varh}
where the inner edge remains stable. There is no vortex that
grows to the critical mass to induce rapid migration. This implies
that the disc temperature must be sufficiently low for vortex-planet
interactions to occur.



\subsection{Effect of planet mass: comparison to Jupiter}

Recently, \cite{peplinski08a,peplinski08b,peplinski08c} performed
numerical simulations of type III migration of a Jupiter-mass
planet. Although \citeauthor{peplinski08b} did not include physical
viscosity, their Cartesian grid is more diffusive and vortex-formation
is suppressed \citep{valborro07}. Hence, they
obtained smooth migration curves. For completeness, the migration of
Jupiter in the presence of vortices is presented below. 

\begin{figure}
  \center
  \includegraphics[scale=.55,clip=true,trim=0cm 0cm 2.1cm
  0cm]{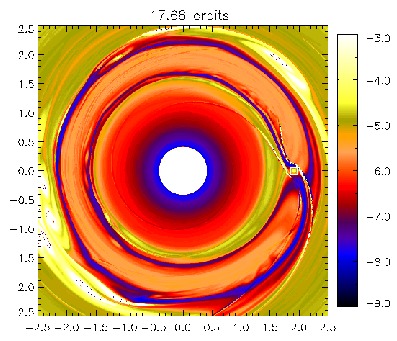}\includegraphics[scale=.55,clip=true,trim=.5cm
  0cm 0cm 0cm]{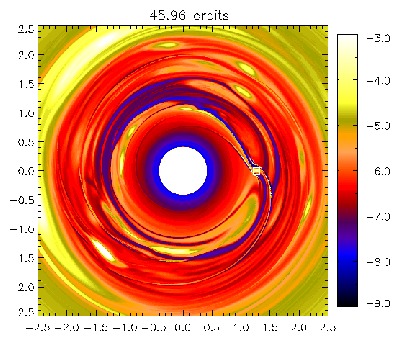}   
  \caption{Vortex-induced migration
    of a Jupiter mass planet in an inviscid disc. Maps of
    $\ln{(\Sigma/\omega)}$ are shown. As with Saturn-mass planets
    , rapid migration is associated with the flow of a
    vortex across the co-orbital region, simultaneously destroying the
    thin vortensity rings. 
    \label{vorten_jup}}
\end{figure}

Fig. \ref{orbit2_jup} 
compares the migration curves $a(t)$ for $M_p=10^{-3}M_*$ in inviscid
and viscous  discs ($\nu_0=1,\,2$) to the previous case of Saturn in a
inviscid disc.  
Vortex-planet 
scattering also occur for Jupiter in the inviscid disc. This
is seen in Fig. \ref{vorten_jup}, where a 
vortex associated  with the inner gap edge builds up and flows across
the co-orbital region.  Jupiter induces stronger shocks and 
higher-amplitude vortensity rings, and are more unstable. The
instability growth timescale is therefore shorter than that of
Saturn, so vortex-induced migration occurs very soon after  
the planet in introduced. 

\begin{figure}[!ht]
  \center
  \includegraphics[width=\linewidth]{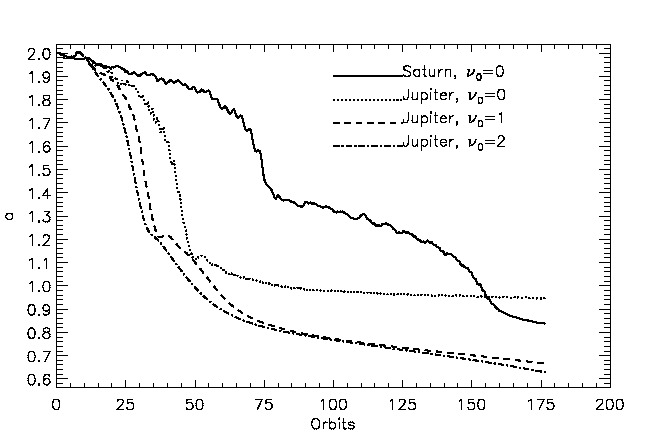}
  \caption{Vortex-induced migration for a Jupiter-mass planet in an
    inviscid disc (dotted) and viscous disc ($\nu_0=1$, dashed;
    $\nu_0=2$, dashed-dot) compared to Saturn-mass planet in a inviscid
    disc (solid).
    \label{orbit2_jup}}
\end{figure}

The $a(t)$ for Jupiter in an inviscid disc
show two migration phases --- fast and slow, and that significant
orbital decay occur within $\sim 50P_0$.  These features were also
observed by \cite{peplinski08b}. However, the fast phase of
\citeauthor{peplinski08b} is almost a linear function of time, 
here migration is clearly accelerating inwards during
$t\lesssim50P_0$. In this case there is an abrupt transition to the
slow phase whereas that of \citeauthor{peplinski08b} is smoother and
no vortices were identified. 

Interestingly, standard viscosity 
$\nu_0=1$ results in a similar behaviour for the fast phase, though
the transition to slow migration is smoother than zero viscosity.
The vortensity distribution for $\nu_0=1$ was found to be much
smoother than $\nu_0=0$, without vortices 
 associated with the inner gap edge. 
This is consistent with \cite{valborro07} who
found that $\nu=O(10^{-5})$ suppress vortex formation. With
$\nu_0=1$, flow-through the co-orbital region is a smoother function
of time.

A case with $\nu_0=2$ is also shown in
Fig. \ref{orbit2_jup}. The orbital decay timescale is again $\sim
50P_0$, consistent with type III migration, but there could be
complications from the fact that a Jovian-mass planet in a viscous
disc can undergo type II migration. The transition from fast to slow
is smoothed (cf. $\nu_0=0,\,1$) and migration is  
 qualitatively closer to \cite{peplinski08b}. These results
suggest that inviscid cases here have lower total viscosity than
those considered by \citeauthor{peplinski08b}, the difference
attributed to different numerical grids employed. 

\section{Additional simulations}\label{addsim}


The simulations above illustrate the physics of vortex-induced
migration. It can be understood in the language of type III
migration. However, its association with 
co-orbital flow  potentially makes it sensitive to numerics near
the planet. 

The locally isothermal equation of state allow 
mass accumulation near the planet, and may lead to 
spurious torques from within the Hill sphere. 
The fiducial simulation above 
was repeated with the equation of state employed by
\cite{peplinski08a}, where the disc temperature increases close to the
planet (Eq. \ref{pepeos}). This reduces mass accumulation onto the planet. 

\begin{figure}[!ht]
  \center
  \includegraphics[width=\linewidth]{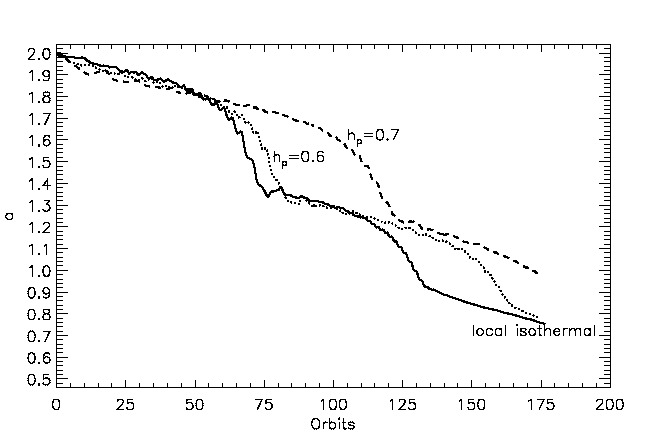}   
  \caption{Vortex-induced migration of a Saturn mass planet in
    inviscid discs with the modified equation of state of
    \cite{peplinski08a}. The parameter $h_p$ is a measure of
    the artificial temperature increase near the planet, relative to
    the locally isothermal equation of state. 
    curves. \label{peplinski_eos}}  
\end{figure}

For the fiducial case, employing this EOS with the
parameter $h_p=0.6$ gives a sound-speed about $18\%$ higher than the
locally isothermal value at the planet's initial position. When
$h_p=0.7$, the increase is about $38\%$. 
Migration curves for these $h_p$ are compared to the locally
isothermal case in Fig. \ref{peplinski_eos}. These again yields 
vortex-planet scattering, which is an indication that significant
torque contribution during the interaction is not sensitive to flow
close to or within the Hill sphere. 


\begin{figure}[!ht]
  \center
  \includegraphics[width=\linewidth]{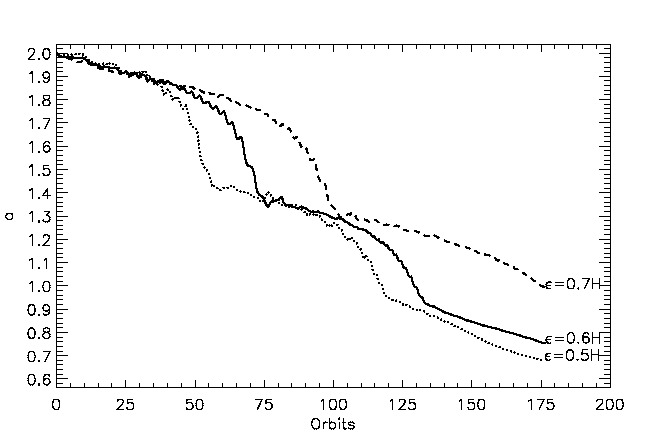}   
  \caption{Vortex-induced migration of a Saturn mass planet in
    inviscid discs with different softening lengths for the planet
    potential.\label{softening}} 
\end{figure}

Finally, there is the issue of softening length. Simulations previous
to this section all used $\epsilon_p=0.6H$. In Fig. \ref{softening},
migrating cases with softening lengths of $0.5H$ and $0.7H$ are
displayed. Both cases display vortex-planet interaction, although at
earlier times when $\epsilon$ is lower. This is expected since 
smaller softening has the same effect as increasing the planet mass,
and thus producing a more unstable disc.  


    \graphicspath{{Conclusions/ConclusionsFigs/EPS/}{Conclusions/ConclusionsFigs/}}

\section{Summary and discussion}\label{conc}
Numerical simulations in this Chapter show that 
migration of a giant planet induced by large-scale 
vortices is a robust phenomenon in low viscosity discs. 
This vortex-planet interaction is best understood as a 
non-smooth form of type III migration. 



Planets more massive than about Saturn induce spiral
 shocks extending into its co-orbital region and 
generate vortensity rings that delineate the gap. 
Local vortensity minima associated with gap edges
are dynamically unstable to non-axisymmetric perturbations.
 Dominant  unstable modes are radially localised. They develop into 
vortices and merge, on dynamical timescales, into a single 
vortex circulating either gap edge. 


The effect of vortices on migration is most significant in low
viscosity discs ($\nu\lesssim 0.25\times10^{-5}$) because the
instability is suppressed at higher viscosity \footnote{In the inviscid
limit only a small numerical viscosity is present. Suppose one
considers a physical viscosity of $\nu=O(10^{-6})$ ,  it is still
large compared to the numerical viscosity, so one expects the latter
to be unimportant. However,  $\nu=O(10^{-6})$ is already unable to
suppress the instability. Results are therefore unaffected by
numerical viscosity of the \fargo code.}.   
The presence of high density vortices at the gap edge produces
nonsmooth migration via vortex-planet scattering 
, with episodes of fast migration where the
planet's orbital radius jumps by a few Hill radii per episode. 
 This is analogous to planet-planet scattering. 
The vortex is also responsible for stalling migration in discs 
with initially flat surface density. In this case there can be 
repeated episodes of vortex-induced
migration. Viscosity smooths the flow across the co-orbital
radius, but has limited effect on the extent of orbital decay via the
type III mechanism.   

The extent of orbital decay in a single episode of
fast migration is independent of surface density scaling. This
suggests a critical vortex mass or surface density is required to interact,
which can be linked to the concept of the co-orbital mass deficit
that drives type III migration. The effect of the background surface
density profile, disc aspect-ratio and planetary mass were
also examined.  The  effect of vortices in these
cases are all consistent with linear analysis of gap stability and
interpretation of vortex-induced migration as a form of type III migration.  


\subsection{Outstanding issues}
The vortex instability requires low viscosity. It is then natural to question their existence as 
long-lived structures in real discs with finite viscosity.  
Recently, \cite{li09}
studied migration in nearly-laminar discs in the type I regime. They
found effects of lowering viscosity, begin to appear for 
$\nu\leq10^{-7}$. 
It is unclear if viscosity in real discs will be low enough for vortices to
develop and affect planets that undergo type I migration. 
For giant planets considered here, non-smooth migration begins 
at $\nu=O(10^{-6})$. 
Vortices can thus interfere with giant planet migration
when viscosity is not more than one order of magnitude smaller than
the standard value $\nu=10^{-5}$. This could be satisfied in `dead
zones' in protoplanetary discs \citep{gammie96}. 

One physical issue is the lack of inclusion of disc self-gravity.  
It may be important when discussing type III migration since
this operates in massive discs \citep{masset03}, and the vortices 
here are over-densities. In any case, self-gravity 
is required for a self-consistent treatment of disc-planet interactions 
\citep{baruteau08}. 
  In the next 
Chapter, the effect of disc self-gravity on the vortex instability is explored.

\chapter{Vortex modes in self-gravitating discs}\label{paper2}
\graphicspath{{Chapter2/}}



A self-consistent protoplanetary disc model should
account for its gravitational potential, or self-gravity. 
Even if the unperturbed disc has low mass and can be considered
non-self-gravitating, if high density structures such as 
vortices develop, it may still be important to include self-gravity. 
In this Chapter, the effects of disc self-gravity on vortex forming
instabilities at planetary gaps are examined. 

The disc-planet models are  first 
described in \S\ref{governing}. The linear stability problem is then
discussed in \S\ref{analytical} and linear numerical
calculations presented in \S\ref{linear}. Vortex modes
are stabilised by self-gravity through  linear perturbations, 
but destabilised by its effect through the background gap structure. 
The above effects cause the most unstable vortex modes to shift to
higher azimuthal wavenumber $m$, and hence more vortices to develop, as
 the disc mass increases. 
Nonlinear  hydrodynamic simulations presented in \S\ref{hydro1} 
confirm this. 
Vortices can become self-gravitating with strong over-surface densities.
This enables pairs of vortices to undergo coorbital dynamics 
with horseshoe trajectories, allowing 
them to survive much longer against merging 
than in the non-self-gravitating limit.
%
 This is strongly suggested by supplementary simulations in
\S\ref{hydro2}, of interacting Kida-like vortices without an embedded
planet. 

Most of this study considers fixed-orbit simulations.  
However,  
some migrating cases are discussed in  \S\ref{migration}. 
As in the non-self-gravitating cases (Chapter \ref{paper1}),  
episodes of rapid migration occur as vortices are scattered by the 
planet, but longer time intervals between them are expected on account
of  stabilisation by self-gravity and slowed vortex merging.  
Finally, results are summarised in \S\ref{conclusions}. 


\section{Disc-planet model}\label{governing} 
The models for disc-planet systems are described here.
Supplementary simulations in \S\ref{hydro2} employ a different set up
from that described below.

The protoplanetary disc is now self-gravitating, but
still considered two-dimensional. The governing fluid equations are 
Eq. \ref{2d_governing}, with the disc gravitational potential $\Phi$
given by the Poisson integral, Eq. \ref{2d_poisson_int}. The softening
length adopted for the disc potential in linear and nonlinear
calculations is $\epsilon_g=0.3H$. The equation of state is locally
isothermal (Eq. \ref{eos_localiso}) with fixed aspect-ratio
$h=0.05$. The disc has a uniform kinematic viscosity $\nu$. 


The disc occupies $r=[r_i,r_o]=[1,10]$. The initial surface density
profile is modified from \cite{armitage99}: 
\begin{align} \label{armitage_disc}
 \Sigma(r) = \Sigma_0 r^{-3/2}\left(1-\sqrt{\frac{r_i}{r+H_i}}\right),
\end{align}
where $H_i=H(r_i)$ is introduced  to prevent a singular
pressure force at the inner boundary because $\Sigma\to 0$
there. 
Determining the disc potential via the Poisson integral
does not require explicit boundary conditions to be imposed, but
because the integral is only taken over the computational domain, it
implicitly assumes no contribution to $\Phi$ from disc material outside the
computational domain ($\Sigma = 0$ there). Hence, the use of
Eq. \ref{armitage_disc}, with surface density tapering towards
radial boundaries, is for consistency with the Poisson integral.

The surface density scale $\Sigma_0$ is chosen via the parameter
$Q_o\equiv Q_\mathrm{Kep}(r_o)$, where  
\begin{align}\label{myQ}  
  Q_\mathrm{Kep}(r) = \frac{c_s\kappa}{\pi G\Sigma}=\frac{hM_*}{\pi
    r^2\Sigma(r)} 
\end{align}
is the Toomre $Q$ parameter for razor thin Keplerian discs with a locally
isothermal equation of state. Note that specifying $\Sigma_0$ this way
renders $Q_\mathrm{Kep}$ independent of $h$.  
$Q_o$ is used to  label  disc models and   
$Q_p\equiv Q_\mathrm{Kep}(r_p)$ is used to indicate gravitational
stability at the planet's initial location. Specifying $Q_o$ also
determines the disc-to-star mass ratio, $M_d/M_*$.    

The disc is initialised with  the azimuthal velocity  required  
by centrifugal balance.  The initial
radial velocity is $u_r = 3\nu/r$, corresponding to the initial radial velocity
profile of a Keplerian disc with uniform kinematic viscosity and surface density
$\propto r^{-3/2}$. Motivated by results in Chapter \ref{paper1}, a viscosity
$\nu=10^{-6}$ in code units is chosen to enable vortex instabilities
in disc-planet interactions. 


Most of this Chapter will focus on the stability of the gap induced by
an embedded planet. Accordingly the planet is fixed
on a circular orbit of radius $ r_p= 5$. 
The planet is introduced at $t=25P_0$ with 
azimuthal velocity that takes account the contribution from the 
gravitational force due to the disc. 
 The planet has mass $M_p=3\times10^{-4}M_*$ and its
 potential is ramped up
over a time period of $10P_0$.     

\section{Linear modes in discs with structured  surface density
}\label{analytical} 
The stability of interest is that of a surface density gap 
 self-consistently opened by a giant planet in a disc in which
self-gravity is not neglected. To simplify linear theory, viscosity,
indirect potentials and the planet potential are ignored. As shown in
Chapter \ref{paper1},   
the planet's role is only to set up the basic state, assumed to be
axisymmetric and defined by $\Sigma(r),\,u_\phi(r)$ and $u_r = 0$. 

As usual, perturbations to the disc state variables are assumed to have
azimuthal and time dependence through a factor  $\exp{i(\sigma t
+m\phi)}$ which is taken as read. 
Denoting locally isothermal perturbations 
by $\dd$, the linearised equations of motion for a spatially fixed
sound-speed profile gives

\begin{align}
 \dd u_r &= -\frac{1}{D}
\left[i\bar{\sigma} \left(c_s^2\frac{dW}{dr} + \frac{d
      \dd\Phi}{dr}\right) + \frac{2im\Omega}{r}\left( c_s^2W +
    \dd\Phi\right)\right]\\  
 \dd u_\phi &= \frac{1}{D}
\left[\frac{\kappa^2}{2\Omega}\left(c_s^2\frac{dW}{dr} +
    \frac{d\dd\Phi}{dr}\right) + \frac{m\bar{\sigma}}{r}\left( c_s^2 W
  + \dd\Phi\right)\right],
\end{align}
where $D = \kappa^2 - \sbar^2$. Recall  $\bar{\sigma}\equiv \sigma
+m\Omega(r)$ and $W=\dd\Sigma/\Sigma$.
These equations are very similar to the non-self-gravitating case
(Eq. \ref{velocities1}---\ref{velocities2}), except now the disc
gravitational potential perturbation $\dd\Phi$ is included, 
and is given by the linearised Poisson integral 

\begin{align}\label{poisson}
&\dd\Phi = -G \int_{r_i}^{r_o} K_m(r,r')\Sigma(r')W(r')r' dr' ,\\
&K_m(r,r') = \int_{0}^{2\pi} \frac{\cos(m\phi)d\phi}{\sqrt{r^2+r'^2 -
    2rr'\cos{(\phi)}+ \epsilon_g^2(r')}}.
\end{align}
Using the linearised equations of motion to eliminate the velocity
component perturbations from  the linearised continuity equation 
 (Eq. \ref{lin_cont}) 
yields the  governing equation 
\begin{align}\label{sggoverning}
  &\frac{d}{dr}\left[\frac{r \Sigma}{D}\left(c_s^2\frac{dW}{dr} 
      + \frac{d\dd\Phi}{dr}\right)\right] +
  \left[ \frac{2m}{\bar{\sigma}} 
    \left(\frac{\Sigma\Omega}{D}\right)
    \frac{dc_s^2}{dr} - r\Sigma \right]W \notag\\
  &+ \left[\frac{2m}{\bar{\sigma}}\frac{d}{dr}
    \left(\frac{\Sigma\Omega}{D}\right)
    -\frac{m^2\Sigma}{rD}\right]
  \left(c_s^2 W +\dd\Phi\right)= 0.
\end{align}


\subsection{Analysis for barotropic discs}\label{perttheory0}
The linear modes of interest are those found to be localised to gap
edges and lead to vortex formation in the nonlinear regime. 
The effect of self-gravity on these modes can be evaluated
analytically with appropriate simplifications. It will be shown that,
paradoxically, in view
of the fact that increasing self-gravity in general destabilise
discs, that self-gravity tends to stabilise vortex modes. 
To do this, one considers a  small change to the  strength of self-gravity 
and  apply perturbation theory to determine the consequence  of the
change  for these modes.  


To make the  analysis tractable, further simplifications are made. The
fluid is considered to be barotropic, so that $p=p(\Sigma)$. This
includes the strictly isothermal case where $c_s$ is a constant. 
There should be no significant
difference between adopting a strictly isothermal,  or
 barotropic equation of state,  and the
fixed $c_s$ profile  as used in nonlinear simulations. This is
because the modes of interest are driven by local features and
disturbances are localised, whereas $c_s(r)$ for the adopted locally
isothermal equation of state varies on a global scale.
For mathematical convenience, the softening prescription is assumed to
be such that $K_m(r,r') = K_m(r', r)$ is symmetric, eg.
$\epsilon_g=\mathrm{constant}$. Although convenient mathematically, 
this is not expected to lead to significant changes for the same
reason as that given above.

Introducing the variable $S$  
\begin{align}\label{s_def}
  S \equiv  c_s^2 W+ \dd\Phi = c_s^2 W - G\int_{r_i}^{r_o}
  K_m(r,r')r'\Sigma(r') W(r')dr', 
\end{align}
the governing equation 
 derived from  Eq. \ref{sggoverning}  simply by taking  $c_s$ to be
 constant is
\begin{align}\label{barotropic}
r\Sigma W = &\frac{d}{dr}\left[\frac{r \Sigma}{D}\left(\frac{dS}{dr}\right)\right] 
+ \left\{\frac{m}{\bar{\sigma}}\frac{d}{dr}
 \left[\frac{1}{\eta (1-\bar{\nu}^2)}\right]
 -\frac{m^2\Sigma}{rD}\right\}S\notag\\
& \equiv  rL (S).
\end{align}
where the expression for vortensity $\eta =
\kappa^2/2\Omega\Sigma$ has been used, and $\bar{\nu}\equiv
\bar{\sigma}/\kappa$. $L$ is a linear operator such that
$L(S)=\dd\Sigma$. In fact, these equations  also hold for a general
barotropic equation of state. 

\subsection{The limit of negligible self-gravity}
When self-gravity is negligible in the linear response, one may set
$G=0$ in Eq. \ref{s_def}, giving $W=S/c_s^2$. 
Substituting this into Eq. \ref{barotropic} gives the  second order 
ordinary differential equation for $S$ that governs 
stability in the limit of zero self-gravity  in the form
\begin{align}\label{NoselfG}
  \frac{r\Sigma S}{c_s^2}  = &\frac{d}{dr}\left[\frac{r \Sigma}{D}\left(\frac{dS}{dr}\right)\right]
  + \left\{\frac{m}{\bar{\sigma}}\frac{d}{dr}
    \left[\frac{1}{\eta (1-\bar{\nu}^2)}\right]
    -\frac{m^2\Sigma}{rD}\right\}S.
\end{align}
This is the same equation studied in Chapter
\ref{paper1} when $c_s$ is regarded constant. If one is concerned with
the effect of weak self-gravity, then the vortex modes should still
have co-rotation radius $r_c$, such that $\sbar(r_c)=0$, at or near a
vortensity minimum, as found in Chapter \ref{paper1}. 


\subsection{Localised low $m$ modes}\label{Lowmm}
When $m$ is small the 
modes described above are localised around co-rotation
and therefore insensitive to distant boundary conditions. These were
the dominant modes found in Chapter \ref{paper1}, where self-gravity
was ignored. It is then appropriate to adopt these simplifications to
calculate the effect of weak self-gravity. 
For these modes, one can neglect $\bar{\nu}$ and
set $D=\kappa^2$ in Eq. \ref{NoselfG} to get the simpler equation 
\begin{align}\label{NoselfG1}
  \frac{r\Sigma S}{c_s^2}  = &\frac{d}{dr}\left[\frac{r
      \Sigma}{\kappa^2}\left(\frac{dS}{dr}\right)\right] 
  + \left\{\frac{m}{\bar{\sigma}}\frac{d}{dr}
    \left[\frac{1}{\eta}\right]
    -\frac{m^2\Sigma}{r\kappa^2}\right\}S.
\end{align}
Localisation occurs because the solutions of Eq. \ref{NoselfG1} can be
seen to decay exponentially away from the co-rotation region, where
there are large background gradients, on a length scale comparable to
$H.$  For Eq. \ref{NoselfG1} to be a good approximation,
one requires $ |\bar{\nu}^2|\ll 1$ in the region of localisation which
is also comparable to $H$ in extent. This in turn requires $H \ll
2r/(3m),$ a condition which is satisfied for low $m$.   
The analysis below assumes localisation so that boundary conditions do
not play a role. It therefore only applies for low $m$.


\subsection{Evaluating the effect of small changes 
to low $m$ modes  using perturbation theory}\label{S3.4}
Eq. \ref{s_def} and Eq. \ref{barotropic} can be used to  
investigate the effect of self-gravity. 
Note that strengthening
self-gravity by scaling up the surface density is equivalent to
increasing $G,$ provided that the background form remains fixed. 
Thus although the effective disc gravity is increased,
the background disc model shall be assumed unchanged.
The background disc model is structured by a perturbing planet
and so its response to changing the disc gravity is non-trivial
to evaluate analytically. Calculations presented in
\S\ref{SGRole}  show that  changes to the background  surface density
profile induced by incorporating the disc gravity tends to make
the vortex modes more unstable. However, the direct effect of
self-gravity through the linear response considered below turns out to
be more important for localised modes with low $m,$ and acts to
stabilise them. 

A solution $S$ corresponding to a neutral
mode ($\gamma=0$) with co-rotation radius $r_c$ located at a
vortensity minimum, is first 
assumed ($\sbar(r_c)= d\eta/dr|_{r=r_c} =0$) . As the associated
$\sigma$ is real,  for this value, 
the operator $L$ is real and regular everywhere. Now
perturb this  solution by altering the strength of  self-gravity via   
\begin{align*}
  G \to G + \dd G
\end{align*}
so that $\dd G>0$ corresponds  to increasing the importance of
self-gravity and vice versa (note that the initial value, $G,$
could be zero). This induces  perturbations
\begin{align*}
   S \to S + \dd S,\quad 
   \dd\Sigma \to \dd\Sigma + \dd\Sigma_1, \quad
   \sigma \to \sigma + \dd \sigma, \quad
   L \to L + \dd L ,
\end{align*} 
with
\begin{align*}
  &\dd \sigma = \dd \sigma_R + i\gamma, \quad
  \dd L = \frac{\p L}{\p\sigma}\dd\sigma,
\end{align*}
where $\dd \sigma_R$ and $\gamma$ are real. 
Noting that $\delta$ denotes a small change and $\gamma$
is small, one can linearise in terms of these quantities  about the
assumed original neutral mode and determine $\gamma.$
The governing equations lead to
\begin{align}
  &L(\dd S) + \dd L(S) = \dd\Sigma_1,\label{perttheory1}\\
  &\dd S = c_s^2 \frac{\dd\Sigma_1}{\Sigma} - G\int_{r_i}^{r_o} 
  K_m(r,r')r'\dd\Sigma_1(r')dr' 
  - \dd G\int_{r_i}^{r_o} 
  K_m(r,r')r'\dd\Sigma(r')dr' .\label{perttheory2}
\end{align}
Next, define the inner product between two functions $U(r),\,V(r)$ as 
\begin{align}
  \langle U, V\rangle \equiv \int_{r_i}^{r_o} rU^*(r)V(r) dr,
\end{align}
Then, assuming localised functions
corresponding to localised modes so that boundary values
can be assumed to vanish  when integrating by parts,
\begin{align}
  \langle U, L(V)\rangle =  \langle L(U), V\rangle,
\end{align}
which used the fact that $L$ corresponding to the neutral mode is
real, making $L$ self-adjoint. Now consider
\begin{align}
  \langle L(S), \dd S \rangle =   
  \langle \dd\Sigma, \dd S \rangle =
  \langle S, L(\dd S) \rangle.
\end{align}
The inner products evaluate to
\begin{align}
  &\langle \dd\Sigma, \dd S \rangle  \notag\\
  &=\int_{r_i}^{r_o} r\dd\Sigma^*(r)\left[c_s^2 \frac{\dd\Sigma_1(r)}{\Sigma(r)} - G\int_{r_i}^{r_o} 
    K_m(r,r')r'\dd\Sigma_1(r')dr' 
     - \dd G\int_{r_i}^{r_o} K_m(r,r')r'\dd\Sigma(r')dr' \right]dr,
\end{align} 
and
\begin{align}
  &\langle S, L(\dd S) \rangle = \langle S, \dd\Sigma_1 - \dd L(S)
  \rangle \notag\\
  &= \int_{r_i}^{r_o} r \left[c_s^2 \frac{\dd\Sigma^*(r)}{\Sigma(r)}
    - G\int_{r_i}^{r_o} K_m(r,r')r'\dd\Sigma^*(r')dr' \right]
  \dd\Sigma_1(r)dr - \langle S, \dd L(S) \rangle.
\end{align} 
These expressions can be equated to yield
\begin{align}\label{perttheory3}
  \langle S, \dd L(S)\rangle &=  \dd
  G\int_{r_i}^{r_o}\int_{r_i}^{r_o}
  rr' K_m(r,r')\dd\Sigma^*(r')\dd\Sigma(r) dr' dr.\notag \\
  &\equiv \dd GE, 
\end{align}
where symmetry of the  kernel $K_m$ was used.
Note that $E>0$ and is  proportional to the magnitude of
the gravitational energy of the mode. Following \cite{papaloizou89},
one separates out the contribution to the perturbed linear operator $\dd
L$  that is proportional to the vortensity gradient (and potentially
singular at co-rotation) and write 
\begin{align}\label{perttheory4}   
  \langle S, \dd L(S)\rangle \equiv \langle S, \dd L_1(S)\rangle +
  \langle S, \dd L_2(S)\rangle,
\end{align}
where $\dd L_2(S)$  contains  the potentially singular  contribution:
\begin{align}
  \langle S, \dd L_2(S)\rangle &= -m
\dd\sigma \left[
\mathcal{P}\left(\int_{r_i}^{r_o}\frac{g(r)}{\bar{\sigma}}dr\right)
+ \int_{r_i}^{r_o}{\rm i}\pi\dd(\bar{\sigma})g(r)dr\right]\label{deltaL2}\\
  g(r)&\equiv\frac{|S|^2}{\bar{\sigma}
}\frac{d}{dr}\left(\frac{1}{\eta}\right).
\end{align}
Note that the usual Landau prescription for negative $\gamma$ (growing
mode) has been used to deal with the pole at co-rotation and that
${\cal P}$ outside the above integral  indicates 
that the principal value is to be taken.
Note too  that  $\dd L_1$ accounts for  the remainder  of $\dd L$ and
is not singular at co-rotation. The contribution from 
 $\dd 
L_2$ is the only one  that can lead to an imaginary contribution
in the limit $\dd\sigma\to 0$ because of the pole at
co-rotation. 
No
such contributions arise from Lindblad resonances where $D=0.$
This is because, as is well known, these do not
constitute effective  singularities  in a gaseous disc
\citep[eg.][]{papaloizou91}.

Recalling that  $\gamma$ is small,  Eq. \ref{perttheory3}- \ref{deltaL2}  can be combined to give
\begin{align}
  \dd GE &=\dd\sigma\left[\frac{\p\left\langle S, 
         L_1(S)\right\rangle}{\p \sigma} -m{\cal P}\left(\int_{r_i}^{r_o}
      \frac{g(r)}{\bar{\sigma}}dr\right) \right. 
    \left. - i m\pi\int_{r_i}^{r_o} 
    \dd(\bar{\sigma})g(r)dr \right]\notag\\  
  & \equiv (\mathcal{A} - im\zeta)\dd\sigma
\end{align}
where $\mathcal{A}$ is the
contribution from the  $ L_1$ term plus the principle  value integral
 and  
\begin{align}
  \zeta \equiv \left.\frac{\pi
      |S|^2}{m\Omega^\prime|
      m\Omega^\prime|}\frac{d^2}{dr^2}\left(\frac{1}{\eta}\right)
  \right|_{r=r_c}.  
\end{align}
In the limit $ \gamma\to 0_-$, $\mathcal{A}$ is real so
\begin{align}\label{stabilisation}
   \gamma = \frac{m\zeta}{\mathcal{A}^2 + m^2\zeta^2}\dd GE.
\end{align}
Vortex forming modes are associated with vortensity
minima (or maxima of $\eta^{-1}$).  Consider a marginally stable mode with co-rotation at
$\mathrm{max}(\eta^{-1})$. Typical rotation profiles have
$\Omega^\prime < 0$, which means $\zeta > 0$ for this mode.
For consistency with the assumption of $\gamma < 0$ in deriving
Eq. \ref{stabilisation}, one requires $\dd G < 0$ since
$E>0$. Accordingly in order  to destabilise this mode, the strength of
self-gravity needs to be reduced. 

This leads to the conclusion that:
increasing self-gravity \emph{stabilises}  low $m$ modes with co-rotation at
a vortensity \emph{minimum} (as has been borne out by
linear and nonlinear calculations presented below);
 while increasing self-gravity would
\emph{destabilise} modes which had  co-rotation at a vortensity
\emph{maximum}. This suggests that  for sufficiently strong self-gravity,
modes associated with vortensity maxima  should  be favoured. This has been
found to be the case but they  are not  modes leading to vortex
formation (see Chapter \ref{paper3}).    

Note $\zeta\propto 1/m^2$, suggesting that $|\gamma|$ decreases
for large $m$, so the stabilisation or destabilisation effect of
self-gravity diminishes for increasing azimuthal wave-number. 
This is only speculative at this point, because there are implicit
dependencies on $m$ through terms in the integrals  and through the original
eigenfunction  $S$. Nevertheless, a weakening effect of self-gravity
through the potential perturbation is 
anticipated because increasing $m$ decreases the magnitude of the
Poisson kernel $K_m$. 

 
\subsection{Association of localised low $m$  normal modes with vortensity minima
for the strength of self-gravity below a threshold}\label{S3.5}
Co-rotation radii for localised neutral modes with low $m$
should be at vortensity minima unless self-gravity (
or an appropriate mean value of $Q$) is above (below) a threshold level.
Consider a disc with localised steep vortensity gradients and a
non-axisymmetric disturbance with co-rotation radius  in this region. Multiply
Eq. \ref{barotropic}  by $S^*$ and integrating over the disc. Assuming 
most  of the contribution is from near co-rotation, as is expected for low $m$ modes
(see \S\ref{Lowmm}),  so that the term
that is potentially singular at corotation
and proportional to the  vortensity gradient is
dominant on the RHS, implies the balance
\begin{align}\label{e_balance}
  \int_{r_i}^{r_o} r c_s^2\frac{ |\dd\Sigma|^2}{\Sigma}dr -
  G\int_{r_i}^{r_o} \int_{r_i}^{r_o} K_m(r,r')rr'\dd\Sigma^*(r')\dd\Sigma(r)dr' dr
  \simeq  
  \int_{r_i}^{r_o}\frac{m|S|^2}{\sbar}\frac{d}{dr}\left(\frac{1}{\eta}\right)dr.
\end{align}
If the LHS of Eq. \ref{e_balance} can be shown to always be positive 
then  co-rotation for a localised 
neutral mode  must lie at a  vortensity minimum, or
$\mathrm{max}(\eta^{-1})$ \footnote{The RHS is positive
  if the dominant contribution to the integral comes from the
  co-rotation ($\sbar=0$) region about a vortensity minimum.}.  
This holds if the maximum possible value of $\Lambda$
for any $\dd\Sigma$ is less than unity, where
\begin{align}\label{lambda_def}
\Lambda=\frac{ G\int_{r_i}^{r_o}\int_{r_i}^{r_o} K_m(r,r')rr'\dd\Sigma^*(r')\dd\Sigma(r)dr' dr}
{\int_{r_i}^{r_o} r c_s^2\frac{ |\dd\Sigma|^2}{\Sigma}dr}
\end{align}
is the ratio of the magnitude of gravitational potential energy to thermal energy
of the disturbance.
Consider re-writing the numerator of $\Lambda$ as
\begin{align}
  &G
  \int_{r_i}^{r_o}\int_{r_i}^{r_o}\left[\frac{K_m(r,r')\sqrt{r}\sqrt{r'}\Sigma^{1/2}(r)
      \Sigma^{1/2}(r')}{c_s(r)c_s(r')}\right]\left[\frac{\sqrt{r}c_s(r)\dd\Sigma(r)}{\Sigma^{1/2}(r)}\right]\left[\frac{\sqrt{r'}c_s(r') 
      \dd\Sigma^*(r')}{\Sigma^{1/2}(r')}\right]drdr' \notag\\
  &=\int_{r_i}^{r_o}
  \frac{\sqrt{r}c_s(r)\dd\Sigma(r)}{\Sigma^{1/2}(r)} \left\{ G
    \int_{r_i}^{r_o} \left[\frac{K_m(r,r')\sqrt{r}\sqrt{r'}\Sigma^{1/2}(r)
      \Sigma^{1/2}(r')}{c_s(r)c_s(r')}\right]\left[\frac{\sqrt{r'}c_s(r') 
      \dd\Sigma^*(r')}{\Sigma^{1/2}(r')}\right]dr'\right\}dr\notag\\
& = \int_{r_i}^{r_o} f_1(r)\left\{f_2^*(r)\right\}dr.
\end{align}
The Cauchy-Schwartz inequality is
\begin{align*}
  \left|\int_{r_i}^{r_o} f_1(r)f_2^*(r)dr\right|^2 \leq
  \int_{r_i}^{r_o} |f_1|^2 dr \times \int_{r_i}^{r_o} |f_2|^2 dr.
\end{align*}
So that Eq. \ref{lambda_def} implies
\begin{align}
&  \Lambda^2 \left(\int_{r_i}^{r_o} r c_s^2\frac{
      |\dd\Sigma|^2}{\Sigma}dr \right)^2 
\leq \int_{r_i}^{r_o}
  \frac{rc_s^2(r)|\dd\Sigma(r)|^2}{\Sigma(r)}dr \notag\\
& \times \int_{r_i}^{r_o} \left | G
    \int_{r_i}^{r_o} \left[\frac{K_m(r,r')\sqrt{r}\sqrt{r'}\Sigma^{1/2}(r)
      \Sigma^{1/2}(r')}{c_s(r)c_s(r')}\right]\left[\frac{\sqrt{r'}c_s(r') 
      \dd\Sigma^*(r')}{\Sigma^{1/2}(r')}\right]dr'\right|^2 dr.
\end{align}
Applying the Cauchy-Schwartz inequality to the second line,
\begin{align}
&  \Lambda^2 \left(\int_{r_i}^{r_o} r c_s^2\frac{
      |\dd\Sigma|^2}{\Sigma}dr \right)^2 
\leq \int_{r_i}^{r_o}
  \frac{rc_s^2(r)|\dd\Sigma(r)|^2}{\Sigma(r)}dr \notag\\
& \times \int_{r_i}^{r_o}  \left\{ G^2
    \int_{r_i}^{r_o} \frac{K_m^2(r,r')rr'\Sigma(r)
      \Sigma(r')}{c_s^2(r)c_s^2(r')}dr'\right\}
\left\{ \int_{r_i}^{r_o} \frac{r'c_s^2(r') 
      |\dd\Sigma(r')|^2}{\Sigma(r')}dr'\right\} dr.
\end{align}
The integrals involving $\dd\Sigma$ on either sides cancel.
It follows that
\begin{align}\label{cauchy1}
\Lambda \le  G\sqrt{\int_{r_i}^{r_o}\int_{r_i}^{r_o} K^2_m(r,r')\frac{\Sigma(r)\Sigma(r')}
{c_s^2(r)c_s^2(r')} rr' dr' dr}.
 \end{align}
Thus if the RHS of Eq. \ref{cauchy1} is less than unity,  corotation of a neutral mode
localised at a vortensity  extremum must be localised at a vortensity minimum.
This condition requires that the strength of self-gravity be below a threshold
and this implies a  lower  bound on an appropriate mean $Q$ value.
The fact that this fails for sufficiently strong self-gravity is consistent
with the discussion above that led to the conclusion that  increasing self-gravity tends
to stabilise the vortex forming instability.

Indeed when self-gravity increases further, instability is transferred
to modes with corotation associated with vortensity maxima.
These are different in character to vortex forming modes being more global
and are referred to as edge modes and they are studied 
in Chapter \ref{paper3}. 


\section{Linear calculations}\label{linear}

Numerical solutions to the linear problem (Eq. \ref{sggoverning}) are
now presented. The locally 
isothermal equation of state $p=h^2GM\Sigma/r\equiv c_s^2\Sigma$ and spatially varying
softening length $\epsilon_g(r)=0.3H(r)$, are restored 
for consistency with numerical simulations used to set up
the basic state. The governing equation is regarded as   
the requirement that  an operator $\mathcal{L}$ acting on $W$ should
be zero, thus 
\begin{align}\label{linearode}
  \mathcal{L}(W) = 0. 
\end{align}
One form of $\mathcal{L}$ is 
\begin{align}\label{operator}
  \mathcal{L} =& rc_s^2\frac{d^2}{dr^2} + r\mathcal{I}_2 
 + \left[\frac{1}{r} + \frac{\Sigma^\prime}{\Sigma} -
  \frac{D^\prime }{D}\right]
\left(rc_s^2\frac{d}{dr} + r\mathcal{I}_1\right) 
 + rc_s^{2\prime} \frac{d}{dr} 
 +\left[\frac{2m\Omega
    c_s^{2\prime}}{\bar{\sigma}} - rD\right]\notag\\
 &+\left\{\frac{2m}{\bar{\sigma}}\left[\frac{\Sigma^\prime\Omega}{\Sigma}
    + \Omega^\prime - \frac{\Omega D^\prime}{D}\right] -
  \frac{m^2}{r}\right\}\left(c_s^2 +\mathcal{I}_0\right).
\end{align}
Primes in Eq. \ref{operator} denote $d/dr$. 
The integro-differential operators $\mathcal{I}_n$ are such that
\begin{align}\label{iiop}
\mathcal{I}_n(W) = \frac{d^n \dd\Phi}{dr^n}.
\end{align}
First and second derivatives of the perturbed potential are  performed by replacing
$K_m(r,r')$ by $\p K_m/\p r$ and $\p^2 K_m/\p r^2$ in the Poisson
integral respectively. 

\subsection{Quadratic approximation}
Vortex modes are localised about co-rotation at a  vortensity minimum. 
They extend over a region  characterised by small $|\bar{\sigma}|$ in the
neighbourhood of $r_c$ and are expected to be
insensitive  to boundary conditions. 
Based on these observations, a crude approximation was adopted in the
analytical discussion in Chapter \ref{paper1}, 
where $\sbar$ is neglected everywhere except when it appears as
$1/\sbar$. This enforces locality. 
However, with self-gravity, the disturbance is potentially more global. 
A better approximation for self-gravitating vortex modes can be made as follows.    

Consider multiplying Eq. \ref{operator}  by $\bar{\sigma}D$ and
expand the resulting equation in powers of $\bar{\sigma}$.
Vortex modes are associated with the region $|\sbar|\sim 0$. 
By neglecting terms proportional to $\bar{\sigma}^3$ and higher powers, one can
derive an equation of the form  
\begin{align}\label{qepapprox}
  \left( \sigma^2 \mathcal{L}_2 + \sigma \mathcal{L}_1 +
    \mathcal{L}_0\right)W = 0, 
\end{align}
where the operators $\mathcal{L}_i$  are real and only depend on the basic
state. This is referred to as the quadratic
approximation. Expressions for $\mathcal{L}_i$ are given in Appendix \ref{lep_qep}. 
The eigenfrequency  $\sigma$ now appears explicitly. 

The above procedure is not expected to be  valid in general, but it should
be valid for modes approximately localised about their corotation radii
with the scale of localisation being much less than $r_c$ itself.
Solution eigenvalues from Eq. \ref{qepapprox} 
can be used as  starting values 
in order to obtain an iterative solution for the eigenvalues for  the full Eq. 
\ref{linearode}. The concept of the vortex forming mode as a localised
mode \emph{is confirmed} if the final solution of Eq. \ref{linearode}
is not significantly changed from the solution of Eq. \ref{qepapprox}. 
Note that since the operators $\mathcal{L}_i$  are real,
eigenfrequencies are either real or  occur in complex conjugate pairs.


\subsection{Distinction between including and excluding
  self-gravity}
It is important to 
clearly define self-gravitating (SG) and non-self-gravitating cases (NSG). 
The
background state on which linear stability analysis is performed were obtained
from nonlinear hydrodynamic simulations (see \S\ref{linearbasic}), which can be run 
with self-gravity (SGBG) or without self-gravity (NSGBG). 
In the linear calculations, self-gravity can be disabled 
by setting $\dd\Phi=0$, corresponding to two
disturbance types: a self-gravitating response (SGRSP) and a
non-self-gravitating response (NSGRSP). 

The analytical discussion in \S\ref{perttheory0} applies to the effect
of self-gravity through the linear response, assuming the form of the background
remains unchanged. In the context of hydrodynamic simulations,
including or neglecting self-gravity simply means whether or not the
disc gravitational potential is included.  If it is included, the
background state set up by simulations also depends on self-gravity. 
Hydrodynamic simulations therefore correspond either  to
the combination  SGBG+SGRSP or to the combination  NSGBG+NSGRSP. These are the fully
self-gravitating and fully non self-gravitating cases. 
The advantage of
performing linear calculations is that one can distinguish between
the effects of self-gravity  arising through its effects on  the background state 
and  the effects resulting from its influence on  the linear  response.

\subsection{Numerical approach}\label{linear_numerics}
The linear operators  are discretised on a
grid that divides the radial range $[r_i, r_o]$ into $n_r=385$
 equally spaced grid points at  which  $W$ is evaluated as $W_j, j=1,2..n_r.$
The governing equation thus becomes a system of
$n_r$ simultaneous equations for the $W_j$: 
\begin{align}\label{sggoverning_matrix}
\sum^{n_r}_{j=1}  \mathcal{L}_{ij}W_j = 0,
\end{align}
where $\mathcal{L}_{ij}$ is the discretised version of the integro-differential 
operator $\mathcal{L}$. Eq. \ref{sggoverning_matrix} is a matrix eigenvalue
problem for $\sigma.$ 
The system of equations incorporates the boundary conditions (replacing the first
and last rows of the matrix $\mathcal{L}_{ij}$). 
For simplicity, $dW/dr = 0$ is imposed at boundaries. 
It is known from other linear calculations and simulations 
that vortex-forming modes are localised and insensitive to
boundary conditions \citep[e.g.][]{valborro07}. Tests have shown  that the  boundary
condition  does not influence  the essential effect of self-gravity on localised
modes. 

The discretised quadratic approximation obtained from  Eq. \ref{qepapprox} leads
to a  quadratic  eigenvalue problem that is equivalent to a $2n_r\times
2n_r$ linear matrix  eigenvalue problem.  This is  solved by
standard matrix software to give all eigenvalues $\sigma$.  The most unstable
eigenvalue is then used as a trial  solution for  the full system (Eq. \ref{sggoverning_matrix}), 
of which non-trivial solutions demand:
\begin{align} 
  \mathrm{det}\,\mathcal{L}_{ij} = 0.
\end{align}
Although the determinant can be computed without much effort, numerically 
testing a zero determinant is ambiguous because the
value of $\mathrm{det}\,\mathcal{L}_{ij}$ can change by
many orders of magnitude depending on how the original equation
is scaled, or equivalently, the choice of units.
To obtain a non-dimensional measure of a zero determinant, independent
of how the matrix elements $\mathcal{L}_{ij}$ are scaled,
$\sigma$ can be considered as a parameter for the matrix $\mathcal{L}_{ij}$ in the
standard eigenvalue problem:
\begin{align}
 \sum^{n_r}_{j=1} \mathcal{L}_{ij}(\sigma)W_j = \lambda W_i. 
\end{align}
Then $\lambda = 0$ is required to be a solution for some $\sigma$.
Given a trial $\sigma$, all the eigenvalues $\lambda$ can be found
using standard methods.  
The actual eigenvalue, $\sigma$, is then solved iteratively 
using the Newton Raphson method to zero the quantity
$\lambda_\mathrm{min}/|\lambda_\mathrm{max}|$, where subscript
`min' and `max' denote the $\lambda$ of smallest and largest magnitudes, 
respectively.  

To check the obtained $\sigma$ indeed makes $\mathcal{L}$
singular, a singular value decomposition (SVD) of
$\mathcal{L}$ is performed, and the reciprocal of the condition number computed. 
This number is zero for a singular matrix. The SVD also
gives the solution vector $W_i$ that corresponds to the zero singular
value, i.e. the eigenfunction $W$ of Eq. \ref{sggoverning}. Numerical solutions
presented below typically have a inverse condition number of
$O(10^{-18})$.



\subsection{Background  state}\label{linearbasic}
The background states used for linear calculations were set up by running
disc-planet simulations for the models described  in \S\ref{governing}.
Numerical details will be  described in \S\ref{hydro1}. The planet is allowed
to open a (stable) gap, then azimuthal averages performed to obtain one-dimensional
profiles.  

\begin{figure}[t]
\centering
\includegraphics[width=0.99\linewidth]{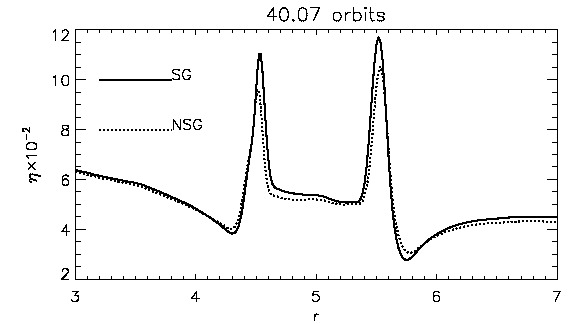}
\caption{
  Gap profiles  opened by a Saturn-mass planet in an initial
  disc model  with $Q_o=4.$ 
The vortensity, $\eta$, obtained with (solid, `SG') and
without (dotted, `NSG') self-gravity is shown. Profiles were obtained 
from azimuthal 
  averages  taken from disc-planet simulation outputs.  
The planet is
  fixed  at 
  $r=5$. 
\label{basic}}
\end{figure}

Fig.\ref{basic} shows the gap profile, in terms of the vortensity $\eta$, 
opened up by a Saturn-mass planet in a disc with $Q_o=4$. 
The formation of vortensity peaks was explained in Chapter \ref{paper1} as
generation of vortensity through shocks 
induced by the planet. 
The profiles set up with and without self-gravity are very similar so
the formation mechanism is not affected by disc self-gravity.
When self-gravity is included, the peaks and troughs have slightly
larger amplitudes due to increased effective planet mass 
(see \S\ref{gapprofiles}). 

Hydrodynamic simulations indicate vortices predominantly develop at the
outer gap edge, henceforth linear modes associated with the outer
vortensity extrema are considered. The outer  vortensity maximum 
is
located at $r\simeq 5.5$ while the outer vortensity minimum
 is located at $r\simeq 5.75.$  These
extrema are  separated by about $0.89H$ and are therefore local
features. 



\subsection{Solution   in the quadratic approximation}
\begin{figure}[!ht]
\centering
\includegraphics[width=0.99\linewidth]{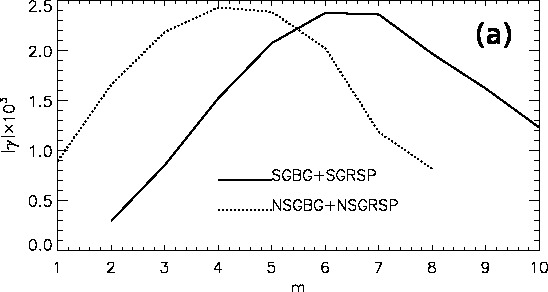}
\includegraphics[width=0.99\linewidth]{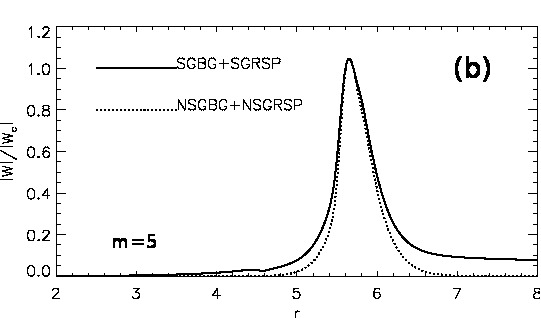}
  \caption{Growth rates (a) obtained from the quadratic approximation to the
    governing equation and the modulus of the $m=5$ eigenfunction
    (b). Here $W_c=W(r=5.7).$  The solid (dotted) lines correspond to the
    case with (without) self-gravity in both  the background  state and response.
    The disc model is  $Q_o=4.$
    \label{QEP_compare1}}
\end{figure}

Solutions for linear modes in  the $Q_o=4$ disc under the quadratic
approximation are shown in  Fig. \ref{QEP_compare1}, which compares  the growth rates
$,\gamma,$ and the eigenfunctions,  $W,$ obtained for $m=5$  for  the fully
self-gravitating case  and the fully  non self-gravitating case.  
Growth rates  are such that the most unstable mode shifts
to higher $m$ when self-gravity  is included. Without self-gravity, 
the dominant mode has
$m=4,$ whereas with self-gravity,  the dominant mode has $m=6$---7.  The combination of
self-gravity acting through the background and the linear  response stabilises modes with
$m \leq 5$ and destabilises modes with $m\geq 6$.  Thus  higher $m$ vortex
modes are enabled by self-gravity.

The eigenfunctions $W$  with and without self-gravity are similar, and are
for the most part  localised about co-rotation. This
suggests the nature of the instability is unchanged in this case.    
 However, the  eigenfunction amplitude in $r>6,\,r<5.2$
is larger when self-gravity is included.
 Note too that the amplitude is larger for
 $r>6$ than for  $r<5.2.$ This is expected since the outer 
disc has smaller Toomre $Q$ values
it is expected to be more responsive.

It is important to remember that the quadratic approximation assumes
modes are localised about co-rotation. The global nature of
some disturbances may invalidate this approximation. Increasing $m$  eventually 
quenches modes in the non self-gravitating case, thus 
localised modes with $m\geq 9$  
  were not found. For the full governing Eq.\ref{sggoverning}, 
high $m$ modes are expected to have
increasing wave-like behaviour and be  less focused around corotation,
contradicting the quadratic approximation and
so it is not surprising that they are  not found here. 

A localised $m=1$  vortex mode was not found in the self-gravitating case  because
it had been stabilised by the inclusion of  self-gravity, according to earlier
analysis. However, there exists other types of low $m$ mode not
captured by the quadratic approximation. Modes with  extreme values of
$m$ are of less relevance since they do not develop in
hydrodynamic simulations, which typically show the number
of vortices in agreement with the most unstable $m.$


\subsection{Solutions to the full governing linear  equation}
\begin{figure}[!ht]
\includegraphics[width=0.99\linewidth]{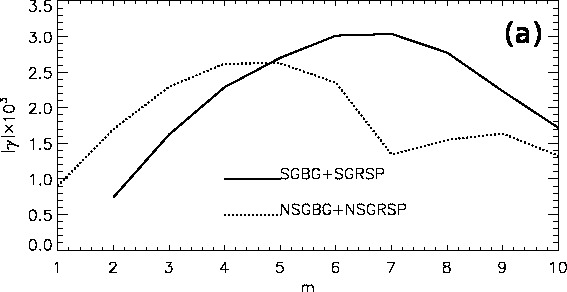}
\includegraphics[width=0.99\linewidth]{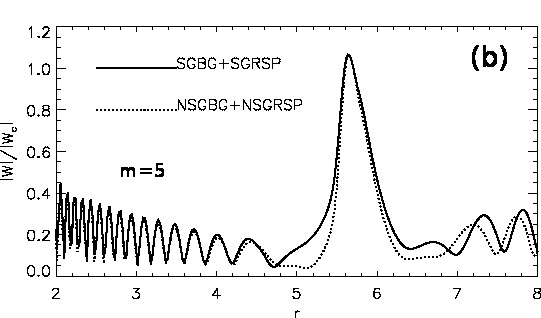}
\caption{Growth rates obtained  for the full governing equation (a) and the eigenfunction
   $|W|$ for $m=5$ (b). Here $W_c = W(5.7)$.  Solid lines are for the fully  
  self-gravitating case and dotted lines are for the fully
  non self-gravitating case. The disc model is  $Q_o=4.$ In
    (a), the local maximum around $m=9$ in the dotted curve is most
    likely caused by a boundary condition 
    effect. Non-self-gravitating modes with
    $m\geq7$ are not seen in nonlinear
    simulations and are not relevant. 
\label{FULL_compare1}}
\end{figure}

Solutions to the full governing Eq. \ref{sggoverning}, for
the fiducial cases  with $Q_o=4$  above are shown in
Fig. \ref{FULL_compare1}. 
Growth rates and the
$m=5$ eigenfunction are compared.  The co-rotation radii for the
self-gravitating and non-self gravitating cases are  $r_c =
5.88,\,5.79$, respectively.  These radii are close to the local
vortensity minimum. 
In the non-self-gravitating case, 
co-rotation almost coincides with the minimum.  

The non-self-gravitating case has maximum growth
rate for  $m=4$ and the self-gravitating case is most unstable for
$m=6$---7. Modes with $m<5$ are stabilised while $m>5$
are destabilised by self-gravity.
 This results in a shift to higher
$m$ modes  when self-gravity is included in nonlinear simulations of the model. 
Modes with the larger $m$ values  are destabilised by fully including
self-gravity. This is  
not in contradiction with with earlier analysis which assumed small $m$
and did not take account of variations in the form of the basic state.
It is shown in \S\ref{SGRole} below that the destabilisation of higher $m$ modes
is in fact due to the change in the background state.

Results for $m < 5$ are in essential agreement with those obtained
in  the quadratic  approximation giving
confidence in view of the importance of the dominance of the corotation (vortensity extremum)
region.  
As with the quadratic approximation, a $m=1$ mode with dominant disturbance about the gap edge
was not found in the self-gravitating case.
Thus low $m$ modes are the ones most effectively   stabilised  by increasing self-gravity.

In the non self-gravitating case, growth
rates for $m\geq 8$ are larger than $m = 7$ and do not follow the
trend of decreasing growth rates seen from $m=5\to7.$  
Non-self-gravitating $m\geq8$  modes found here
are unlikely to be the same type of vortex modes as $m\leq7$
because they have significant wave-like regions 
in $W$ and are not concentrated near co-rotation as for  $m \leq 7.$  In addition, $m\geq
8$ is also where the quadratic approximation  appears to 
fail. These wave-dominated modes demand radiative boundary conditions
rather than the simplistic conditions applied here. 
The $m\geq8$ non-self-gravitating  modes identified here are thus
likely to be artifacts of the boundary condition.  
Fortunately, these modes  
are irrelevant because they are not the
most unstable, nor are they observed in the corresponding
nonlinear simulations. 
By considering  the behaviour for  $m\leq7,$ one expects a cut-off for vortex modes
around $m=8$ for this model.    


The $m=5$ eigenfunctions in shown in  Fig. \ref{FULL_compare1} are similar. Both
have dominant disturbance around  co-rotation. Behaviour in the  region $r\leq4.6$ is 
essentially identical (since self-gravity weakens with decreasing radius for this disc model). 
The largest difference is found  in 
$r\geq6.4$, consistent with the general picture that self-gravity is
usually more important in outer parts of accretion discs. 
The  disturbance around co-rotation is also somewhat
wider in the self-gravitating case, signifying the global nature of
gravity. Comparing with Fig. \ref{QEP_compare1}, 
the quadratic approximation captures the main feature of the mode in 
the co-rotation region. However, it removes the
wave-like behaviour in the extended disc. 


\subsection{The role of self-gravity}\label{SGRole}
The calculations above can be  compared to hydrodynamic
simulations where self-gravity is either enabled or not.  
The linear problem allows one to examine separately the effect 
of self-gravity through its influence on the basic state and through
its influence on the
linear response. 

Below, growth rates for the  $Q_o=4$ disc model are compared for
a pair of cases, one with and the other without 
 self-gravity in setting up the basic state
(i.e. the nonlinear simulations were run with and without disc self-gravity enabled)
but both without self-gravity implemented in the
 linear mode calculation (i.e. setting
$\dd\Phi = 0$). Another pair of cases is also compared,
one  with and the other without self-gravity
implemented  in the
linear response, but with  both having  the background 
state calculated with the disc self-gravity incorporated.

\begin{figure}[!ht]
\centering
\includegraphics[width=0.99\linewidth]{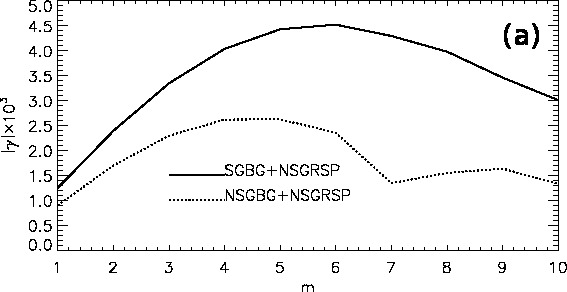}
\includegraphics[width=0.99\linewidth]{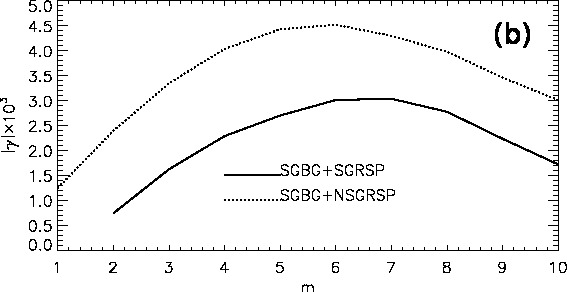}
\caption{The effect of self-gravity on growth rates 
 of vortex-forming instabilities at gap edges, through the background (a) and through the response (b).  
The disc model is $Q_o=4$. In
    (a), the local maximum around $m=9$ in the dotted curve is most
    likely caused by a boundary condition 
    effect. Modes with
    $m\geq7$ without self-gravity are not seen in nonlinear
    simulations and are not relevant. 
\label{sgrole}}
\end{figure}

Fig. \ref{sgrole}(a) examines the influence of self-gravity through its
modification of  the
background state. It shows that this modification  is destabilising.
This is because a deeper gap is set up when
self-gravity is included in the simulation. 
In going from the case without self-gravity
to the one with self-gravity, the most unstable mode shifts
from $m=4$---5 to $m=6.$ 
The difference in growth rates decreases  as $m$ decreases so the
effect due to the modification of the background 
is smallest for small $m.$ In other words,  high $m$ modes are made more unstable  
by including self-gravity in the basic state. 


Fig.\ref{sgrole}(b) compares the growth rates obtained from linear
calculations for identical background states but
 with and without self-gravity
implemented in the response.  The analysis
in \S\ref{perttheory0} applies to this comparison.  Consistent with
that, enabling self-gravity in the response decreases
$|\gamma|$ and stabilises the system
against modes with co-rotation associated with a  vortensity
minimum for \emph{ any value of $m$}.
 Unlike the effect acting  through the background, 
the influence of self gravity through the linear response is more
significant for low $m$ 
and can lead to complete stabilisation (e.g. $m=1$). 
The most unstable mode  without self-gravity has  $m=5$  and
including self-gravity 
 shifts it to $m=6$. 

The fact that self-gravity acting in the linear 
response and background  state has
opposite effects on growth rates is consistent with the comparison between fully
self-gravitating and fully non-self-gravitating cases 
(see Fig. \ref{FULL_compare1}, where peak growth rates are approximately equal). 
The effect of self-gravity through changes to the background
 and through direct influence on the linear response both
contribute to favouring higher $m$. However, the background effect is
achieved by increasing high $m$ growth rates, whereas the effect via
the response works by  stabilising low $m$ modes in accordance
with the discussion in \S\ref{S3.4}.
Thus  low $m$ vortex modes become  disfavoured. 
Overall, one expects more vortices to form,
corresponding to increasing $m,$  with
increasing self-gravity.


\subsection{Models with different $Q_o$ }
In the calculations below,
self-gravity is always included in setting up the basic state. 
The $Q_o=4$ disc above is compared 
to $Q_o=3$ (a more massive disc) and $Q_o=8$ 
(a less massive disc).

\begin{figure}[!ht]
\centering
\includegraphics[width=0.99\textwidth]{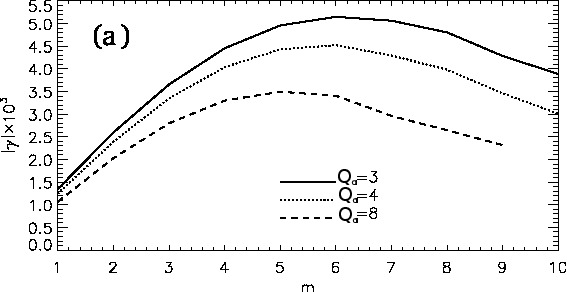}
\includegraphics[width=0.99\textwidth]{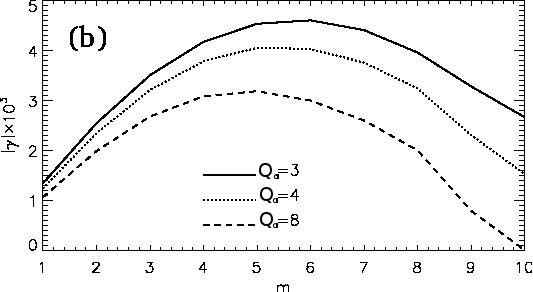}
\caption{Growth rates for unstable modes in background discs with
  $Q_o=3,\,4,\,8$.  (a) is obtained from solving the full linear equation, while (b)
  is from the quadratic approximation. Self-gravity is only
  incorporated in the background and 
  not in the linear response. The similarity between (a) and (b)
  indicate these modes are localised disturbances.  
  \label{compareQs_BG}}
\end{figure}

Fig. \ref{compareQs_BG} show growth rates obtained from the full 
equation and the quadratic approximation, \emph{without} self-gravity in the
linear response. The two sets of curves are similar. As with switching
self-gravity on or off, increasing self-gravity in the background
destabilises all modes, but this effect diminishes with decreasing
$m$.  Although lowering $Q_o$ shifts the most unstable modes to
slightly higher $m$, the low $m$ modes still exist.  

\begin{figure}[!ht]
   \centering
   \includegraphics[width=0.99\linewidth]{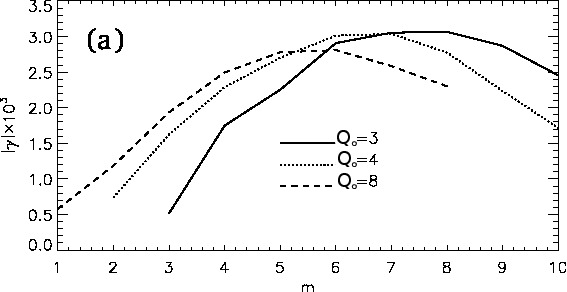}
   \includegraphics[width=0.99\linewidth]{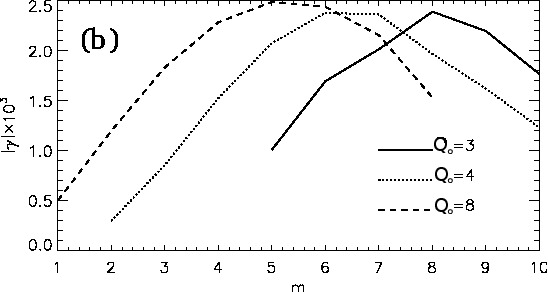}
   \caption{Growth rates for unstable modes in background discs with
     $Q_o=3,\,4,\,8$.  (a) is obtained from solving the full linear equation, while (b)
     is from the quadratic approximation. Self-gravity is incorporated
     in the basic state and the linear response. 
     \label{compareQs}}  
 \end{figure}

 Growth rates shown in Fig. \ref{compareQs} fully incorporates 
 self-gravity. 
 The plot
 demonstrates a clear shift of the most rapidly 
 growing mode  to higher $m$ as $Q_o$ is lowered
 . For $Q_o=8$ the most unstable mode has
 $m=5-6$ and for $Q_o=3$ it shifts 
 to $m=7-8$.  The shift is accompanied by the stabilisation (or loss) of low $m$
 modes. For example, when the disc mass is increased  as $Q_o$ changes
 from  $Q_o=4$ to $Q_o=3,$  the
 $m=3$ growth rate decreases by a factor of 3. Modes with the  lowest $m$  are
 stabilised by self-gravity in the response, as $m=1,\,2$ modes were not found for
 $Q_o=3$.   

 Fig. \ref{compareQs}(b) shows a similar dependence 
 of growth rates on  $Q_o$ in the quadratic approximation. 
 The shift to higher $m$ is more
 apparent. For $Q_o=3$ no modes with $m\leq 4$ were found.   Since
 the approximation reinforces the localised property of vortex-forming
 modes, it means that localised modes for low $m$ become increasing
 unlikely as $Q_o$ is lowered.  
 Hence, for massive discs only high $m$
 vortex forming  modes can develop. The agreement between 
 results obtained using the quadratic
 approximation and the  full governing 
 equation indicates a lack of sensitivity to boundary conditions and
 so is reassuring for the modes of interest. 

\clearpage
\section{Nonlinear hydrodynamic simulations}\label{hydro1}
Hydrodynamic simulations of vortex formation and evolution
at edges of gaps opened by a giant planet were performed for discs of 
varying masses.  
The disc-planet  models were described in \S\ref{governing}. 
The planet mass is $M_p=3\times10^{-4}M_*$ and 
the disc has uniform kinematic viscosity $\nu=10^{-6}$ 
corresponding to an $\alpha$
viscosity of $\alpha = \nu/(c_sH) = 1.8\times10^{-4}$ at the planet's
fixed orbital radius at  $r_p=5.$


\subsection{Numerical scheme}
The hydrodynamic equations are evolved using an extended version of
the \fargo code \citep{masset00a,masset00b} which includes disc self-gravity.    
A 2D Poisson solver for \fargo was implemented 
 by \cite{baruteau08}. The gravitational acceleration due to the  disc
is calculated directly using Fast Fourier Transforms in both azimuth
and radius. 

The disc is divided into $N_r\times N_\phi = 768\times2304$
computational grid cells  in  radius 
and azimuth. The  grid 
in the  radial (azimuthal) direction is logarithmically (evenly) spaced.
An open boundary is imposed at $r=r_i$ and non-reflecting boundary as used by
\cite{zhang08} at $r=r_o$ \citep{godon96}.    
Since vortices are localised features, as 
long as  gap edges are  far from boundaries, boundary conditions can
only  have a limited effect. Test  simulations with
damping boundary conditions \citep{valborro07} and  open outer
boundaries yield similar results to those presented below.


\subsection{The effect of self-gravity}\label{sg_nsg}
The standard distinction between self-gravitating and 
non-self-gravitating disc-planet simulations is either including or
excluding 
the disc gravitational potential in the total potential calculation
\citep{nelson03a,zhang08}. A pair of such cases for the $Q_o=4$ disc
are compared. The disc mass is $M_d = 0.024M_*$. Note the
response of a non-self-gravitating disc to the planet is independent 
of $Q_o$.  

\begin{figure}[t]
  \centering
  \includegraphics[scale=.67,clip=true,trim=0cm 0cm 1.5cm 0cm]{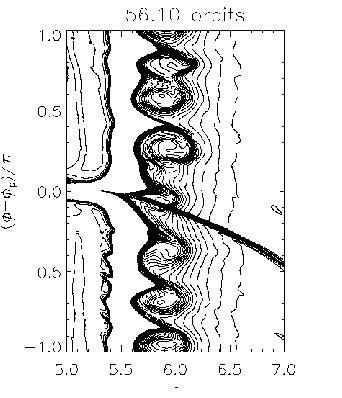}\includegraphics[scale=.67,clip=true,trim=1.1cm 0cm 0cm 0cm]{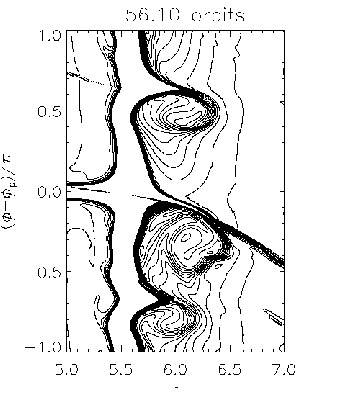}
  \caption{Vortensity contours for $Q_o=4$ with (left) and without (right)
    self-gravity. The planet is located at $r = 5,\,\phi =   \phi_p$. The
    thick lines crossing the outer radial boundaries
    correspond to spiral shocks induced
    by the planet. Local vortensity maxima ($r\sim 5.5$) are also
    produced by such shocks. 
    \label{vortices_sg_nsg}}
\end{figure}

Fig. \ref{vortices_sg_nsg} shows vortensity contours at the
onset of vortex formation at the outer gap edge. The local 
vortensity maximum at about $r=5.5$ 
remains  largely undisturbed  indicating that instability 
is associated with structure outside it and associated
with  the vortensity minimum, consistent with earlier analytical
arguments. More vortices are excited when self-gravity is included. In
that case  the $m=6$ vortex mode dominate whereas  the $m=3$ mode
dominates in the non-self-gravitating case. This contrast is
consistent with linear calculations.

Fig. \ref{vortices_sg_nsg} also shows that vortices have radial
length-scales comparable to 
 the local scale-height ($H(6) = 0.3$). 
The  vortices  have radial sizes of  $\sim 2.2H(6)$ without self-gravity and
$\sim 1.3H(6)$ with self-gravity. The  
vortices are of smaller radial extent  when self-gravity is included
 because of the preference for
higher $m.$ A decrease in radial size can be expected from the
decrease in width of the WKBJ evanescent zone centred on corotation.
This is determined by the condition 
\begin{align}
  (\sigma + m\Omega)^2 = \kappa^2(1 - 1/Q^2). 
\end{align}
 Writing $\sigma = -m\Omega(r_c)$, it
is straight forward to show that the radial width of the evanescent zone 
approximately scales as $1/m$.  
Since without self-gravity
the preferred $m$ is a factor of two smaller,
vortices in this case are expected to be approximately double the size
of those in the self-gravitating case.

With self-gravity, the  vortices are approximately centred  along the
radius $r=5.9$, close to the 
corotation radius expected from linear calculations  $r_c=5.88$. 
In the non-self-gravitating case,
 linear calculation gives $r_c=5.81$ for $m=3$, but 
the  vortices are approximately centred along $r=6.$  
However, perturbations here are already in the nonlinear regime
and interaction between vortices  or with the spiral shock
may shift the vortices around. 
In this regard  there is 
more variation in the radial locations of the vortices as compared
to the case with self-gravity. 

\begin{figure}[t]
  \centering
  \includegraphics[scale=.67,clip=true,trim=0cm 0cm 1.5cm 0cm]{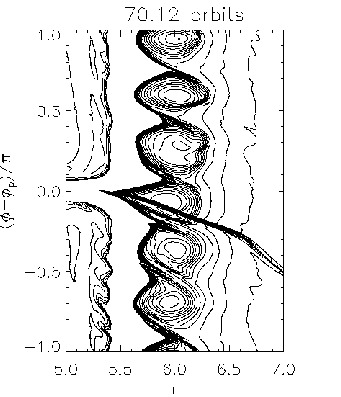}\includegraphics[scale=.67,clip=true,trim=1.1cm 0cm 0cm 0cm]{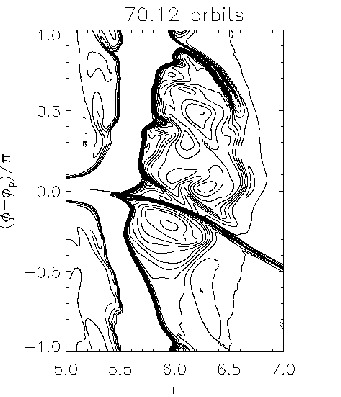}
  \caption{Same as  Fig. \ref{vortices_sg_nsg} but at a later time. The
     left panel includes self-gravity whereas the right panel is
     non-self-gravitating.\label{vortices_sg_nsg_later}}
\end{figure}

Evolving the systems further, Fig. \ref{vortices_sg_nsg_later} shows
that in the non-self-gravitating case, a single vortex has formed through
merging. This is a standard result for gap edge vortices 
\cite[e.g][]{valborro07}. However, in the self-gravitating case 6
vortices remain. Resisted merging is 
explored in more detail later on, but this is not too surprising if
one regards these vortices as planets in mean motion resonance.       


\subsection{Varying disc mass: gap profiles}\label{gapprofiles} 
Self-gravitating disc models with $ 1.5 \leq Q_o \leq
8$ are presented below. These are equivalent to $ 2.6 \leq Q_p \leq 14$
or  $0.063 \geq M_d/M_* \geq 0.012$. The equilibrium gap profiles have
a range of $Q$ values with local extrema of $1.5\leq
\mathrm{min}(Q)\leq 9.5$ and $4.8\leq \mathrm{max}(Q)\leq 21.6$ near
the outer gap edge.  The $Q$ profiles behave similarly to 
the vortensity profiles (see Chapter \ref{paper3} for an explicit
illustration).

\begin{figure}
\centering
\includegraphics[width=0.99\linewidth]{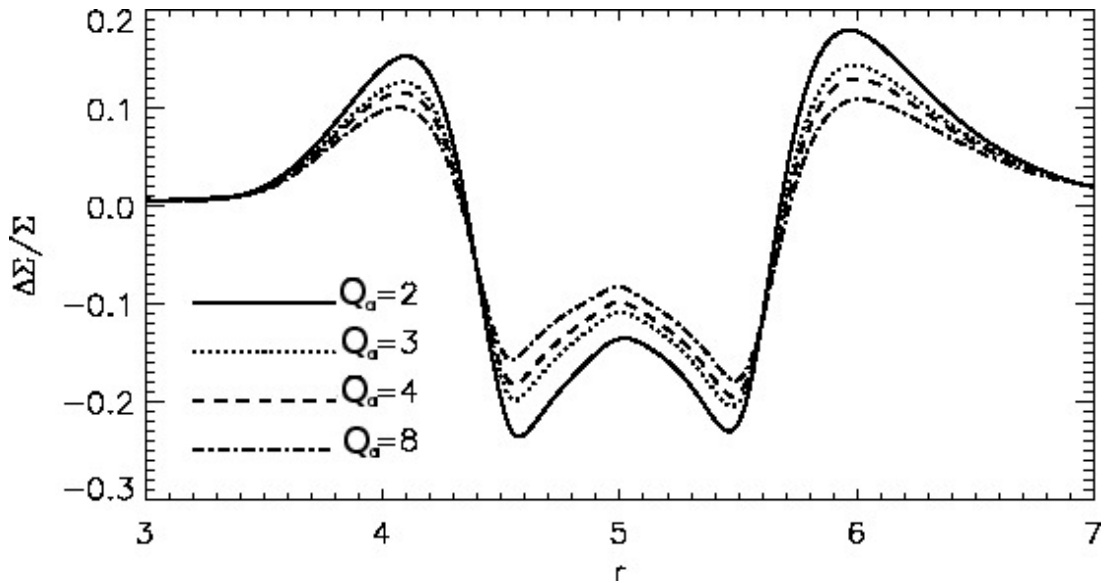}
\caption{Gap profiles opened by a Saturn mass planet in
  self-gravitating discs as a function of disc mass, parametrised by
  $Q_o$. 
  The azimuthally averaged relative surface density
  perturbation is shown. The planet located at $r = 5 $.  
\label{gap_profiles}}
\end{figure}

The gap profiles opened by the planet are shown  in
Fig. \ref{gap_profiles} for the range of $Q_o$ that develop
vortices. Outside the plotted region 
the curves are indistinguishable. 
The gap  deepens with decreasing $Q_o$ and gap edges become steeper. 
In going from $Q_o=8\to2$ the
gap depth $|\Delta\Sigma/\Sigma|$ increases by about 0.05---0.08, similarly the
bumps near gap edges increase by 0.05---0.06. 

Self-gravity affects gap structure on  a local scale by increasing
the effective planet mass so that $M_p \rightarrow M_p^\prime$.
 A straightforward estimate, based on the unperturbed disc model,
 of the expected  mass within  the Hill
radius $,r_h,$ of a
point mass planet with $M_p=3\times10^{-4}M_*$ is $M_H=0.047M_p$ for $Q_o=8$ and
$M_H=0.17M_p$ for $Q_o=2.$ It is likely that
 $M_H$, or at least some significant  fraction of it
adds to the effective mass of the planet acting  on the disc when
self-gravity is self-consistently included.
 Thus $M_p^\prime,$ and therefore also the gap depth and 
steepness of gap edges, 
 are expected to  increase with disc
mass. 

Without carefully tuning $M_p$,
self-gravitating disc-planet calculations with different surface
density scales will always differ in $M_p^\prime$. Since the gap profiles
differ, stability is affected also.
Note that the gap
width $w$ in Fig. \ref{gap_profiles} does not change greatly with $Q_o$, 
which is consistent with the scaling
$w\propto r_h\propto M_p^{\prime1/3}.$ However, the 
peaks/troughs for $Q_o=2$  do lie slightly closer to the orbital radius  $r_p$
than other cases. This
is because shocks are induced closer to the planet due to increased 
$M_p^\prime$ (Chapter \ref{paper1}).

\subsection{Varying disc mass: gap stability} 
\begin{landscape}
  \begin{figure}
    \centering
    \includegraphics[scale=0.39]{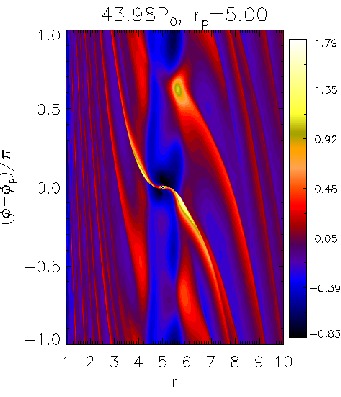}
    \includegraphics[scale=0.39,clip=true,trim=2.3cm 0cm 0cm 0cm]{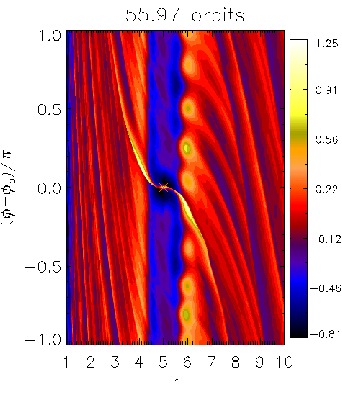}
    \includegraphics[scale=0.39,clip=true,trim=2.3cm 0cm 0cm 0cm]{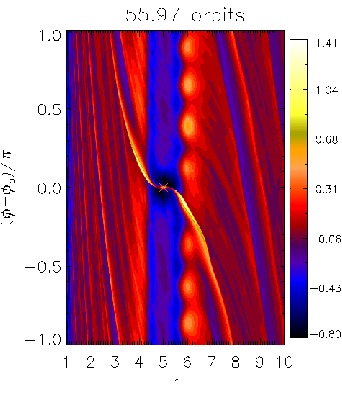}
    \includegraphics[scale=0.39,clip=true,trim=2.3cm 0cm 0cm 0cm]{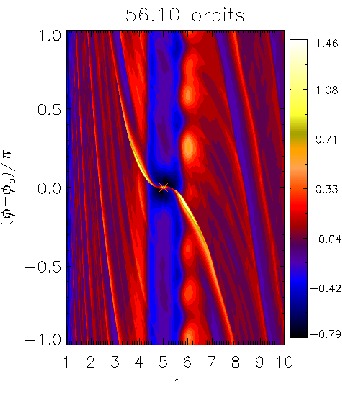}
    \includegraphics[scale=0.39,clip=true,trim=2.3cm 0cm 0cm 0cm]{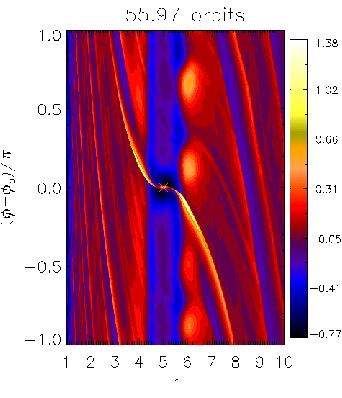}\\
    \caption{Instability at the outer gap edge of a Saturn-mass planet, in
      self-gravitating discs with minimum Toomre parameter, 
      from left to right, of $Q_o = 1.5,\, 2,\, 3,\, 4,\, 8$.
      The relative surface density perturbation is shown.
      \label{compare_v}}
  \end{figure}
\end{landscape}

Fig. \ref{compare_v} shows snapshots of the  relative surface density
perturbation as instability sets in. Consider  $Q_o\geq 2$ first. 
The instability is associated with the outer gap edge while the inner
edge remains relatively stable. In the least massive
disc $Q_o=8$, 3---4  vortices form at the outer gap edge,
similar to what happens in the non-self-gravitating disc in
\S\ref{sg_nsg}. As $Q_0$ is lowered, more vortices develop. By $Q_o=2$, 
8 surface density maxima can be identified \footnote{Note that a
vortex may be obscured by the planetary wake.}.

In
moving from $Q_o=8\to Q_o=2,$  vortices become smaller and their centres
move inwards. When  $Q_o=8 $ local surface density maxima lie just
outside $r=6$ while for $Q_o = 2 $ they  lie just interior to
$r=6$. The shift to smaller radial location is consistent with
analytical expectations that a vortensity maximum is favoured against a 
vortensity minimum as self-gravity is increased, and for planetary outer gap edges the 
local vortensity maximum lies closer to the planet than the 
vortensity minimum (Fig. \ref{basic}).  
\vspace{-0.1cm}
Increasing $M_d$  makes the perturbation more
global. Vortices in massive discs can gravitationally perturb parts of
the disc further out, similar to a planet. Lowering $Q_o$, vortices
develop longer, more prominent trailing  spirals exterior to them. This
is most notable with $Q_o=2$, where the vortex spirals can have
comparable amplitudes as the planet wake. However, lowering $Q_o$
even further, a global instability 
eventually develops as the edge disturbances perturb the disc strongly via
self-gravity.  

The $Q_o=1.5$ disc does not develop vortices. This is
consistent with the stabilisation effect of self-gravity on the vortex
instability through the linear response. Instead, $Q_o=1.5$ develops
global spiral instabilities associated with the gap edge. 
These are referred to as edge modes and are discussed in detail in
Chapter \ref{paper3}. One important difference from vortex modes is
that edge modes  are associated with local vortensity \emph{maxima},
as expected from energy arguments (\S\ref{S3.5}).  

\begin{figure}
  \centering
  \includegraphics[width=0.99\linewidth]{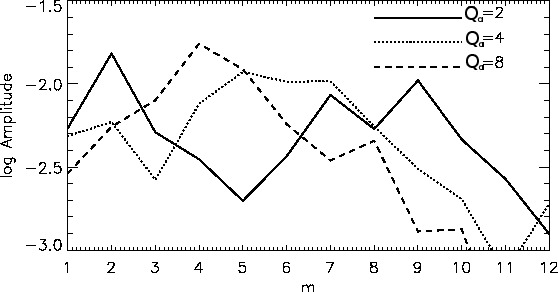} 
  \caption{Fourier amplitudes of the surface density in the outer disc
  $r\in[5,10],$ normalised by the axisymmetric component,
    for some of the  models illustrated in Fig. \ref{compare_v}. The
    plotted quantity is $\log|C_m/C_0|$, where $C_m=\int a_m(r)dr$ and
    $a_m(r)=\int \Sigma\exp{(-im\phi)}d\phi$.   
  \label{compare_amps}}
\end{figure}

The Fourier amplitudes of the surface density averaged over the region
$r\in[5,10]$, as a function of $m,$  is shown in
Fig. \ref{compare_amps} for $Q_o=2,\,4,\,8$. The snapshot is taken 
at $t=56P_0$. The shift to 
higher $m$ vortex modes with increasing disc mass is evident as
expected from linear calculations.  For $Q_o=8,4,2$ the preferred
vortex modes have  respectively  $m=4,\, 5$---7
and 7---9, with average amplitudes that decrease with  $Q_o$. The
latter is consistent the stabilisation effect of self-gravity
on  linear modes with low $m$.

An important region in Fig. \ref{compare_amps} is $m\leq
3$. There is a peak in amplitude at $m=2$ for $Q_o=2,\,4$. These are
the edge modes described above. They were not seen in linear
calculations because those calculations focused on finding vortex modes,
which have corotation at or close to local vortensity minima. The
loss of low $m$ vortex modes with increasing self-gravity observed in
linear calculations, is replaced by the increasing prominence of
global edge modes. Fig. \ref{compare_amps} shows the $m=2$ amplitude
becomes more significant with increasing self-gravity.   

\begin{figure}
    \centering
    \includegraphics[scale=0.39,clip=true,trim=0cm 0cm 1.68cm
    0cm]{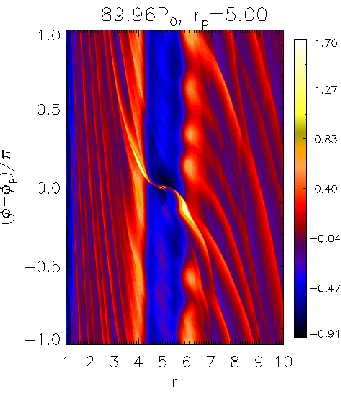} 
    \includegraphics[scale=0.39,clip=true,trim=2.3cm 0cm 1.68cm
    0cm]{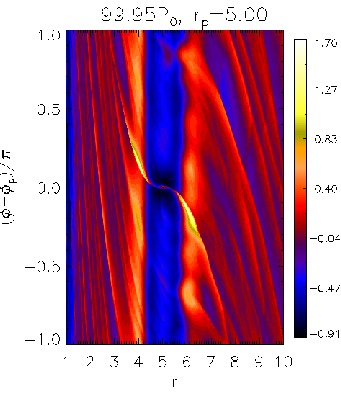} 
    \includegraphics[scale=0.39,clip=true,trim=2.3cm 0cm 1.68cm
    0cm]{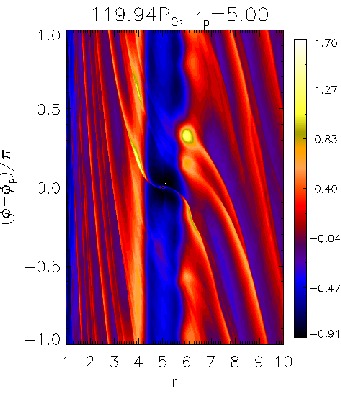}
    \includegraphics[scale=0.39,clip=true,trim=2.3cm 0cm 0.0cm
    0cm]{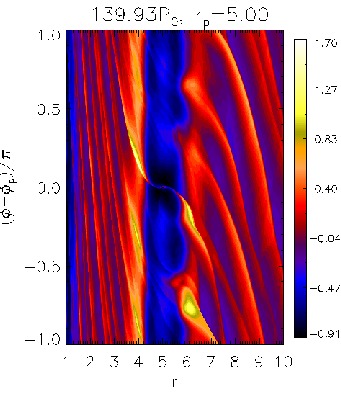} 
    \caption{Evolution of gap edge instabilities for the $Q_o=2$ case,
      for which both vortex and global spirals develop. Although
       Fourier amplitudes suggest $m=2$ global modes to be present
      early on, 3---4 global spirals can be identified in the last
      snapshot. These higher $m$ spiral modes may have resulted from
      nonlinear evolution of vortices perturbing the disc globally, rather than growth
      from linear instability of edge modes.
      \label{Q2_transition_case}}
  \end{figure}

The $Q_o=2$ simulation does display evidence of global disturbances
hindering vortex evolution. This transition case is difficult to
analyse since both vortices and spirals 
develop, as indicated by evolutionary snapshots of $Q_o=2$ in 
Fig. \ref{Q2_transition_case}.   
Fig. \ref{compare_amps} shows the 
low $m$ and high $m$ modes have comparable amplitudes in $r>5$. However,
if only averaged over the region $r\geq 7$, then in $Q_o=2$ the
$m=2$ mode was found to be dominant, because this spiral mode is
global  whereas vortex mode disturbances are localised to the 
edge ($r\simeq6$). Increasing self-gravity further, eventually edge
modes become dominant, this is seen in the $Q_o=1.5$ case in
Fig. \ref{compare_v}.  

\subsection{Evolution  and merging of self-gravitating vortices} 
\begin{landscape} 
   \begin{figure*} 
     \centering 
     \includegraphics[scale=.39]{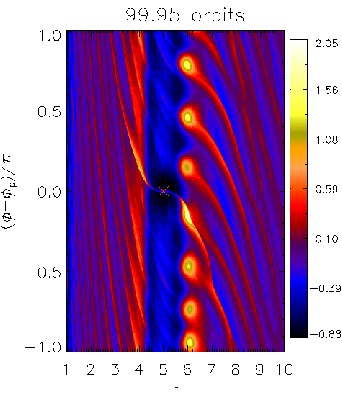} 
     \includegraphics[scale=.39,clip=true,trim=2.3cm 0cm 0cm
     0cm]{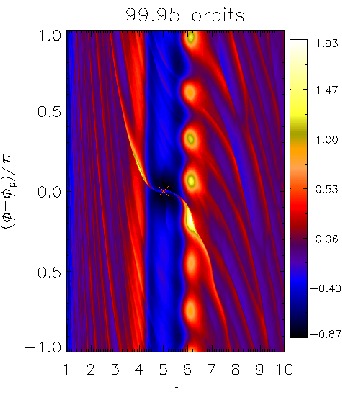} 
     \includegraphics[scale=.39,clip=true,trim=2.3cm 0cm 0cm
     0cm]{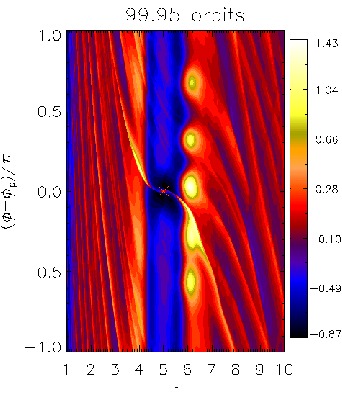} 
     \includegraphics[scale=.39,clip=true,trim=2.3cm 0cm 0cm
     0cm]{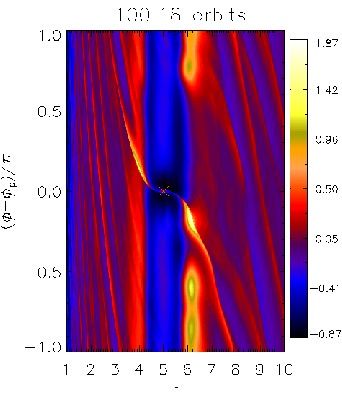}  
    \includegraphics[scale=.39,clip=true,trim=2.3cm 0cm 0cm
     0cm]{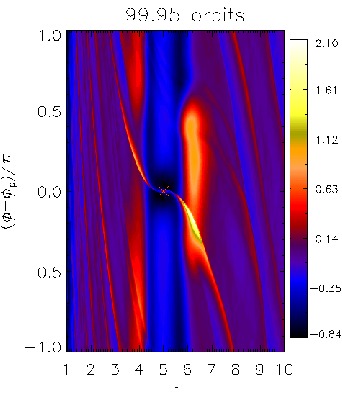} 
     \caption{Non-linear evolution of  vortex instabilities at the
       outer gap edge of a Saturn mass planet as a function of 
       disc mass, parametrised by the minimum Toomre parameter $Q_o$,
       from left to right, of $Q_o=2.5,\,3.0,\,3.5,\,4.0,\,8.0$. 
       \label{compare_v_nonlin}} 
\end{figure*}
\end{landscape}

Disc self-gravity also affects nonlinear interaction between vortices.  
Fig. \ref{compare_v_nonlin} compares surface density 
perturbations for a range of $Q_o$ at $t=100P_0.$  Since the
vortices emerge at roughly $t=56P_0$, they have evolved for a  similar
time. Typical growth times for vortex modes 
are  $t_g\sim 6P_0$, so the vortices have evolved for
about $7t_g$, well into the nonlinear regime.   

The snapshot for $Q_o = 8$ shows  a single vortex resulting from 
merging of the initial vortices. The vortex disturbance is largely
confined to within  a local scale-height of the gap edge.
 A single large vortex is the typical result for simulations with no
 self-gravity \citep[e.g.][, and in Chapter \ref{paper1}]{valborro07}
 , although the self-gravitating vortex
 here is thinner than the completely non-self-gravitating simulation
 (Fig. \ref{vortices_sg_nsg_later}). The $Q_o = 4$ disc shows vortex
 merging taking place, as   
individual surface density maxima can still be identified within the
large vortex behind the shock. 
 For  $Q_o=4$,
 a single vortex  forms at $t\simeq 110P_0$.

Fig. \ref{compare_v_nonlin} shows  that increasing the disc mass delays
vortex merging. For $Q_o=3.5$, 5 vortices remain and for $Q_o=3$ and $Q_o=2.5$, 
7 vortices remain. They have not merged into a single vortex as happened when
$Q_o=8$. A  25\% increase in disc mass as  $Q_o=4\to3$ causes
merging to be  delayed by $50P_0$. This suggests that increased gravitational
interaction between vortices opposes merging.

As $Q_o$ is lowered  
the inter-vortex distance increases. When  $Q_o=2.5$ their
azimuthal separation can be larger than the vortex itself. Also, vortices
become less elongated and more localised being symptomatic 
of  gravitational condensation. Trailing wakes 
from vortices become more prominent as the vortices more strongly
perturb the disc via their  self-gravity. In fact, the planetary wake becomes less
identifiable among the vortex wakes. Notice  vortex wakes are
mostly identified with vortices upstream of the planetary wake, rather
than just  downstream. Vortex wakes appear to detach
from the vortex after passing through the planetary shock. 
There is a sharp contrast in gap structure between the $Q_o=2.5$ and $Q_o=8$
cases. The single vortex that results when 
 $Q_o=8$ is aligned along the outer gap edge, which is
still approximately identified as circle  $r=6.$
 However, when  $Q_o=2.5$ 
the vortices and  their wakes intersect the circle $r=6$ 
 making the radius of the outer gap edge less well-defined.  

\begin{figure}
  \centering
  \includegraphics[width=0.99\linewidth]{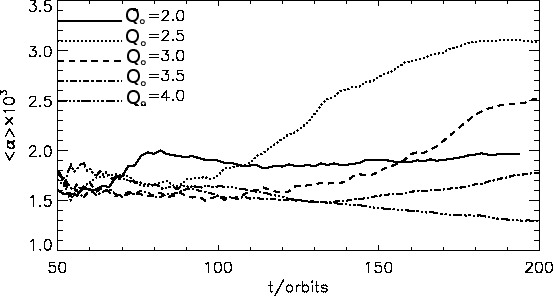}
  \caption{Running time average of the 
    viscosity parameter $\alpha$ averaged over  the vortex region
    for a range of disc models.  
    \label{alpha_visc}}
\end{figure}

An important feature of vortices is their wakes are 
associated with the excitation of density waves
which transport  angular momentum transport outwards
\citep[see e.g.][]{paardekooper10}, even without self-gravity.  
A measure of this can be made through the $\alpha$ viscosity
parameter. Here, it is defined as

\begin{align}
  \alpha(r,\phi) \equiv \Delta u_r\Delta u_\phi/c_s^2(r)
\end{align}
where $\Delta$ denote deviation from azimuthal averaged values. For
diagnostics, 
$\alpha$ is spatially averaged over the vortex region $r\in [5.7,7.1]$
and its running-time average $\langle \alpha \rangle$ is plotted in
Fig. \ref{alpha_visc}. 
The parameter $\langle \alpha\rangle $  associated with vortices
is $O(10^{-3})$, an order of magnitude larger than the imposed value 
associated with $\nu.$ 
When $Q_o=4$, $\langle\alpha\rangle$ decays steadily
 after an initial transient growth. The case
$Q_o=2$ has a roughly constant  $\langle\alpha\rangle \simeq 1.9\times10^{-3}$
for $t\gtrsim 80P_0$ after a brief growth around $t\sim
70P_0$. Interestingly, for intermediate strengths of self-gravity,
$Q_o=2.5$---3.5, there is growth in $\langle\alpha\rangle$ over
several tens of orbits. 
  
The behaviour of $\langle\alpha\rangle$ as a function of $Q_o$ is
consistent with the general picture of
vortex formation and evolution (Fig. \ref{compare_v_nonlin}).  
The decay in $\langle\alpha\rangle$ is associated with vortex merging
leading to fewer vortices. This is the case for $Q_o=4$. 
Vortex merging happens more readily  
for lower disc masses, hence although multiple vortices  develop
from the instability, this phase does not last long enough for the
multi-vortex configuration to significantly transport angular
momentum. Increasing self-gravity to  $Q_o=3.5\to2.5$,
 vortices become less prone to merging and the multi-vortex phase
 lasts longer. 
 They have time to evolve into compact objects
that further perturb the disc.
 This is consistent with fact that $\langle\alpha\rangle$-growth is prolonged 
with increased disc mass, as merging is delayed.
 However, if self-gravity is too strong, such as when  $Q_o=2$, growth is
again limited, because the $m=2$ global spiral mode develops and hinders vortex
evolution.


\subsection{ Long term evolution of gap edge vortices}
\begin{figure}[!ht]
\centering
\includegraphics[width=.99\linewidth]{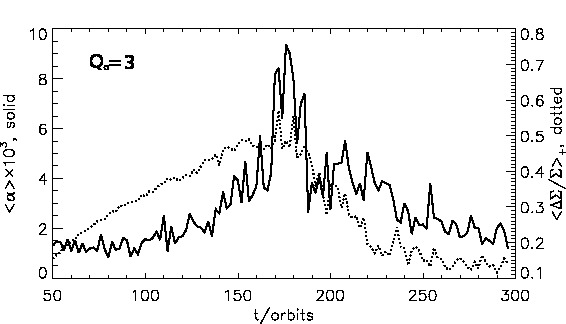}\\
\includegraphics[scale=0.45]{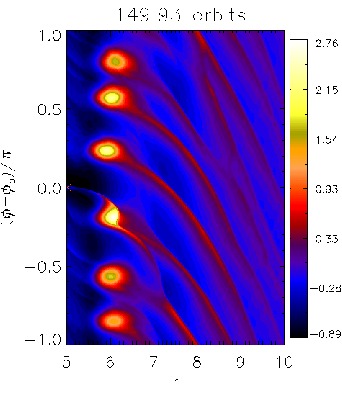}
\includegraphics[scale=0.45,clip=true,trim=2.3cm 0cm 0cm
0cm]{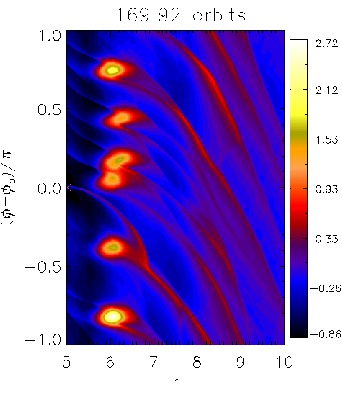} 
\includegraphics[scale=0.45,clip=true,trim=2.3cm 0cm 0cm
0cm]{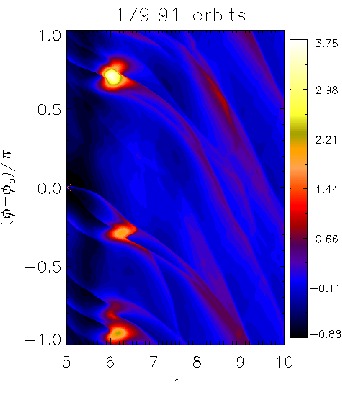} 
\caption{Vortex evolution  for the disc with $Q_o=3$. Line plot: instantaneous
  $\alpha$ viscosities (solid) and average over-density
  (dotted). Contour plots:  relative surface density
  perturbations before and after vortex merging.
\label{Qm3_alpha}}
\end{figure}

The long term evolution of vortices in the $Q_o=3$ disc is shown 
in Fig. \ref{Qm3_alpha}. It shows the  instantaneous $\alpha$ viscosity
parameter, average over-density (defined in  regions where the
relative surface density perturbation is positive) and contour plots
of relative surface density perturbation in the vortex region.  

The viscosity parameter 
 $\alpha$ grows from $t=100P_0$ to $t=170P_0$ with $\mathrm{max}(\alpha) \simeq
8\times 10^{-3}$, which is $\sim50$ times larger than the background
viscosity. At $t=150P_0$ there remains 6 distinct vortices (one vortex just 
passing through the planetary wake). The multi-vortex
configuration has been maintained for a further$\sim 50P_0$
since Fig. \ref{compare_v_nonlin}. With weak self-gravity, a single
vortex would have formed through 
merging. Delayed merging allows individual vortices to evolve and
become planet-like. The typical
over-density in a vortex is $\sim 1$ at $t=100P_0$ and increases to
$>1.5$ by $150P_0.$  As a consequence one expects increased
disturbance in the surrounding disc. The phase of $\alpha$-growth
correlates with a linear increase in the average over-density in the
vortex region.  

\begin{figure}[!ht]
\centering
\includegraphics[width=0.49\linewidth]{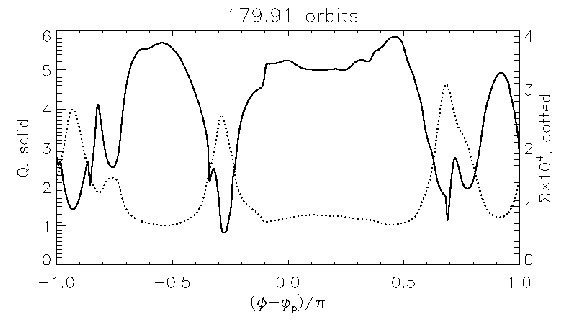}
\includegraphics[width=0.49\linewidth]{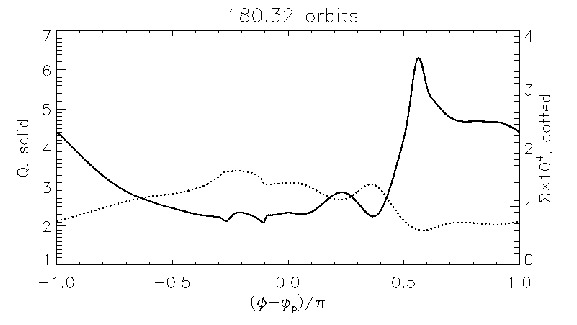}
\caption{Comparison of post-merging vortices for discs with  $Q_o=3$ (left) and
  $Q_o=4$ (right). The Toomre $Q$ (solid) and surface density (dotted)
  averaged over $r\in[6,6.5]$ is shown as a function of azimuth relative
  to the planet. 
\label{compare_merged_vortex}}
\end{figure}

At $t=170P_0$, Fig. \ref{Qm3_alpha} shows 6 vortices still remain, 
 and the parameter $\alpha$ is a  maximum.
However, after a  burst of vortex merging events, 
$\alpha$ decreases rapidly. The snapshot at $t=180P_0$ shows a quieter
disc with only 3 vortices, but of similar size to those before
merging. This is unlike cases with weak self-gravity where a
larger vortex results from
merging. Fig. \ref{compare_merged_vortex} compares post-merging
vortices in discs with  $Q_o=3$ and $Q_o=4.$  
For $Q_o=3$, the post-merging vortices are localised in azimuth,
with  $Q\sim 1$ and the contour plot shows the densest vortex has
an over-density of $\simeq 3.75$. However,  for  the disc with  $Q_o=4$,
a single vortex forms that extends about half the total azimuth and
has $Q>2$. This hints at gravitational collapse of vortices in the
$Q_o=3$ disc. 

Note that the mass of the  $Q_o=3$ disc is only $M_d=0.031M_*$,
usually considered  insufficient for fragmentation into bound objects by
classic Toomre instability (though it is still massive compared to
more typical protoplanetary disc masses of $M_d\sim 0.01M_*$).  
However, the comparison above shows that adding self-gravity to
the vortex instability  enables collapse into compact objects that 
survive against shear.   

\begin{figure}[!ht]
   \centering
   \includegraphics[width=0.99\linewidth]{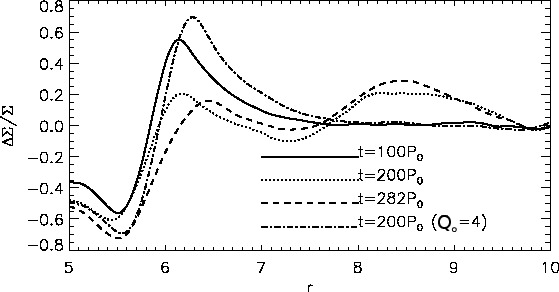}
 \caption{The  
gap profile for the $Q_o=3$ disc at different times 
 (solid, dotted and dashed) as shown through
	relative surface density perturbation. A snapshot of the
profile for the  disc 
with  $Q_o=4$ (dash-dot) is also 
	plotted for comparison. \label{v3_gap_evol}}
 \end{figure}

Vortices in $Q_o=3$ noticeably affect the gap structure. 
Fig. \ref{v3_gap_evol} shows several snapshots
of the gap structure from $t=100P_0,$ when there were
 multiple vortices, to the end of the simulation 
when a vortex pair remained.
 The quasi-steady gap profile for $Q_o=4$ is also shown for
 comparison. Neither  the single 
vortex in  $Q_o=4$ nor multiple vortices with $Q_o=3$
affect the one-dimensional 
gap profile  at  $t=100P_0$  because the disturbances only
redistribute mass in the azimuthal direction. 
However, in the $Q_o=3$ disc, the original bump at the outer
gap edge is diminished after vortex  merging takes place at $t=200P_0.$   
A surface density depression of  $\Delta\Sigma/\Sigma \simeq -0.1$ then
develops at $r=7.3$ and a bump develops at $r=8.5.$  These features
last to the end of  the simulation.

Self-gravitating vortices behave like planets. Assuming vortices lie near the surface
density maximum at $r=6,$ the  creation of the surface density deficit  
in  $r\in[7,7.5]$ and the new surface density maximum  at $r=8.5$
could  be induced,  in a similar manner as  it would be by  a planet, 
through  the outward transport of angular momentum by the density
waves launched by the vortices. 

Roughly speaking, treating $m$
vortices at $r=r_v$ as a gravitational perturbation on the exterior
disc, it exerts a positive torque at $r_L =
r_v(1+1/m)^{2/3}$. Inserting $r_v\sim6$ and $m\sim3$
(Fig. \ref{Qm3_alpha}---\ref{compare_merged_vortex}) gives
$r_L\sim7.3$. The  material removed from the region of the deficit,
which has gained angular momentum, ends up  contributing to the new
surface density maximum at $r=8.5.$   

\subsection{Anticyclonic vortices}

\begin{figure}
  \centering
  \subfigure[Gap vortices, $\Delta
  \Sigma/\Sigma$\label{v3_vortex}]{\includegraphics[width=0.32\linewidth]{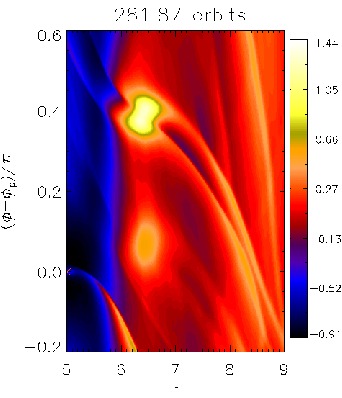}} 
  \subfigure[Gap vortex, $\eta$
  \label{v3_vorten}]{\includegraphics[width=0.32\linewidth]{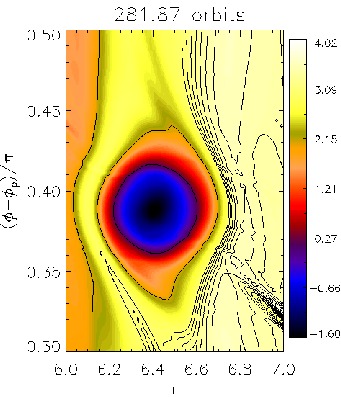}} 
  \subfigure[Artificial vortices, $\Delta \Sigma/\Sigma$
  \label{kida_vortex}]{\includegraphics[width=0.32\linewidth]{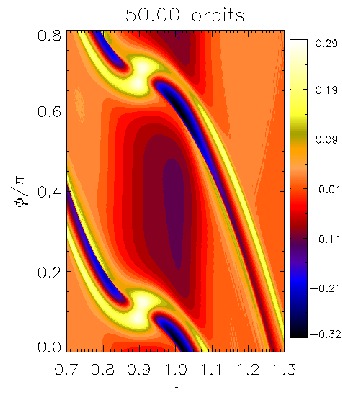}} 
\caption{The final vortex pair at the end of the  disc-planet
  simulation for $Q_o=3:$ (a) Relative surface density perturbation contour plot (b)Vortensity contours for the upper vortex. 
 Shown in (c) is the relative surface density perturbation for  a pair of  vortices formed by imposing
Kida vortex solutions as a perturbation in 
  a standard power-law disc. \label{Qm3_final_vortex}
}
\end{figure}

For the $Q_o=3$ disc, a vortex-pair forms at $t\simeq 230P_0$ and
lasts until the end of the simulation ($t\simeq 300P_0$) and is
reminiscent of co-orbital planets. The pair survives on a timescale
well beyond which merging occurs in non-self-gravitating simulations. 
A snapshot is shown in Fig. \ref{v3_vortex}. Two blobs can be
identified along the gap edge, the upper vortex being more over-dense
than the lower one. They lie at a radius $r_v = 6.4$, corresponding to
local surface density maximum  in azimuthally
averaged  one-dimensional profiles, which is expected to be
neutral  for vortex migration \citep{paardekooper10}.

The upper vortex in Fig. \ref{v3_vortex} is different to  the pre-merging
vortices or those with weak self-gravity. It
has two spiral disturbances extending from the vortex to $(r,\phi -
\phi_p) = (9,-0.2\pi)$, whereas the pre-merging vortices have one trailing
spiral.  Its vortensity field is shown in Fig. \ref{v3_vorten}.  The
vortex core has $\eta < 0$, whereas the final large vortex in the disc
with  $Q_o=4$
has $\eta >0$ in its core (though it is still a local minimum).

The region with $\eta < 0$ has a
mass of  $M_v\simeq 5.46\times10^{-5}M_*\simeq 0.18M_p$ and average radius
$\bar{r} \simeq 0.92H(r_v)$ about the vortex centroid. It has  semi-major
and semi-minor axes of $\sim 1.57H(r_v)$ and $\simeq 0.55H(r_v)$ 
respectively, corresponding to aspect-ratio of 2.9.     
This is about 18 Earth masses if
$M_*$ is Solar. Denoting the mean square relative velocity (to
the vortex centre) as $\Delta v^2$,  $GM_v/\bar{r}\Delta v^2
\sim 3.9$. This region is gravitationally bound. Although planets of
such mass are not expected to open significant gaps, the surface
density deficit in $r=[7,7.5]$ in Fig. \ref{v3_vortex}  indicates
vortices may do so. The process could be assisted by
the fact that a vortex of size $H$ produces a  perturbation of a
magnitude similar what would be produced by Saturn when  $h=0.05$ 
\citep{paardekooper10}. 


Finally, Fig. \ref{kida_vortex} show $\Delta\Sigma/\Sigma$ for a pair
of Kida-like vortices placed in a disc. Note the similarity between
the Kida-like vortex and the upper vortex in 
Fig. \ref{v3_vortex}, particularly the double wake structure, the
tilted core and the surface density deficit just outside the
vortices. This is a surprising coincidence,
given that the simulation setups that produced them  were completely
different (for more details see below).  

\section{Simulations of co-orbital Kida vortices}\label{hydro2}

One of the effects of self-gravity shown in simulations above is to
delay vortex merging. This effect is significant when vortices develop
into compact structures. Furthermore, the similarity between the final
vortex in the disc with $Q_o=3$  and a Kida-like vortex \citep{kida81}
motivates one to consider the effect of self-gravity on the interaction
between two Kida vortices. 

Simulations presented here are supplementary and focus on the
interaction between vortices without a planet. 
This isolates effects due to vortex-vortex interactions
and the influence of self-gravity without interference from the
planet. 

\subsection{Numerical setup}
A standard power law disc is adopted, with initial surface density profile
$\Sigma=\Sigma_0 r^{-1/2},$ where $\Sigma_0$ is a scaling
constant. The disc occupies $r\in [0.25,2.5]$. The value of $\Sigma_0$
is chosen to provide  a specified  $Q_\mathrm{Kep}$ at a specified radius as
before. The initial 
azimuthal velocity is chosen so that the disc is initially 
in hydrostatic equilibrium. The initial radial velocity is zero. The disc is
locally isothermal with $h=0.05$. The viscosity is 
$\nu=10^{-9}$ which is essentially the inviscid limit. No planets are
introduced and the notation $(r_p,\phi_p)$ is used to denote a
vortex centroid.  

Artificial vortices are set up by 
following \cite{lesur09b}.  
Details of the implementation
is described in Appendix \ref{kidasetup}.
Two vortex perturbations are imposed with initial angular
separation $\theta = \pi/2$\footnote{Not to be confused with the polar angle in 3D.}. Table \ref{experiments} summarises the
numerical experiments, which considers switching self-gravity on or off, a
range of disc masses (here specified   through $Q_1\equiv Q_\mathrm{Kep}(1)$) 
and initial radial separations  $X_0$ of the two vortices (so that for
one of the vortices, $r_p\to r_p + X_0$).     

The fluid equations are evolved using the \fargo code with resolution
$N_r\times N_\phi = 800\times 2400$ giving $\simeq 17$ grid points per
scale height.  The vortices are typically of that size. Damping
boundary conditions are applied \citep{valborro06}. 

\begin{table}
  \centering
    \caption{Parameters for two-vortex interaction simulations.
     The first column gives the nomenclature for the simulations. 
     The  initial radial separation is $X_0$ and  $Q_1$ is the
    Keplerian Toomre parameter $Q_\mathrm{Kep}$ at unit radius.
     The fourth column indicates whether self-gravity was included.} 
  \begin{tabular}{lcccl}
    \hline
    Case & $X_0/H $ & $Q_1$ & self-gravity \\ 
    \hline\hline
    Fnsg  & 0  &  8.0 &  NO \\
    Fsg   & 0  &  8.0 &  YES \\
    M1  &  0 & 15.9 & YES \\
    M2  &  0 & 5.3 & YES \\
    S1  &  0.1 & 8.0 & YES \\
    S2  &  0.2 & 8.0 & YES \\
    S3  &  0.3 & 8.0 & YES \\
    S4  &  0.6 & 8.0 & YES \\
    \hline
  \end{tabular}
  \label{experiments}
\end{table}

\subsection{Vortex pairs}
\begin{figure}
  \centering
  \includegraphics[width=0.5\textwidth]{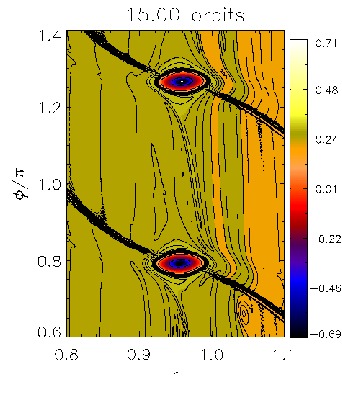}
  \caption{Vortices formed by imposing Kida vortex solutions as a 
    perturbation to a global disc. The scaled vortensity field is
    shown. 
    \label{nsg_vortices}}
\end{figure}

As an example of vortices formed by the above procedure, 
Fig. \ref{nsg_vortices} shows  the vortensity field for the case Fnsg.  
The vortex pair circulates at $r\simeq 0.96$ and are localised in
radius and azimuth.   
The vortex centroids have vorticity $-1.57$ (dimensionless units) in
the non-rotating frame, 
 close to the local Keplerian shear ($-1.5r_p^{-3/2}\simeq -1.59$).  
 The upper vortex has a long tail reaching
the outer part of the lower vortex rather than its centroid.
The wakes associated with each vortex are
similar to those induced by a planet and is responsible for vortex
migration  \citep{paardekooper10}.  

Focusing on one vortex, the region with negative absolute vorticity has half
width in radius and azimuth of $\simeq 0.6H$ and $1.3H$
respectively.
 This region has mass $m_v \sim
1.2\times 10^{-5}M_*$, or about $4M_\earth$ if $M_*=1M_\sun$.    
The estimate is not far from the assumption that the final vortex has
size $H$ with average density that of the unperturbed
disc at the location it was set up, which gives 
$m_v\sim 1.6\times10^{-5}M_*$. If self-gravity is enabled, one can expect
gravitational interaction between vortices to behave like that between  two
co-orbital planets of at least a few Earth masses. 

\subsection{Simulation results}
\begin{figure}[ht]
  \centering
  \includegraphics[width=.24\linewidth,clip=true,trim=1.8cm 2.1cm 1.9cm 1.1cm]{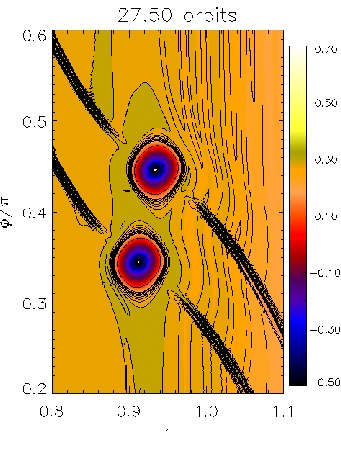}
  \includegraphics[width=.24\linewidth,clip=true,trim=1.8cm 2.1cm 1.9cm 1.1cm]{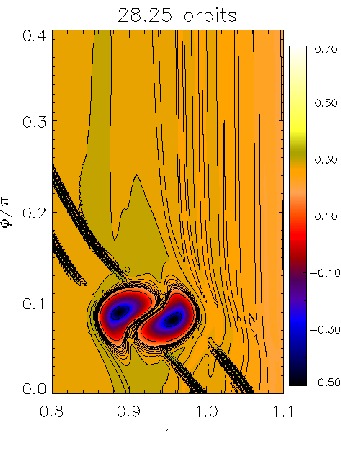}
  \includegraphics[width=.24\linewidth,clip=true,trim=1.8cm 2.1cm 1.9cm 1.1cm]{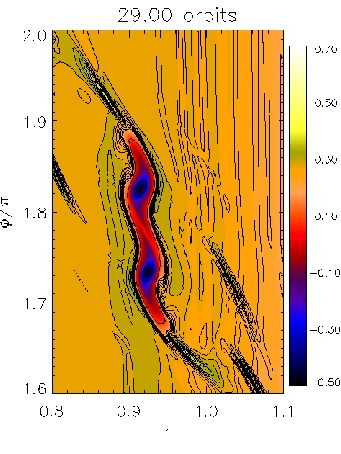}
  \includegraphics[width=.24\linewidth,clip=true,trim=1.8cm 2.1cm 1.9cm 1.1cm]{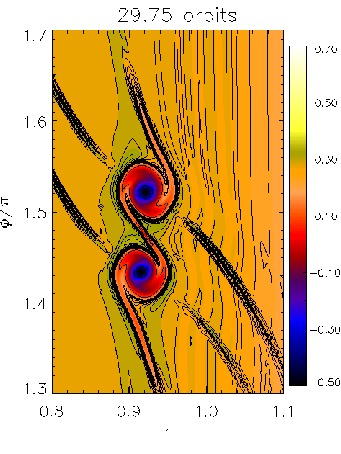}\\
  \includegraphics[width=.24\linewidth,clip=true,trim=1.8cm 2.1cm 1.9cm 1.1cm]{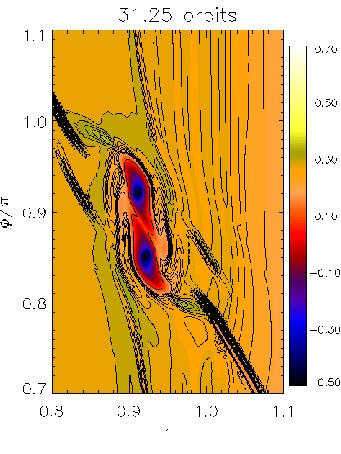}
  \includegraphics[width=.24\linewidth,clip=true,trim=1.8cm 2.1cm 1.9cm 1.1cm]{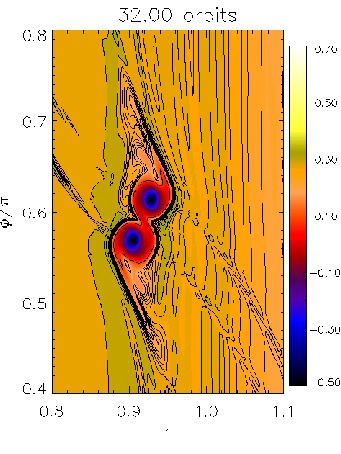}
  \includegraphics[width=.24\linewidth,clip=true,trim=1.8cm 2.1cm 1.9cm 1.1cm]{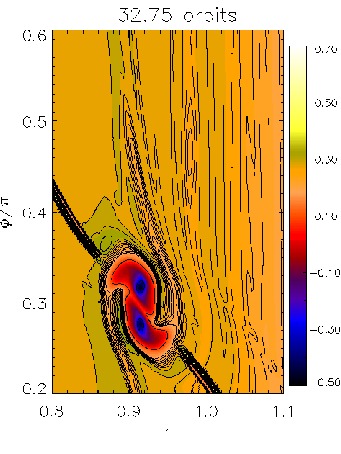}
  \includegraphics[width=.24\linewidth,clip=true,trim=1.8cm 2.1cm 1.9cm 1.1cm]{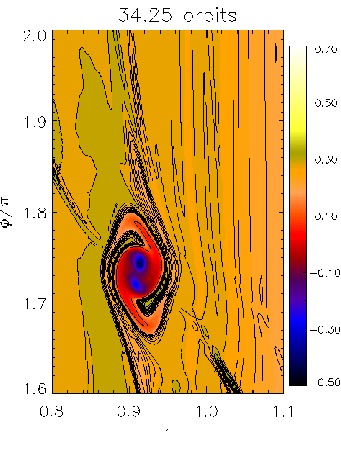}
  \caption{Typical merging of a vortex pair. There is no
  qualitative difference in this process between self-gravitating or 
  non-self-gravitating vortices. The scaled vortensity field is
  shown. The vertical direction is azimuthal, and horizontal is
  radial. The vortices have size $\sim H$. 
  \label{nsg_vortices_merging}}
\end{figure}

\begin{figure}[ht]
  \centering
  \includegraphics[width=0.5\linewidth]{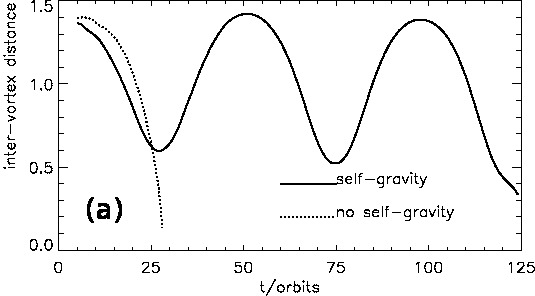}\includegraphics[width=0.5\linewidth]{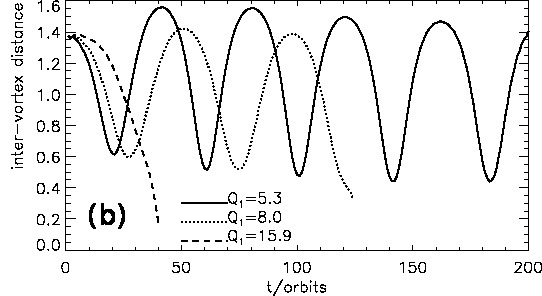}\\
  \includegraphics[width=0.5\linewidth]{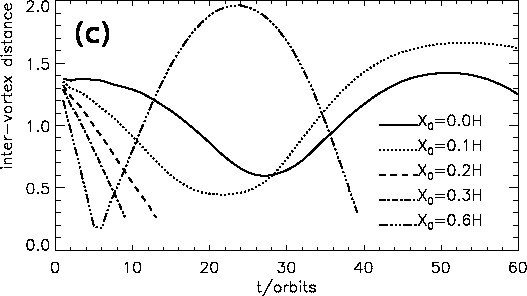}
  \caption{Inter-vortex distances for vortex pair simulations: (a) With and 
    without self-gravity for $Q_1=8$ ( simulations  Fsg and  Fnsg) (b) For  discs of different masses
    characterised  by $Q_1$, the Keplerian Toomre parameter at $r=1,$ self-gravity being included
    (simulations Fsg, M1, and M2).
    (c) Vortices input with  varying initial radial separation, $X_0,$ with self-gravity included
    for disc models with $Q_o=8$( simulations Fsg, S1,S2,S3, and S4).
    \label{kida_expts}}
\end{figure}

Consider first the cases Fsg and Fnsg. These are for the disc with
$Q_1=8$ with and without self-gravity respectively.  The distance
between vortex centroids as a function of time is shown in
Fig. \ref{kida_expts}(a). It takes almost 5 times as long for the
self-gravitating vortices to merge. A merging event is
shown in Fig. \ref{nsg_vortices_merging}.

By examining the vortensity distribution,
non-self-gravitating vortices were observed to begin merging at $t=28P_0,$
but self-gravitating vortices at $t=126P_0.$   For Fnsg,  
vortices  approach each other  within $\sim 23P_0$ of formation and  merge. 
Assuming they move relative to each other because of  Keplerian shear, 
the time taken to merge implies that 
their radial separation at vortex formation must be 
$\simeq 0.14H.$ This length-scale is only resolved by about 2.5 grid
cells  indicating that a non-zero initial separation is generated by
grid effects, despite the vortex perturbations being imposed at the
same radius. This is actually not unphysical because 
vortex formation through instability is not expected to occur at exactly the
same radii.    


In the self-gravitating case Fsg, the inter-vortex distance
oscillates with period $50P_0$ and there are two close-encounters
before merging. 
This is analogous to the survival of  the vortex pair towards the end of the 
disc-planet simulation for $Q_o=3$. Given a  maximal separation of $\simeq 1.4$ and
that the vortices  are circulating  near unit radius, the maximal angular
separation is  $\simeq \pi/2$. 
 The minimal separation during the first two close encounters  is $\simeq 0.6$, or about $12H$
which implies  that the vortices are on tadpole
orbits.  
As the minimum separation is much larger than the typical vortex size,
 merging does not occur during the first two close encounters.

The experiment in Fig. \ref{kida_expts}(b) varies
the strength of self-gravity via 
$Q_1$. The reference case has $Q_1=8$.  
Doubling $Q_1$ to $Q_1=15.9$ (case M1) weakens self-gravity and vortices merge within few tens
of $P_0.$  For $Q_1=5.3$ (case M2) the
oscillation period is $\simeq 40P_0$ and maximal
separation is 1.55, larger than for  $Q_1=8$, as is the first minimum
separation: the increased self-gravity has enhanced the mutual repulsion of
vortices. The two-vortex configuration in M2 lasts until the end of the
simulation, but
there is a secular decrease in the minimum
separation (of $0.6$ at $t=20P_0$ and $0.45$ at $t=180P_0$) due to
vortex migration. Merging will likely occur eventually. Consistent with
the behaviour seen for  gap edge vortices, increasing self-gravity
delays merging.   

The final experiment  varies the  initial
radial separation of the vortex perturbations.
Results are  shown in Fig. \ref{kida_expts}(c). Increasing $X_0$
to $0.1H$ from the reference case, the first minimum separation decreases to 0.45
from 0.6: vortices approach each other more closely, but still repel and
undergo co-orbital dynamics. The increased oscillation amplitudes  imply a larger tadpole
orbit. For $X_0=0.2H$ and $0.3H$, vortex separation decreases more rapidly
and become small enough for vortex merging. However, for $X_0=0.6H$,
vortices simply circulate past by each other and no merging occurs,
despite reaching a similar minimal separation  as when
$X_0=0.2H,\,0.3H$. Vortices should
collide head on if they are to merge. 

\subsection{Vortices as co-orbital planets}\label{vortices_planets}
Simulations above indicate that self-gravitating vortex pairs behave 
like co-orbital planets. This means there exists an initial radial 
separation (or impact parameter) below which  vortices execute 
horseshoe orbits relative to each other. Analytical and numerical work indicates that vortices 
merge if their centroids are within $d\sim 3s$
of each other \citep[e.g.][]{zabusky79,melander88}, where $s$ is the
vortex size. Although one cannot assume the same critical $d/s$ apply
here because the situations  are very different, it is still 
reasonable to expect merging if vortices can reach within some
critical distance of one another. 

The results above can  be anticipated from existing treatments 
of co-orbital  dynamics. The first is   
\cite{murray00}'s  model of the co-orbital satellites of Saturn, the
Janus-Epimetheus system.  The governing equation from \cite{murray00}
gives a relationship between two configurations,  the  `final'
configuration (subscript $f$) and the   `initial' configuration
(subscript $i$) in the form 
\begin{align}\label{murray_model}
  &\left(\frac{\Delta r_\mathrm{i}}{r_0}\right)^2 - \left(\frac{\Delta
      r_\mathrm{f}}{r_0}\right)^2 =
  -\frac{4}{3}q\left[H(\theta_\mathrm{i}) - H(\theta_\mathrm{f})\right],\\
  &H(\theta) = \left[\sin{(\theta/2)}\right]^{-1} - 2\cos{\theta} - 2 \notag
\end{align}
where $\Delta r$ is the radial separation of the satellites,
 $r_0$ the average orbital
radius (assumed fixed, thereby ignoring migration), $\theta$ their
angular  separation and $q=\mathcal{M}/M_*,$  with $\mathcal{M}$ being
the sum of the satellite masses. It is assumed $q\ll1.$  

Eq. \ref{murray_model} should apply to vortex pairs if there is
significant mutual gravitational interaction. Their
final radial separation is assumed zero, $\Delta r_\mathrm{f}=0$, when
the vortices are at their minimum angular separation.  
Inserting $\theta_i = \pi/2,$
$\Delta r_\mathrm{i} = 7.1\times10^{-3}$ (corresponding to the initial
conditions deduced for Fnsg), $r_0=1$ and
$q=2.3\times10^{-5}$ (corresponding to the measured mass of the
negative absolute vorticity region in Fsg), Eq. \ref{murray_model} gives    
$\theta_\mathrm{f} \simeq 0.41$ or a minimal angular separation of $\sim
8H$ so merging is not expected if vortices have size $H.$ 
This estimate is lower than the observed value of 0.6, but 
this $q$ is a lower limit on the vortex mass. Inserting $q=1.2\times10^{-4}$ gives
$\theta_f = 0.6$ assuming all other parameters remain the same. This
implies the effective gravitational mass of the  vortex should be 20 Earth
masses. 

\begin{figure}[ht]
  \centering
  \includegraphics[width=0.99\linewidth]{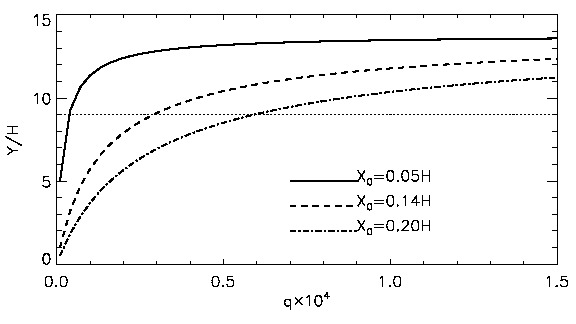}
  \caption{Merging conditions based on
    \citeauthor{murray00}'s model of co-orbital satellites. $Y$ is the
    minimum inter-vortex   separation and the horizontal line is a
    hypothetical critical   separation below which merging occurs.
    Thus merging is less likely for larger $q$. \label{merge_cond_q}}
\end{figure}

Eq. \ref{murray_model} is illustrated in
Fig. \ref{merge_cond_q}.  For fixed $X_0$ and increasing $q,$
(e.g. as a by product of increasing the disc mass)
the minimal separation $Y$ increases, eventually
becoming too large for merging to occur.
For fixed $q,$  $Y$ decreases
as $X_0$ increases. This is  similar  to a test particle on a
horseshoe orbit in the restricted three body problem.
The larger its impact parameter, 
 the closer it approaches the secondary mass.  

\begin{figure}[ht]
  \centering
  \includegraphics[width=0.99\linewidth]{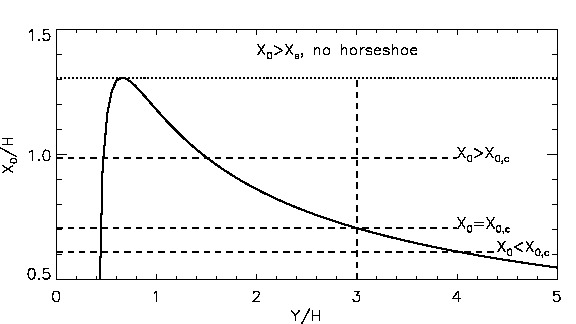}
  \caption{Merging conditions based on shearing
    sheet dynamics. For a given $X_0$, the allowed values
    of minimum separation $Y$ lies between the intersection of the horizontal
    line $X_0/H =$ constant
    and the solid curve. 
    The vertical dashed line is a hypothetical critical
    separation, below which merging occurs.\label{min_sep}}
\end{figure}

Mutual horseshoe turns may also be analysed in a shearing sheet
approximation. This is just the generalisation of the result 
for a test particle and a planet, to the
two-planet case (Appendix \ref{coorb_vortices}). Since the vortices
have characteristic sizes of the local scale-height, it is convenient
to non-dimensionalise lengths by $H$, in the expression for the
maximum initial radial separation of vortices that allow horseshoe
turns: 
\begin{align}\label{min_sep2}
  \hat{x}_0 < \frac{1}{h}\left(\frac{8q}{3\hat{y}h} -
    \frac{q^2}{3\hat{y}^4h^4}\right)^{1/2},
\end{align}
where $\hat{x}_0 = X_0/H,\, \hat{y}=Y/H$. Given $X_0$,
Eq. \ref{min_sep2}  can be inverted to give the range of possible $Y.$ 
This inequality is displayed in Fig. \ref{min_sep}. There exists a
critical $X_0=X_s$ beyond which  
vortices do not execute horseshoe turns and instead
circulate past  each other. There may also exist a value $X_{0,c}$
such 
that   $X_{0,c}< X_s$ and  for $X_{0,c}<X_0<X_s$, vortices execute
U-turns  
and the values of $Y$ attained are sufficiently small to allow
merging.  
This happens  only when the vortices have small enough mass. 
When $X_0<X_{0,c}$, $Y$  can be larger 
than critical, so that merging does not occur.

These simple models give qualitatively the same results as vortex pair and
disc-planet simulations, and serve  to explain the effect of
self-gravity delaying merging by interpreting vortices as co-orbital
planets that execute mutual horseshoe turns, thereby imposing a
minimum inter-vortex separation (mainly in the azimuthal direction). 
If this minimum separation is still larger than a critical separation known to exist
for vortex merging, then merging cannot occur. Hence, multiple-vortex
configurations can be sustained longer  as the strength of
self-gravity is increased.  


\section{Implications on vortex-induced migration}\label{migration}

Chapter \ref{paper1} established that vortices at gap edges can lead to
brief phases of rapid inwards migration of giant
planets.  In summary, gap edges become unstable to
vortex modes leading to  vortex 
mergers  on dynamical timescales, 
in turn resulting in a large-scale vortex circulating at  either gap edge.
Upon approaching the planet, the inner vortex can
execute a horseshoe turn and  move  outwards 
across the gap, gaining angular momentum.
Thus, the planet loses angular momentum and is 
scattered inwards.

\begin{figure}[ht]
   \centering
   \includegraphics[width=0.99\linewidth]{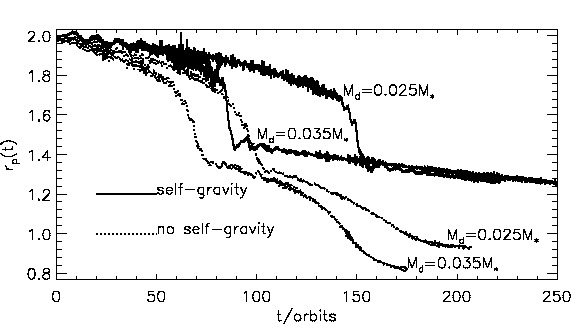}
   \caption{Vortex-induced migration with (solid) and without (dotted)
     disc self-gravity. The discs are inviscid and initially with
     uniform surface density.
     \label{sgNsg}} 
\end{figure}

\begin{landscape}

  \begin{figure}
    \centering

    \includegraphics[scale=0.39,clip=true,trim=0cm 1.8cm 1.96cm 0cm]{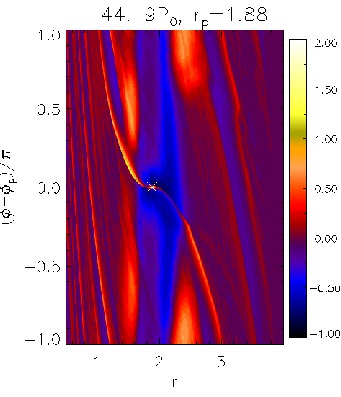}
    \includegraphics[scale=0.39,clip=true,trim=2.3cm 1.8cm 1.96cm 0cm]{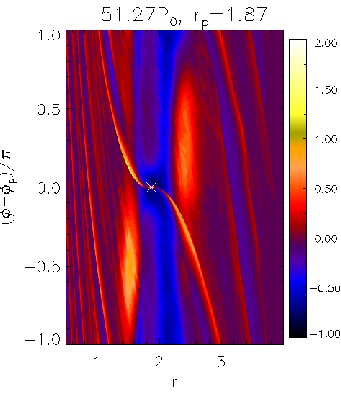}
    \includegraphics[scale=0.39,clip=true,trim=2.3cm 1.8cm 1.96cm 0cm]{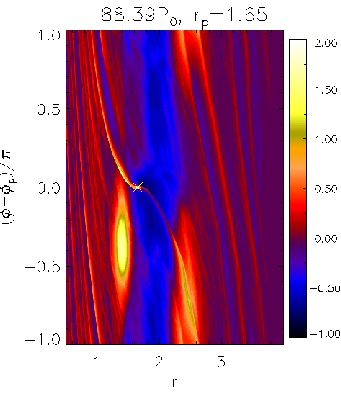}
    \includegraphics[scale=0.39,clip=true,trim=2.3cm 1.8cm 1.96cm 0cm]{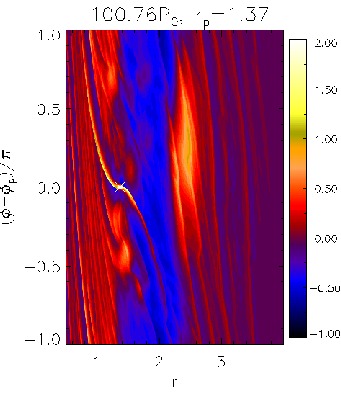}
    \includegraphics[scale=0.39,clip=true,trim=2.3cm 1.8cm 1.96cm
    0cm]{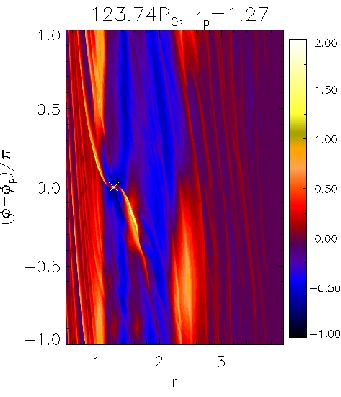}
     \includegraphics[scale=0.39,clip=true,trim=2.3cm 1.8cm 0cm
    0cm]{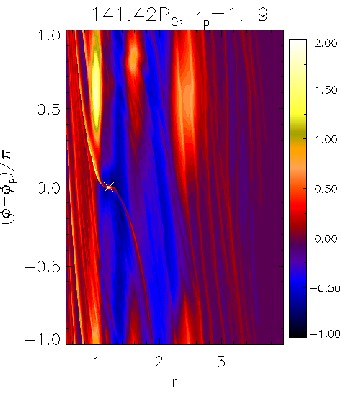}\\

    \includegraphics[scale=0.39,clip=true,trim=0cm 1.8cm 1.96cm 0cm]{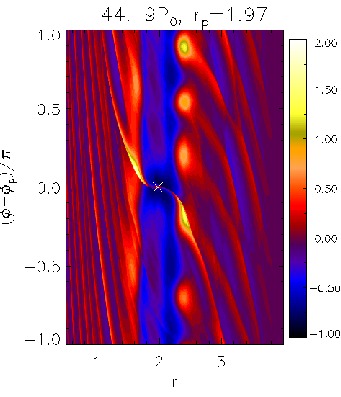}
    \includegraphics[scale=0.39,clip=true,trim=2.3cm 1.8cm 1.96cm 0cm]{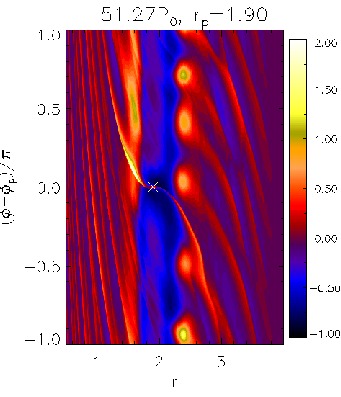}
    \includegraphics[scale=0.39,clip=true,trim=2.3cm 1.8cm 1.96cm 0cm]{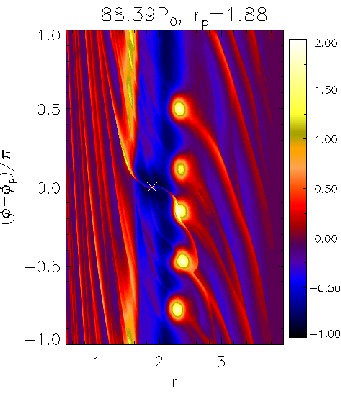}
    \includegraphics[scale=0.39,clip=true,trim=2.3cm 1.8cm 1.96cm 0cm]{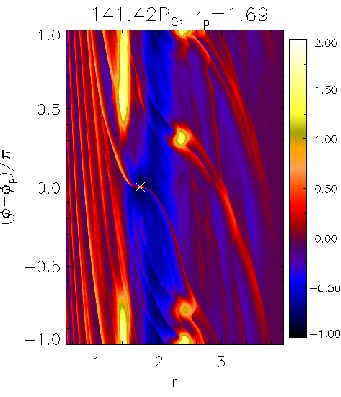}
    \includegraphics[scale=0.39,clip=true,trim=2.3cm 1.8cm 1.96cm
    0cm]{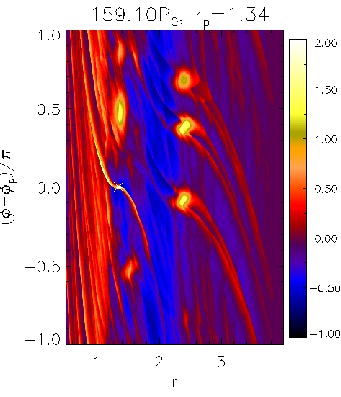}
     \includegraphics[scale=.39,clip=true,trim=2.3cm 1.8cm 0cm
    0cm]{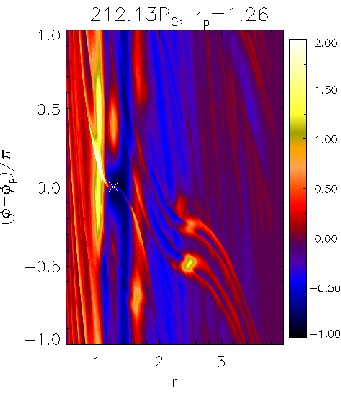}\\
    \caption{Vortex-induced migration in the $M_d=0.025M_*$ disc (see
      Fig. \ref{compare_v}), without self-gravity (top)
      and with self-gravity (bottom). The single vortex at the inner
      gap edge causes the planet to jump inwards
      when it scatters the planet. The formation of this vortex is
      slower in the self-gravitating case.
      \label{compare_v2}}
  \end{figure}
\end{landscape}

For comparison purposes, the non-self-gravitating 
simulations performed in Chapter \ref{paper1} were repeated  
with disc self-gravity. Fig. \ref{sgNsg} shows
the orbital migration of a Saturn-mass planet in discs with
total mass $M_d=0.035M_*$ and $0.025M_*$. 
For $M_d=0.025M_*$, the surface density perturbations
are also compared in Fig. \ref{compare_v2}. 
The $M_d=0.035M_*$
non-self-gravitating disc was the fiducial run in  
Chapter \ref{paper1}. 
Including self-gravity  delays vortex-induced migration from 
$t=60P_0$ to $t=85P_0.$ 
The delay is consistent with both the stabilisation  of low $m$ modes, 
and the slower vortex merging induced by self-gravity.  

Consider $M_d=0.025M_*$.
In the 
self-gravitating disc, it takes longer 
for the inner vortex to build up (e.g. comparing $t=88.4P_0$ in
Fig. \ref{compare_v2}), disrupt 
the coorbital region and flow across the gap. This delays              
vortex-induced rapid migration, by about $50P_0$ compared to the
non-self-gravitating case.

The extent of rapid migration is unaffected by self-gravity.
Notice also the increased oscillations in $r_p(t)$ when self-gravity is included.
This is because of the sustained multi-vortex configuration at the
outer gap edge causing large oscillations in disc-planet torques,  
whereas without self-gravity these vortices would have merged. 

After the first scattering event, migration stalls while the planet 
opens a gap at its new orbital radius and vortex formation recurs. 
With self-gravity included,  the re-formation of a single vortex takes longer
 and it is narrower in radial extent than 
the non self-gravitating case. When the thinner self-gravitating vortex
passes by the planet, little vortex material splits off from 
the main vortex and flows across the gap, unlike the 
non-self-gravitating vortex 
 where material breaks off more easily (Chapter \ref{paper1}).  
This is probably due to 
the  self-gravity of the vortex. Hence, there is a longer stalling
period when self-gravity is included.  Thus the net effect of
self-gravity is to slow the migration in this  
example.

\begin{figure}
  \centering
  \includegraphics[width=\linewidth]{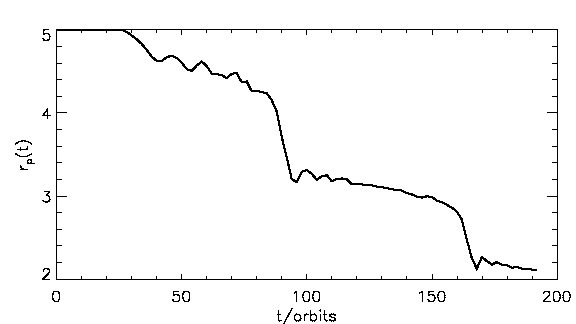}
    \caption{Repeated episodes of vortex-induced migration in a
      self-gravitating disc, for the $Q_o=4$ disc model used in
      previous sections.
      \label{M2HR2}}
\end{figure}

For the setup used in Chapter \ref{paper1}, a second fast migration episode
was not seen within the simulated time when self-gravity was included, but it may eventually occur. The
total practical simulation time of a few hundred orbits is still very
short compared to disc lifetimes.
However, for the $Q_o=4$ disc model used in previous parts of this
Chapter, two episodes of rapid migration occurs
(Fig. \ref{M2HR2}). Self-gravity does not change the physical  nature
of vortex-induced, non-smooth migration.

\section{Summary and discussion}\label{conclusions}

The effect of disc self-gravity on vortex-forming
instabilities associated with planetary gaps were explored. 
Analysis shows vortex  
 modes are stabilised by self-gravity through  its effect on
the linear mode when the background remains fixed. This aspect has
been confirmed by linear calculations. 
 Linear calculations showed that  the vortex forming modes with the
highest growth rate  shift to higher $m$ with increasing disc mass. 
This is due to the combined effect of self-gravity through the
response and through the background state.

Hydrodynamic simulations 
with and without  self-gravity for a range of disc masses were
performed. More vortices form as the disc mass
increased in accordance with linear calculations. However, for 
sufficiently strong self-gravity, the vortex modes are
suppressed and global spiral  modes develop instead. 
Self-gravity delays vortex merging and
 multi-vortex configurations can
be sustained for longer with increasing disc mass, allowing vortices
to evolve individually.

The nature of post-merging
vortices is also affected by self-gravity. With weak self-gravity ( i.e. $M_d
\leq 0.024M_* $), a single vortex, extended in azimuth forms and circulates
at the outer gap edge. For $M_d=0.031M_*$ the final
configuration is a vortex pair at the outer gap edge, each localised
in azimuth. In this case a vortex, containing on the
order of $20M_{\oplus}$, is gravitationally bound. 
The  effect of their gravitational influence on the rest of the disc
is to redistribute mass radially whereas for lower disc masses,  
redistribution is restricted  to being azimuthal. 

The internal flow in these self-gravitating, localised
vortices adjusts so that they are not destroyed by the background shear, 
taking on a structure similar to that of a vortex resulting from
perturbing the disc with the Kida solution \citep{kida81}. 
Such vortices form in discs much less massive than 
that required for direct disc fragmentation. 
Vortices can trap dust particles due to their 
association with pressure maxima \citep{bracco99,chavanis00}.
Increasing   self-gravity leads to vortices of stronger density contrast 
(Fig. \ref{compare_merged_vortex}), so self-gravitating 
vortices may be more effective at collecting solid particles, and
assist planetesimal formation.  


Supplementary simulations of Kida-like vortices were performed to understand
the effect of self-gravity on inter-vortex interactions. 
In a self-gravitating disc, coorbital 
vortex pairs  behave like coorbital planets 
and can execute mutual horseshoe turns. 
As a consequence  there exists a minimum inter-vortex
distance. 
Merging is  then avoided if this minimal separation
is still larger than a critical separation 
below which vortex merging occurs.


The effect of self-gravity found here, should 
also applicable to other types of  structured features  in a protoplanetary
disc that could  support vortex forming  instabilities.
Such a possibility is  the boundary region  
between  a dead zone  and active region of the disc \citep{lyra09}.


The basics of vortex-induced migration remains unchanged by
self-gravity. With self-gravity,  
the resistance to forming a single large vortex 
results in such migration being delayed. 
Self-gravitating  vortices are
are less effective in scattering the planet because they
 do not disrupt the co-orbital region as significantly 
as their non-self-gravitating  counterparts.
It can be said that in the regime of 
disc masses where vortices form and are significantly
affected by  self-gravity, vortex-induced migration is slowed down.

\chapter{Edge modes in self-gravitating discs}\label{paper3}
\graphicspath{{Chapter3/}}

Numerical experiments Chapter \ref{paper2} showed that global spiral
disturbances associated with gap edges can develop in massive
discs. 
They arise from a linear instability associated with both
the self-gravity of the disc and  local vortensity maxima which
coincide with  gap edges.  They can be physically interpreted as
disturbances localised around a  vortensity maximum
that further perturb gravitationally the smooth parts of the disc by
exciting waves at Lindblad resonances. The angular momentum carried
away reacts back on the edge disturbance so as to destabilise it.

These edge modes are studied in this Chapter. 
The disc-planet models are identical to that used in Chapter \ref{paper2},
but briefly reproduced in \S\ref{basic_equations} for ease of reference.  
In \S\ref{motivation}, the existence of edge modes and its basic features 
are illustrated with a numerical simulation. 
An analytic discussion and interpretation of edge modes is then presented in
\S\ref{analytic1}---\S\ref{analytic2}. 
In   \S\ref{linear_paper3}, linear calculations for various
disc models confirm the existence of edge dominated modes for
low values of the azimuthal mode number. The physical picture of edge
modes presented in the analytical discussion is consistent with these
numerical solutions.

In \S\ref{hydro} results from hydrodynamic simulations 
for a range of disc masses are discussed. These are all stable
in the absence of the planet. However, they exhibit low $m$
edge modes once a planet-induced gap is present.
The form and behaviour of these is found to be in accord with linear
theory. 
The spiral arms associated with the edge instabilities are shown to produce
fluctuating torques acting on the planet. Fast inwards type III
migration may occur, but \emph{outwards} migration
due to interaction with spiral arms is also observed. 
\S\ref{conclusions_paper3} concludes this Chapter. 

\section{Disc  model}\label{basic_equations}
To apply angular momentum conservation in the analytical description
of edge modes, three-dimensional disc models
are required. This is because although the disc material may be
considered to be confined to a thin sheet, the gravitational potential
it generates is still three-dimensional. Thus, it is important to 
consider the self-gravity part as fully three-dimensional. 
This is governed by the Poisson equation   
\begin{align*}
  \nabla^2\Phi=4\pi G\rho,
\end{align*}
where $\rho$ is the three dimensional mass density and $\Phi$ is the
associated gravitational potential which can be found via
\begin{align}\label{Poissonint}
\Phi(\bm{r})= -G\int_{\cal{D}}
\frac{\rho(\bm{r}^\prime)d^3\bm{r}^\prime}{|\bm{r} - \bm{r}^\prime|}, 
\end{align}
where ${\cal{D}}$ is the domain where $\rho$ is non zero. This
coincides with the disc domain when this is not separated from
external material. The full 3D hydrodynamic equations are listed in
\S\ref{3d_models} and is the basis for analytical work. In analytic
discussions, fluids with a general barotropic equation of state  are
considered, for which $p=p(\rho)$ and $dp/d\rho =c_s^2.$

\subsection {Razor thin discs}
For numerical work, the disc is considered two-dimensional and
only its gravitational potential in the plane is used. The
governing equations are therefore those listed in
\S\ref{2d_approx}, which were used throughout Chapter \ref{paper2}. In
numerical experiments, the locally isothermal equation of state is used 
with fixed aspect-ratio $h=0.05$. 
 The physical setup and notation is the same as Chapter 
\ref{paper2}, but simulations will be presented for a range of
viscosity and softening parameters. 

The discs occupy $r=[r_i,r_o]=[1,10]$ and are parametrised by $Q_o$, 
the Keplerian Toomre $Q$ value at the outer boundary. The initial surface
density profile is given by Eq. \ref{armitage_disc} (Chapter \ref{paper2}).  
Recall that 2D self-gravity requires the use of a
gravitational softening length (see \S\ref{2d_approx},
Eq. \ref{2d_poisson_int}) to prevent a singularity and approximately
account for the vertical dimension. The softening length
prescription $\epsilon_g(r) = \epsilon_{g0} h r$ is used, and 
$\epsilon_{g0}=0.3$ set as a fiducial value, but values up to 1.0
will be considered. 

The initial azimuthal velocity is found by assuming the
 centrifugal force balances forces due to 
stellar gravity,  the disc's self-gravity and pressure gradient. 
Thus the  initial azimuthal velocity is given by
\begin{align}
  u_{\phi}^2 = \frac{r}{\Sigma}\frac{d p}{dr} + \frac{GM_*}{r} + r\frac{d\Phi}{dr},
\end{align}
For the  local isothermal equation of state adopted, the contribution
due to the pressure is given by
\begin{align}
  \frac{r}{\Sigma}\frac{d p}{dr} = c_s^2 
  \left\{-\frac{5}{2} + \frac{r\sqrt{r_i
      }(r+H_i)^{-3/2}}{2\left[1-\sqrt{r_i/(r+H_i)}\right]}\right\}.
\end{align}
At $r=r_i,$ for $h=0.05,$  this is $\sim 20c_s^2$  which is
approximately $5\%$ of the square of the local Keplerian speed
so that the  contribution  of the pressure force is indeed minor  when
compared to that arising from the gravity of the central star. 

The initial radial velocity is set to $u_r = 3\nu/r$, corresponding
to  the  initial radial velocity velocity of an one-dimensional
Keplerian accretion disc with  $\Sigma\propto r^{-3/2}$ and uniform
kinematic viscosity.  The fiducial value $\nu = 10^{-5}$ in
dimensionless units is adopted, but results for $\nu=10^{-7}$ to
$\nu=5\times10^{-5}$ will be presented. The disc is evolved  for 
$\sim 280$ Keplerian orbits at $r_i$ before introducing the 
planet.  

When a planet of mass $M_p$ is introduced,  it is inserted in a circular
orbit, under the gravity of both the central star and disc,
at a distance  $r_p = 5$ from the central star. Its
gravitational potential is then ramped up  over a time interval
$10P_0.$ Recall $P_0$ is the circular Keplerian orbital period at 
the planet's initial radius. 


\section{Properties of edge modes}\label{motivation} 
The existence of spiral modes associated with 
planetary gaps is first demonstrated through numerical simulations. 
Numerical details are given
in \S\ref{hydro} where additional hydrodynamic simulations 
of this type as well as others,  where the planet is allowed 
to migrate,  will be investigated
more fully.  

\subsection{Example simulation}\label{Firstsim}

Described here are results obtained  
for the $Q_o=1.5$ disc model, corresponding to
a total disc mass of $M_d= 0.063M_*$, in  which there is an                                
embedded planet with mass $M_p=3\times10^{-4}M_*.$ 
 The  uniform kinematic viscosity was taken to be  $\nu=10^{-5}$ in
dimensionless units. For this simulation
the planet was held on a fixed circular orbit. 
Without a perturber the  disc is gravitationally stable because
the $Q_\mathrm{Kep} = 2.62,\, 1.5$ near the planet and the outer boundary, respectively.  
General non-axisymmetric gravitational instabilities occur in smooth discs
when $Q_\mathrm{Kep}\lesssim1.5$.    

\begin{figure}[t]
\centering
\includegraphics[width=0.99\linewidth]{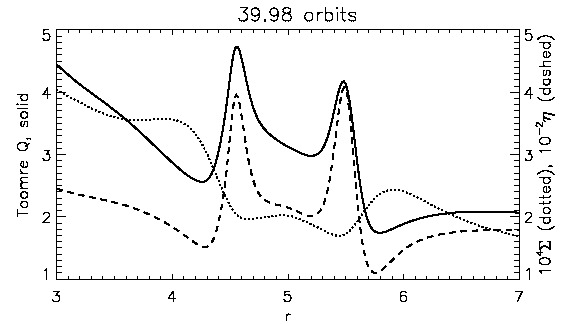}
\caption{Gap profile  produced  by a Saturn-mass planet  embedded 
  in a disc with
  $Q_o=1.5.$ The  azimuthally averaged surface
  density (dotted line), vortensity (dashed line) and  Toomre $Q$ parameter (solid line)
  are plotted. Note that the $Q$ profile shows a maximum and minimum
  associated with each gap edge. 
  The planet is in a fixed circular orbit
  located at $r=5.$ 
  \label{Qm1.5_basic}}
\end{figure}

The profile of the gap  opened by the planet is shown in
Fig. \ref{Qm1.5_basic} in terms of the azimuthally averaged surface
density $\langle \Sigma\rangle_{\phi}$, vortensity $\langle
\eta\rangle_{\phi}$ 
and Toomre $Q$. The latter is calculated using the 
azimuthally averaged epicycle frequency. 
 The surface density gap corresponds to  a
decrease of surface density by $\simeq 20\%$ relative to the
initial profile.  

Notice that the vortensity and Toomre $Q$ profiles
are qualitatively similar. The neighbourhoods of the inner and outer
gap edges both contain maxima and minima of $Q$ and $\eta.$ For 
typical  disc models structured by  disc-planet interactions, such
as those  illustrated in Fig. \ref{Qm1.5_basic}, local maxima/minima
in $Q$ and $\eta$ approximately coincide. Vortensity 
generation  as material flows through planet-induced shocks   
leads to vortensity maxima. Since the axisymmetric Toomre parameter can
be directly related to vortensity, $Q=(c_s/\pi G)(2\Omega \eta/
\Sigma)^{1/2}$, the resulting $Q$ and $\eta$
profiles are very similar in their extrema locations.

Fig. \ref{Qm1.5_basic} shows that in the neighbourhood of the outer
gap edge, the  maximum  and minimum  values of $Q$ occur at $r=5.45$  and $r=5.75,$
and  are $Q=4.2$ and $Q=1.75$ respectively. In the region of  the
inner gap edge,  the  maximum  and minimum  values of $Q$ occur at
$r=4.55$  and $r=4.25.$ These are $Q=4.75$ and $Q=2.55$
respectively. 
The characteristic width of the gap edge is the local scale-height.  
On average $Q\sim 2$ for $r>6$.  Since $Q>1$ everywhere, the disc is
stable against local  axisymmetric perturbations. Furthermore, since 
$Q>1.5$ in the gap region, gravitational instabilities should not be
directly associated with the local value of $Q$. Instead, instability
is associated with the global $Q$-profile.    

    
\begin{figure}[ht]
  \centering
  \includegraphics[width=0.99\linewidth]{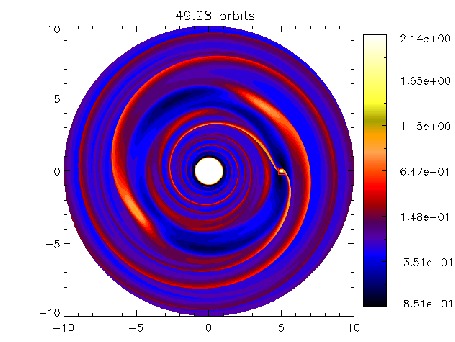}
  \caption{Relative surface density perturbation for $Q_o=1.5$ at
    $t=50P_0$.\label{Qm1.5_t50_contour}}
\end{figure}

The simulation shows that both gap edges become unstable. The
surface density contours  at $t=50P_0$  are   shown in
Fig. \ref{Qm1.5_t50_contour}. A mode with azimuthal wavenumber $m=3$ and
relative surface density perturbation $\Delta\Sigma/\Sigma\simeq 0.4$
is associated with the inner edge and a  mode with $m=2$ and
$\Delta\Sigma/\Sigma\simeq 
0.9$ associated with  the outer edge. The modes have become nonlinear, with the development of
shocks, on dynamical time-scales. Spirals do not extend across the gap indicating
effective gap opening by the planet and disc self-gravity is not strong enough to
connect the spiral modes on either side of $r_p.$

\begin{figure}[ht]
\centering
\includegraphics[width=0.99\linewidth]{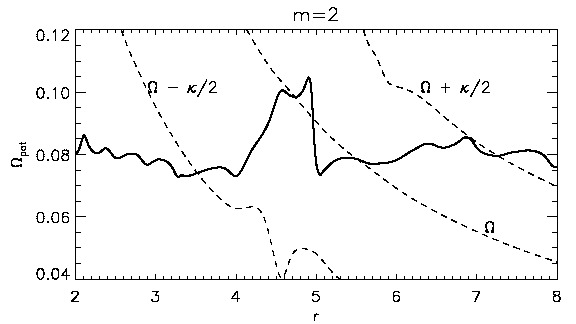}
\caption{Same case as Fig. \ref{Qm1.5_t50_contour}, but shown here is the
  the radial form of the  pattern speed found from  the $m=2$ component
  of the Fourier transform of the surface density
  (solid). Note that this mode is associated with the outer disc
  and, apart from within the gap,
  this mode appears to have a  stable pattern speed of $\Omega_{pat} \sim 0.078$
  corresponding to a  corotation 
  point at $r=5.5,$ corresponding to vortensity maximum in the gap profile. 
  Plots of  $\Omega$ and $\Omega\pm\kappa/2$ are also given  
  (dashed).
  \label{Qm1.5_t50_pattern}}
\end{figure}

Plots of the radial  dependence of  the pattern rotation speed 
\footnote{The pattern speed is $(1/m)\p\varphi/\p t$,
  where $\tan{\varphi} = -\mathrm{Im}(a_m)/\mathrm{Re}(a_m)$. Recall $a_m=\int
  \Sigma e^{-im\phi}d\phi$ is the Fourier
  amplitude. Calculating the pattern speed thus require
  $\p\Sigma/\p t$. However, $\p\Sigma/\p
  t=-\nabla\cdot(\Sigma\bm{u})$ from the continuity
  equation. Instantaneous time-derivatives can therefore be calculated
  from spatial derivatives. This gives the pattern speed from a single data
  output.}, 
$\Omega_\mathrm{pat},$
obtained  from the $m=2$ component of 
the Fourier transform of  the surface density,  together with
the azimuthally
averaged  values of $\Omega$ and $\Omega\pm\kappa/2$   
 are  presented  in Fig. \ref{Qm1.5_t50_pattern}.
The mode associated with the outer gap edge is of particular interest, 
because  the $m=2$ spiral mode has more
than twice the relative  surface density amplitude compared to  the
inner spiral mode. 

By necessity, the Fourier transform includes features due to  the planetary wakes and
inner disc spiral modes as well as the  dominant outer disc mode. 
On average, $\Omega_\mathrm{pat}\sim 0.08$. This value was also
explicitly checked by measuring the angle through which
the spiral pattern rotates  in a given time interval.  
There is a co-rotation point in the outer disc  at $r=5.5$ where
$\Omega=\Omega_\mathrm{pat}\sim 0.078 $, i.e. at  the local vortensity
 maximum (surface density minimum) of the basic state, which is just
within the gap (see Fig. \ref{Qm1.5_basic}).
This is consistent with pattern speeds obtained from linear calculations presented
later. Fig. \ref{Qm1.5_t50_pattern} indicates another co-rotation
point at $r\simeq 4.7$ but this is due to  inclusion of features
associated with the planetary wakes.  

Unstable spiral modes identified here are referred to as
\emph{edge modes}. Their properties  can 
be compared to those of \emph{groove modes} in particle discs \citep{sellwood91} or 
fluid discs \citep{meschiari08}. 
\citeauthor{sellwood91} describe groove modes as  gravitational instabilities
associated with  the  coupling of disturbances associated with two edges 
across the gap between them.  However,
the analytic description of the edge mode instability, presented later, is a coupling between a
disturbance at a single   gap edge and the disturbance it excites in
the adjoining  smooth  disc  away from  co-rotation. 
The other gap edge plays no role in this description.  
Although both types of mode  are associated with vortensity maxima,
the groove modes described above would have co-rotation at the gap
centre midway between the edges, whereas for the planetary gap 
co-rotation is at the gap edge.  

This is significant because it implies differential rotation between the
spiral pattern and the planet. In addition, unstable modes on the
inner and outer gap edges need not have the same  $m$
(Fig. \ref{Qm1.5_t50_contour} shows $m=3$ on the inner edge and $m=2$
on the outer edge). If there was a groove mode with co-rotation at the
gap centre, then the spiral pattern  would  co-rotate  with the planet 
and the coupled edges must have disturbances dominated by the same
value of $m$.

\subsection{Comparison to vortex instabilities}

It is interesting to compare
edge modes to the localised vortex-forming instabilities
studied in Chapter \ref{paper1}---\ref{paper2}, which
 develop in non-self-gravitating or weakly self-gravitating discs
near gap edges.  However, the vortex instability requires low
viscosity. For comparison purposes, here a kinematic viscosity of
$\nu = 10^{-6}$ 
is
imposed. 
The behaviour of disc models with $Q_o=1.5$ and $Q_o=4.0$ are compared
below, the former having strong self-gravity and the latter weak
self-gravity.

\begin{figure}
  \centering
\includegraphics[scale=.45,clip=true,clip=1.6cm 0cm 0cm
0cm]{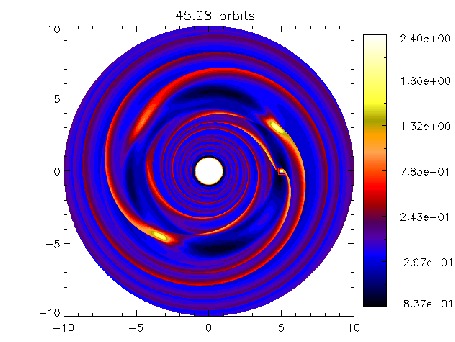}\includegraphics[scale=.45,clip=true,trim=1.6cm
0cm 0cm 0cm]{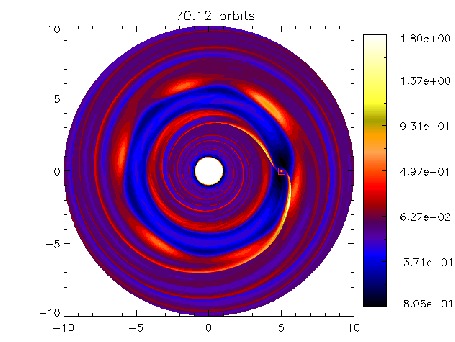} 
 \caption{Relative surface density perturbation for the $Q_o=1.5$
    disc (left) and the $Q_o=4.0$ disc (right) with
    $\nu=10^{-6}$. These plots show that the gap 
    opened by a Saturn-mass planet supports vortex instabilities in low
    mass discs and spiral instabilities  in sufficiently  massive discs. 
    \label{vortex_edge}}
\end{figure}

Fig. \ref{vortex_edge} shows two types of instabilities depending on
disc mass. The lower  mass disc with $Q_o=4$ develops 6 vortices localised 
in the vicinity of the outer gap edge. 
The more massive disc with $Q_o=1.5$  develops  edge modes of the type identified
in the earlier run with $\nu=10^{-5}.$ Here, the $m=3$ spiral mode is
favoured because of the smaller viscosity coefficient used
\footnote{This results
in a sharper gap profile, which generally supports higher $m$.}

Vortex modes are said to be local because instability is associated
with flow in the vicinity of co-rotation \citep{lovelace99} and does
not require the  excitation of waves. Edge modes are said to be global
because  instability requires interaction between  an edge disturbance 
and waves launched at Lindblad resonances, away from co-rotation 
in the smooth part of the disc. Vortices perturb the disc even without 
self-gravity \citep{paardekooper10}.  Although the vortex mode in
Fig. \ref{vortex_edge} shows  significant wave perturbations  in the
outer parts of the disc,  these should be seen as  a consequence  
of unstable vortices  forming around co-rotation, rather than the
cause of  a linear instability.

Vortex modes have co-rotation close to local vortensity
{\it minima} in the undisturbed gap profile. The 
self-gravitating vortex modes are localised with higher $m$ being dominant ($m>5$). On
the other hand the spiral modes found here are global with 
$m\leq 3$. 
 Edge modes make the gap less identifiable than vortex modes do
because the former, with corotation closer to $r_p$, 
protrudes the gap edge more.  While the edge mode only takes a few
$P_0$ to become non-linear with  spiral shocks attaining comparable
amplitudes to the planetary wakes, the vortex mode takes a
significantly longer time (the plot for $Q_o=4$ is $30P_0$ after gap
formation). This can be understood as waves emitted by the vortex
modes tends to stabilise them, whereas wave emission drives the edge
mode instability.  

\section{Analytical discussion of edge modes}\label{analytic1}
The fiducial disc simulated above  is massive with
$M_d\sim 0.06M_*$ but it is still stable against gravitational
instabilities in the sense that $Q\geq 1.5$ everywhere
and the disc without an embedded planet, which has
a smoothly varying surface density profile,  does not
exhibit the spontaneous growth of spiral instabilities. A
planet of Saturn's mass ($M_p=3\times10^{-4}M_*$) 
is expected to only open a partial
gap  ($\sim 30\%$ deficit in surface density), 
but even this is sufficient to trigger a $m=2$ spiral
instability. 

The case above thus suggests spiral modes
may develop under conditions that are not as extreme as those
considered by \cite{meschiari08}. They used a
prescribed gap profile with a gap depth of $90\%$ relative to the background,
which is three times deeper than the simulation above. Their gap
corresponds to $M_p=0.002M_*$, although a planet potential was not
explicitly included.  In both their model and the fiducial case,
the gap width is $\simeq 2r_h$, 
but since \citeauthor{meschiari08} effectively  used a  two Jupiter
mass planet, at the same $r_p$ their  gap is about  1.9 times wider
than for a Saturn mass planet. 

This section and the next are devoted to a theoretical discussion of
spiral modes associated with planetary gap edges. This  work is based
on the analysis of  the governing equations for linear
perturbations.  As angular momentum balance is important for  enabling
small perturbations to grow unstably,  one begins by formulating the
conservation of angular momentum for linear perturbations. 

\subsection{The conservation of angular momentum for a  perturbed disc}\label{barcons}
A conservation law can be derived for angular momentum associated with
the perturbations of a disc. This enables the angular momentum density
and flux to be identified within the framework of linear perturbation
theory. The behaviour of these quantities is found to be important for 
indicating the nature of the angular momentum balance in a system
with a neutral or weakly growing normal mode and how positive (negative) angular momentum fluxes
associated with wave losses may drive the instability of a disturbance
that decreases (increases)  the local angular momentum density.

\subsection{Barotropic discs}
The linearised equation of motion in three dimensions for the
Lagrangian displacement $\bm{\xi}$ 
for a differentially rotating  fluid with a  barotropic equation of state  and  with
 self-gravity included is given by 
\begin{align}\label{lagrangian}
\frac{D^2\bm{\xi}}{D t^2} + 2\Omega\hat{\bm{k}}\wedge\frac{D\bm{\xi}}{D
 t}  + r\hat{\bm{r}}\left(\bm{\xi}\cdot\nabla\Omega^2\right) = -\nabla S 
 - \nabla\Phi_\mathrm{ext},
\end{align}
where
\begin{align}
  S = &c_s^2 \frac{\dd\rho}{\rho}+ \dd\Phi 
\end{align}
\citep[see eg.][]{lbo67,lin93b}. A cylindrical polar coordinate system
$(r,\phi,z)$ is adopted here. Perturbations to quantities are denoted with
$\dd$. The operator
$D/Dt\equiv \p/\p t +\Omega\p/\p\phi$  is the convective derivative
following the unperturbed motion and $\hat{\bm{k}}$ 
is the unit vector in the $z$  direction  which is normal to the disc
mid-plane. For  perturbations that depend on $\phi$ through  a factor 
$\exp({\rm i}m\phi),$  $m$ being the azimuthal mode number, 
assumed positive, the operator $D/Dt$ 
reduces  to  the operator $\left(    \partial/\partial t  + {\rm
    i}m\Omega\right).$ 
In addition, for a barotropic equation of state, $\Omega =\Omega(r)$
depends only on the cylindrical radius. 
Then Eq. \ref{lagrangian} becomes
\begin{align}\label{lagrangianm}
 \frac{\partial^2\bm{\xi}}{\partial t^2} + 
2\Omega\hat{\bm{k}}\wedge\frac{\partial\bm{\xi}}{\partial t}
+2{\rm i}m\Omega\frac{\partial\bm{\xi}}{\partial t}
+2{\rm i}m\Omega^2\hat{\bm{k}}\wedge{\bm{\xi}} 
+ r\hat{\bm{r}}\left(\bm{\xi}\cdot\nabla\Omega^2\right)
-m^2\Omega^2\bm{\xi} = -\nabla S
- \nabla\Phi_\mathrm{ext}.
\end{align}
The Eulerian density perturbation is given by
\begin{align}\label{3Dcont}
\dd\rho=-\nabla\cdot(\rho \bm{\xi}).
\end{align}
The perturbation to the disc's gravitational potential is given
by linearising Poisson's equation to give
\begin{align}\label{3DPoisson}
\nabla^2\dd\Phi=4\pi G\dd\rho.
\end{align}
An external potential perturbation  $\Phi_\mathrm{ext}$  has been
added to the equations of motion to assist with  the identification of
angular momentum flux later on.


By taking the scalar product of Eq. \ref{lagrangianm} with ${\rho
  \bm{\xi}}^*,$ and taking the imaginary part, after making use of  
Eq. \ref{3Dcont} and Eq. \ref{3DPoisson}, a conservation law
that expresses the conservation of angular momentum for 
the perturbations can be derived:
\begin{align}\label{3DCONS}
  \frac{\p\rho_J}{\p t} + \nabla\cdot(\bm{F}_A + \bm{F}_G +
  \bm{F}_\mathrm{ext}) =  T,
\end{align}
where
\begin{align}
  \rho_J &\equiv -\frac{m\rho}{2} 
  {\rm Im}\left(\bm{\xi}^*\cdot\frac{\p\bm{\xi}}{\p t} +
  \Omega{\hat{\bf k}}\cdot\bm{\xi}\wedge\bm{\xi}^* +
  im\Omega |\bm{\xi}|^2\right)\label{rhoJ} \\
\bm{F}_A &= -\frac{m\rho}{2}{\rm Im}\left(\bm{\xi}^*S\right)\label{FA} \\
\bm{F}_G &= -\frac{m}{2}{\rm Im}\left(\frac{1}{4\pi
      G}\dd\Phi\nabla \dd\Phi^*\right)\label{FG} \\
\bm{F}_\mathrm{ext} &=
-\frac{m\rho}{2}{\rm Im}\left(\bm{\xi}^*\Phi_\mathrm{ext}\right)\label{Fext}
\end{align}
and
\begin{align}\label{extTorque}
T=\frac{m}{2}{\rm Im}\left(\Phi_\mathrm{ext}\dd\rho^{*}\right).
\end{align}
Here $\rho_J$ is the angular momentum density and the angular momentum 
flux is split into three contributions, $\bm{F}_A$ being 
proportional to the Lagrangian displacement, is the advective
angular momentum flux, $\bm{F}_G$ is the flux associated with
the perturbed gravitational stresses and $\bm{F}_\mathrm{ext}$
is the flux associated with the external potential
which is inactive for free perturbations.

The quantity $T$ is a torque density associated with the external
potential as can be seen from that fact that
the real torque integrated over azimuth and divided by $2\pi$ is
\begin{align}\label{extTorque1}
-\frac{1}{2\pi}\int^{2\pi}_0{\rm Re}(\dd\rho^{*}e^{\mathrm{i}m\phi}){\rm Re}\left(\frac{\p(\Phi_\mathrm{ext}e^{\mathrm{i}m\phi})}{\p \phi}\right)d\phi
 =T=\frac{m}{2}{\rm Im}\left(\Phi_\mathrm{ext}\dd\rho^{*}\right).
\end{align}
The $e^{\mathrm{i}m\phi}$ dependence has been reinstated on the LHS for clarity. 
This justifies the scalings used for the angular momentum density
and fluxes. 

\subsection{Two dimensional discs} 
The form of the conservation law for a  razor thin disc
is obtained by integrating Eq. {\ref{3DCONS} over the vertical coordinate.
For terms $\propto \rho$ the integrand is non zero only within the
disc. $\bm{\xi}$ 
may be assumed to depend only
on $r$ and $\phi.$ The effect of the integration is simply to replace
$\rho$ and $\dd\rho $ by the surface density $\Sigma$  and its perturbation 
 $\dd\Sigma $ respectively. The fluxes  become vertically integrated fluxes.
Vertically integrating the term associated with the gravitational
stresses, assuming the potential perturbation vanishes at large
distances, gives 
\begin{align}\label{vintFG}
\int^{\infty}_{-\infty}\nabla\cdot{\bf F}_Gdz
= -\frac{m}{2}{\rm Im}\left(\dd\Phi(r ,\phi,0)\dd\Sigma ^{*}\right)
=\frac{1}{r}\frac{\p\left( r F_{Gr}\right)}{\p r}, 
\end{align} 
where $\nabla^2\dd\Phi=4\pi G \dd\Sigma \dd(z)$ has been used. 
The angular momentum flux associated with
self-gravity is 
\begin{align}\label{FGrdef}
  F_{Gr}
  =-\frac{m}{8\pi G}{\rm Im}\int^{\infty}_{-\infty} {\dd\Phi}\frac{\p
    \dd\Phi^*}{\p r}dz . 
\end{align}
Thus for a two dimensional  razor thin disc the conservation law is of the
same form as Eq. \ref{3DCONS} but with $\rho_J$,
${\bf F}_A,$ and ${\bf F}_\mathrm{ext}$
given by Eq. \ref{rhoJ}, Eq. \ref{FA} and Eq. \ref{Fext} with $\rho$
replaced by $\Sigma$ respectively.  As the vertical direction has been
integrated over only the radial components of the fluxes contribute. 
From the above analysis the vertically integrated angular momentum
flux due to gravitational stresses becomes 
\begin{align}\label{FG2D}
  {\bf F}_G \to F_{Gr}{\hat {\bf r}},
\end{align}
with $F_{Gr}$ given by Eq. \ref{FGrdef}. Accordingly this term, unlike
the others involves an integration over the vertical direction. 
Explicit expressions for the fluxes in terms of the
perturbation $S=\dd\Sigma  c_s^2/\Sigma +\dd\Phi$, appropriate to the
razor thin disc, shall be derived below. 


\subsection{Locally isothermal equation of state}
It is possible to repeat the analysis of \S\ref{barcons} when
the disc has a locally isothermal equation of state. 
In this case $p=\rho c_s^2,$ where $c_s$ is a prescribed function of position and
the linearised equation of motion (Eq. \ref{lagrangian}) is modified to read
\begin{align}\label{lagrangianli}
\frac{D^2\bm{\xi}}{D t^2} +
2\Omega\hat{\bm{k}}\wedge\frac{D\bm{\xi}}{D 
 t}  + r\hat{\bm{r}}\left(\bm{\xi}\cdot\nabla\Omega^2\right) = -\nabla
S +\frac{\dd\rho}{\rho}\nabla(c_s^2) 
 - \nabla\Phi_\mathrm{ext}.
\end{align}
An expression of the form of Eq. \ref{3DCONS} may be derived 
but now the torque density $T$ takes the form 
\begin{align}\label{extTorque2} 
  T=\frac{m}{2}\left[{\rm Im}\left(\Phi_\mathrm{ext}\dd\rho^{*}\right) - 
    {\rm Im}\left(\dd\rho\bm{\xi}^*\cdot\nabla(c_s^2)\right) \right].
\end{align}
When $c_s$ is constant corresponding to a strictly isothermal equation of state, 
the fluid is again barotropic and previous expressions recovered. Eq. \ref{extTorque2}
contains terms in addition to  external contributions. In general 
these imply the possibility of an exchange of angular momentum between perturbations
and the background. A very similar discussion applies to the razor
thin limit. The analysis above can be applied to determine 
the angular momentum balance for the  linear perturbations 
of razor thin discs 
and thus determine  properties of and conditions for  unstable normal
modes. 
 
\subsection{Properties of the linear modes of a razor thin disc}
Linear perturbations are assumed to have $\phi$ and $t$
dependence through a factor  of the form  $e^{i(\sigma t +
  m\phi)}$ where $\sigma$ is the complex eigenfrequency and $m$ is the
azimuthal mode-number.  From now on this factor is taken as read. 
Linearising the hydrodynamic equations 
for the inviscid case and ignoring external potentials, gives
\begin{align}
  \ {\rm i}\sbar\dd\Sigma  & = -\frac{1}{r}\frac{d(\Sigma r \dd u_r )}{dr}
  -\frac{{\rm i}m \Sigma \dd u_\phi}{r}\label{cl}\\
  \ {\rm i}{\bar{\sigma}}\dd u_r  -2\Omega \dd u_\phi &= -\frac{dS}{dr}
  +\frac{\dd\Sigma }{\Sigma}\frac{d c_s^2}{dr}\label{mrl}\\
  \ {\rm i}{\bar{\sigma}}\dd u_\phi + \frac{\kappa^2}{2\Omega}\dd u_r  &=-\frac{{\rm i}m S}{r},
  \label{mphil}\end{align}
Recall $\bar{\sigma} = \sigma + m\Omega$ is the
shifted mode frequency. These equations were also used in Chapter
\ref{paper1} and Chapter \ref{paper2} in terms of $W\equiv
\dd\Sigma/\Sigma$. However, it is convenient for the analytical 
discussion later on, to derive a governing equation in a slightly
different form than previous Chapters. Using Eq. \ref{mrl} and
Eq. \ref{mphil} to eliminate $\dd u_r $ and $\dd u_\phi$ 
in Eq. \ref{cl} gives
\begin{align}\label{barotropic2}
 r\dd\Sigma  = & \frac{d}{dr}\left[\frac{r\Sigma}{D}
\left(\frac{dS}{dr} +
    \frac{2m\Omega\bar{\sigma}S}{r\kappa^2}\right)\right]
-\frac{2m\Omega\bar{\sigma}\Sigma}{\kappa^2D}\left(\frac{dS}{dr} +
  \frac{2m\Omega\bar{\sigma}S}{r\kappa^2}\right)
+ \frac{mS}{\bar{\sigma}}\frac{d}{dr}\left(\frac{1}{\eta}\right)\nonumber \\ 
&-\frac{4m^2\Omega^2\Sigma S}{r\kappa^4}
 -\frac{d}{dr}\left[\frac{r\dd\Sigma }{D}\frac{dc_s^2}{dr}\right]
+\frac{2m\Omega\dd\Sigma }{D\bar{\sigma}}\frac{dc_s^2}{dr}.
\end{align}
Recall $D\equiv \kappa^2 - \bar{\sigma}^2$ and  $\eta =
\kappa^2/2\Omega\Sigma$. This form is similar to that used by
\cite{papaloizou91} and explicitly shows that a co-rotation singularity,
where  $\bar{\sigma}=0$, can be avoided if co-rotation corresponds to a
vortensity extremum ($d\eta/dr=0$). The potential singularity at
co-rotation associated with the term involving $dc_s^2/dr$ has
negligible effect in practice, explained below. 

The above expressions may be applied to a disc, either  with a locally
isothermal equation of state, or with a barotropic 
 equation
of state, where the integrated pressure
$p=p(\Sigma)$ and the sound-speed $c_s^2\equiv dp/d\Sigma$. 
To obtain the latter case , the quantity  $dc_s^2/dr$
is simply replaced by zero. 
Inclusion of additional terms dependent on this 
has been found numerically to produce only a slight modification of
the discussion that applies when they are neglected. 
This is because the locally isothermal $c^2_s$
varies slowly compared to either the linear perturbations
or the surface density in the vicinity of an edge, thus
$(r/c_s^2)dc_s^2/dr$ shall be neglected in the analysis from now
on. Also recall that $ S = \dd\Sigma c_s^2/\Sigma+\dd\Phi $   
and  the gravitational potential is given by the vertically integrated 
Poisson integral as 
\begin{align*}
  \dd\Phi & = S-\frac{\dd\Sigma c_s^2}{\Sigma} = - G\int_{\cal D} K_m(r,r')\dd\Sigma (r')r'dr' 
\end{align*}
where
\begin{align*}
  K_m(r,r') &= \int^{2\pi}_0  \frac{\cos(m\phi)d\phi}{\sqrt{r^2+r'^2 -
      2rr'\cos{(\phi) + \epsilon_g^2}}}. 
\end{align*}
Here the domain of integration ${\cal D}$ is that of the disc provided 
there is no external material. However, in a situation where density
waves propagate beyond the disc 
boundaries ${\cal D}$ either should be formally extended or the 
form of $K_m$ changed to properly reflect the boundary conditions.


The governing equation above, together with angular momentum
conservation, can be used to model angular momentum flux balance for
normal modes. For simplicity only barotropic case are considered in
the analysis.  The analytic discussion concerns a disc model with one
boundary being a sharp edge with  arbitrary  propagation of density waves 
directed away from this boundary being allowed.
When applied to a gap opened by a planet in a global disc, 
the analytic models corresponds to the section of the disc from the inner disc boundary to the
inner gap edge or the section from the outer gap edge to the outer disc boundary. 
These two sections are assumed to be decoupled 
in the analytic discussion of stability but not in the numerical one.
In addition, although it produces the
background profile,
the planet is assumed to have no effect on the linear stability analysis
discussed here, but it is fully incorporated in nonlinear simulations.
The correspondence between the various approaches adopted indicates
the above approximations are reasonable.
The discussion  of the outer and inner disc sections is very similar,
accordingly  these  are considered below  together.


\subsection{Angular momentum flux balance for normal
  modes}\label{Fluxbal} 
The vertically integrated angular momentum flux associated with
perturbations can be related to the background vortensity
profile. 
A barotropic disc model is assumed, so terms
involving $dc_s/dr$ are neglected. Multiplying
Eq. \ref{barotropic2} by $S^*$ and integrating:  
\begin{align}\label{angmom_balance}
&{\cal {F}}(S,\sigma) \equiv -\int_{r_i}^{r_o} r\dd\Sigma  S^* dr 
\pm\left. \left[ \frac{r\Sigma}{D}\left(\frac{d S}{d r}  +
 \frac{2m\Omega\bar{\sigma}}{r\kappa^2}S\right)
 S^*\right]\right|_{r_b} \nonumber \\
&- \int_{r_i}^{r_o} 
 \frac{r\Sigma}{D}\left(\frac{d S}{d r}  +
 \frac{2m\Omega\bar{\sigma}}{r\kappa^2}S\right)
 \left(\frac{d S^*}{d r}  +
 \frac{2m\Omega\bar{\sigma}}{r\kappa^2}S^*\right)dr
 \nonumber\\ 
& -\int_{r_i}^{r_o}
 \frac{4m^2\Omega^2\Sigma|S|^2}{r\kappa^4}dr + \int_{r_i}^{r_o}
 \frac{m|S|^2}{\bar{\sigma}}\frac{d}{dr}\left(\frac{1}{\eta}\right) dr 
=0,\end{align}
where, here and below,  the positive sign alternative applies to the case where the disc domain
has an inner sharp edge where $r=r_i$ and then $r_b=r_o,$ the outer boundary radius.
The negative sign alternative applies to the case where the disc domain
has an outer sharp edge where $r=r_o$ and then  $r_b=r_i,$ the inner boundary radius.
In both cases waves may propagate through $r=r_b.$
Note also  that there is no contribution from edge boundary terms
as the surface density is assumed to be negligible  there.

One may now express the terms in the above integral relation,
which for convenience has been taken to define a functional of $S$
that depends on  $\sigma,$ 
in terms of quantities related to the transport of angular momentum
by taking its imaginary part.
One begins by assuming $\sigma$ is real or, more precisely,
the limit that marginal stability is approached.

By making use of Eq. \ref{vintFG}, one finds that
\begin{align}
  &{\rm Im}\left[\int_{r_i}^{r_o} r\dd\Sigma  S^* dr\right]=
  {\rm Im}\left[\int_{r_i}^{r_o} r\dd\Sigma \left( \frac{\dd\Sigma ^*c_s^2}
      {\Sigma} +\dd\Phi^*\right)dr\right]
  ={\rm Im}\left[\int_{r_i}^{r_o} r\dd\Sigma \dd\Phi^*dr\right]\nonumber\\
&=\pm\frac{2}{m}
  \left. [rF_{Gr}]\right|_{r_b}.
\end{align}
Recall that $F_{Gr}$ is the vertically integrated 
 angular momentum flux transported by gravitational stresses. 
Note that $F_{Gr}$ at the sharp edge is ignored. Eq. \ref{vintFG},
together with the requirement that $F_{Gr}$ is regular for $r\to0$ and
vanishes 
more rapidly than $1/r$ for $r\to\infty$, implies that $F_{Gr}$ is
zero at the sharp 
edge separating the disc domain and the exterior (assumed) vacuum or
very low density region.  

In addition, from Eq. \ref{mrl} and Eq. \ref{mphil},  
the radial Lagrangian displacement is given by 
\begin{align}
  \xi_r &= \frac{\dd u_r}{{\rm i} \sbar}= -\frac{1}{D}\left(
    \frac{d S}{d r} +
    \frac{2mS\Omega}{r\bar{\sigma}}\right).
\end{align}
From this it follows that the  imaginary part of the second term on the RHS of
Eq. \ref{angmom_balance}) is
\begin{align*}
\pm\left.{\rm Im} \left[ \frac{r\Sigma}{D}\left(\frac{d S}{d r}  +
 \frac{2m\Omega\bar{\sigma}}{r\kappa^2}S\right)
 S^*\right]\right|_{r_b}=  -~\pm~{\rm Im}~[r\Sigma\xi_rS^*]|_{r_b} =
-\pm(2/m)[rF_{Ar}]|_{r_b}. 
\end{align*}
Using the above and taking the imaginary part of
Eq. \ref{angmom_balance} yields the remarkably simple   
expression
\begin{align}\label{angmom_balance2}
  \pm\left[r(F_{Gr}+F_{Ar})\right]|_{r_b}  
  = \frac{m}{2}{\rm Im}\left[\int_{r_i}^{r_o} \frac{m|S|^2}{\bar{\sigma}}
    \frac{d}{dr}\left(\frac{1}{\eta}\right)
    dr\right].
\end{align}
In the above, it has been assumed that $\sigma$ is real
and  that the mode is at marginal stability.
Then the RHS of Eq. \ref{angmom_balance2}  is apparently real. 
However, the integrand is potentially singular at corotation  where 
$\sbar=0.$ Thus the approach to marginal stability 
has to be taken with care. 

Setting $\sigma=\sigma_R+{\rm i}\gamma,$ where
$\sigma_R,$ being  the real part of $\sigma,$
defines the corotation radius, $r_c,$  through $\Omega(r_c)=-\sigma_R/m$
and $\gamma$ is the imaginary part of $\sigma$, so that the mode is
unstable for $\gamma <0$ with growth rate $-\gamma$. 
Marginal stability can be approached by  
assuming that  $\gamma$ has a vanishingly small magnitude
but  is negative. Then one can apply the Landau prescription and
replace $1/\sbar$ by $  
{\cal{P}}(1/\sbar) + i\pi\dd(\sbar),$
where  ${\cal{P}}$ indicates that the principal value
of the integral is to be taken and $\dd$ denotes Dirac's delta
function. 
Adopting this, Eq. \ref{angmom_balance2} becomes
\begin{align}\label{angmom_balance3}
  \left[r(F_{Gr}+F_{Ar})\right]|_{r_b}
  = \pm\frac{\pi |m|}{2}\left.\left[\frac{|S|^2}{|d\Omega/dr|}
\frac{d}{dr}\left(\frac{1}{\eta}\right)\right]\right|_{r_c}.
\end{align}
The above relation can be viewed as stating that at marginal stability,
either corotation is at a vortensity extremum and both terms in Eq. 
\ref{angmom_balance3} vanish, or
angular momentum losses as a result of waves
passing through $r=r_b$ (the smooth boundary), are balanced by torques   
exerted at corotation. The latter torque is proportional to the gradient
of the vortensity $\eta $ \citep{goldreich79}.  
Note that $\eta^{-1}$ scales with $\Sigma$.


The propagating waves  quite generally
carry negative angular momentum   when located in an inner disc
($r<r_c$) 
and positive angular momentum  when located in  an outer disc
($r>r_c$) \citep{goldreich79}.  
Thus  a balance   may be possible between  angular momentum losses resulting from
the propagation of these waves out of the system  and
angular momentum gained or lost by  an  edge disturbance (near
co-rotation radius) as expressed by Eq. \ref{angmom_balance3}.   
For an edge disturbance in a region of increasing (decreasing) surface 
density, such as an inner (outer)  edge to an outer (inner) disc, the
edge disturbance loses (gains) angular momentum.  

In practice, an inner (outer) edge disturbance
may be associated with a positive (negative) surface density
slope  near a vortensity maximum  or near a vortensity minimum
occurring as a result of the variation of both  $\kappa^2$ and
$\Sigma$ \citep[e.g.][]{lovelace99,papaloizou89}.    
The former case is associated with discs for which self-gravity is important
and spiral density waves are readily excited. The latter is associated
with weakly or non self-gravitating  discs and is associated with
vortex formation at gap edges as has already been discussed in
Chapters \ref{paper1}---\ref{paper2}. 

The  boundary condition at $r=r_b$ (the non-sharp boundary) may
correspond to stipulating outgoing waves or a radiation condition,
so that  the fluxes in Eq. \ref{angmom_balance3} are non zero. 
Another possibility is to have a surface density taper to zero 
which would mean wave reflection and removal of the  boundary flux
terms. In fact, the issue of stability, for the particular modes of
interest, is not sensitive to precise boundary conditions at the
smooth boundary.  
In the case of non self-gravitating discs and the low $m$ vortex
modes,  the  regions away from the edge  are evanescent and amplitudes 
are small there, the disturbance being localised in the vicinity of the edge.
For self-gravitating discs, as indicated below, instability can appear
because of an unstable interaction between {\it outwardly propagating
  positive} ({\it inwardly propagating negative}) angular momentum
density waves  and a {\it negative} ({\it positive})  angular momentum
edge disturbance localised  near an {\it inner} ({\it outer}) 
disc gap edge. As long as these waves are excited it should 
not matter whether they are reflected or transmitted at large distance,
as long as their angular momentum density remains in the wider disc
and is not fed back to the exciting edge disturbance.

For these reasons and also simplicity it shall be assumed that at the
boundary where $r=r_b,$  either the  surface density  has tapered 
 to zero, or there is reflection such  that  
the  boundary terms in expressions like Eq. \ref{angmom_balance} may
be dropped. Eq. \ref{angmom_balance3} then simply states that marginal
stability occurs when corotation is at a vortensity extremum. 
It is found that the vortensity {\it maximum}  case 
is associated with an instability in self-gravitating discs that is
associated with strong spiral waves. On the other hand
the   vortensity {\it minimum}  case is associated with 
the vortex-forming instability in  non-self gravitating discs
that has been previously studied. 


\section{Description of the edge mode instability}\label{analytic2} 

The edge mode instability in a  self-gravitating disc can be
interpreted as
due to a disturbance associated with the 
edge causing the excitation of spiral waves  that propagate away from it 
 resulting  in destabilisation. This occurs because the emitted waves
carry away  angular momentum which has the opposite
sign to that associated with the edge disturbance 
 ( see \S\ref{Fluxbal} and Eq. \ref{angmom_balance3}). 
Accordingly the excitation process is expected to lead to the growth
of this disturbance.

Thus  one can consider a disturbance localised in the vicinity of the
edge and the potential perturbation it produces to excite spiral density
waves in the  bulk of the disc. 
Numerical calculations show that edge dominated modes of this type
occur for low $m$ and are dominant in the nonlinear regime.
The analysis below  assumes the mode is weakly growing corresponding
to the back reaction of the waves on the edge disturbance being weak.
Thus  in the first instance one calculates a neutral edge
disturbance (zero growth rate) and then calculate the wave emission as a perturbation.
It should be emphasised again that an 
important feature is that corotation is at a vortensity {\it maximum}
(in contrast to the non-self-gravitating case for which corotation
is located at a vortensity {\it minimum}).
Such modes require self-gravity. In an average sense they  require a
sufficiently small $Q$ value and so  
cannot occur in the non self-gravitating limit ($Q\to\infty$).  


\subsection{Neutral edge disturbances with corotation at a vortensity
  maximum}\label{Neutedge} 
Consider the expression for  $\dd\Sigma $ expressed by 
Eq. \ref{barotropic2} and 
assume that only the vortensity gradient term (third term on the RHS) need be retained.
This is because this term should dominate near corotation in the presence
of large vortensity gradients that are presumed to occur near the edge
and the disturbance is assumed localised there 
\footnote{See also the discussion of groove modes in collisionless
particle discs of \cite{sellwood91}, where similar assumptions are made.}.
Thus 
\begin{align}\label{sigma_bigS}
  \dd\Sigma  = \frac{S}{r{\bar{\omega}}}\frac{d}{dr}\left(\frac{1}{\eta}\right),
\end{align}
where $\bar{\omega} = \sbar/m$. The associated gravitational potential is given by
\begin{align}\label{PPOT}
\dd\Phi =-
G\int_{r_i}^{r_o} 
K_m(r,r')\frac{S(r')}{{\bar{\omega}}(r')}\frac{d}{dr'}\left[\frac{1}{\eta(r')}\right]
dr'.
\end{align}
The relation $S =\dd\Sigma  c_s^2/\Sigma+\dd\Phi$
gives the integral equation
\begin{align}
&S(r)\left[ 1-
  \frac{c_s^2}{r\Sigma{\bar{\omega}}}\frac{d}{dr
}\left(\frac{1}{\eta}\right)\right] 
=- \int_{r_i}^{r_o}GK_m(r,r')\frac{S(r')}{{\bar{\omega}}(r')}\frac{d}{dr'}\left[\frac{1}{\eta(r')}\right]
dr'.\label{Inteq}
\end{align}
For a disturbance dominated by self-gravity, pressure should be negligible. 
Under such  an approximation, $S\sim\dd\Phi.$ 
Eq. \ref{sigma_bigS} then implies
$\mathrm{sgn}{(\bar{\omega})}=\mathrm{sgn}(d\eta/dr)$ in order 
to obtain a negative (positive) potential perturbation
for a  positive (negative) surface density perturbation. 
At co-rotation where $\bar{\omega}=0$ and there is a vortensity extremum, 
this requirement becomes $\mathrm{sgn}{(d\bar{\omega}/dr)}=\mathrm{sgn}(d\Omega/dr)
=\mathrm{sgn}(d^2\eta/dr^2)$. Since typical rotation profiles have $d\Omega/dr\equiv\Omega' 
< 0$, the physical 
requirement that potential and surface density perturbations have opposite signs implies that
$d^2\eta/dr^2<0$ at co-rotation, i.e. the vortensity is a maximum. The
discussion thus applies to vortensity maxima only. 
For the  class of vortensity  profiles for which 
$\mathrm{sgn}(d\eta/dr) = \mathrm{sgn}(\bar{\omega}),$ the integral
Eq. \ref{Inteq} may be transformed into a Fredholm  integral
equation with symmetric kernel. This is demonstrated below. 

Noting that corotation is located at a vortensity maximum,
the requirement above implies the vortensity  
increases (decreases) interior (exterior) to corotation 
\footnote{The discussion may also be extended to the case when that is true 
with $d^2\eta/dr^2$ vanishing at corotation}.
As the disturbance is expected to be localised around corotation,
it is reasonable to assume this holds and that one may contract
the integration domain $(r_i,r_o)$ to exclude regions which do not conform
without significantly affecting the problem. 

Next, introduce the function $\mathcal{H}$ such that
\begin{align}  S(r) =
  \mathcal{Z}(r) 
  \mathcal{H}(r),
\end{align}
where
\begin{align}
  \mathcal{Z}(r)=
  \left[-\frac{d}{dr}\left(\frac{1}{\eta}\right)\frac{1}{\bar{\omega}}\right]
  ^{-1/2} 
  \left[  1-\frac{c_s^2}{r\Sigma\bar{\omega}}
    \frac{d}{dr}\left(\frac{1}{\eta}\right)\right]^{-1/2}.
\end{align}
Note that for $\mathrm{sgn}(d\eta/dr) = \mathrm{sgn}(\bar{\omega})$,
$\mathcal{Z}$ is real. Defining the new symmetric kernel
$\mathcal{R}(r,r')$ as 
\begin{align}
\mathcal{R}(r,r') \equiv  &  GK_m(r,r')\mathcal{Y}(r)\mathcal{Y}(r')
\end{align}
where
\begin{align}
\mathcal{Y}(r)=
\left[-\frac{d}{dr}\left(\frac{1}{\eta}\right)\frac{1}{\bar{\omega}}\right]
^{1/2}
\left[  1-\frac{c_s^2}{r\Sigma\bar{\omega}}
\frac{d}{dr}\left(\frac{1}{\eta}\right)\right]^{-1/2}
\end{align}
one obtains the integral equation
\begin{align}
  \mathcal{H} = \int_{r_i}^{r_o}\mathcal{R}(r,r')\mathcal{H}(r')dr'.
\end{align}
In other words, $\mathcal{H}$ is required to be
the solution to
\begin{align}\label{fredholm}
  \lambda \mathcal{H} = \int_{r_i}^{r_o}\mathcal{R}(r,r')\mathcal{H}(r')dr',
\end{align}
with $\lambda=1$. However, Eq. \ref{fredholm} is just a standard
homogeneous Fredholm integral  equation of the second kind. Thus the
existence of a nontrivial solution ($\mathcal{H}\neq 0$) implies  $\mathcal{H}$ is an
eigenfunction of the kernel $\mathcal{R}$ with unit eigenvalue.

The kernel $\mathcal{R}$ is  positive and  symmetric.  Accordingly the
Fredholm 
equation above has the property that the maximum eigenvalue  $\lambda > 0$
 and is the maximum of the quantity $\Lambda$ given by 
\begin{align}\label{CauSCh}
  \Lambda = \frac{\int_{r_i}^{r_o}\int_{r_i}^{r_o}\mathcal{R}(r,r')
    \mathcal{H}(r')\mathcal{H}^*(r)drdr'}{\int_{r_i}^{r_o}|\mathcal{H}(r)|^2 dr}
\end{align}
over  all $\mathcal{H}$. Thus, if it is possible
to show that $\mathrm{max}(\Lambda) < 1,$ then it
is not possible to have a non-trivial solution corresponding
to a neutral  mode. Upon application of the
Cauchy-Schwarz inequality (twice), one can deduce
\begin{align}
\Lambda^2 \leq
\int_{r_i}^{r_o}\int_{r_i}^{r_o}|\mathcal{R}(r,r')|^2drdr'
 = \Lambda_\mathrm{est}^2 \ge \mathrm{max}(\Lambda^2) .
\end{align}
Thus the existence  of a neutral mode
of this type is not possible if $\Lambda_{est}^2$  is less than unity.

To relate the necessary condition for mode existence to physical
quantities, one can estimate $\Lambda_{est}^2$.
The Poisson kernel $K_m(r,r')$ is largest at $r = r'$. Other factors in
the numerator of  $\mathcal{R}$ involve the factor
$1/\bar{\omega}$. Assume these factors have their largest contribution
at  corotation where $r=r_c$ and $\bar{\omega} = 0.$ 
Then $ \Lambda_{est}^2 \sim
\int_{r_i}^{r_o}\int_{r_i}^{r_o}|\mathcal{R}(r_c,r_c)|^2drdr'
$. Assuming the edge region has width of order $L_c,$ taken to
be much less than the local radius but not less than  the local scale-height,
gives the estimate
\begin{align}\label{lambda_est}
\Lambda_{est} \sim &|\mathcal{R}(r_c, r_c)|L_c 
 =
GK_m(r_c,r_c)\left|\frac{1}{\Omega^\prime}\left(\frac{1}{\eta}\right)^{\prime\prime}\right|L_c
\left/ \left| 1 - \frac{c_s^2}{r_c \Sigma
      \Omega^\prime}\left(\frac{1}{\eta}\right)^{\prime\prime}\right|\right..
\end{align}
All quantities are evaluated at co-rotation  and the  double prime denotes $d^2/dr^2$.
Because the edge is thin
by assumption,  the Poisson kernel can be further approximated by
\begin{align}
  K_m(r,r') \sim \frac{2}{\sqrt{rr'}}K_{0}\left(\frac{m}{\sqrt{rr'}}\sqrt{|r
      - r'|^2 + \epsilon_g^2} \right),
\end{align}
where $K_{0}$ is the  modified Bessel function of the second kind of order
 zero  and recall that
$\epsilon_g$ is the softening length. If pressure 
effects are negligible ($c_s^2$ being small) then 
\begin{align}\label{maxL}
  \Lambda_\mathrm{est} \sim
  \frac{2GK_{0}(m\epsilon_g/r_c)}{r_c}
  \left|\frac{1}{\Omega^\prime}
    \left(\frac{1}{\eta}\right)^{\prime\prime}\right|_{r_c}L_c.
\end{align}
Since $\eta = \kappa^2/2\Omega\Sigma$, the requirement that $\Lambda^2_\mathrm{est}$
exceeds unity, leads  to the  expectation that modes of the type under
discussion can exist  only for sufficiently large surface density
scales. Setting $\Omega^\prime\sim \Omega/r_c$ and suppose the vortensity
varies on a length scale $L_c$ implies 
\begin{align}\label{maxL1}  \Lambda_\mathrm{est} \sim
\frac{4G\Sigma K_{0}(m\epsilon_g/r_c)}{\Omega^2 L_c}. 
\end{align}
Note $\kappa\sim\Omega$ has been used to obtain the final expression.  
Thus taking $L_c\sim H$ and $\epsilon_g$ being a fraction $\epsilon_{g0}$ of the local
scale height, gives $ \Lambda_\mathrm{est} \sim 4\ln(1/(m\epsilon_{g0}h))/(\pi Q).$
On account of the logarithmic factor,
the condition $\Lambda_\mathrm{est}>1$  suggests
that edge disturbances which lead to instabilities can be present
for $Q$ significantly larger than  unity, as is confirmed numerically.
However, a surface density threshold must be exceeded.
Thus such modes associated with vortensity
{\it maxima} do not occur in a non self-gravitating disc.
 The fundamental quantity is the vortensity profile, 
Eq. \ref{maxL} indicates that if it is not sufficiently peaked,  there
can be no mode. Similarly, mode existence becomes less favoured  for
increasing softening and/or increasing $m$.  

It is important to note an analysis of the type discussed
above does not work for vortensity minima. An equation
 similar to Eq. \ref{fredholm} 
could be derived but in this case the  corresponding kernel $\mathcal{R}$ 
would be negative, implying inconsistent negative eigenvalues $\lambda.$ 
This situation results from the surface density perturbation
leading to a gravitational potential perturbation with the
wrong sign to satisfy the condition Eq. \ref{sigma_bigS}.  
Thus the type of gravitationally-dominated edge disturbance considered 
in this section is not expected to be associated 
with vortensity minima. 
The fact that edge modes associated with vortensity maxima
require a threshold value of $Q^{-1}$ to be exceeded separates
them from vortex forming modes which are  modes  associated with vortensity
minima and require $Q^{-1}$ not to be above a threshold in order
to be effective. For most  
disc models considered in this Chapter, with minimum $Q\le 2$, vortex
modes do not occur. Discs with minimum $Q\ga 2$ were considered in
Chapter \ref{paper2}.


\subsubsection{Implications for disc-planet systems}\label{imp_disc_planet}
For the locally isothermal disc models adopted in this Chapter, with a
small aspect-ratio 
$h=0.05$, edge modes could appear more easily as the disc is cold. 
This is because if  the disc thickness sets the length  scale 
of the edge,  Eq. \ref{lambda_est} and Eq. \ref{maxL}---\ref{maxL1} 
indicate that, for fixed $Q$, lowering sound-speed and hence the disc
aspect ratio increases $\Lambda_\mathrm{est}$. 

Equations above can be applied to gaps opened by a Saturn mass
planet, which is considered in linear and nonlinear numerical calculations later on.   
Without instabilities, 
these gaps  deepen with time.  There is also vortensity mixing in the
co-orbital region.  
In a fixed orbit,  these two effects respectively reduce the gap surface density 
and the magnitude of the edge vortensity peaks. 
 In this way  the conditions for edge modes 
become less favourable with time. In this case  edge modes 
are expected to  develop early on during gap formation, if at all. 
However, this effect may be less pronounced if the orbit evolves  because the planet migrates.

For larger planetary masses such as Jupiter, a deeper gap will be opened
but stronger shocks are also induced, which may lead to stronger vortensity peaks. 
Thus, in view of potentially competing effects, the conditions for edge modes 
as a function of planet migration and planetary mass must be investigated 
numerically. 


\subsection{Launching of spiral density waves}\label{GTexcite} 
Although localised at the  gap  edge, the edge disturbance
perturbs the bulk of the  disc through its gravitational potential
exciting density waves. This is expected to be through torques
exerted at Lindblad resonances \citep{goldreich79}. 
When the disc is exterior (interior) to the edge 
the wave excitation can be viewed  as occurring at the
the outer (inner) Lindblad resonance respectively. These resonances
occur  where $\sigma/m=-\Omega\mp\kappa/m.$
Here the negative (positive) sign alternative applies to the outer (inner) 
Lindblad resonance respectively. 
The perturbing potential is given by Eq. \ref{PPOT} and below
 this potential is given the symbol $\Phi_{edge}.$

The total conserved  angular momentum  flux   associated 
with the launched waves, when they are assumed to propagate out 
of the system is given by \cite{goldreich79} as 
\begin{align}
  F_T = \frac{\pi^2 mr\Sigma }{\beta}\left|\frac{d\Phi_{edge}}{dr} + \frac{2m\Omega\sbar
      \Phi_{edge}}{r\kappa^2}\right|^2.
\end{align}
where $\beta = 2\kappa(\mp \kappa^\prime -  m\Omega^\prime).$
These waves carry positive angular momentum outwards when excited by an inner edge,
or equivalently negative angular momentum inwards when excited by an outer edge. 
Thus they  will destabilise
 negative angular momentum disturbances at an inner disc edge
or   positive angular momentum disturbances  at an outer disc edge that cause their emission.
They will lead to an angular momentum flux at the boundary where $r=r_b$  given by
$(2\pi)\left[r(F_{Gr}+F_{Ar})\right]|_{r_b}=F_T$


\subsection{ Spiral density waves and the growth of edge modes}\label{spirals}
For weakly growing modes, the effects of  wave losses at the non-sharp
boundary can be considered as a perturbation on the edge disturbance. 
To do this, the assumption that the surface density tapers to zero at
the  boundary where $r=r_b$ is relaxed.  Instead, a small value is adopted there such that 
self-gravity is not important for the density waves.
Thus the domain ${\cal D}$ is retained for evaluating the gravitational
potential ($r\in [r_i,r_o]$). 
Then, for real frequencies,  all the terms in Eq. \ref{angmom_balance}
apart from the term inversely proportional  
to $\bar{\sigma}$ and the  boundary  term  associated with the advective angular momentum flux
at $r=r_b$  are real.  
Assume that the mode is close to marginal stability (subscript `ms' below) with corotation
at a vortensity maximum where $\sigma=\sigma_r$ and  write
Eq. \ref{angmom_balance} in the form 
${\cal {F}}(S,\sigma )={\cal {F}}(S,\sigma_r +\delta\sigma)=0.$
For small $\delta\sigma$, one can  expand to first order in $\delta\sigma,$  obtaining
\begin{align}\label{FINM}
  \left.\frac{\partial {\cal {F}}}{\partial \sigma }\right|_{ms}\delta \sigma
  + {\cal {F}}(S,\sigma_r)\equiv (D_r+ {\rm i}D_i)\delta\sigma +{\cal {F}}(S,\sigma_r) =0.
\end{align}
Differentiating Eq. \ref{angmom_balance} gives 
\begin{align}\label{margnx4a}
  D_r=& \left.\left(
      {\cal P}\int_{r_i}^{r_o}
      \frac{m|S|^2}{\eta^2\bar{\sigma}^2}\frac{d\eta}{dr} dr\right)
  \right|_{\sigma=\sigma_r}
   -\left.{\frac{\partial}{\partial\sigma}\left(\int_{r_i}^{r_o}
        \frac{r\Sigma}{D}\left|\frac{d S}{d r}  +
          \frac{2m\Omega\bar{\sigma}}{r\kappa^2}S\right|^2
        dr\right)}\right|_{\sigma=\sigma_r}
\end{align}
and
\begin{align}\label{margnx5a}
  D_i=\left.-
    \frac{\pi |S|^2}{m\Omega'|\Omega'|}
    \frac{d^2}{dr^2}\left(\frac{1}{\eta}\right)\right|_{ms}.
\end{align}
Setting $\delta\sigma = \delta\sigma_r+i\gamma,$ where $-\gamma$ is the growth rate
and taking the imaginary part of Eq. \ref{FINM}, noting that
the imaginary part of 
the first two terms on the RHS
 of Eq. \ref{angmom_balance}
contribute to the imaginary part of $\mathcal{F}$  giving this as  
$\mathrm{Im}(\mathcal{F})= -\pm (2/m)\left[r(F_{Gr}+F_{Ar})\right]|_{r_b}= \mp (1/(m\pi))F_T$
with $F_T$ given above, 
\begin{align}\label{margnx6}
-\gamma=\mp\frac{F_T}{m\pi D_r}+\delta\sigma_r \frac{D_i}{D_r}.
\end{align}
Similarly the real part of (\ref{FINM}) gives
\begin{align}\label{margnx6real}
 D_r\delta\sigma_r = \gamma D_i -\mathrm{Re}(\mathcal{F}),\end{align}
where $\mathrm{Re}(\mathcal{F})$ is the real part of $\mathcal{F}$.
Suppose that 
the waves emitted by the edge disturbance result in $\gamma\neq0,$
but that the back reaction   does not change the location of the
co-rotation point from that of  the marginally
 stable mode (as indicated by numerical results). Thus 
co-rotation remains at the original vortensity maximum, implying   that
conditions  adjust  so that  $\delta\sigma_r=0.$ Then: 
\begin{align}\label{margnx7}
-\gamma=\mp\frac{F_T}{m\pi D_r},
\end{align}
where the upper (lower) sign applies to an inner (outer) edge. The
inner edge case is the one considered in linear and nonlinear
calculations. Instability occurs when  wave emission
causes negative (positive) angular momentum to be  transferred to a negative (positive)
angular momentum edge disturbance located at an inner (outer) sharp edge. 

According to Eq. \ref{margnx7}, to 
obtain instability $(-\gamma>0)$ for an inner edge disturbance one
requires  
$D_r<0$ (since $F_T>0$). For a disturbance concentrated at corotation near  an inner sharp edge,
with the disc lying beyond, the first term on the  RHS 
of Eq. \ref{margnx4a} dominates. As corotation is at a  vortensity
maximum, one expects the contribution to the first integral of $D_r$ from
the region just beyond corotation,  
where $d\eta/dr < 0$, to be negative
and the contribution from the region just interior to corotation,  
where $d\eta/dr>0$, to be positive. 
Because the region exterior to an inner edge has higher 
surface density than that interior to it, and the integrand for the first term 
in $D_r$ is proportional to $\Sigma$, one may expect the region exterior to co-rotation
dominates the contribution to $D_r$, making it negative and therefore unstable. 
Indeed, it was found that $D_r<0$ for the numerical fiducial case
presented in \S\ref{num_fiducial}.  
In this case there is instability due to the reaction of 
the emitted outwardly propagating positive angular  momentum wave on
the negative angular momentum inner gap edge  disturbance.
A corresponding discussion applies to a disc with an outer sharp edge.

The above considerations depend on the excitation
of waves at a Lindblad resonance that were transported with
a conserved action or angular momentum flux towards a boundary
where they were lost. However, linear calculations and simulations
described below indicate lack of sensitivity to  such a boundary condition.
This can be understood if it is emphasised that the significant issue
is that after emission, the wave action density should not return
to the edge disturbance responsible for it. This is possible for
example if the waves become  trapped in a cavity in the wider disc in which case a radiative
boundary would not be needed.

\section{Linear calculations}\label{linear_paper3}

Numerical solutions to the linear problem are now presented. These
calculations are a continuation of those presented in Chapter
\ref{paper2} to massive disc models where edge modes are the dominant
disturbance. 
The basic state is obtained  from hydrodynamic disc-planet  
simulations by azimuthally averaging the surface density and azimuthal 
velocity component fields to obtain one dimensional profiles. The
background is thus axisymmetric.  The radial velocity is assumed to
be zero. For consistency with nonlinear simulations, the local
isothermal equation of state  $p= h^2 GM_*\Sigma/r$
with $h=0.05$ is adopted, and the self-gravity softening prescription used
is that presented in \S\ref{basic_equations}. This prescription makes
the Poisson kernel asymmetric, but this is of no practical
significance since the softening length is small (compared to the
local radius and characteristic width of the edge region).  
 

Rather than solve
in terms of the quantity $ S=\dd \Sigma c_s^2/\Sigma +\dd \Phi,$
which was convenient for analytic discussion, it is more convenient
here to work in terms of the relative surface density perturbation,
$W=\dd\Sigma/\Sigma$ for which a single governing equation may be
written down explicitly:
\begin{align}\label{localiso}
&\frac{d}{dr}\left[\frac{r \Sigma}{D}\left(c_s^2\frac{dW}{dr} +
    \frac{d\dd\Phi}{dr}\right)\right] 
+ \left[ \frac{2m}{\bar{\sigma}}
 \left(\frac{\Sigma\Omega}{D}\right)
 \frac{dc_s^2}{dr} - r\Sigma \right]W \notag\\
& + \left[\frac{2m}{\bar{\sigma}}\frac{d}{dr}
 \left(\frac{\Sigma\Omega}{D}\right)
 -\frac{m^2\Sigma}{rD}\right]\left(c_s^2 W +
 \dd\Phi\right )
\equiv {\mathcal L}(W) = 0.
\end{align}
This is the same linear equation as that used in Chapter \ref{paper2}
. Since this is a homogeneous equation, one can multiply
through by, e.g. $\sbar$, without changing the problem. However,
numerical convergence was more easily attained with
this additional factor. The term in $dc_s^2/dr$ has been kept for consistency
with simulations used to setup the basic state. By solving the linear problem
with $dc_s^2/dr=0$, this term can be verified to have 
negligible effect on the results obtained below.  
This is because $c_s$ varies on a global scale, whereas the edge
disturbance is associated with strong local gradients.

Eq. \ref{localiso} is converted to a matrix equation
$\sum_{j=1}^{N_r}\mathcal{L}_{ij}(\sigma)W_j = 0$ by discretising 
it on  $N_r=1025$ equally spaced grid points in the domain
$r=[1,10]$. The same iterative method of solution was used as in 
Chapter \ref{paper2} (\S\ref{linear_numerics}), except that  
the quadratic approximation used to study localised vortex modes in Chapter
\ref{paper2}, which gave trial eigenvalues for the full problem, 
is not appropriate here since edge modes are intrinsically global. 

However, because simulations suggest the important
disturbances are associated with the outer gap edge, it is reasonable
to search for values of  $\sigma$ such that co-rotation lies near  the outer gap
edge. The reciprocal of the final matrix condition number was verified
to be small (at the level of  machine precision) in 
order for results to be accepted. For simplicity,  the boundary 
condition $dW/dr= 0$ was applied at $r=r_i=1$ and $r=r_o=10.$
As discussed in \S\ref{Fluxbal}, modes are  found to be 
driven by the back reaction of emitted
density waves on  a disturbance located at an outer gap edge,
a process  expected to be insensitive to boundary  conditions.
Indeed, results are insensitive as  to whether
the above or other boundary conditions are used (see \S\ref{edge_mode_bc}). 



\subsection{Fiducial case}\label{num_fiducial}
As a fiducial case, consider the disc model with  $Q_o=1.5$ and with a
gap opened by a Saturn mass planet. This is adopted  in \S\ref{hydro}, 
However, simulations described in \S\ref{hydro} from which the model was extracted had 
a viscosity $\nu=10^{-5},$ whereas the linear calculations
described below are for inviscid discs.
 
The basic state gap profile is illustrated
in  Fig. \ref{linear_basic} where 
the surface density $\Sigma,$ the relative deviation
of the  angular velocity from the Keplerian value  $\Omega/\Omega_k -1$
and the vortensity $\eta$ are plotted.
As expected from analytic discussions
of modes near to marginal stability,  local extrema in
$\eta$ in the vicinity of gap edges are closely
associated with  instability. This is manifest through
mode co-rotation points being very close to them.  

\begin{figure}[!ht]
  \centering
  \includegraphics[width=1.0\textwidth]{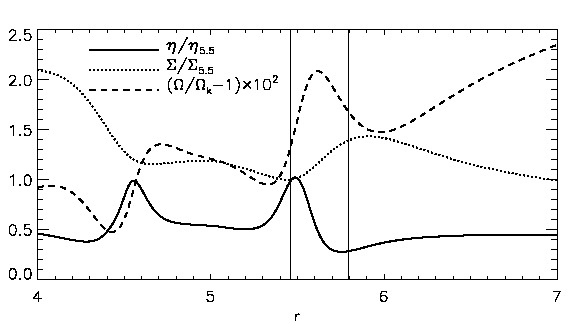}
  \caption{ Gap profile produced  by a Saturn-mass planet in the $Q_o=1.5$
    disc with viscosity $\nu=10^{-5}$. The surface
    density $\Sigma$ and vortensity $\eta$ are scaled by
    their values at $r=5.5.$ The relative  deviation of the angular velocity
     from the Keplerian value is also plotted. Vertical lines indicate
     co-rotation radii $r_c$ (where
    ${\mathrm{Re}}(\bar{\sigma})=0$) for the $m=2$ mode. With 
    self-gravity, $r_c = 5.5$, close to a vortensity maximum. When
    self-gravity is neglected $r_c = 5.8$, 
    close to a vortensity minimum. 
\label{linear_basic}}
\end{figure}

The magnitude of relative surface density perturbation, $|W|$, for
self-gravitating (SG) and non-self-gravitating (NSG) responses are
shown in Fig. \ref{Qm1.5_linear}. NSG means that  $\dd\Phi =
0$. However, the background profile is identical for both
calculations. The eigenfrequencies
for SG (NSG) modes are given by  $-\sigma = 0.1587 +
i0.4515\times10^{-2}$ ($-\sigma = 0.1458 + i0.2041\times10^{-2}$).
These correspond to co-rotation points at  $r_c=5.4626,\,5.7959$ for the
SG and NSG modes, respectively. The SG growth rate corresponds to $\sim
3$ times the local orbital period. Thus although $|\gamma|/\sigma_R \sim
0.03,$ is relatively small, the  instability grows  on a
dynamical timescale.

The co-rotation points of SG and NSG modes are very close to a local
maximum and minimum  of  $\eta$, respectively
(Fig. \ref{linear_basic}). The SG mode grows twice as fast as the NSG mode, consistent with
the observation made when comparing edge modes and vortex
modes in \S\ref{motivation}, where the former became non-linear
sooner. The SG and NSG eigenfunctions are similar around co-rotation
($r\in [5,6]$), although 
the SG mode has a larger width and is shifted slightly to the left 
(Fig.\ref{Qm1.5_linear}).
As expected from the discussion in \S\ref{GTexcite}, 
the SG mode has significant wave-like region
interior to the inner Lindblad resonance ($r\leq 3.4$) and exterior to
the outer Lindblad resonance at ($r\geq7.2$). By contrast, the  NSG
mode has negligible amplitude outside $r\in[5,7]$ compared
to that at co-rotation, whereas the
SG amplitude for  $r\in [8,10]$ can be up to $\simeq 56\%$ of
the peak amplitude near co-rotation. 
Increasing $m$ in the NSG calculation increases the amplitude in the wave regions,
but even for $m=6$, it was found that  waves in $r\in[8,10]$ for the NSG case 
have an amplitude of about 33\% of that
at co-rotation, a smaller value than that pertaining to
the SG case with  $m=2.$

The  behaviour  in the wider disc, away from co-rotation, shows that the
for low $m,$ the NSG mode is a  localised vortex mode whereas the  SG mode
corresponds to  an edge mode  with global spirals. Noting that
maxima in the vortensity and Toomre $Q$ nearly  coincide,
it makes sense that the SG mode
is global because away from co-rotation, the background Toomre $Q$ is
decreasing, which makes it easier to excite density waves, because
the evanescent zones between corotation and the Lindblad resonances
that are expected from WKBJ theory,  narrow accordingly. 
\begin{figure}[!ht]
  \centering
  \includegraphics[width=1.0\textwidth]{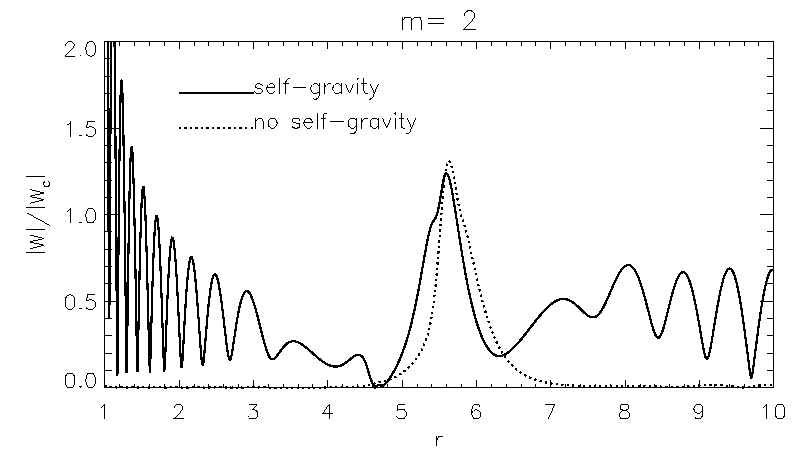}
    \caption{Unstable $m=2$ modes found from linear stability analysis 
	for the fiducial model, with
      self-gravity (solid) and without self-gravity (dotted) in the
      linear response. $|W|$ has been scaled by its value at co-rotation.  
      \label{Qm1.5_linear}}
\end{figure}

Comparing the SG and NSG modes in 
Fig. \ref{Qm1.5_linear} shows that
including self-gravity in the linear response enables  additional waves
in the disc at low $m$. Fig. \ref{Qm1.5_linear_pot} shows the gravitational 
potential perturbation for the 
SG mode. A comparison with $|W|$ in Fig. \ref{Qm1.5_linear} shows that
the surface density perturbation around $r_c$ has an
associated potential perturbation that varies on a more global scale.
The peak in $|W|$  about $r_c$ is confined to $r\in [4.6,6.2]$ whereas
that for $|\dd\Phi|$ is confined to $r \in [3.2,7.5],$ 
overlapping Lindblad resonances (see Fig. \ref{Qm1.5_t50_contour}}) in the
latter case. Thus $\dd\Phi$ is less localised  around co-rotation
than $|W|$. This is qualitatively similar to a planet --- a localised
mass with globally varying potential. 

Rapid oscillations seen in $|W|$  for $r>7$ are  not observed for
$|\dd\Phi|$ in the same region. Furthermore, $|\dd\Phi|(r>7)$ is at most
$\simeq 20\%$ of the co-rotation amplitude, which is a smaller ratio
than that for $|W|$. This leads to the notion that the disturbance about
the outer gap edge is driving the disturbance in the outer disc as in
the analytical discussion given  above. In effect, the disturbance at 
co-rotation  acts like an external perturber (e.g. a planet) to drive
density waves in the wider disc, through its gravitational field. 
Clearly, this is only possible in a self-gravitating disc.   


\begin{figure}[!ht]
  \centering
  \includegraphics[width=1.0\linewidth]{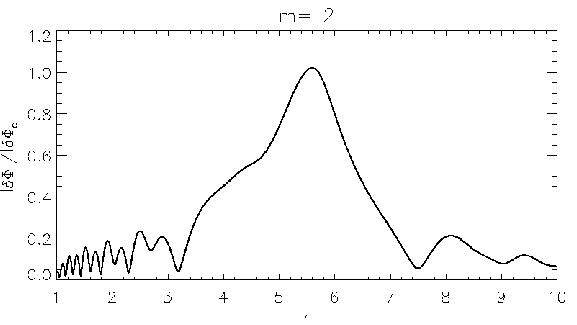}
  \caption{Gravitational potential perturbation $\dd\Phi$
    ($\equiv\delta\Phi$ in the plot),  
      scaled by its value at $r=5.5$ for the $m=2$ mode with  self-gravity. 
\label{Qm1.5_linear_pot}}
\end{figure}


\subsection{Energy balance of edge and wavelike disturbances} \label{energy_balance}
The analysis in \S\ref{analytic2} describes the edge mode
as being associated with a coupling between disturbances associated
with  vortensity extrema  near an edge 
and the smooth regions interior or exterior to the edge. 
This interpretation can be validated by applying energy balance 
to the reference case with self-gravity. 
Specifically, the region $r\in [5,10]$ is considered  because
hydrodynamic simulations indicate the spiral arms are more prominent
in the outer disc (\S\ref{motivation} and \S\ref{hydro}).  
Multiplying the governing Eq. \ref{localiso}  by 
$S^*$, integrating over $[r_1,r_2]$ and then taking the  real part
gives 
\begin{align}\label{balance}
  {\mathrm{Re}}\int_{r_1}^{r_2}{\varrho}dr = 
  {\mathrm{Re}} \int_{r_1}^{r_2}( \varrho_\mathrm{corot} + \varrho_\mathrm{wave})dr,
\end{align}
where
\begin{align}
  &\varrho \equiv r \Sigma W S^* = r(c_s^2
  |\dd\Sigma|^2/\Sigma + \dd\Sigma\dd\Phi^*) \label{totalenerg},\\
  &\varrho_\mathrm{corot} \equiv \frac{2m}{\bar{\sigma}}\frac{d}{dr}
  \left(\frac{\Sigma\Omega}{D}\right)|S|^2,\\
  &\varrho_\mathrm{wave} \equiv S^*\frac{d}{dr}\left[\frac{r \Sigma}{D}\left(c_s^2\frac{dW}{dr} +
      \frac{d\dd\Phi}{dr}\right)\right] 
  \left(\frac{\Sigma\Omega}{D}\right)
  \frac{dc_s^2}{dr}WS^*   -\frac{m^2\Sigma}{rD}|S|^2.\label{rho_wave}
\end{align}
$\varrho$ has dimensions of
energy per unit length and, for a normal mode,
 its real part is  four times the energy   per unit length associated with
pressure perturbations (the term  $\propto c_s^2$ ) and gravitational potential
perturbations (the term  $\propto \dd\Phi^{*}$).  As the factor of four 
is immaterial to the discussion, the integral of 
$\varrho$ over the region concerned will simply be referred to as the
thermal-gravitational energy (TGE).  
Note that when self-gravity dominates, the TGE is negative 
and when pressure dominates, it is positive.

When the integral is performed, the TGE is balanced by various terms on the RHS of
Eq. \ref{balance}.  For simplicity, terms on the RHS are split into just
two parts that are integrals of  $\varrho_\mathrm{corot}$ and
$\varrho_\mathrm{wave}$ over the region of interest.   
This is of course not  a unique procedure. The 
vortensity term $\varrho_\mathrm{corot}$ has been isolated because it
contains the potential  co-rotation  singularity  which can be amplified by 
large vortensity gradients at the gap edge. 
The rest  of the RHS
is collected into $\varrho_\mathrm{wave}$. 
Note that the term in $dc_s^2/dr$ is included in 
$\varrho_\mathrm{wave}$, despite being proportional to $1/\sbar$. 
This is motivated by trying to keep $\varrho_\mathrm{corot}$ as close to
the vortensity term identified in the analytical formalism 
as possible. However, it has been numerically checked that attributing
the $dc_s^2/dr$ term to  $\varrho_\mathrm{corot}$ or neglecting it
altogether makes negligible difference to results below. 
Again, this is because $c_s^2(r)\propto 1/r$ varies on a much larger scale than
vortensity gradients.  

It is important to note that while the real part of the
\emph{combination} $ \varrho_\mathrm{corot} + \varrho_\mathrm{wave}$ gives
rise to the TGE, one cannot interpret ${\mathrm{Re}}(\varrho_\mathrm{corot})$ as
an energy density of the co-rotation region. It
contains a term contributing to the TGE
that is proportional to the vortensity gradient (see below) 
and potentially associated with a corotation singularity. 
Similar arguments apply to
$\varrho_\mathrm{wave}$ which, in the strictly isothermal case, can
be seen to be associated with density waves
(see \S\ref{Fluxbal} above).  This splitting is used to show that 
for the modes of interest, the vortensity term contributes  most to
the TGE.  



\begin{figure}[!ht]
  \centering
  \includegraphics[width=1.0\linewidth]{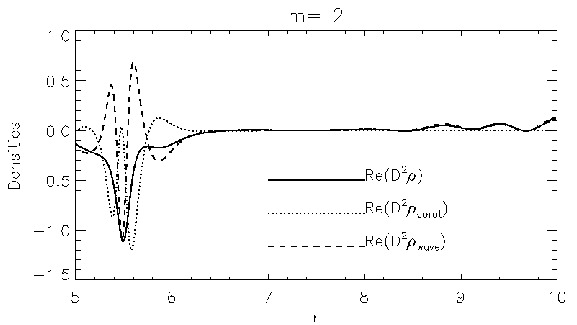}
  \caption{One-dimensional modified energy densities computed from the
    eigenfunctions $W,\, \dd\Phi$ for the fiducial case $Q_o=1.5$. 
    \label{energy_den}}
\end{figure}

Numerically however, quantities defined  above 
are inconvenient because  of the vanishing of  $D$
for neutral modes at Lindblad resonances.  
To circumvent this, one can work with  $D^2\varrho,\,
D^2\varrho_\mathrm{corot},$ and $D^2\varrho_\mathrm{wave}.$
These are called modified energy densities.
This change does not disrupt the purpose of this section, because
the most important balance  turns out to be between  
the terms involving ${\mathrm{Re}}(D^2\varrho)$ and
${\mathrm{Re}}(D^2\rho_\mathrm{corot})$, both focused near co-rotation, where $D\sim\kappa^2$. 
In the region near Lindblad resonances and beyond, 
${\mathrm{Re}}(\varrho)$ and ${\mathrm{Re}}(\varrho_\mathrm{corot})$ are  small.  
Thus the incorporation of the $D^2$ factor does not influence 
conclusions about the TGE balance. 
The  modified energy densities are
plotted in Fig. \ref{energy_den} for the $m=2$ mode in the fiducial case.
The curves share essential features 
with  those obtained  using the original definitions without the additional
factor of $D^2$ (these are 
presented and discussed in  Appendix \ref{energy}).

Fig. \ref{energy_den} shows that ${\mathrm{Re}}(D^2\varrho)$ is negative around co-rotation 
which is located near the
outer gap edge, $r\in[5,6]$). Accordingly  ${\mathrm{Re}}(\varrho)<0$
here, and  must be dominated by the
gravitational energy contribution since the pressure contribution is positive
definite.  This supports the interpretation of the disturbance as an
edge mode where self-gravity is important. If it were the 
vortex mode then  ${\mathrm{Re}}(D^2\varrho)>0$ near corotation,
because in that case, self-gravity is unimportant compared to
pressure.

For $r\geq 8.4$, ${\mathrm{Re}}(D^2\rho)$ becomes positive due to pressure effects  and
oscillates towards the outer boundary as the perturbation becomes
wave-like. Similarly, the pressure perturbation dominates towards the
inner boundary (not shown). This signifies that self-gravity  
becomes unimportant relative to pressure within these regions. This is
consistent with the behaviour of the gravitational potential perturbations, which are 
largest near co-rotation. 

Fig. \ref{energy_den} shows  that ${\mathrm{Re}}(D^2\rho_\mathrm{corot})$ and 
${\mathrm{Re}}(D^2\rho_\mathrm{wave})$ have their largest amplitudes for
$r\in[5,6]$.
Integrating over $r\in[5,10]$ gives $\tilde{U}<0$, where
\begin{align*}
  &\tilde{U}\equiv {\mathrm{Re}} \int_5^{10} D^2\varrho dr. 
\end{align*}
Using a normalisation such that $\tilde{U} = -|\tilde{U}|$ gives
\begin{align*}
  &\tilde{U}_\mathrm{corot}\equiv{\mathrm{Re}} \int_5^{10}
  D^2\varrho_\mathrm{corot} dr \simeq -0.822|\tilde{U}|,\\ 
  &\tilde{U}_\mathrm{wave}\equiv{\mathrm{Re}} \int_5^{10} D^2\varrho_\mathrm{wave} 
  dr \simeq -0.179|\tilde{U}|.  
\end{align*}
This implies the TGE is negative, and hence a
gravitationally-dominated disturbance. As
$|\tilde{U}_\mathrm{corot}/\tilde{U}|\sim 0.8$, this suggests the TGE is 
predominantly balanced by the vortensity  term which from
Fig. \ref{energy_den}, is localised in the gap edge
region. It balances the gravitational energy of the mode. Because
vortensity gradients are largest near the gap edge, gravitational
energy is most negative here  overtaking  pressure, resulting in a
negative TGE.

\subsubsection{ Further analysis of the vortensity  term}\label{furtherdecomp}
The vortensity term  $\rho_\mathrm{corot}$ defined above differs from  that
naturally identified in the alternative form of the governing equation
in \S\ref{analytic1}. However, this difference is not significant.
The vortensity term in the linear calculations can be further
decomposed as 
\begin{align*}
  &\varrho_\mathrm{corot} = \varrho_\mathrm{corot,1} +
  \varrho_\mathrm{corot,2}\\
  &\varrho_\mathrm{corot,1}  =
  \frac{m}{\bar{\sigma}(1-\bar{\nu}^2)}\frac{d}{dr}\left(\frac{1}{\eta}\right)|S|^2,\\ 
  &\varrho_\mathrm{corot,2}  =
  \frac{2m}{\kappa\eta(1-\bar{\nu}^2)^2}\frac{d\bar{\nu}}{dr}|S|^2,
\end{align*}
where $\bar{\nu}=\bar{\sigma}/\kappa$. For the time being, consider
$\varrho_\mathrm{corot,1}$ as the new vortensity term, so that
$\varrho_\mathrm{corot} \to \varrho_\mathrm{corot,1}$ and attribute
$\varrho_\mathrm{corot,2}$ to the wave term so that 
$\varrho_\mathrm{wave}\to \varrho_\mathrm{wave} + \varrho_\mathrm{corot,2}$. 
This new vortensity term is proportional  to the vortensity gradient
explicitly.  
The new definitions give
\begin{align*}
  |\tilde{U}_\mathrm{corot}/\tilde{U}|\sim 0.690.
\end{align*} 
Thus  the new co-rotation
term alone accounts for $\sim 70\%$ of the (modified) TGE and
giving  almost the same result as the previous one. In addition as
most of the contribution comes from  around co-rotation, where $|\bar{\nu}|\ll 1$. Replacing the
factor $(1-\bar{\nu}^2)$ by unity has a negligible effect on 
the results. However, making this replacement gives 
\begin{align*}
  \rho_\mathrm{corot,1}\to
  \frac{m|S|^2}{\sbar}\frac{d}{dr}\left(\frac{1}{\eta} \right),
\end{align*}
which  is exactly the vortensity term used in the analysis given in
\S\ref{analytic1}. 
Whether this term contributes positively or negatively to the TGE depends on 
the sign of $\int_{r_1}^{r_2}\varrho_\mathrm{corot,1}dr$. Most of the contribution
to this integral is expected from the co-rotation region where $\sbar\simeq0$. 
Then $\mathrm{sgn}\left(\int_{r_1}^{r_2}\varrho_\mathrm{corot,1}dr\right)
=\mathrm{sgn}(\left.\varrho_\mathrm{corot,1}\right|_{r_c})$.  
Evaluating this for a neutral mode with co-rotation at a
vortensity extremum gives  
$\left.\varrho_\mathrm{corot,1}\right|_{r_c} =
\left.-\eta^{-2}|S|^2(d^2\eta/dr^2)/\Omega^\prime\right|_{r_c}$.
Since quite generally,  $\Omega^\prime\equiv d\Omega/dr < 0,$
if co-rotation occurs at $\mathrm{max}(\eta)$
then the vortensity term contributes negatively to the TGE.
As stated above, when self-gravity dominates pressure, the TGE is negative,
so it can then be mostly accounted for  by the vortensity term. 
This is precisely the type of balance assumed for the neutral edge
mode modelled  in \S\ref{Neutedge}.

However, a consequence of the above discussion is that  if co-rotation occurs at
a minimum value of $\eta,$ then the vortensity
term contributes positively to the TGE.
This can only give the dominant contribution when the TGE is
positive, which can only be the case when the pressure contribution
dominates over that from self-gravity (see also \S\ref{S3.5}).  
In the limit of weak self-gravity, the TGE will  be $>0$ and it can be
balanced by a localised disturbance with co-rotation at a  vortensity
minimum  as happens  for the vortex modes.  


\subsection{Dependence on $m$ and disc mass} \label{varm_varQ}
Solutions to the linear eigenmode problem for discs with $1.2\leq Q_o \leq
1.6$ are presented below. 
The basic state is set up from hydrodynamic simulations as
described in \S\ref{motivation}, with $\nu=10^{-5}$. 
The local $\mathrm{max}(Q)$ near the outer gap edge ranges 
between $Q=3.3$ and $Q=4.4$. Self-gravity is included in the
response ($\dd\Phi\ne 0$).

\begin{figure}[!ht]
  \centering
  \includegraphics[width=1.0\textwidth,clip=true,trim=0cm 0.5cm 0cm 0cm]{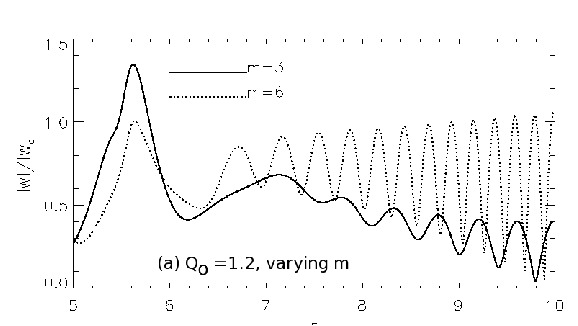}
  \includegraphics[width=1.0\textwidth,clip=true,trim=0cm 0cm 0cm
  1cm,keepaspectratio=false]{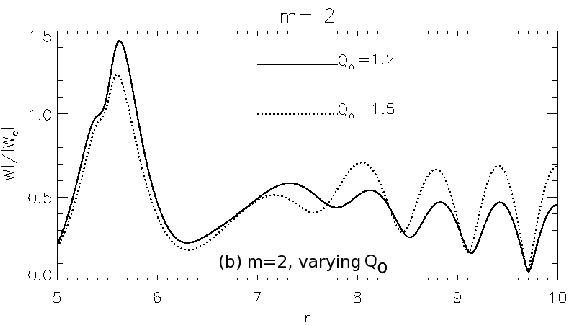} 
  \caption{The effect of azimuthal wave-number $m$ (a) and the effect
    of disc mass, parametrised by $Q_o$ (b) on linear edge
    modes. \label{varm_and_Q}} 
\end{figure}

Fig \ref{varm_and_Q}(a) compares eigenfunctions $,|W|,$ for different $m$ for the
$Q_o=1.2$ disc. Increasing $m$ increases the amplitude in the
wave region relative to that around co-rotation ($r_c \simeq
5.5$), resulting in $m=6$ being more global than $m=3$. 
High $m$ modes here do
not fit the description above of edge modes in the region of interest ($r>5$), 
because the disturbance at co-rotation becomes comparable to  
or even smaller than that in  the wave region beyond the outer Lindblad resonance.
It is then questionable to interpret the waves
as a secondary phenomenon induced by the
edge disturbance, even though the physics of the modes,
as being entities driven by an unstable interaction between edge and wave
disturbances, may be
more or less  the same.  
Hence, the edge-dominated modes described above were only found with
$m\la 3$ in simulations where the surface density perturbation is maximal  near
the gap edge.  For fixed $m=2$, hence in that  regime,
Fig. \ref{varm_and_Q}(b) shows  that lowering $Q_o$ makes the co-rotation
amplitude larger relative to that in the wave region. This can be expected because
increasing the level of self-gravity means the necessary condition for 
edge disturbance is more easily satisfied (\S\ref{Neutedge}). 


\begin{figure}[!ht]
  \centering
  \includegraphics[width=1.0\textwidth,clip=true,trim=0cm 0.5cm 0cm
  0cm]{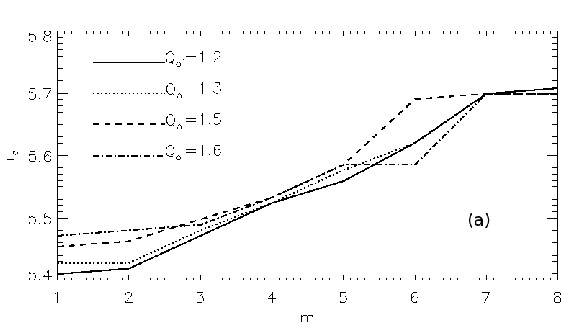}
  \includegraphics[width=1.0\textwidth,clip=true,trim=0cm 0cm 0cm
  1cm]{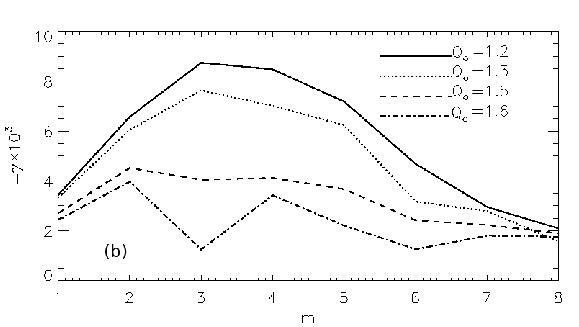}
  \caption{Co-rotation radii (a) and growth rates (b)
    as a function of azimuthal wave-number $m$, for a range of $Q_o$
    values.
    \label{compare_gamma_and_corot}}
\end{figure}

Fig. \ref{compare_gamma_and_corot} shows the
corotation radii and growth rates 
for the unstable edge modes as a function of the azimuthal mode
number, $m$ and a range of disc masses (parametrised by $Q_o$). Note
that changing $Q_o$ affects both the background state and  the linear
response, while $m$  acts as  a parameter of the linear response only. 
The plots in   Fig. \ref{compare_gamma_and_corot}(a) show
that the co-rotation radius tends to move outwards with
increasing $m$ and/or $Q_o.$ 
However, corotation is always located in the edge region where the vortensity
is either decreasing or maximal.
The largest growth rates are found for low $m$ ($\la 3$) edge-dominated modes.

Referring to the basic state (Fig. \ref{linear_basic}), one sees that
increasing $m$ and/or $Q_o$, which weakens self-gravity, 
shifts co-rotation towards $\mathrm{min}(\eta)$,
which disfavours edge dominated modes.  For sufficiently low mass or
non-self-gravitating discs, co-rotation is associated with
vortensity minima  so vortex modes dominate
(e.g. $Q_o=4$ in \S\ref{motivation}).  Thus one only expects edge
dominated  modes for low $m$ and sufficiently large  disc mass.    
An  exception to the general trend above is that  the $Q_o=1.6,\,m=6$
mode, has a smaller co-rotation radius  than the corresponding 
mode  with $Q_o < 1.6$. This 
may be a boundary effect because for  this value of  $m$ and $Q_o$
the modes are distributed through the outer disc  and
require the implementation of  accurate  radiative
boundary conditions, rather than the simplified  boundary conditions 
actually applied.  However, such high $m$ modes typically have growth rates
smaller than  those for lower $m$, accordingly
they are not expected to dominate, nor are they observed in the
nonlinear simulations discussed later.

Fig.\ref{compare_gamma_and_corot}(b) shows that
growth rates increase as $Q_o$ is 
lowered (increasing surface density scale). 
Note that decreasing $Q_o$ results in deeper gaps due to 
increased effective planetary mass (see \S\ref{gapprofiles}).      
Steeper gaps are expected to be more
unstable, therefore there is a contribution to the increased growth
rate  from changes to the background profile as the disc mass is
increased. This also contributes to the larger disturbance
amplitude around co-rotation relative to the wave region as $Q_o$ is
lowered, as observed in Fig. \ref{varm_and_Q}(b).


One expects that the effect of lowering $Q_o$, through the linear response $\dd\Phi$, is 
to destabilise edge dominated modes as they rely on self-gravity. 
However, the effect of self-gravity through the response is weaker for
larger $m$ values because the size of the Poisson kernel decreases. 
Hence, the increase of the growth rates with disc mass 
for $m=4$---6  is likely more attributable to the effect of
self-gravity on the basic state. Growth rates for  the most unstable
mode approximately doubles as the  disc mass is increased by 
25\% ($Q_o=1.5\to Q_o=1.2$). Notice for fixed $Q_o\le1.5$, the
$m=3$---5 growth rates are similar but $m>3$ are not edge modes.

For $Q_o=1.2,\,1.3$ the most
unstable mode has  $m=3,$ whereas for $Q_o=1.5,$ the $m=2$ mode has
the largest growth rate. As discussed in \S\ref{analytic2}, edge dominated 
modes require sufficient disc mass 
and/or low $m$. For fixed $m$, they are stabilised with increasing $Q_o$. 
Hence, for edge modes to exist with decreasing disc mass, they must shift to 
smaller $m$. The higher $m$ modes are less affected by self-gravity
through $\dd\Phi$.  
This may explain the double peak in growth rate plotted as a function of $m$ 
for $Q_o=1.6$ in moving from $Q_o=1.5$ (for $Q_o=1.5$ the $m=3$ growth rate
is also slightly smaller than $ m= 2,\,4$).



\subsubsection{Energy partition}

  The integrals of the modified energy densities  for each
azimuthal wavenumber are given for the
fiducial case with $Q_o=1.5$ in Table \ref{Qm1.5_varm}.
 Only the  modes with $m\leq2$ are edge dominated modes
 because  for these sum of the integrals contributing to the (modified) TGE 
 is negative, corresponding to a gravitationally dominated disturbance
and over 50\% of the energy is accounted by the vortensity edge term. 
The $m=2$ mode had the largest growth rate 
with corotation radius at $r=5.46.$ This is fully consistent
with the nonlinear simulation of the fiducial case performed 
in \S\ref{Firstsim}. There the dominant mode in the outer disc had $m=2$
and corotation radius at $r\simeq 5.5.$

For $m>2$, the total (modified) energy  becomes positive, which
must be due to the pressure term ($c_s^2|\dd\Sigma|^2/\Sigma$) 
becoming dominant. The vortensity term alone cannot
balance this  because it contributes negatively.  Increasing $m$
shifts co-rotation to increased radius, away from the local vortensity
maxima, and its contribution becomes smaller in magnitude. However, it
does not change sign because $r_c$ never reaches the local vortensity minimum.  
For $m\geq 4$, the vortensity term contribution is small consistent with high 
$m$ solutions being predominantly wave-like(Fig. \ref{varm_and_Q}). 
However, note that in spite of this the vortensity term may still play some role
in driving the modes through corotation torques. 
These trends have also been found for other models 
so they appear to be general. However, high $m$ global modes are unimportant 
as they do not develop in nonlinear simulations.  

\begin{table}
  \centering
  \caption{Eigenfrequencies, 
       corotation radii and 
     modified energy integrals  for
    different  azimuthal mode numbers, $m$, for the fiducial  disc with
    $Q_o=1.5$. The modified energy integrals are taken over  the outer disc 
    $r\in[5,10]$.\label{Qm1.5_varm}}   
  \begin{tabular}{ccccll}
    \hline\hline
    $m$ & $-\sigma_R$ & $-\gamma\times10^{2}$ & $r_c$ &
    $-\tilde{U}_\mathrm{corot}/|\tilde{U}|$ & $-\tilde{U}_\mathrm{wave}/|\tilde{U}|$\\ 
    \hline
    1 & 0.079 & 0.270 & 5.454  &
    0.65 & 0.35 \\
    2 & 0.159 & 0.452 & 5.463  &
    0.82 & 0.18 \\
    3 & 0.237 & 0.404 & 5.498  & 
    0.34 & -1.24 \\
    4 & 0.313 & 0.412 & 5.533  &
    0.026 & -1.03 \\
    5 & 0.386 & 0.366 & 5.585  & 
    0.0048 & -1.004 \\
    6 & 0.462 & 0.242 & 5.603  &
    0.0015 & -0.998 \\
    7 & 0.524 & 0.223 & 5.699  & 
    0.0010& -0.997 \\
    8 & 0.599 & 0.189 & 5.699  & 
    0.00099 & -0.995 \\
   \hline
  \end{tabular}
\end{table}


\subsection{Softening length}
Gravitational softening  has to be used in a
two-dimensional calculation.  It prevents a singularity at $r=r'$ in
the Poisson kernel and approximately accounts for the disc's vertical
dimension but is associated with some uncertainty.  
Apart from the linear response, softening also has an effect through 
the gap profile set up by nonlinear simulations. A fully self-consistent 
treatment requires a new simulation with each new softening considered.

For reasons of numerical tractability though,
experiments were only performed using two values
of softening parameter $\epsilon_{g0,\mathrm{e}}$ in simulations
to setup the gap profile
and range of values $\epsilon_{g0,\mathrm{l}}$ 
used in the linear response. In nonlinear simulations a single disc
potential softening parameter 
$\epsilon_{g0}$ is used. Results are summarised in Table
\ref{soft_expts}. The integrated total 
modified energy values indicate edge modes 
since the major contribution is  due to the co-rotation/vortensity
term.  

Comparing growth rates for fully self-consistent cases
$\epsilon_{g0,\mathrm{e}}=\epsilon_{g0,\mathrm{l}} =0.3,\,0.6$ shows
that increasing the softening length stabilises edge modes, which is
expected since they are driven by self-gravity. In addition  all
growth rates for $\epsilon_{g0,\mathrm{e}}=0.6$ are smaller than those
for $\epsilon_{g0,\mathrm{e}}=0.3$ independently of
$\epsilon_{g0,\mathrm{l}}$. This indicates  stabilisation of  edge
modes via the basic state when softening is increased (since the
effective planet mass decreases with increasing
$\epsilon_{g0,\mathrm{e}}$, resulting in a shallower profile). 
For fixed $\epsilon_{g0,\mathrm{e}}$, $|\gamma|$ decreases as
$\epsilon_{g0,\mathrm{l}}$ increases. Eventually, 
edge modes for $\epsilon_{g0,\mathrm{l}} \geq
0.6$ with $\epsilon_{g0,\mathrm{e}}= 0.3$ were not found, nor for
$\epsilon_{g0,\mathrm{l}} \ga 0.83$ when $\epsilon_{g0,\mathrm{e}}=
0.6$. An upper limited is expected from analytical considerations
(see \S\ref{Neutedge}) because if self-gravity in the linear response is
too weak, a self-gravitating vortensity/co-rotation disturbance
cannot be maintained. And because the edge mode requires gravitational
interaction between gap edge and the smooth disc, if self-gravity is
too weak, then this interaction is also weak so edge dominated modes are
suppressed for large softening lengths. 

\subsubsection{Convergence issues}\label{linear_convergence}
For fixed $\epsilon_{g0,\mathrm{e}}=0.3$, the energy ratios in
Table \ref{soft_expts} indicate convergence as
$\epsilon_{g0,\mathrm{l}}\to0$. Perhaps surprisingly, 
$|\tilde{U}_\mathrm{corot}/\tilde{U}|$ decreases with softening. 
This is because $\varrho_\mathrm{wave}$, as defined  by equation
(\ref{rho_wave}), includes a term proportional to $d^2\dd\Phi/dr^2$
which was found to be the dominant 
contribution to $|\tilde{U}_\mathrm{wave}|$, and its contribution mostly comes from 
the co-rotation region (as the potential perturbation is largest there).  
As the strength of self-gravity is increased by decreasing softening, 
the edge disturbance is modified, but the subsequent effect of the
$d^2\dd\Phi/dr^2$ term on the energies cannot be anticipated a priori. 
 
For $\epsilon_{g0,\mathrm{l}}=0.03$, the vortensity term does not
dominate the modified total energy as much. 
However, this case is in fact not self-consistent 
because the softening used to setup the gap profile is an order of 
magnitude larger than that used in linear calculations. 
The limit of zero softening is also physically irrelevant because 
the disc has finite thickness. For the self-consistent cases with 
reasonable softening values ($0.3,\,0.6$), the vortensity
term dominates the energy balance, so the analytic description developed 
above works reasonably well for more realistic setups.  

\begin{table}
  \centering
  \caption{As for Table \ref{Qm1.5_varm}
    but for linear calculations with different softening
    parameters, for the $Q_o=1.5$  fiducial case and azimuthal mode number
    $m=2$. Subscripts `l' denote the softening parameter used in the linear
    response, and `e' denotes the  softening  parameter used in setting up the basic
    state. Note that although growth rates vary
    by about a factor of two, the location of the corotation
    radius does not vary  significantly. \label{soft_expts}}   
  
    \begin{tabular}{cccccl}
    \hline\hline
    $\epsilon_{g0,l}$ & $\epsilon_{g0,e}$ &
    $-\sigma_R$ & $-\gamma\times10^{2}$  &
    $-\tilde{U}_\mathrm{corot}/|\tilde{U}|$ &  $-\tilde{U}_\mathrm{wave}/|\tilde{U}|$\\ 
    \hline
   0.03& 0.3 & 0.159 & 0.560 & 0.66 &    
    0.33\\    
    0.05& 0.3 & 0.159 & 0.544 & 0.67 &
    0.32\\ 
    0.1 & 0.3 & 0.159 & 0.508 & 0.69 &
    0.30\\ 
    0.2 & 0.3 & 0.159 & 0.461 & 0.74 &
    0.23 \\ 
    0.3 & 0.3 & 0.159 & 0.452 & 0.82 & 
    0.18 \\
    0.5 & 0.3 & 0.158 & 0.435 & 0.87 &
    0.14 \\
    0.3 & 0.6 & 0.159 & 0.420 & 0.91 &
    0.09 \\
    0.5 & 0.6 & 0.158 & 0.410 & 0.95 &
    0.064 \\
    0.6 & 0.6 & 0.158 & 0.375  & 0.99 &
    0.0017 \\
    0.8 & 0.6 & 0.158 & 0.276  & 1.15 &
    -0.14 \\
   \hline
  \end{tabular}
\end{table}


\subsection{Edge mode boundary issues}\label{edge_mode_bc}
The boundary condition $W^\prime=0$ was applied for
simplicity. Vortex modes in low mass or non-self-gravitating discs are
localised and insensitive to boundary conditions
\citep{valborro07}. This is explicitly demonstrated in Table
\ref{bound_expts} where the fiducial problem was solved with
self-gravity switched off in the linear response, for various boundary
effects described below. The edge dominated modes rely on self-gravity
and are therefore intrinsically global, boundary effects cannot be
assumed unimportant a priori.

\begin{table}
  \centering
  \caption{Eigenfrequencies for miscellaneous linear
    calculations with different boundary conditions. These calculations
    neglect self-gravity and are therefore vortex
    modes. In the `none' BC case, derivatives at the boundaries
    are approximated with interior grid points. Solutions are essentially independent of boundary
    effects. \label{bound_expts}}    
  \begin{tabular}{cccccc}
    \hline\hline
    $r_0$ & $r_m$& BC &$-\sigma_R$ & $-\gamma\times10^{2}$ & $r_c$ \\
    \hline
    1 & 10 & $W^\prime = 0$ & 0.1458 & 0.2041 & 5.7959  \\
    2.5 & 10 & $W^\prime = 0$ &  0.1458  & 0.2041 &  5.7930 \\
    1 & 10 & none          &  0.1458   & 0.2041 & 5.7959 \\
    1.1 & 9.8 & $W^\prime = 0$ &    0.1458  &  0.2041  & 5.7941 \\
    \hline
  \end{tabular}
\end{table}

Additional experiments were performed with varying boundary
conditions for the fiducial  case ($Q_o=1.5$). These include 
re-locating the inner boundary to $r=1.1,\,2.5$ or the outer boundary to $r=9.8$ and
approximating/extrapolating boundary derivatives using interior grid points
\citep{adams89}. In the last case, the linear ODE is applied at endpoints of the grid
so that boundary conditions are not explicitly set. This is strictly a numerical procedure
to generate a linear system to solve, which may or may not correspond to 
physical solutions. However, this numerical experiment gave similar results to
cases where boundary conditions are explicity imposed. 

\begin{table}
  \centering
  \caption{Same as Table \ref{bound_expts} but now with
    self-gravity, so these are the edge modes. \label{bound_expts2}}
  \begin{tabular}{cccccc}
    \hline\hline
    $r_0$ & $r_m$& BC &$-\sigma_R$ & $-\gamma\times10^{2}$ & $r_c$ \\
    \hline
    1 & 10 & $W^\prime = 0$ & 0.1587 & 0.4515 & 5.4626   \\                         
    2.5 & 10 & $W^\prime = 0$ &  0.1587  & 0.4689 & 5.4640 \\
    1 & 10 & none          & 0.1591 & 0.4904 &  5.4538    \\
    1.1 & 9.8 & $W^\prime = 0$ &  0.1591  &  0.4794 & 5.4542 \\
    \hline
  \end{tabular}
\end{table}

\begin{figure}[!h]
   \centering
   \includegraphics[width=1.0\textwidth]{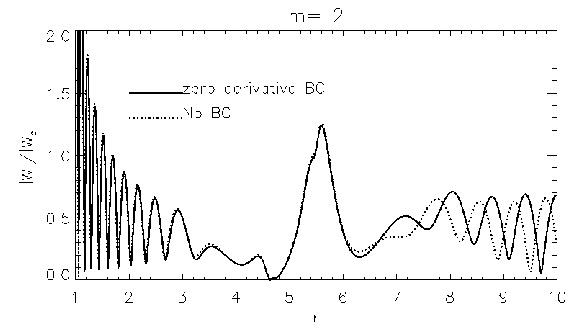}
   \caption{Effect of boundary conditions on edge modes. The solid
     curve denote the reference solution with $W^\prime=0$ applied at
     boundaries, the dotted line indicate evaluation of linearised
     equations at boundaries by approximating boundary derivatives
     using interior points. 
     \label{Qm1.5_varBC}}
 \end{figure}

For self-gravitating disc calculations giving edge dominated modes,
Table \ref{bound_expts2} show these various conditions gave
co-rotation radii  varying by about 0.2\% and growth rates varying by
at most 10\%. Overall the eigenfunctions  $W$  are
similar, e.g. Fig. \ref{Qm1.5_varBC}, showing  the essential features
of the edge mode, including relative amplitude of outer disc wave
region and the co-rotation region. The existence of edge dominated
modes is therefore not too sensitive to boundary effects. Physically
this is because edge modes are  mainly driven by the local vortensity
maximum or edge, which is an internal feature away from boundaries.  


\section{Nonlinear hydrodynamic simulations}\label{hydro}
Having understood the basic physics of linear edge modes, their
nonlinear evolution can be studied with hydrodynamic simulations.  
Disc models with $1.2 \leq
Q_o \leq 2$, corresponding to disc masses $0.079M_* \geq M_d \geq
0.047M_*$, are simulated. Most simulations use a viscosity of
$\nu=10^{-5}$ which has been typically adopted for protoplanetary
discs. This value has  also been 
found to suppress vortex instabilities \citep{valborro07}. 
The fiducial self-gravitational softening  parameter 
is set to $\epsilon_{g0}=0.3$. The planet has mass $M_p/M_* = 3\times
10^{-4}$ and is set on a fixed circular 
orbit at $r_p=5$ in order  to focus on gap stability. The
planet potential is introduced at $25P_0$ and its gravitational
potential ramped up  over $10P_0$, where $P_0$ is the Keplerian period
at $r=5$. 
The effects of varying 
viscosity and softening lengths are explored later in this section. 
Planetary migration is briefly considered in the next section and
explored in more detail in Chapter \ref{edge_migration}.

\subsection{Numerics} 
The numerical setup and methods are the same as those employed in
Chapter \ref{paper2}. The hydrodynamic equations are evolved with the
\fargo code including self-gravity \citep{masset00a, baruteau08}.  
The disc is divided into $N_r\times N_\phi = 768\times2304$  grid
points in radius  and azimuth giving a resolution of $\Delta r/H =
16.7$ cells per scale-height. The grid cells 
are nearly square, with $\Delta  r/r\Delta\phi = 1.1$. 
Open boundaries are imposed at $r_i$ and non-reflecting boundaries at
$r_o$ \citep{godon96}\footnote{Here, open boundary means the flow out of
the domain is allowed while flow into the domain is inhibited. The non-reflecting
boundary applies a similar condition, but 
to flow characteristics (Riemann invariants), rather than to the primitive flow variables directly. 
Waves then propagate out of the domain but not into it.}. 
The latter has also been used in the  self-gravitating
disc-planet simulations of \cite{zhang08}. As argued above, 
because edge modes are physically driven by an internal edge, boundary  
conditions are not expected to significantly affect whether or not
they exist. Indeed, simulations with open outer boundaries, or damping
boundaries \citep{valborro06} all show development of edge modes.

\subsection{Overview}
\begin{figure}
\centering
\includegraphics[width=0.32\textwidth,clip=true,trim=1.5cm 0cm 0cm 0cm]{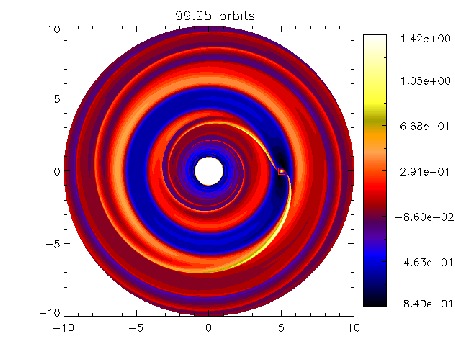}
\includegraphics[width=0.32\textwidth,clip=true,trim=1.5cm 0cm 0cm 0cm]{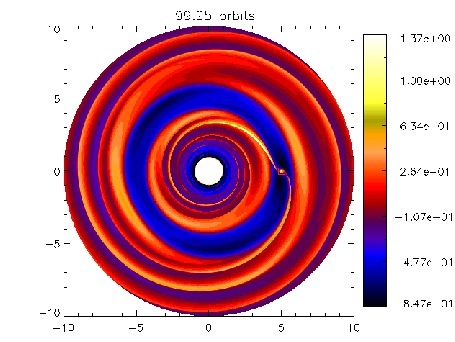}
\includegraphics[width=0.32\textwidth,clip=true,trim=1.5cm 0cm 0cm 0cm]{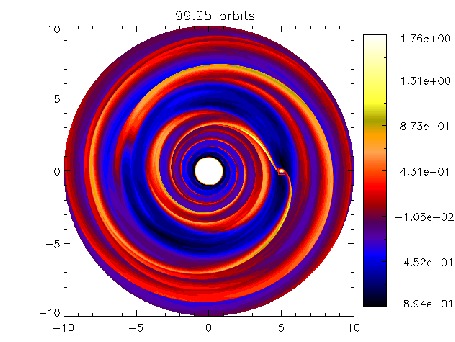}
\caption{Surface density perturbations (relative to $t=0$) at
  $t= 100P_0$ as a function the minimum value of Toomre $Q$ in the
  disc, $Q_o$. From left to right: $Q_o=2.0,\,1.5,\,1.2$. 
\label{qm_density}}
\end{figure}

Fig. \ref{qm_density} shows the relative surface
density perturbation ($\Delta\Sigma/\Sigma$) for $Q_o=1.2,\,1.5$ and
2.0. The gap profile formed in these cases have local $\mathrm{max}(Q)=
3.3, \, 4.2,\,5.3$ near the outer gap edge and the average $Q$ for
$r\in [6,r_o]$ is 
$1.6,\,2.0,\,2.6$. Although  the edge mode is associated with the  local
vortensity maximum (located close to  $\mathrm{max}(Q)$ for gap profiles),
it requires coupling to the external smooth disc via
self-gravity. Hence, edge modes only develop if $Q$ in these smooth regions
is sufficiently small, otherwise global disturbances cannot be
constructed.

The case $Q_o=2.0$ gives a standard result for gap-opening planets,  
typically requiring planetary masses comparable or above that of Saturn. 
For that mass a partial gap of depth $\sim
50\%$ is formed,  a  steady state attained and remains stable to the
end of the simulation. This serves as a stable case for comparison
with more massive discs. This simulation was repeated with
$\nu=10^{-6}$, in which case we identified  both vortex and edge
modes. The standard viscosity $\nu=10^{-5}$ thus suppress both types
of instabilities for $Q_o=2$.  Although  $\nu=10^{-5}$ was adopted to avoid 
complications from vortex modes, note that they are stabilised 
by sufficiently massive  discs anyway (e.g. $Q_o=1.5,\, \nu=10^{-6}$
do not show vortices, see \S\ref{motivation} and Chapter
\ref{paper2}).

Gap edges become unstable for the fiducial case $Q_o\leq 1.5,\, \nu= 10^{-5}.$
As implied by linear analysis, the $m=2$ edge mode 
dominates when  $Q_o=1.5$ and it  saturates by $t=100P_0.$
It is more prominent in the outer disc, and over-densities deform the gap edge
into an eccentric ring. 
The inner disc remains fairly circular, though weak non-axisymmetric
disturbance ($m=3$) can be identified. 
Note the global nature of the edge modes in the outer
disc: spiral arms can be traced back to disturbances at the gap edge
rather than the planet. Consider the spiral disturbance at the outer
gap edge just upstream of the planet where $\Delta\Sigma/\Sigma$ is
locally maximum. Moving along this spiral outwards,
$\Delta\Sigma/\Sigma$ reaches a local minimum at $(x,y)=(6,-4)$ 
before increasing again. This is suggestive of the outer disc spirals
being induced by the edge disturbance.   

For the $Q_o=1.2$ case,  there is significant
disruption to the outer gap edge, it is no longer clearly
identifiable. The $m=1$ edge mode  becomes  dominant in the outer
disc, although overall it  still appears to carry  an $m=2$
disturbance because of the perturbation due to the planet. The inner
disc becomes visibly eccentric with an $m=2$ edge mode
disturbance. The shocks due to edge modes can comparable strength to
those induced by the planet.  


\subsection{Development of edge modes for  $Q_o=1.5$}
 
\begin{landscape}
\begin{figure}
  \centering
  \includegraphics[scale=0.56          ,clip=true,trim=0cm
  1.83cm 1.645cm 0cm]{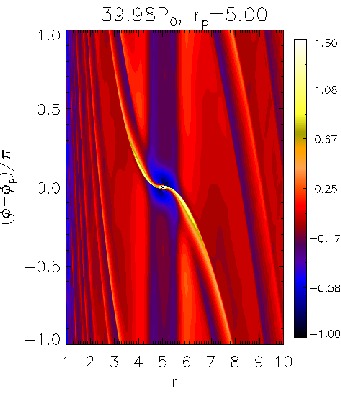}
  \includegraphics[scale=0.56               ,clip=true,trim=2.34cm
  1.83cm 1.645cm 0cm]{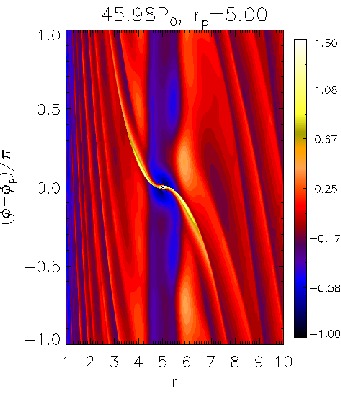}
  \includegraphics[ scale=0.56                  ,clip=true,trim=2.34cm
  1.83cm 1.645cm 0cm]{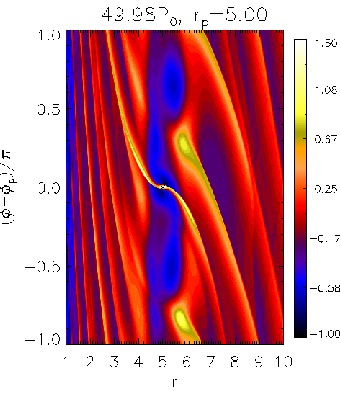}
  \includegraphics[ scale=0.56                 ,clip=true,trim=2.34cm
  1.83cm 0cm 0cm]{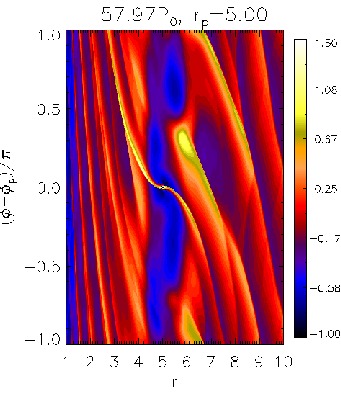}\\
  \includegraphics[scale=0.56    ,clip=true,trim=0cm
  0cm 1.645cm 0cm]{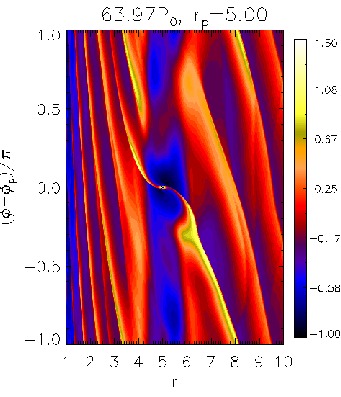}
  \includegraphics[scale=0.56                ,clip=true,trim=2.34cm
  0cm 1.645cm 0cm]{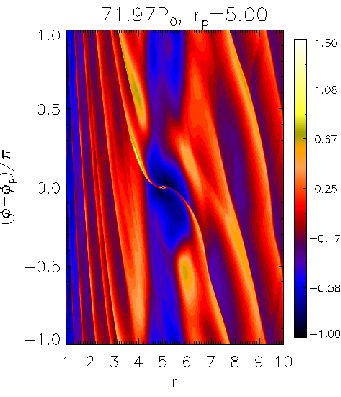}
  \includegraphics[ scale=0.56                  ,clip=true,trim=2.34cm
  0cm 1.645cm 0cm]{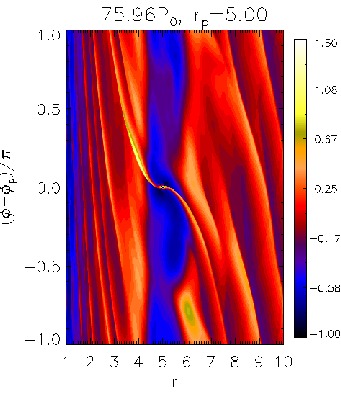}
  \includegraphics[ scale=0.56                 ,clip=true,trim=2.34cm
  0cm 0cm 0cm]{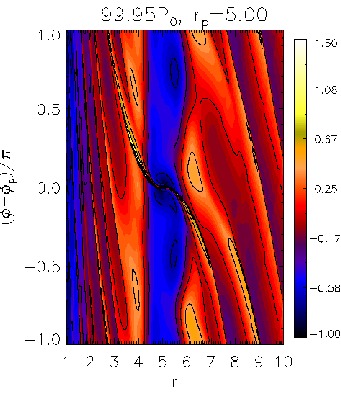}
  \caption{Evolution of  the surface density perturbation 
(relative to $t=0$ ) for the $Q_o=1.5$
    case from $t\sim 40P_0$ to $t\sim 100P_0$.   
    \label{Qm1.5_nonlin1}}
\end{figure}
\end{landscape}

Fig. \ref{Qm1.5_nonlin1} shows the development of the
edge mode in the fiducial disc $Q_o=1.5$. The final dominance of $m=2$
is consistent with linear calculations. The planet opens a
well-defined gap by $t\sim 40P_0$, or  
$\sim 5P_0$ after the planet potential has been  fully 
ramped up. At  $t\sim 46P_0,$ two over-dense blobs, associated with the 
edge mode  can be  identified at the outer gap
edge. At this time the gap only has $\Delta\Sigma/\Sigma \simeq -0.3$,
implying edge modes can develop \emph{during} gap formation, since the
steady state stable gap in $Q_o=2$ reaches $\Delta\Sigma/\Sigma \simeq
-0.6$. This can be expected from the discussion in
\S\ref{imp_disc_planet}.   
Edge modes are global  and are associated with
long trailing spiral waves  with density perturbation  amplitudes
that are  not small compared to
that at the gap edge.
This is clearly seen at  $t\sim 50P_0$ when spiral
shocks have already developed.
The edge modes have a growth rate consistent with predictions from
linear theory and   become nonlinear within the time frame  $46P_0 <
t < 50P_0$, i.e. dynamical time-scales.

The edge perturbations  penetrate the 
outer gap edge and trail an angle  similar to the planetary wake.
Notice the $m=3$ disturbance near $r=4$ in 
the snapshots taken at $t=50P_0$ and $t=56P_0.$  Two of the three
over-densities, seen as a function of azimuth,
correlate to the $m=2$ mode in the outer disc. The third  over-density
adjacent to the planet is probably caused by the outer planetary wake
because of the disc self-gravity. It is also
possible that different $m$ modes have developed for the inner and
outer discs, since the analytic description of edge modes instability only
involves one edge. 

The presence of both planetary wakes (with pattern speed $\simeq
\Omega_k(5)$) and edge modes (with pattern speed $\simeq 
\Omega_k(5.5)$) of comparable amplitude enable direct interaction
between them. 
Passage of edge mode
spirals through the planetary wake may disrupt the former, explaining some of the 
apparently split-spirals in the plots. 
At 
$t\sim 64P_0$  Fig. \ref{Qm1.5_nonlin1} indicates  that a 
spiral density wave feature coincides with the
(outer) planetary wake. 
The enhanced outer planetary wake implies an  increased negative torque 
exerted on the planet at this instant. 
This effect occurs during the  overlap of positive
density perturbations. Assuming that  at a fixed radius the
edge mode spiral has azimuthal thickness $nh$ where $n$ is a
dimensionless number, then the time taken for this spiral to cross the
planetary wake is $  \Delta T = nh/|\Omega_k(5.5) -
  \Omega_k(5)| \simeq 0.06nP_0$. This is $\ll P_0$ for $n=O(1)$. Such 
crossing spiral waves will induce sharply peaked oscillations in the 
disc-planet torque.

The nonlinear evolution of edge modes in disc-planet systems can give
complicated surface density fields. The gap becomes  highly 
deformed.  Fig. \ref{Qm1.5_nonlin1} shows that  large voids develop to
compensate for the over-densities  in spiral  
arms, leading to  azimuthal gaps ($t=64P_0$). 
At $t=72P_0$, the gap width ahead of the planet is narrowed by the
spiral disturbance at the outer edge trying to connect  to that at the 
inner edge. 
However, this gap-closing effect is opposed by the planetary  torques
which tend to open it.

A quasi-steady state is reached by $t\geq 76P_0$  showing an eccentric
gap. For $t=99.5P_0$ additional contour lines are
plotted to indicate  two local surface density maxima along the edge mode
spiral adjacent to the planet. This spiral is disjointed around
$r\sim7.5$ where on traversing an arm,  there is a minimum in $\Delta\Sigma/\Sigma$.
 Taking co-rotation of the spiral at $r_c=5.5$ gives the
outer Lindblad resonance of an  $m=2$ disturbance at $r=7.2,$ assuming  a
Keplerian disc. This is within the disjointed region of the spiral
arm, which supports the interpretation that the disturbance at the
edge is driving activity in the outer disc by perturbing it
gravitationally and launching waves at outer Lindblad resonances. 

\subsubsection{Eccentric gaps}
\begin{figure}
   \centering
\includegraphics[width=0.32\textwidth,clip=true,trim=1.5cm 0cm 0cm 0cm]{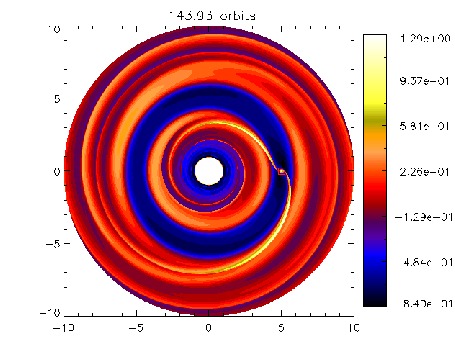}
\includegraphics[width=0.32\textwidth,clip=true,trim=1.5cm 0cm 0cm 0cm]{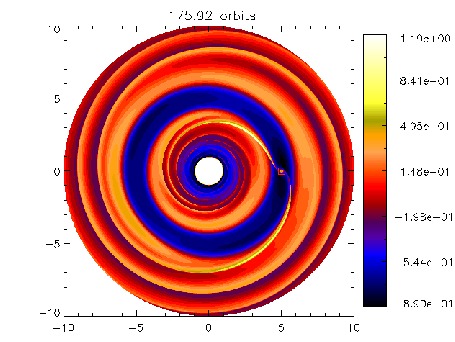}
\includegraphics[width=0.32\textwidth,clip=true,trim=1.5cm 0cm 0cm 0cm]{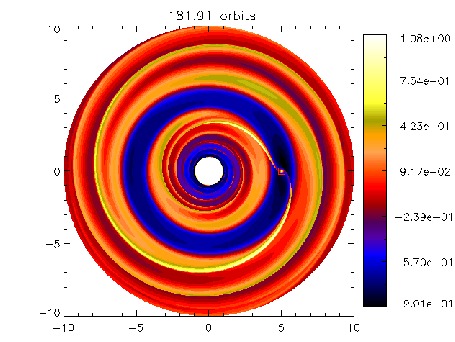} 
   \caption{Precession of an eccentric gap in the
     presence of an $m=2$ edge mode for a disc with  $Q_o=1.5.$ 
     \label{Qm1.5_ecc}}
 \end{figure}

Several snapshots of the  $Q_o=1.5$ run  
taken during  the second half of the simulation are
shown in Fig. \ref{Qm1.5_ecc}, which  display  the precession of an
eccentric outer gap edge. Deformation of a  circular
gap into an eccentric one requires an $m=1$  disturbance.
By inspection of the form of the relative surface
density perturbation for  $Q_o=1.5$
  (see Fig. \ref{qm_density}), edge modes are seen 
 associated with an  eccentric outer  gap edge. 

In their non-self-gravitating disc-planet simulations, \cite{kley06}
found  eccentric discs can result for planetary masses $M_p > 0.003M_*$
fixed on circular orbit. Simulations here show that edge modes in a
self-gravitating disc can provide an additional perturbation to deform
the gap into an eccentric shape for lower mass planets.

\subsection{  Evolution of the gap structure  for  $Q_o=1.2$}

\begin{figure}[!ht]
   \centering
   \includegraphics[scale=0.5,clip=true, trim=1cm 0cm 3.3cm
   0cm]{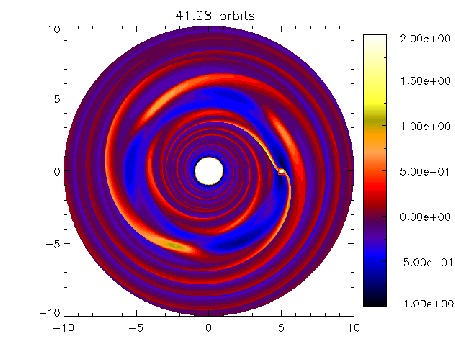}\includegraphics[scale=0.5,clip=true, trim=1cm 0cm 0cm
   0cm]{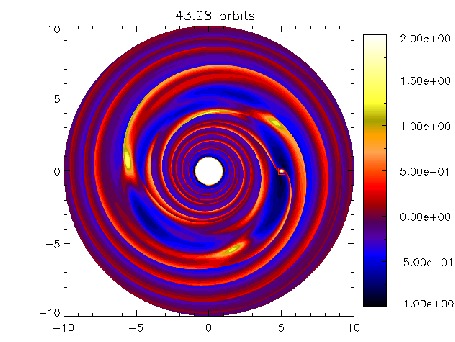}
   \caption{ The disc model with  $Q_o=1.2$: emergence of the $m=3$ edge mode.
     \label{Qm1.2}}
 \end{figure}

With increasing disc mass,  the gap profile evolves away from its original form
. Fig. \ref{Qm1.2} illustrates  the emergence of the $m=3$ edge
mode, expected from linear calculations, and subsequent
evolution of the gap profile. 
By $t\sim 42P_0$ the $m=3$ edge mode has already become nonlinear,
whereas for  $Q_o=1.5$ the $m=2$ mode only begins to emerge at
$t\sim 44P_0,$ being consistent with linear calculations that the former has
almost twice the growth rate of the latter. In Fig. \ref{Qm1.2} 
there is a  $m=3$ spiral at $r<r_p$ in phase with the
mode for  $r>r_p.$ The gap narrows where they almost touch. These are
part of a single mode.
By  $t=44P_0,$   shocks  have formed   and  
extend continuously across the gap.  The   self-gravity of the disc
 has overcome the planet's gravity which is responsible for
gap-opening. Edge mode spirals can be more prominent than planetary wakes. 

\begin{figure}[!ht]
   \centering
   \includegraphics[width=1.0\textwidth,clip=true,trim=0cm 0cm 0cm
   1cm]{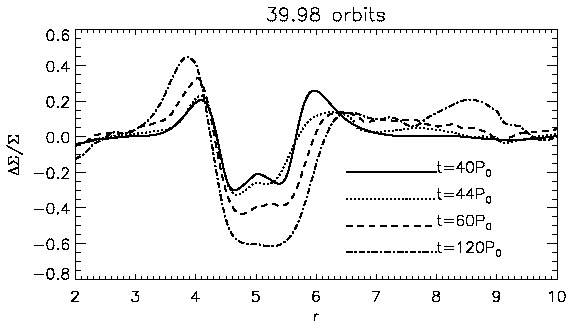}
   \caption{ The disc model with  $Q_o=1.2$:
     evolution of azimuthally averaged surface density perturbation. 
     \label{Qm1.2_aziavg}}
 \end{figure}

The evolution of the form  of the gap  is shown  in  the one-dimensional  averaged profiles
plotted in Fig. \ref{Qm1.2_aziavg}. The quasi-steady  profile before the onset 
of the edge mode  instability is 
$t\sim 40P_0.$ 
By $t=44P_0$, the original bump at $r=6$  has
diminished. This bump originates from the planet expelling
material as it opens a gap but is subsequently `undone' by the edge
mode (with associated over-density around co-rotation at $r\simeq
5.5$) as it grows and attempts to 
connect across the gap. This gap 
filling effect is  possible as non-axisymmetric 
perturbation  of the inner disc  may be induced via the outer  disc self-gravity.
Notice also in Fig. \ref{Qm1.2_aziavg}
at $t=44P_0$ the increased surface density at $r\simeq7.5$ implying
radial re-distribution of material. 
By $t=120P_0$, the
bump at the inner gap edge ($r=4$) is similar to that produced by a
standard gap-opening planet. By 
contrast, the outer bump cannot be maintained. The gap widens and
there is an overall increase in surface density for $r>6.5$ 

\subsubsection{Vortensity field}

\begin{figure}[!ht]
   \centering
   \includegraphics[width=1.0\textwidth,clip=true,trim=0cm 0cm 0cm
   1cm]{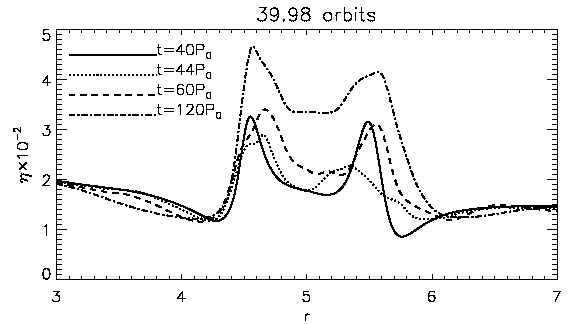} 
   \caption{ The disc model with  $Q_o=1.2$: evolution of the
     azimuthally averaged vortensity $\eta$ profile in the region of the gap. 
     \label{Qm1.2_vortensity}}
 \end{figure}

Fig. \ref{Qm1.2_vortensity} illustrates
the evolution of the vortensity profile within the
gap. The  development of edge modes temporarily reduces the
amplitude of the vortensity maxima set up by the perturbation of the planet.
This is particularly noticeable when the profiles at 
$t=40P_0$ and  $t=44P_0$ are compared.  However, vortensity
is generated through material passing through shocks
induced by the planet.
This provides a vortensity source that enables the amplitudes of
vortensity maxima to increase again, as can be seen from the profile  at $t=60P_0.$ 
The maxima remain roughly symmetric about the planet's location at
$r_p\pm 2r_h$.  Note a general increase in  the co-orbital vortensity level
caused by  fluid elements repeatedly
passing  through  shocks.  

\begin{figure}
  \centering
  \includegraphics[scale=0.39          ,clip=true,trim=0cm
  0.0cm 1.88cm 0cm]{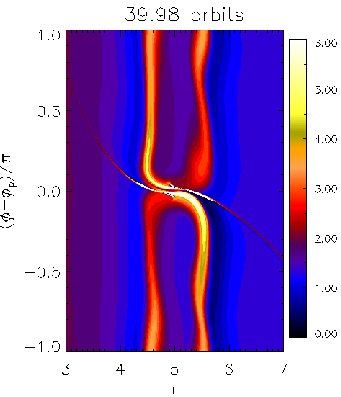}
  \includegraphics[scale=0.39                 ,clip=true,trim=2.34cm
  0cm 1.88cm 0cm]{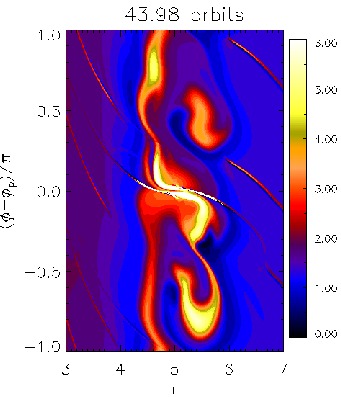}
  \includegraphics[ scale=0.39                  ,clip=true,trim=2.34cm
  0cm 1.88cm 0cm]{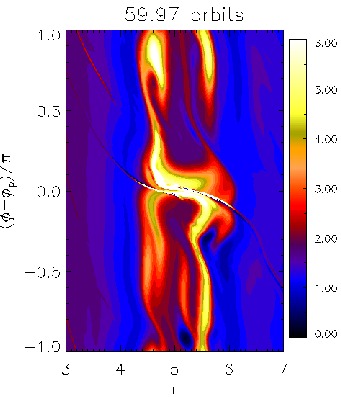}
  \includegraphics[ scale=0.39                 ,clip=true,trim=2.34cm
  0cm 0cm 0cm]{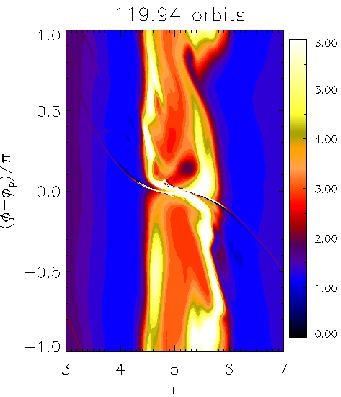}
  \caption{Evolution of the non-axisymmetric vortensity field for the
    $Q_o=1.2$ disc model. 
    \label{Qm1.2_vortxy}}
\end{figure}

The corresponding non-axisymmetric
vortensity field is shown in 
Fig. \ref{Qm1.2_vortxy}. The important point here is that, 
edge modes destroy the axisymmetric rings of
vortensity maxima, because they are associated with
$\mathrm{max}(\eta)$. However, for the outer gap edge, local vortensity 
minima resides exterior to maxima. Hence, global edge modes in the
outer disc affect vortensity minima too. This contrasts to vortex
modes, which are localised and associated with vortensity minima and
their development do not affect the vortensity  maxima rings.   

\subsection{Similarity between an edge mode and a planet} 
The similarity between the effects induced by an  edge mode and an
external perturber can be further illustrated   by the
induced  mass transport. 
Just as when a  planet opens a gap, the development of an  edge
disturbance   results in a  radial re-distribution of  mass.   
Angular momentum transport can be expressed in terms of an $\alpha$ 
viscosity parameter, defined through  $\alpha = \Delta u_r\Delta
u_\phi/c_s^2$, where $\Delta$ here denotes deviation from the
azimuthal average. 

\begin{figure}[!ht]
   \centering
   \includegraphics[width=0.99\linewidth]{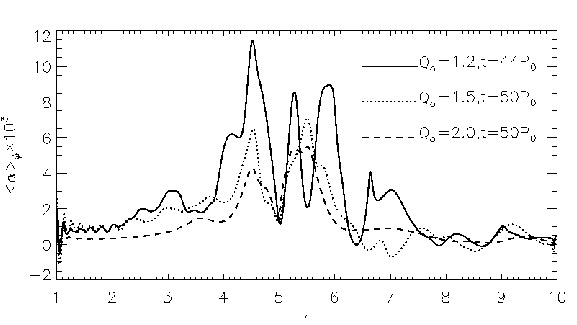}  
   \caption{Azimuthally averaged Reynolds stresses associated with edge modes 
     when they become nonlinear ($Q_o=1.2,\,1.5$), the $Q_o=2$ case
     is shown as a stable case with no edge modes for
     comparison.
     \label{masstrans}}
 \end{figure}

Angular momentum is also transmitted  through  gravitational torques.
However, in practice this was found to be a  small effect  compared to 
transport  due to Reynolds stresses. The discs here are not
gravito-turbulent. 
The azimuthally averaged $\alpha$ parameters 
$\langle\alpha\rangle$  are shown in
Fig. \ref{masstrans}. The stable disc with $Q_o=2$ has
non-axisymmetric features entirely induced by the planetary
perturbation and $\langle\alpha\rangle$ is localised about the
planet's orbital radius giving a 
two-straw feature about $r_p$. This is typical of disc-planet interactions,
and it provides  a  useful case
for comparison with  more massive discs to see the effect
of edge modes and their  spatial dependence.  

Increasing self-gravity to $Q_o=1.5$, angular momentum  transport is 
enhanced around $r_p$ and for  $r<4$, signifying the global nature of 
edge modes. Towards the inner boundary, $\alpha\simeq 10^{-3}$
being   twice as large as for  the case with  $Q_o=2.$
The $Q_o=1.5$ case  also has enhanced
wave-like behaviour for  $r>7$ as compared to $Q_o=2.0.$ 

The $Q_o=1.2$ case is more dramatic, with $\langle\alpha\rangle$ reaching
a factor of 2--3 times larger than the imposed physical viscosity
   around the planet's location ($\nu/h^2 = 4\times10^{-3} $). 
This case displays a three-straw feature, with an additional 
trough at  about $r=5.5$, i.e. close to edge mode co-rotation, as if another
planet were placed there. 
Although the $Q_0=1.5$ case also develops edge modes, there is no such 
 additional
trough at co-rotation. This may because the  $Q_o=1.2$ case  is more
unstable,  developing  an  $m=3$ mode ( which produces  3 additional 
effective coorbital  planets
placed at the gap edge), whereas the  $Q_o=1.5$ case  develops a less prominent
$m=2$ mode. However, compared to $Q_o=1.5$,  $Q_o=1.2$ has no enhanced
transport away from the gap region.

\begin{figure}[!ht]
   \centering
   \includegraphics[width=0.99\linewidth]{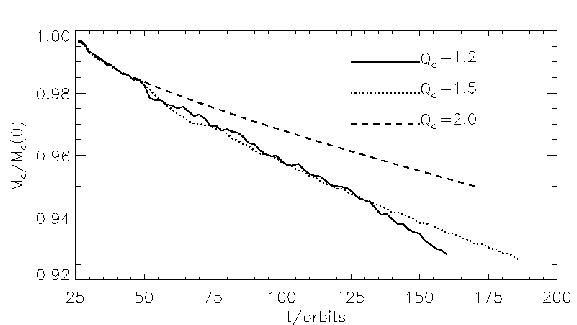} 
   \caption{Evolution of the total disc mass for the three cases in
     Fig. \ref{masstrans}, scaled by initial disc mass.   
     \label{masstrans_md}}
 \end{figure}

Fig. \ref{masstrans_md} shows the  evolution of disc mass as a
function of time, $M_d(t)$.  
Viscous evolution leads to mass loss ($Q_o=2.0$) and there is clearly
enhanced mass loss if edge modes develop ($Q_o=1.2,\,1.5$). For $t\leq
50P_0$ the curves are identical, this interval includes gap formation and
the emergence of edge modes. When edge modes become nonlinear,
mass loss is enhanced ($t\simeq 50P_0$), but there is no significant
difference between $Q_o=1.2,\,1.5$, which is consistent with the
previous $\langle \alpha\rangle $ plots near the inner boundary. The $Q_o=1.2$ curve
is non-monotonic, though still decreasing overall, possibly because
of boundary effects, and mass loss is enhanced once more around
$t=125P_0$. As $Q_o$ is lowered, $M_d(t)$ becomes less smooth,
showing the dynamical nature of edge modes. 

\subsection{Additional effects}
Edge modes are affected by softening and viscosity. To analyse mode
amplitudes in these cases, the surface density field is Fourier
transformed in azimuth giving $a_m(r)$ as the (complex) amplitude of
the $m^{\mathrm{th}}$ mode. $a_m$ is then integrated over the outer
disc $r>5$ to obtain $C_m =\int a_m dr$. 

\subsubsection{Softening}
In linear theory, edge modes could not be constructed for
gravitational softening parameters that are too large. The $Q_o=1.5$
disc has been simulated with $\epsilon_{g0}=0.6,0.8,\,1.0$ to confirm
this. 

\begin{figure}[!ht]
   \centering
   \includegraphics[width=1.0\textwidth]{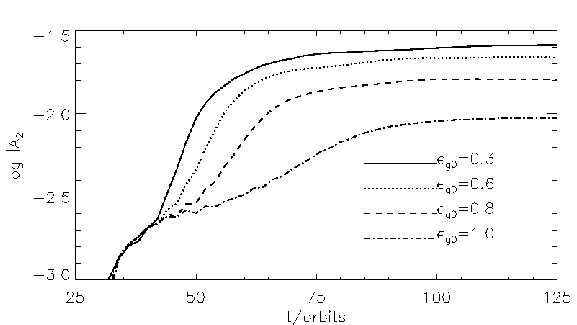} 
   \caption{Running-time average of the surface density $m=2$ Fourier amplitude, integrated over
     $r\in[5,10]$ and scaled by the axisymmetric component, 
     for the $Q_o=1.5$ disc, as a function of the  gravitational softening
     length. 
     \label{vsoft}}
 \end{figure}

Fig. \ref{vsoft} shows the evolution of the running-time averaged $m=2$ amplitudes. 
The evolution for all cases is very similar for $t\leq
40P_0$. Unstable modes develop thereafter, with increasing growth rates
and saturation levels as $\epsilon_{g0}$ is lowered.
$\epsilon_{g0}=1$ displays either no growth or  a  much reduced growth rate. 
The gap for $Q_o=1.5,\,\epsilon_{g0}=1.0$ reaches a steady state
close to the stable case with $Q_o=2.0,\,\epsilon_{g0}=0.3$.  This is
consistent with the fact that as softening increases, edge disturbances
can no longer perturb the remainder of the disc via self-gravity. The
edge itself is also stabilised against gravitationally dominated
disturbances, according to linear theory. The residual
non-axisymmetric structure is due to the planetary perturbation.  

Interestingly, the decrease in
saturation level in going from $\epsilon_{g0}=0.3\to0.6$ is smaller than going from
$\epsilon_{g0}=0.6\to0.8$, which is in turn smaller than $\epsilon_{g0}=0.8\to1.0$,
despite a larger relative increase in softening in one pair than the next.  
This is suggestive of convergence for the nonlinear saturation amplitude  
as $\epsilon_g\to0$. This can be expected 
from convergence in linear theory (\S\ref{linear_convergence}).   
However,
  at $\epsilon_{g0}=0.3$ there are only 5 cells per softening
  length. To probe even smaller $\epsilon_{g0}$ requires prohibitively high
  resolution simulations.  
  Furthermore, since $\epsilon_g$ approximately accounts for 
  the disc's non-zero vertical extent, very small softening lengths 
  are not physically relevant to explore.  


\subsubsection{Viscosity}
An important difference between edge dominated  modes and the already
well-studied vortex modes in planetary gaps is the effect of
viscosity on them. The standard viscosity $\nu=10^{-5}$ prevents vortex mode
development, but \S\ref{motivation} showed that such
viscosity values still allow the $m=2$ edge modes to develop. Vortices are
localised disturbances and more easily smeared out than global spirals
in a given time interval. Hence, vortex growth is inhibited more
easily by viscosity than spirals. 

\begin{figure}[!hb]
   \centering
   \includegraphics[width=1.0\textwidth]{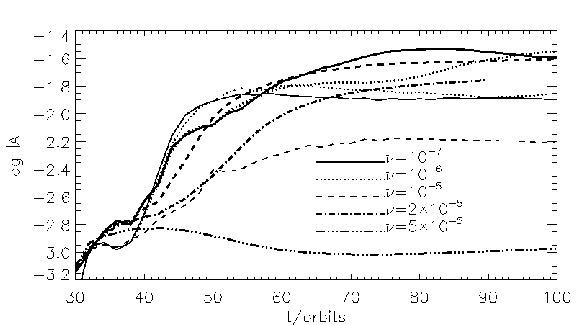} 
   \caption{Mode amplitudes 
     as a function of applied viscosity $\nu$ for $Q_o=1.5$. 
     The Fourier amplitudes are scaled by the axisymmetric component
     and its running-time average plotted. Thick lines indicate the $m=2$ mode and is
     plotted for all $\nu$. Thin lines indicate the $m=3$ mode and are
     plotted for $\nu=10^{-7}$---$10^{-5}$ only. 
     \label{vvisc}}
 \end{figure}

The $Q_o=1.5$ run was repeated with a range of viscosities. Results are
shown in Fig. \ref{vvisc}. Generally, lowering viscosity 
increases the amplitudes of the non-axisymmetric modes.
For $\nu\leq 10^{-6}$, the $m=3$  mode emerges first 
(note that this is not in conflict with linear calculations 
which used $\nu=10^{-5}$ in its basic state), rather than $m=2$.  
This is because lowering viscosity allows a sharper vortensity peak to develop in the background model, 
and thus has the same effect as increasing the disc mass. 
The latter generally enable higher $m$ edge modes.

Unlike vortex modes, even with twice the standard viscosity
($\nu=2\times10^{-5}$) the edge mode develops and grows. It is only
 suppressed when $\nu\geq5\times
10^{-5}$. However, this is because no vortensity maxima could be setup
at the gap edges due to vortensity diffusion in the co-orbital region
 making an almost uniform vortensity distribution in the gap (e.g. the
 final plot in Fig. \ref{Qm1.2_vortensity}). Since 
the necessary condition for edge modes is not achieved, no linear
modes exist. This differs  from the nature of the suppression of
vortex modes, because in that case vortensity extrema can still be
setup.

\section{Applications to disc-planet interactions}\label{disc-planet}
In order to focus on the issue of gap stability, 
the planet was held on a fixed orbit. 
However, torques exerted by the non-axisymmetric disc
on the planet will induce migration that could occur
on a short time scale \citep[e.g.][]{peplinski08b}.
Such torques  will be affected by the presence
of edge modes.

\subsection{Disc-planet torques}
The presence of large-scale edge mode spirals of comparable
amplitude as the planetary wake will significantly modify
disc-on-planet torques. 
Here, this torque is measured while still keeping the planet on fixed
orbit. This eliminates torques \emph{due} to migration. 

\begin{figure*}
   \centering
\includegraphics[width=0.49\textwidth]{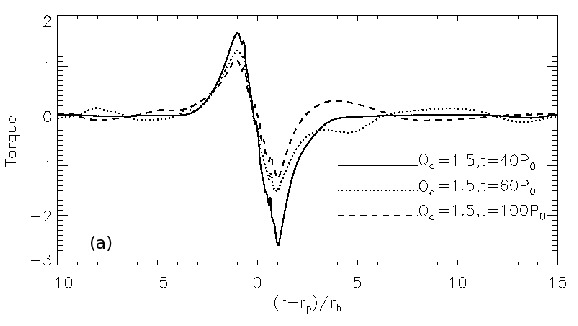}
\includegraphics[width=0.49\textwidth]{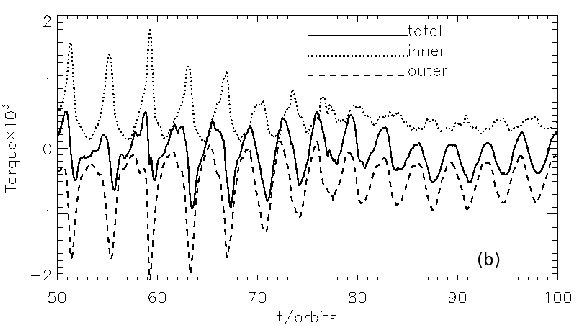}
\includegraphics[width=0.49\textwidth]{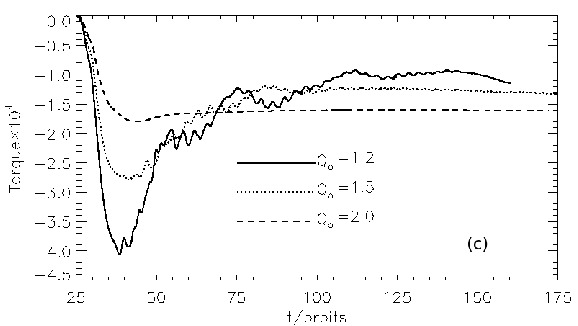} 
   \caption{(a) Disc  torque per unit radius acting on the planet.
     (b) Time dependent evolution of the disc planet torques, the  contribution
        from the inner disc (dotted curve), the
     outer disc (dashed curve) and the  total torque (solid curve).
     (c) The running  time average of the total
      torque acting on the planet  as a function of $Q_o.$  
     \label{Qm1.5_torques}}
 \end{figure*}

Fig. \ref{Qm1.5_torques}(a) shows
the evolution of the 
torque per unit length  for the $Q_o=1.5$ fiducial case. 
The
torque profile at $t=40P_0,$ before the edge modes
have developed significantly,  shows the outer torque is larger in
magnitude than inner torque, implying inward  migration. This is a
typical result for disc-planet interactions and serves as a  reference. 
At $t=60P_0$ and $t=100P_0$, edge modes mostly modify the  torque contributions
exterior to the planet, though some disturbances are seen in the
interior disc ($r-r_p<-5r_h$, recall $r_h$ is the Hill radius). 
The original outer torque at $+r_h$, is reduced in magnitude as material
re-distributed to concentrate around the  co-rotation radius $r_c$ of
the edge mode 
($r_c$ is $\sim2r_h$ away from the planet). 

Edge modes can both enhance or reduce disc-planet
torques associated with planetary wakes. Consider 
$r-r_p\in [3,5]r_h$ in 
Fig. \ref{Qm1.5_torques}(a).
 Comparing the situation at $t=60P_0$ to that at 
$t=40P_0$, the torque contribution from this region is seen to be  more
negative. This is because  an edge mode spiral overlaps the
planetary wake and therefore contributes an  additional negative torque on
the planet.
 However, at $t=100P_0$, an edge mode spiral is just upstream of the
planet,  exerting  a positive torque on the planet, hence the torque
contribution from  $r-r_p\in [3,5]r_h$ becomes positive. 

Differential rotation between edge modes and the planet produces
oscillatory torques, shown in Fig. \ref{Qm1.5_torques}(b). Unlike typical
disc-planet interactions, which produce inward  migration, the total
instantaneous torques in the presence of edge modes can be of either
sign. For a disturbance with $m$ fold symmetry the time interval between
encounters with the planet is $\Delta T = (2\pi/m)/\dd\Omega$, where
$\dd\Omega$ is the difference in angular velocity of the planet and
the disturbance pattern.  Approximating $\delta \Omega =
|\Omega_k(r_p) - \Omega_k(r_c)|$ with $r_c=5.46$ obtained from linear
theory gives $\Delta T = 4.04P_0$. Indeed, the oscillation period in
Fig. \ref{Qm1.5_torques} is $\simeq 4P_0$. 
Both inner and outer torques oscillate, but since the edge mode is
more prominent in the outer disc, oscillations in the outer torque are
larger, particularly after mode saturation ($t>70P_0$).

Time-averaged torques are compared in Fig. \ref{Qm1.5_torques}(c). In
all three disc masses,  on average a negative torque acts on the planet and its magnitude
largest at $t=40P_0$. 
Up to this point, the  torque becomes more negative as the disc mass increases, as expected.
However, development of edge modes makes the averaged torques more
positive, and eventually reverses the trend with $Q_o$.  While the disc
with $Q_o=1.5$ also attains steady torque values as in $Q_o=2$, the disc with
$Q_o=1.2$ has significant  oscillations
even after time-averaging and is remains  non-steady at the end of the run. 

Before the
instability sets in, torques arise from wakes induced by the planet,
and the outer (negative) torque is dominant. Edge modes concentrate
material into spiral arms, leaving voids in between. Therefore, except
when spiral arms  cross the planet's azimuth, the surface density in 
the planetary wake is reduced because it resides in the void in between edge mode
spirals. Hence the torque magnitude is
reduced. If the reduction in surface density due to voids in between edge mode spirals is
greater than the increase in surface density scale produced 
by decreasing $Q_o,$ the  presence of edge modes will, 
on average make the torque acting on the planet more positive.


\subsection{Outward scattering by spiral arms}\label{edge_scatter}
If edge modes develop, planetary migration may be affected by them.
Their association with gap edges inevitably affects co-orbital
flows. The more long term effect of edge modes on migration is explored in
Chapter \ref{edge_migration}. 
This subsection  highlights an
interesting effect found when edge modes are present. This is
\emph{outward} migration induced by spiral
arms  analogous to vortex-induced rapid migration
described in Chapter \ref{paper1}.

\begin{figure}
   \centering
   \includegraphics[width=1.0\textwidth]{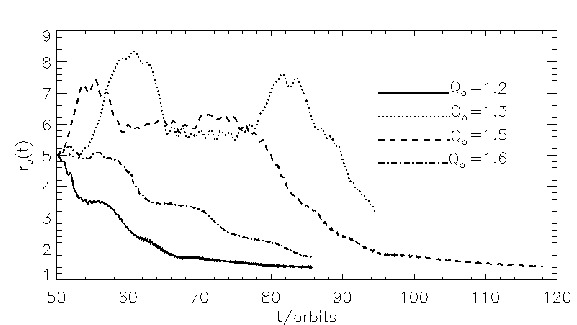} 
   \caption{Orbital radius evolution in discs with edge mode
     disturbances. The instantaneous orbital radius $r_p$ is shown as a
     function of time.
     \label{orbit3_rp}}
 \end{figure}

Some of the simulations described  above were restarted 
at $t=50P_0$  allowing the planet 
to move in response to the gravitational forces due to the disc.
 Fig. \ref{orbit3_rp} shows the orbital radius of the
planet in discs with edge modes present. 
No obvious trend with
$Q_o$ can be seen. This is because of oscillatory torques (positive or negative) 
 due to edge modes and
the partial gap associated with a Saturn mass planet, which is associated
with  type III migration \citep{masset03}. The initial direction of migration can
be inwards or outwards depending on the relative positioning  of the  edge mode
spirals with respect to the planet at the time
the planet was first allowed to move. 

Outward  migration is seen for $Q_o=1.3$ and $Q_o=1.5.$  
For $Q_o=1.5$, the planet migrates by $\Delta r = 2$, or 8.6 times its
initial Hill radius, within only $4P_0.$  This is essentially the result of  a
scattering event. 
Repeating this run with  tapering applied to  contributions 
to disc torques from  
within $0.6r_h$ of the planet, outward  scattering still 
occurs, but limited to $\Delta r = 1$ and the planet remains at
$r\simeq 6$ until $t=128P_0$. Hence, while the subsequent inward
migration seen for  $Q_o=1.5$ may be associated with
conditions close to the planet, 
the initial outward scattering is due to an exterior edge mode spiral.

For $Q_o=1.3$ the planet is scattered to $r_p\simeq 8$.  
It is interesting to note that the subsequent
inward migration for $Q_o=1.5$ and $Q_o=1.3$ stalls at $r\simeq6$, i.e. the
original outer gap edge. The planet remains there for sufficient time for
both gap and edge mode development and for $Q_o=1.3$ a second episode of
outward scattering occurs.  
Interaction with edge modes can affect
the disc well beyond the original co-orbital region of the planet. In the
case of outward  migration, it may promote gravitational activity in
the outer disc. However, boundary effects may be important after
significant outward scattering.  

\begin{figure}[!hb]
   \centering
   
\includegraphics[scale=0.6, clip=true, trim=0cm 1.85cm 0cm
   0cm]{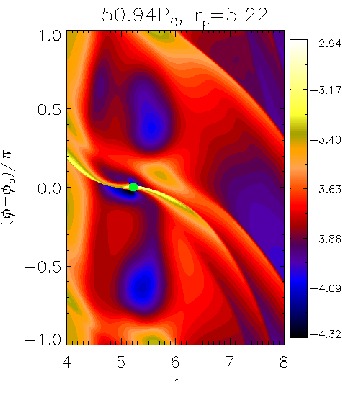}\includegraphics[scale=.6,clip=true,trim=2.2cm 1.85cm 0cm 0cm]{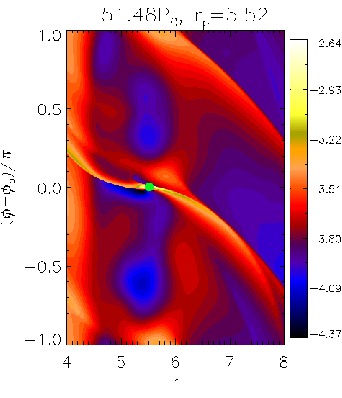}\\
   \includegraphics[scale=.6,clip=true,trim=0cm 0cm 0cm 0cm]{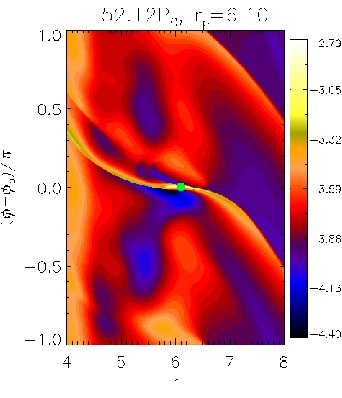}\includegraphics[scale=.6,clip=true,trim=2.2cm 0cm 0cm 0cm]{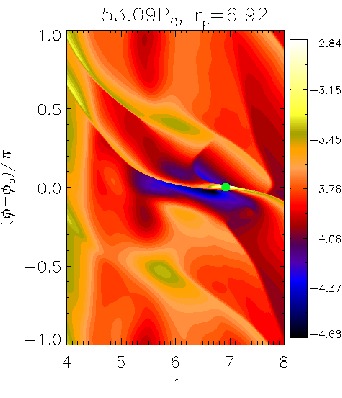}

   \caption{Interaction between an edge mode spiral and the planet (green dot), 
	causing the latter to be scattered outwards. $\log\Sigma$ is shown. 
     \label{Qm1.5_scatter}}
 \end{figure}

The spiral arm-planet interaction for $Q_o=1.5$ above is detailed in
Fig. \ref{Qm1.5_scatter}.   
A spiral arm approaches the planet from the
upstream direction ($t=50.9P_0$) and exerts  positive torque on the planet,
increasing $r_p$. The gap is asymmetric in the azimuthal direction
with the surface density upstream  of the planet being larger
than that downstream of the planet ($t=51.5P_0$).
 As the spiral arm  passes through the
planetary wake ($t=51.5P_0$---$52.1P_0$) the gap surface density just
ahead of the planet builds up as an edge mode is setup across the
gap. Notice the fluid blob at $r=5,\,\phi-\phi_p=0.3\pi$. This
signifies material executing horseshoe turns ahead of the planet. 
At the later time $t=53.1P_0$ the planet has exited the original gap, in
 which the
average surface density is now higher than before the scattering.
 Material that was originally outside
the gap loses angular momentum and moves into the original gap. This
material is not necessarily that composing the spiral.

Migration through  scattering by spiral arms  differs from
vortex scattering. Vortices form about vortensity minima, which
lie further from the planet than vortensity maxima.
Vortensity maxima are  stable in the case of vortex formation. This
means the ring of vortensity maxima must be disrupted in order for
vortices to flow across the planet's orbital radius for direct
interaction. Edge modes themselves correspond to a  disruption of
vortensity maxima, hence direct interaction is less hindered than in  the
vortex case. Furthermore, a vortex is a material volume of fluid. Vortex-planet 
scattering results in an  orbital radius change of both objects.
However, a spiral pattern is generally not a material volume of fluid.
It can be seen in Fig. \ref{Qm1.5_scatter} that the local
$\mathrm{max}(\Sigma)$ in the spiral remains  at 
approximately the same radius before and after the
interaction. The spiral pattern as a whole does not move inwards. In this
case, the spiral first increases  the  planet's orbital radius, thereby
encouraging it to interact with the outer gap edge and scatter that
material inwards. 

\section{Summary and discussion}\label{conclusions_paper3}
The investigation of planetary gap stability has been extended to   
massive protoplanetary discs. 
Vortex instabilities,
associated with local vortensity minima, have previously 
been found to occur in weakly or non
self-gravitating discs. However,  edge modes  that are  associated with local
vortensity maxima and require sufficiently strong  self-gravity
were  found in the more massive discs in this Chapter. These have very
different properties in that they are global 
rather than local and are associated with large scale spiral arms
as well as being less affected by viscous diffusion.

For the disc-planet models here, with fixed planet mass
$M_p=3\times10^{-4}M_*$ and fixed disc aspect-ratio $h=0.05$, edge
modes develop for gap profiles with an average Toomre $Q\la 2$ exterior 
to the planet's orbital radius. In the unperturbed smooth disc, 
this corresponds to $Q\la 2.6$ at the planet's location
and $Q\la 1.5$ at the outer boundary, or equivalently a disc mass
$M_d\ga 0.06M_*$.  For fixed $M_d$, nonlinear simulations show 
edge modes develop readily for uniform kinematic
viscosity $\nu\la 2\times10^{-5}$ and self-gravitational softening
$\epsilon_g\la 0.8H$. 

A theoretical description  of edge modes was developed. The edge mode
is interpreted as a disturbance associated with an interior vortensity
maximum that requires self-gravity to be sustained, and which
further perturbs the remainder of the disc through its
self-gravity causing the excitation of density waves. These density
waves acts to destabilise the edge disturbance. 
Linear calculations confirm this picture and also show
the expected strengthening of the instability
when self-gravity is increased via the  disc mass and the expected
weakening of the instability through  increasing 
gravitational  softening.

Hydrodynamic simulations of disc-planet interactions confirm earlier
suggestions by \cite{meschiari08}, who found gravitational instabilities
associated with prescribed gap profiles without a planet. Indeed,
edge modes can develop for
gaps self-consistently opened by introducing a planet
into the disc. However, models here showed their development
\emph{during} gap formation of a Saturn-mass planet, when the gap
only consists of a $20$---$30\%$ deficit in surface density relative
to the unperturbed disc. 
This is much shallower than the Jovian-mass planetary gaps 
considered in \cite{meschiari08}. 
 Edge modes exist for the disc models considered here 
with masses $M_d\ga 0.06M_*$, again this is less massive than 
required in \cite{meschiari08}, but  more massive than those typically
used in modelling protoplanetary discs. 


%
%

Simulations show edge modes with  $m=2$ and $m=3.$  They have
surface density maxima localised near the outer gap edge, but the
disturbance they produce  
extends throughout the entire disc, consistent with analytical 
expectation and linear calculations.
One important difference between edge modes and
the vortex forming  modes in non-self-gravitating discs
is that the latter require low viscosity. The typical
viscosity values adopted for protoplanetary discs, $\nu=10^{-5}$, will
suppress vortex formation but not edge modes around a 
gap opened by a Saturn-mass planet.

Edge modes produce large oscillations in
the disc torques acting on the planet, 
which can be positive or negative. The presence of edge modes
reverses the trend of the time 
averaged disc-on-planet torque as a function of
disc mass: the torque being more positive as disc mass increases.   
Direct interaction between the planet and  a
spiral arm associated with an  edge mode is
possible. These  should be more prominent in the disc 
 section with the lower
$Q.$ In practise this corresponds to $r>r_p,$  so the planet tends to
interact with the outer spirals,  which results  in a scattering of the planet 
outwards.

\chapter{Planetary migration in the presence of edge modes}\label{edge_migration}
\graphicspath{{Chapter4/}}

Edge modes significantly modify disc-planet torques from the
classic picture of monotonic inwards migration in low-mass discs. 
Hence, migration of gap-opening planets in massive protoplanetary
discs is expected to be qualitatively different to that in gravitationally
stable discs. For example, \S\ref{edge_scatter} indicate the
possibility of direct gravitational interaction with an edge mode
spiral arm. This is analogous, although distinct, from vortex-planet 
scattering.    

This Chapter explores the effect of edge modes associated with
gap edges on planetary migration averaged over timescales much longer
than that for scattering. Torque measurements from fixed-orbit simulations in
Chapter \ref{paper3} show the time-averaged disc-on-planet torques
become more positive with disc mass. For Saturn-mass planets this
effect is due to modification of Lindblad torques. However, for the
Jovian mass planets adopted here, edge modes were found to supply
torques from a region very close to the planet's orbital radius.

This study is fully numerical and is organised as
follows. The disc-planet model is described in \S\ref{setup} and their
stability  discussed in \S\ref{model_stability}. Simulation results
are presented in \S\ref{overview} and a fiducial case is analysed in
\S\ref{Q1d7}. \S\ref{concl} concludes this Chapter with a discussion of
the implications of the results.

\section{Simulation setup}\label{setup}
The class of disc profiles are identical to that
used in Chapter \ref{paper2}---\ref{paper3}. The discs are
two-dimensional and are parameterised by their Keplerian Toomre value at
the outer boundary, 
$Q_o$, or equivalently the disc-to-star mass ratio $M_d/M_*$. These
are chosen via the surface density scale. The
discs have uniform kinematic viscosity $\nu=10^{-5}$. 

The disc spans $r\in[1,25]$ and a larger planetary mass $M_p/M_* =
2\times10^{-3}$ is considered. This $M_p$ opens a deep gap 
so that strong scattering with spiral arms leading to large jumps 
in orbital radius on orbital timescales, do not occur. 
It allows one to examine the long term
effect of edge modes averaged 
over many orbits. Numerically, a larger $M_p$ is advantageous because there are
more cells per Hill radius than smaller $M_p$.  

The planet is introduced  on a circular orbit at $r_p=10$, this is
further from disc boundaries than in disc-planet models used previously. 
This attempts to reduce boundary effects, should the planet move
significantly.

\subsection{Equation of state}
The equation of state is locally isothermal, so the vertically
integrated pressure is $p=c_s^2\Sigma$ where $c_s$ is a specified
function. However, as noted in Chapter \ref{paper2}, adopting $c_s^2 =
h^2 GM_*/r$ (as used in most of the previous simulations) in a
self-gravitating disc means the planet mass is no 
longer a free parameter. The effective planet mass is $M_p$ plus disc
material gravitationally bound to the planet. This effect will be
significant for Jovian mass planets. To reduce it, one
can artificially increase $c_s$ near the planet by using the EOS
described in \S\ref{pepeos}. 
The parameter $h_p$ in this EOS controls the temperature
increase relative to the $1/r$ profile, and is fixed at $h_p=0.5$. 
Far away from the planet, the disc aspect-ratio is fixed at $h=0.05$
and the temperature profile returns to $\propto 1/r$.     
The motivation in using this EOS is entirely numerical. 

\subsection{Numerics}
The disc is divided into $N_r\times N_\phi = 1024\times 2048$ zones in
radius and azimuth, logarithmically and uniformly spaced
respectively. The resolution is approximately  16 cells per
 $H$ (or 28 cells per $r_h$). The cells are nearly square ($\Delta r/r\Delta\phi
= 1.02$). Open boundaries are applied at the inner boundary and
non-reflecting at the outer boundary, as before. The self-gravity softening is
$\epsilon_g = 0.3H$.    

The planet is introduced with zero mass at $t=20P_0$. 
Its mass is then switched on smoothly over 
$10P_0$ to its final value $M_p$. The planet is then allowed to
respond to disc gravitational forces for $t>30P_0$. A standard fifth
order Runge-Kutta scheme is used to integrate its equation of
motion. The softening length for the planet potential is $\epsilon_p =
0.6H$. 

\section{Disc models and their stability}\label{model_stability}
Four discs are presented here to assess the stability of
the adopted models: $Q_0=1.3,\,1.5,\, 1.7,\, 2.0$.  
These correspond to a Toomre parameter at the planet's initial radius
of $Q_p = 2.40, \, 2.77,\, 3.14, \, 3.70 $ and total disc masses of
$M_d/M_*= 0.092,\, 0.080,\, 0.071, 0.060$, respectively. The discs are
gravitationally stable to axisymmetric perturbations. 
For smooth radial profiles, they are also expected to be
gravitationally stable to non-axisymmetric perturbations near $r_p$
because $Q_p > 1.5$.

\begin{figure}[!ht]
  \centering
  \includegraphics[scale=.35,clip=true,trim=0cm 0.0cm 0cm
  0cm]{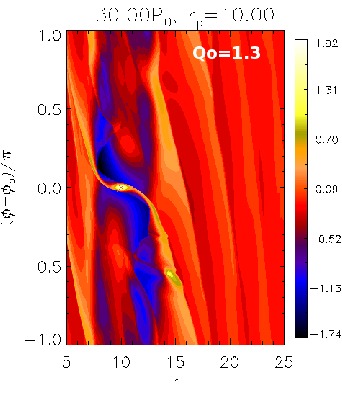}\includegraphics[scale=.35,clip=true,trim=2.13cm
  0cm 0cm
  0cm]{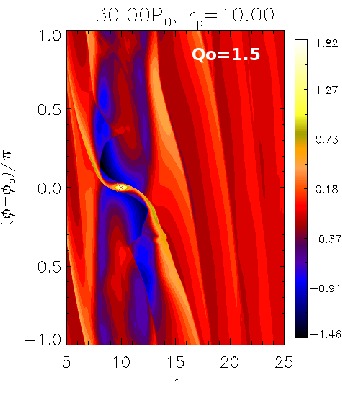}\includegraphics[scale=.35,clip=true,trim=2.13cm
  0cm 0cm 0cm]{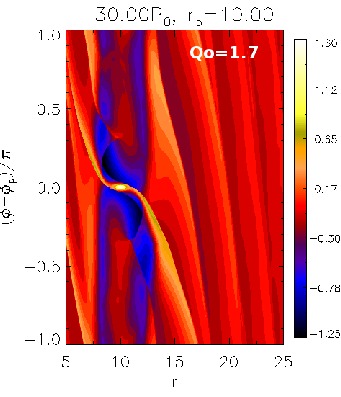}\includegraphics[scale=.35,clip=true,trim=2.13cm
  0cm 0cm 0cm]{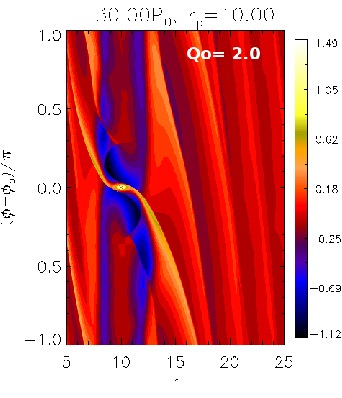}   
  \caption{Logarithmic relative surface density perturbation
    $\log{\left[\Sigma/\Sigma(t=0)\right]}$ when the planet reaches
    its final mass.  The colour bar 
    corresponds to the maximum and minimum perturbations within each
    plotted region, which occur near the planet.    
    \label{ch4_polarxy}}
\end{figure}

Fig. \ref{ch4_polarxy} shows the surface density perturbation for the
disc models when the planet mass reaches its final value. The
gap structure is similar and relatively stable. 
The over-density at the outer gap edge upstream of the planet
($r\simeq14,\,\phi-\phi_p\simeq 0.5\pi$) resembles an edge mode. 
However, this is likely associated with the introduction of a
massive planet over a short time-scale, rather than instability of an
axisymmetric gap. This is because it is present in all disc masses with
the same azimuthal wavenumber ($m=1$), whereas the most unstable edge
mode is expected to be of higher $m$ with increasing disc
mass. This transient feature increases in
magnitude with $M_d$, and for $Q_o=1.3$ it visibly makes the gap less
clean.


\begin{figure}[!ht]
\centering
\includegraphics[width=0.49\linewidth]{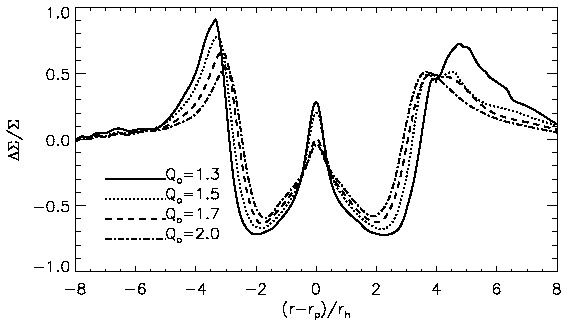}\includegraphics[width=.49\linewidth]{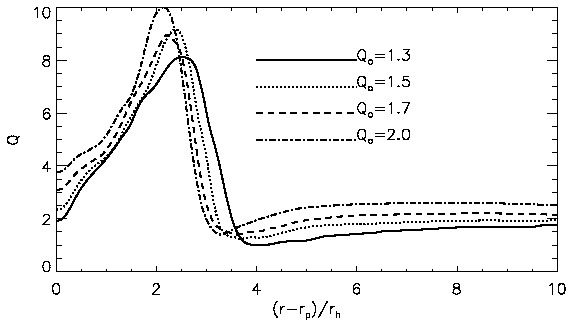}    
\caption{One dimensional gap profiles in terms of relative
  surface density perturbation (left) and the Toomre $Q$ profile of
  the outer disc. The horizontal axis is the displacement from the
  planet in units of its Hill radius. The snapshots are taken at
  $t=30P_0$. 
\label{ch4_1D_profile}}
\end{figure}

For stability discussion, one dimensional gap profiles should be used, 
because linear analysis is usually applied to axisymmetric 
backgrounds. 
Fig. \ref{ch4_1D_profile} compares the azimuthally
averaged surface density perturbation and Toomre $Q$ profiles at
$t=30P_0$. The gap is almost cleared ($\Delta\Sigma/\Sigma \simeq
-0.7$). The surface density perturbation 
is similar to that induced by a Saturn-mass planet, apart from the
spike near $r_p$ due to disc material bound to the planet. As
expected, this causes the gap edges to become more steep with lowering
$Q_o$. 

The fundamental quantity for stability is the
vortensity $\eta$. The edge mode is associated with
$\mathrm{max}(\eta)$, which for gap profiles coincide with
$\mathrm{max}(Q)$. Fig. \ref{ch4_1D_profile} shows that edge mode
corotation $r_c$  should occur inside the gap. Assuming
the coorbital region of a giant planet is $|r - r_p|\lesssim 2.5r_h$
\citep{artymowicz04b,paardekooper09}, $r_c$ is just within the outer gap edge.
This suggests the development of edge modes could
bring over-densities into the coorbital region of the planet.

Edge modes
also require coupling to the outer smooth disc, which is easier with
lower $Q_o$. Interestingly, in Fig. \ref{ch4_1D_profile} the $Q$ profile beyond the outer gap
edge becomes smoother as $Q_o$ is reduced. The local
$\mathrm{min}(Q)$ at the outer gap edge is most pronounced for
$Q_o=2.0$, which favour 
vortex modes. The profiles here indicate edge modes will develop more
readily with increasing disc mass and therefore have more significant
effect on migration as $Q_o$ is lowered.


\subsection{Fragmentation of high mass discs}
Giant planets can induce fragmentation in marginally stable
discs. This phenomenon was previously identified in smooth 
particle hydrodynamic simulations by \cite{armitage99} and
\cite{lufkin04}. However, grid-based codes are generally better at
resolving shocks and therefore planetary wakes.
Fig. \ref{ch4_polarxy} confirms that planetary wakes may become
gravitationally unstable as a clump forms along the outer wake in
$Q_o=1.3$. 

\begin{figure}[!ht]
\centering
\includegraphics[width=0.49\linewidth]{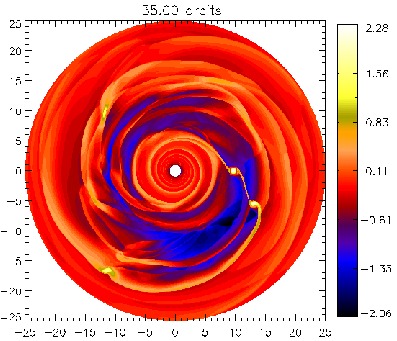}\includegraphics[width=.49\linewidth]{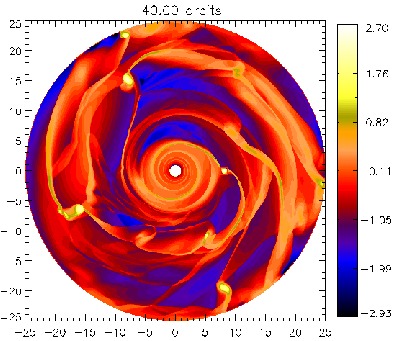}  
\caption{Fragmentation induced by the planet in the most massive
  disc model $Q_o=1.3, \, M_d/M_* =
  0.092$. $\log{\left[\Sigma/\Sigma(t=0)\right]}$  is shown.  
\label{ch4_Q1d3_frag}}
\end{figure}

Fig. \ref{ch4_Q1d3_frag} shows that fragmentation indeed occur for
$Q_o=1.3$ within $10P_0$ of releasing the planet, and there is no
annular gap. This means that for 
very massive discs ($M_d\gtrsim 0.1M_*$), migration will be significantly
affected by these clumps, rather than large-scale spirals associated 
with edge modes. 

\section{Migration and disc structure}\label{overview}
Edge modes lead to  non-axisymmetric over-densities inside the
gap. This is a potential source for torques to originate from within the
planet's coorbital region. Because $r_c$ is expected near
the outer edge of the coorbital region, edge mode over-densities
may undergo horseshoe turns upon approaching the planet.
Furthermore, fluid elements just inside the separatrix
will traverse close to the planet when it
executes the U-turn, and this may provide significant torque.
 
\begin{figure}[!ht]
  \centering
  \includegraphics[width=\linewidth]{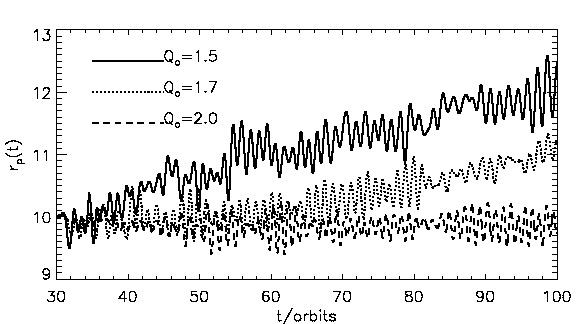} 
  \caption{Migration of a 2-Jupiter mass planet ($M_p/M_* = 0.002$) in
    massive discs. The instantaneous orbital radius is shown. 
    \label{ch4_migration}}
\end{figure}

Fig. \ref{ch4_migration} shows the instantaneous orbital radius of the
planet in non-fragmenting disc models. Migration is non-monotonic
and can be inwards or outwards on short time-scales ($\sim P_0$). However,
with increasing disc mass, outwards migration is favoured on
timescales of a few 10's of orbits. For $Q_o=1.5$, $r_p$ increases by $20\%$
in only $70P_0$. 

\begin{figure}[!ht]
\centering
\includegraphics[scale=.5,clip=true,trim=0cm 0.0cm 1.65cm
0cm]{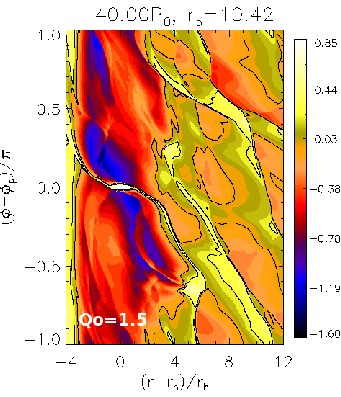}\includegraphics[scale=.5,clip=true,trim=2.13cm
0cm 1.65cm
0cm]{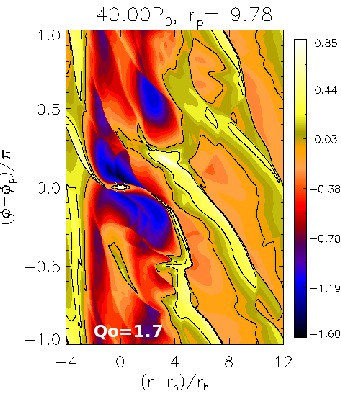}\includegraphics[scale=.5,clip=true,trim=2.13cm
0cm 0cm 0cm]{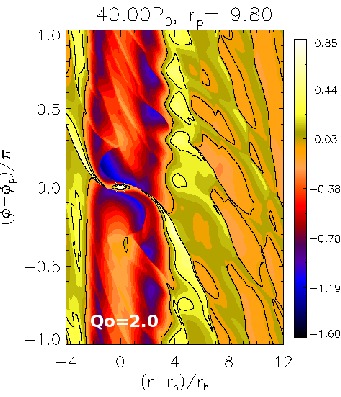}  
\caption{Relative surface density perturbation on a logarithmic scale,
   $\log{\left[\Sigma/\Sigma(t=0)\right]}$, showing gap
   instability. \label{ch4_polarxy_t40}} 
\end{figure}

\begin{figure}[!ht]
\centering
\includegraphics[width=\linewidth]{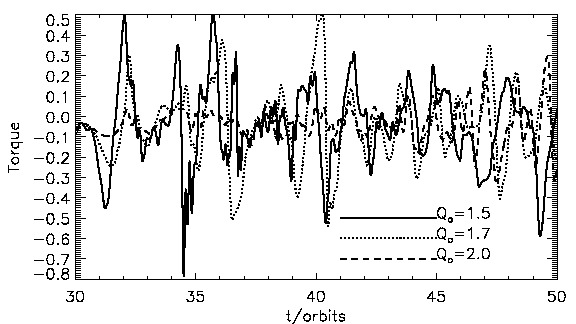}  
\caption{Instantaneous disc-on-planet torques.
  Tapering has been applied to contributions from the Hill sphere to make this plot
  clearer. This procedure does not affect the torque's rapid oscillatory behaviour, or
 its behaviour as a function of $Q_o$. Migration was evolved without tapering since the simulations
are fully self-gravitating \citep{crida09}.  The snapshots in 
  Fig. \ref{ch4_polarxy_t40} 
  are taken at $t=40P_0$. \label{ch4_torque_instant}} 
\end{figure}

The behaviour of $r_p(t)$ correlates with the coherence of the gap
structure. Fig. \ref{ch4_polarxy_t40} shows that $Q_o=2.0$ first 
develops vortex mode (with $m>6$). Although a viscosity of
$\nu=10^{-5}$ suppressed vortex modes in previous studies (for $M_p\leq 0.001M_*$), 
increasing $M_p$ has the same effect as lowering $\nu$,
hence vortex modes are not completely suppressed. $Q_o=2.0$ 
does develop spiral arms later on, but these probably 
result from the vortices perturbing the disc, rather than 
the linear edge mode instability. 
The important feature with $Q_o=2.0 $  is
that instability leaves the gap edge intact and identifiable, unlike more massive discs below. 
Related to this is that Fig. \ref{ch4_torque_instant} shows up to
$t\leq 40P_0$, the torque magnitudes for $Q_o=2.0$ are typically
smaller than in that $Q_o=1.5,\,1.7$.  

$Q_o=1.7$ develops a $m=2$---3 edge mode, making
the gap less well-defined. The chosen snapshot is just
prior to a small outwards jump in $r_p$ for $Q_o=1.7$
(Fig. \ref{ch4_migration}) and Fig. \ref{ch4_torque_instant}  
shows a clear positive disc-on-planet torque at this instant. 
Fig. \ref{ch4_polarxy_t40} shows that
the over-density inside the gap and associated with the edge mode,
just upstream of the planet, could provide such positive torque. 
The outer gap edge in $Q_o=2.0$ is not as disrupted and this results in no secular
increase in $r_p$ over the simulated timescale.

In Fig. \ref{ch4_polarxy_t40}, large-scale spirals can still be seen
for $Q_o=1.5$. Weak fragmentation occurs but collapse into
clumps are not seen. Like $Q_o=1.7$, 
the gap is unclean, but now with significant disruption to the outer gap
edge. It is no longer a clear feature like in $Q_o=2.0$. 

\subsection{Gap evolution}\label{gap_evolution}
\begin{figure}[!ht]
\centering
\includegraphics[width=0.49\linewidth,clip=true,trim=0cm 0cm 0cm 0.74cm]{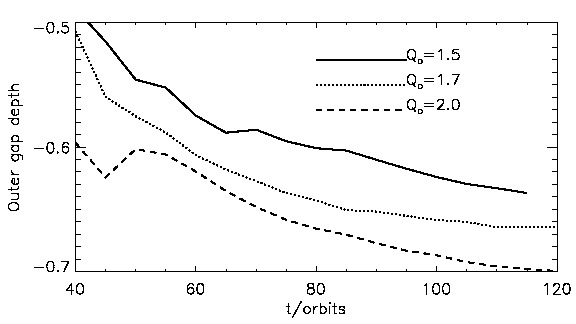}\includegraphics[width=0.49\linewidth,clip=true,trim=0cm 0cm 0cm 0.74cm]{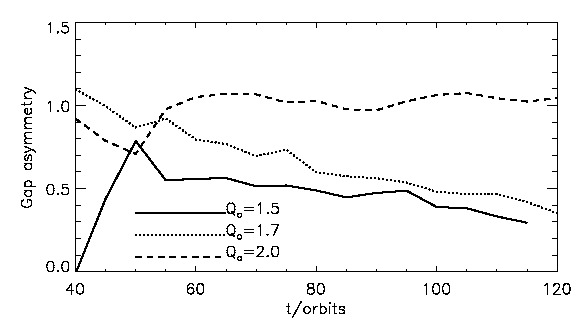}    
\caption{Running-time averages of gap properties: the dimensionless
  outer gap depth  (left) and the gap asymmetry in units of Hill
  radius (right). These quantities are defined in
  \S\ref{gap_evolution}. 
\label{ch4_gap_evolution}}
\end{figure}

The migration curves in Fig. \ref{ch4_migration} results
from modification to the gap structure by instabilities. The following
prescription was used to monitor gap structure. A one-dimensional profile
of the relative surface density perturbation $\Delta\Sigma/\Sigma$ is
first calculated (as in Fig. \ref{ch4_1D_profile}). The \emph{outer gap
  edge} is defined as $r_{e,\mathrm{out}}>r_p$ such that
$\Delta\Sigma/\Sigma = 0$ there, and similarly for the \emph{inner gap
  edge} $r_{e,\mathrm{in}}<r_p$. The 
\emph{outer gap depth} is $\Delta\Sigma/\Sigma$ averaged over
$r\in[r_p,r_{e,\mathrm{out}}]$. The \emph{outer gap width}, in units of
the Hill radius, is $w_\mathrm{out} = |r_{e,\mathrm{out}} - r_p|/r_h$
and the \emph{inner gap width} is $w_\mathrm{in} = |
r_{e,\mathrm{in}} - r_p|/r_h$. The \emph{gap asymmetry} is
$w_\mathrm{out} - w_\mathrm{in}$. Running time-averaged ($t\geq
40P_0$) plots are shown in Fig. \ref{ch4_gap_evolution}

Edge modes have a gap-filling effect. The case $Q_o=1.5$, in which the
planet immediately migrates outwards, has the smallest gap depth
throughout. $Q_o=2.0$ has the deepest gap and is non-migrating
over the simulation timescale. In $Q_o=1.7$, the magnitude of gap depth 
decreases relative to that for $Q_o=2.0$ after $t=80P_0$. 
Note that this corresponds to outwards migration in $Q_o=1.7$. 
These observations suggests that material brought into the gap by edge modes, 
is responsible for outwards migration.

Gap asymmetry was defined as the distance
from the planet to the outer gap edge minus the distance between the planet
and the inner gap edge. It follows that the more negative gap asymmetry is,
the closer the planet is to the outer gap edge.
Fig. \ref{ch4_gap_evolution} shows the planet is located
closer to the outer gap edge in discs with lower $Q_o$ than in discs with higher $Q_o$.
Furthermore, the non-migrating $Q_o=2.0$ reaches a constant asymmetry,
whereas in the outwards-migrating cases asymmetry decreases with time.

The trend above suggests that the planet is moving
material associated with the outer gap edge, which is unstable to edge modes, inwards via horseshoe turns
(this would also be consistent with
shallower gaps with decreasing $Q_o$). This provides a
positive torque on the planet.
If, on average, this effect dominates over sources
of negative torques, e.g. negative Lindblad torques and the enhancement of which by
the coincidence of an edge mode spiral arm with the outer planetary wake, then 
the planet should on average migrate outwards, i.e. closer to the outer gap edge.


\section{A fiducial case in detail}\label{Q1d7}

\begin{figure}[!ht]
\centering
\includegraphics[width=.81\linewidth,clip=true,trim=0cm 0cm 0cm 0.4cm]{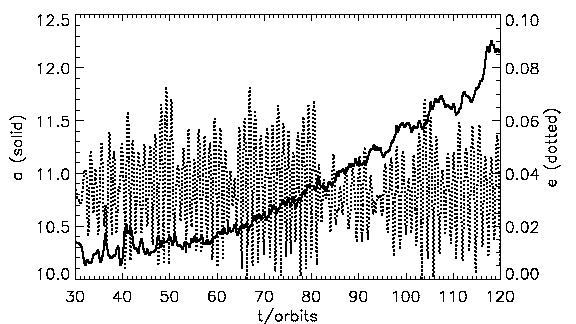}  
\caption{Orbital evolution of the planet in the $Q_o=1.7$ disc, in
  terms of Keplerian semi-major axis $a$ (solid) and eccentricity $e$
  (dotted). These have been calculated without accounting for the disc
  potential.   
\label{ch4_Q1d7_migration}}
\end{figure}

\begin{figure}[!ht]
\centering
\includegraphics[scale=.38,clip=true,trim=0cm 1.8cm 1.65cm
0cm]{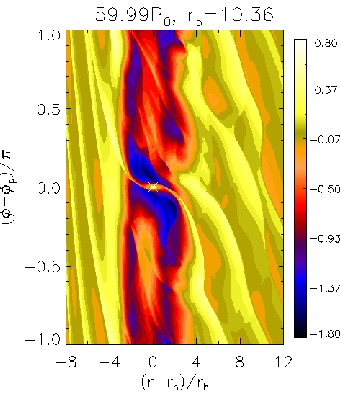}\includegraphics[scale=.38,clip=true,trim=2.13cm
1.8cm 1.65cm
0cm]{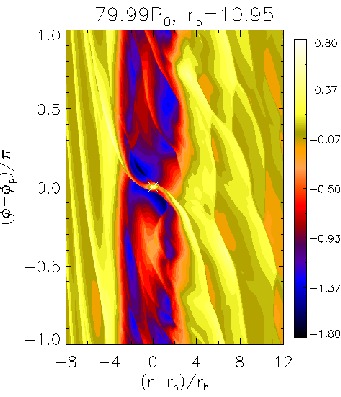}\includegraphics[scale=.38,clip=true,trim=2.13cm
1.8cm 1.65cm 0cm]{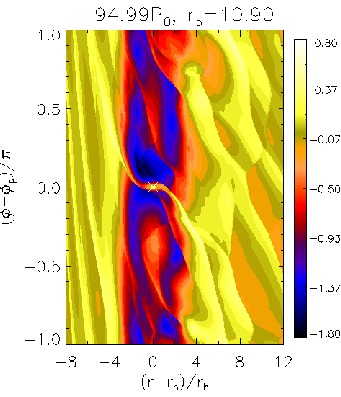}\includegraphics[scale=.38,clip=true,trim=2.13cm
1.8cm 0cm 0cm]{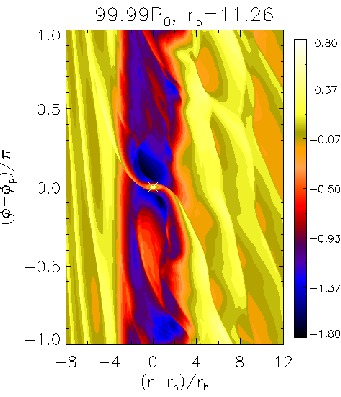}\\
\includegraphics[scale=.38,clip=true,trim=0cm 0.0cm 1.65cm
0cm]{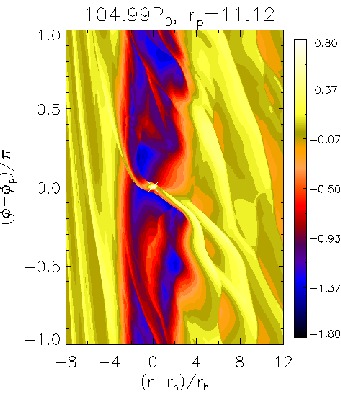}\includegraphics[scale=.38,clip=true,trim=2.13cm
0.0cm 1.65cm
0cm]{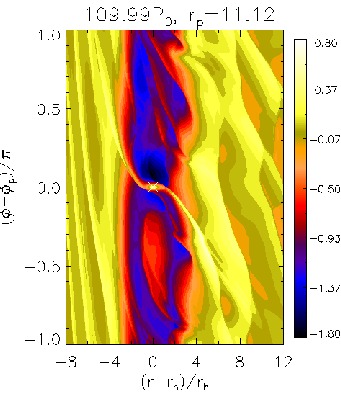}\includegraphics[scale=.38,clip=true,trim=2.13cm
0.0cm 1.65cm 0cm]{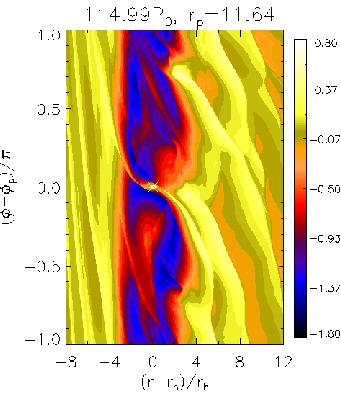}\includegraphics[scale=.38,clip=true,trim=2.13cm
0.0cm 0cm 0cm]{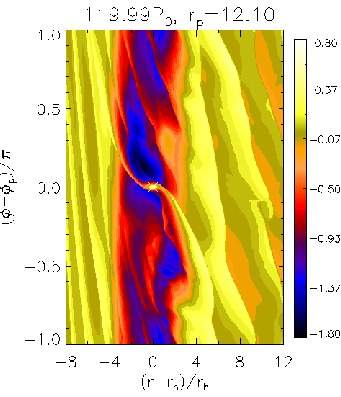}
\caption{Evolution of the $Q_o=1.7$ case. The
  logarithmic relative surface density perturbation
   $\log{\left[\Sigma/\Sigma(t=0)\right]}$ is shown.
\label{ch4_Q1d7_polarxy}}
\end{figure}

Fig. \ref{ch4_Q1d7_migration} and Fig. \ref{ch4_Q1d7_polarxy}
presents a summary of the simulation with $Q_o=1.7$, 
in terms of the evolution of the planet's orbit 
and the disc's surface density perturbation. 
The orbit remains fairly circular $(e< 0.06)$ with no
secular change in eccentricity. However, the semi-major axis $a$ increases
more noticeably than orbital radius $r_p$.

Fig. \ref{ch4_Q1d7_polarxy} shows the outer disc ($r>r_p$) is
unstable to edge modes whereas the inner disc ($r<r_p$) remains
relatively stable. Large-scale spirals are maintained throughout
the simulation. The gap is unclean with material
brought into it by the edge modes. Notice in the $t=105P_0$
 and $t=115P_0$ snapshots the over-density just upstream and exterior to
 the planet. It resides within the planet's coorbital region ($2.5r_h\gtrsim r-r_p>0$)
 so this material is expected to execute inwards horseshoe turns and exert a positive
 torque on the planet.  If this torque is sufficiently large, it may trigger type III migration because
it promotes the planet to move closer to the higher density gap edge and interact with it. This can
explain the faster-than-linear increase in $a$. In this case the 
 edge mode over-densities provides the coorbital mass deficit.

\subsection{Torques and mode amplitudes}

\begin{figure}[!hb]
\centering
\includegraphics[scale=0.68,clip=true,trim=0cm 1.5cm 0cm 0cm]{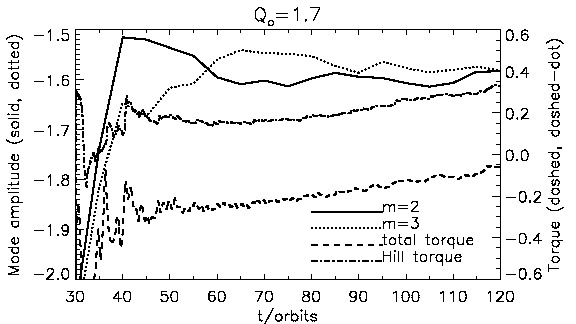}
\includegraphics[scale=0.68]{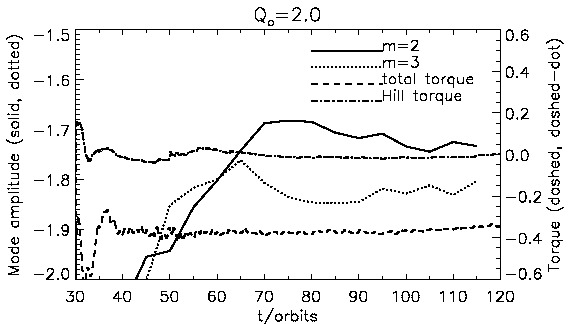}  
\caption{Evolution of the amplitude of the $m=2$ (solid) and $m=3$
  (dotted) modes in surface density in the outer disc
  $r\in[r_p,r_o]$ in comparison to the total disc-on-planet torque
  (dashed) and the torque contribution from the Hill sphere
  (dashed-dot). All quantities are running-time averaged. Disc modes
  have been normalised by the axisymmetric amplitude and is plotted on
  a logarithmic scale. The upper
  plot is $Q_o=1.7$, in which the planet migrates outwards. The lower
  plot is the non-migrating $Q_o=2.0$ case. 
\label{ch4_Q1d7_mode_torque}}
\end{figure}

Fig. \ref{ch4_Q1d7_mode_torque} shows the  time-averaged evolution of
non-axisymmetric modes in the outer disc, together with evolution
of disc-on-planet torques. The effect of edge modes can be seen by
comparison to the non-migrating case $Q_o=2.0$. The
modes have larger amplitudes in $Q_o=1.7$, and the total torque
becomes more positive with time\footnote{The total torque values
are negative, despite overall outwards migration, because the running
time-average is plotted. The average includes the time at planet release, when the disc-planet
torque is large negative. }. 
The figure shows this 
is attributed to torques from within the Hill radius, which is positive. 
By contrast,
although $Q_o=2.0$ also develop large-scale spirals later on, they are of
smaller amplitude and do not make the Hill torque positive. The
total time-averaged torque remains fairly constant. 

Fig. \ref{ch4_Q1d7_mode_torque}, along with Fig. \ref{ch4_Q1d7_polarxy}, confirms that large-scale
spirals which extend into the gap can provide significant torques. Edge modes naturally fit this 
description since they are associated with vortensity maxima just inside the gap.  
Because the outer disc is 
more unstable, the corotation radius $r_c$ of edge modes lie beyond $r_p$ so
its pattern speed is smaller than the planet's rotation. 
Thus, over-densities associated with corotation approach the planet from upstream, 
but because $r_c$ is actually within the planet's coorbital region, the associated
fluid elements are expected to execute (inwards) horseshoe turns upon 
approaching the planet. This provides a positive torque on the planet, and if
the edge mode amplitude is large enough, it may reverse the usual
tendency for inwards migration. 

The picture outlined above is consistent with the fact that edge modes
have corotation at vortensity maxima. A giant planet can induce 
shocks very close to its orbital radius (Chapter \ref{paper1}) and  
vortensity is generated as fluid particles execute horseshoe turns across
the shock. That is, the vortensity rings are associated with particles  
on horseshoe orbits. So when non-axisymmetric disturbances associated 
with vortensity rings develop (i.e. edge modes), the associated over-densities can be 
expected to execute horseshoe turns. 


\subsection{Torque distribution}
\begin{figure}[!ht]
    \centering
    \includegraphics[scale=0.35,clip=true,trim=0cm 1.75cm 0cm
    0cm]{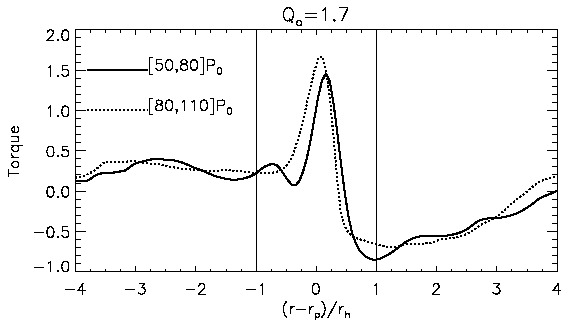}\includegraphics[scale=0.35,clip=true,trim=0cm 1.75cm 0cm
    0cm]{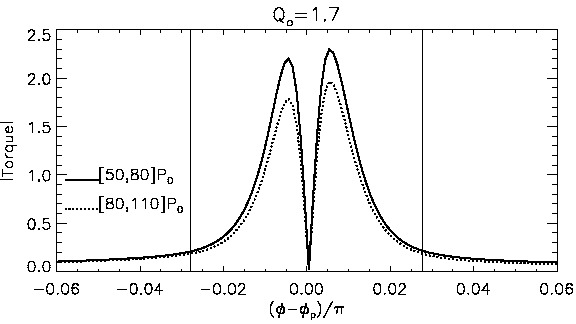}\\
    \includegraphics[scale=0.35]{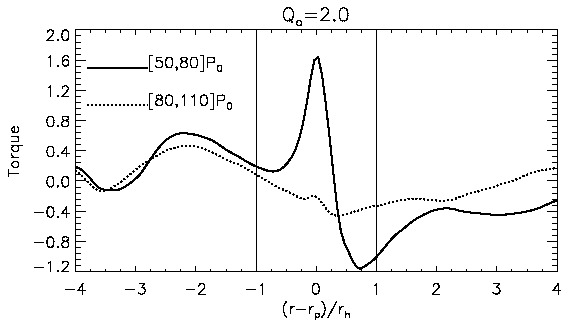}\includegraphics[scale=0.35]{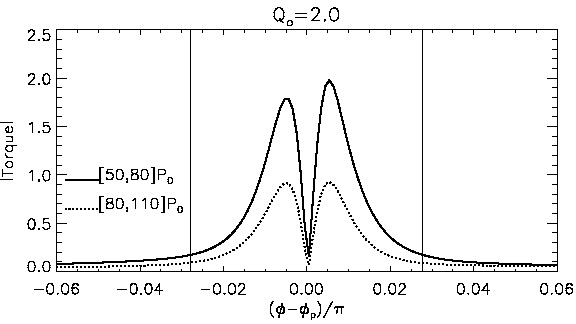}    
    \caption{Torque densities. Radial plot (left) and azimuthal plot 
      (right), averaged over $t\in[50,80]P_0$ (solid) and
      over $t\in[80,110]P_0$ (dotted). The upper plots are from the $Q_o=1.7$ case,
      which display outwards migration and the lower plots are from the
      non-migrating $Q_o=2.0$ case. Vertical lines in each plot indicate
      the Hill radius.
      \label{ch4_torque_contrib}}
  \end{figure}

The analysis above suggests that torques from within the gap are responsible for
the gradual increase in $r_p$ or $a$. 
Fig. \ref{ch4_torque_contrib} compares torque densities in
the fiducial case to the non-migrating case. The torques have been
averaged over $30P_0$ at two time intervals. For $t\in[50,80]P_o$
radial plots for both cases show a positive torque at
$r_p$. By $t\in[80,110]P_0$ this torque has
diminished for $Q_o=2.0$. This is expected for Jovian planets as
they open a clean gap, leaving little material near $r_p$ to torque the planet. 
However, in $Q_o=1.7$, this torque is maintained 
(and even slightly increased) by the edge modes because they bring 
material into the co-orbital region.  

Azimuthal plots in Fig. \ref{ch4_torque_contrib} only account for
material in between the gap edges (quantitatively defined in
\S\ref{gap_evolution}). Both cases have similar distributions for
$t\in[50,80]P_0$. For $t\in[80,110]P_0$ though, the torque becomes
symmetric about the planet's azimuth for $Q_o=2.0$, whereas for
$Q_o=1.7$ asymmetry is maintained, leading to a
positive torque.  
This torque is caused by over-density
ahead in comparison to that behind the planet, i.e. material flowing
radially inwards across the planet's orbital radius. 
However, unlike the single scattering events by vortices or spirals,
which dominate most of the migration, here migration occurs on much 
longer timescales in comparison.  

\section{Summary and discussion}\label{concl}
The numerical experiments above have shown that edge modes can
induce outwards migration over timescales of a few tens of initial orbital
periods. In the planet's frame, an edge 
mode spiral, with corotation at vortensity maximum exterior to $r_p$, 
approaches the planet from $\phi > \phi_p$. 
However, the mode extends into the planet's 
coorbital region, so there are associated over-densities 
that execute inwards horseshoe turns, which
provides a positive torque. 

The torque responsible for outwards migration can be
identified as positive bump at $r=r_p$ in the time-averaged radial torque
plots in Fig. \ref{ch4_torque_contrib}, resulting from azimuthal gap
asymmetry about the planet. 
This torque diminishes with time in the
non-migrating $Q_o=2.0$ case, thus migration itself helps to maintain
this torque. Note that such torque was also absent in the Saturn-mass
case when held on fixed orbit (Fig. \ref{Qm1.5_torques}). Another
difference to Saturn is Jovian mass planets induce larger
perturbations and the gap-filling effect of edge modes tend to be
stronger. 

Edge modes must be sustained in order to provide this 
torque, this in turn requires the  existence of vortensity maximum,
despite the development of edge modes destroying them. 
However, this is easily regenerated by giant planets because they
induce shocks which act as a source of vortensity. Furthermore, since
edge modes also require gravitational coupling to an exterior 
smooth disc, outwards migration of this type is
probably more likely to occur than inwards, because typically discs
become more self-gravitating with increasing radius.

Planetary migration observed here closely resembles classic type
III migration \citep{masset03} because it relies on 
torques from the coorbital region and the accelerated increase in
semi-major axis indicate positive feedback. The main difference to
type III migration originally described by \citeauthor{masset03} is
that the flow across the planet's orbit is non-smooth, because edge
modes present distinct fluid blobs, rather than a
continuous flow across $r_p$. Also, where type III migration usually apply to
partial gaps and therefore Saturn-mass planets, the Jovian mass planet
used here reside in a much deeper gap. 
Radial mass flux across the planet is still possible, despite the
gap-opening effect of Jovian planets, because the unstable edge modes
brings over-densities inside the gap to interact with the planet. 

A consequence of the above is that formation of a clean gap is expected to
become increasingly difficult in massive discs. Planetary gaps
correspond to existence of vortensity maximum, but
these become unstable to edge modes in massive discs and instability
tends to fill 
the gap.  In
very massive discs giant planets induce fragmentation, their
planetary wakes become gravitationally unstable and no annular gap forms. 
Therefore classic type II migration, where the planet moves smoothly with the
viscous gap evolution, probably does not apply in massive discs. 





\subsection{Outstanding issues}
It should be noted that the adopted disc model is biased
towards outwards migration because the Toomre $Q$ decreases outwards
(approximately like $r^{-1/2}$) so edge modes associated with the
outer disc develops preferentially. If a disc model such that the
inner disc is more unstable was used, then inwards migration should
occur. And if independent edge modes of comparable amplitude develop on either side of the gap, 
 there should be little migration overall.

Significant torques originate from material close to the
planet, so numerical treatment of the Hill sphere is a potential
issue. This concern also applies to standard type III migration, but
note that the adopted equation of state was originally designed for numerical
studies of type III migration \citep{peplinski08a}. Furthermore,
fully self-gravitating discs are suited for type III migration \citep{crida09}.
Lower resolution simulations have been performed 
which also resulted in outwards migration.


This EOS mimics heating near the planet but is not a quantitative model for
the true thermodynamics. If this EOS under-estimates the heating, then the positive
torque could be over-estimated and vice versa. On the other hand, edge mode corotation
reside outside the Hill radius but still inside
the gap. So the numerical treatment of the Hill radius does not
affect existence of edge modes and their associated over-densities
should affect torques originating from the coorbital region
in the way described in this Chapter.



\chapter{Three-dimensional simulations}\label{3d}
\graphicspath{{Chapter5/}}

Razor-thin discs have been used in all the previous numerical
simulations. This approximation can be justified for gap-opening planets,  
whose definition can be taken as $M_p/M_*\gtrsim h^3$, so that the
planet's Hill radius exceeds the local disc thickness. Nevertheless,
three-dimensional discs should be considered because instabilities
studied above are associated with radial structures with characteristic
length scales comparable to the local thickness. 

In this Chapter, 
numerical simulations are developed to confirm properties of gap 
stability studied previously in 2D, persist in 3D geometry. The 3D disc-planet
models are presented in \S\ref{3d_disc_model}. \S\ref{adapt_zeus}
describes adaptation of the \zeus 
software to perform disc-planet simulations with and without disc
self-gravity.  Three-dimensional counterparts to previous 2D results
are presented in \S\ref{3d_stability}, including  vortex formation and
vortex-induced 
migration in non-self-gravitating discs, vortex modes in
self-gravitating discs and edge modes in massive discs.
\S\ref{summary} concludes this Chapter.  

\section{Three-dimensional models}\label{3d_disc_model}
The governing equations for three-dimensional discs are
listed in Chapter \ref{intro}. The disc model is extended from that
used in Chapter \ref{paper2}---\ref{edge_migration} by introducing
vertical structure. The disc is 
described in spherical polar co-ordinates $(r,\theta,\phi)$
with origin on the central star,  $r$ is now the  
spherical polar radius, and the cylindrical radius denoted by 
$R=r\sin{\theta}$ and the cylindrical $z=r\cos{\theta}$.


The disc occupies $r\in[r_i,r_o]$, $\theta\in[\theta_\mathrm{min},
\theta_\mathrm{max}]$ and  $\phi\in[0,2\pi]$. The vertical extent
is such that $|\pi/2 - \theta_\mathrm{min}|$ corresponds to $n_H$ number
of scale-heights. The initial density profile is
\begin{align}\label{initial_density}
  \rho_0(R,z) &= \frac{\Sigma_0 R^{-p}}{\sqrt{2\pi
      H^2}}\exp{\left(-\frac{\Phi_*}{c_s^2} - \frac{1}{h^2}
    \right)}
  \times\left[1 - \sqrt{\frac{r_i}{R + hr_i}}\right]\notag\\
  & \equiv f(R)\exp{\left(-\Phi_*/c_s^2\right)},
\end{align}
where $H=hR$ is the scale-height with $h=0.05$ the fixed aspect-ratio,
$\Sigma_0$ is a constant parameterising the density scale. The
power-law index in surface density is fixed at $p=3/2$ for all the
simulations performed. Recall $\Phi_*$ is the stellar potential.  
Eq. \ref{initial_density} results from exact vertical hydrostatic 
equilibrium assuming an isothermal equation of state with sound-speed 
$c_s=c_s(R)$. As before, the density scale $\Sigma_0$ is 
chosen via the Keplerian Toomre parameter $Q_o$ at the outer boundary: 
\begin{align}\label{3d_q0}
  Q_o \equiv \left. \frac{c_s\Omega_k}{\pi G
    \int_{z_\mathrm{min}}^{z_\mathrm{max}} \rho_0 dz}\right|_{R=r_o}.
\end{align}
Note that vertical integration in Eq. \ref{3d_q0} corresponds to the
computational domain so exterior mass is either zero or assumed
negligible. Thus models with fixed $Q_o$ but increasing $n_H$ will
have decreasing midplane density scale represented by $\Sigma_0$.  
 
The disc is initialised without meridional
velocity ($u_r=u_\theta=0$) and azimuthal velocity given by 
centrifugal balance between stellar potential and pressure: 
\begin{align}\label{initial_vphi}
  u_\phi^2 = R\frac{\p c_s^2}{\p R} +
  \Phi_*\frac{\p\ln{c_s^2}}{\p\ln{R}}  + Rc_s^2\frac{\p\ln f}{\p R}.
\end{align}
Exact balance implies $\p u_\phi / \p z \neq 0$, but  since the disc is thin
($h\ll 1$), the azimuthal velocity is well approximated to be
Keplerian ( $u_\phi \simeq R\Omega_k$).


\subsection{Self-gravity}
The three-dimensional disc potential
$\Phi$ is obtained by solving the differential Poisson equation
$\nabla^2\Phi = 4\pi G \rho$.  This removes the need for a softening
length used for 2D self-gravity. However, boundary conditions must be
explicitly imposed and will be described in \S\ref{adapt_zeus}. 

Instead of solving for the initial vertical structure consistent with
self-gravity, a simpler numerical approach is taken. All
calculations are initialised with the non-self-gravitating equilibrium
described above. To account for radial self-gravity, the initial
azimuthal velocity is reset to $v_\phi^2\to v_\phi^2 + rg_r$.  
The vertical self-gravity force, $g_\theta$, is initially set to
zero and then smoothly increased to its full value over time. 



\subsection{Planet treatment and equation of state} 
The planet potential is softened by a fraction of its Hill radius
\begin{align}
  \Phi_p = -\frac{GM_p}{\sqrt{|\bm{r} - \bm{r}_p|^2 + (\epsilon_{p0}
      r_h)^2}}. 
\end{align}
The softening is set to $\epsilon_{p0}=0.1$.
In 2D, softening by a fraction of the local scale-height approximately
accounts for vertical structure, whereas softening in 3D is purely to
avoid numerical divergence.  The Hill radius is calculated with
$|\bm{r}_p|$, but this is very close to the planet's cylindrical
radius from the star because the orbit inclination is negligible for all
the migrating cases presented later. The planet is introduced on fixed circular
orbit of radius $r_p$ in the midplane with Keplerian rotation speed
and its mass is smoothly switched on. 

The disc-on-planet forces are tapered by multiplying  all components
of the force by the factor $1-\exp{\left[-\frac{|\bm{r} -
    \bm{r}_p|^2}{2(\epsilon_cr_h)^2}\right]}$.   
The cut-off parameter $\epsilon_c$ is set to unity
unless otherwise stated. Tapering is applied whether or not disc
self-gravity is enabled. Without self-gravity, the physical motivation
for tapering is that material close to the planet should be
gravitationally bound to it and should be excluded in the
disc-on-planet force. There is no physical need for tapering in a
self-gravitating disc, but limited numerical resolution in a 3D
calculation prevents accurate determination of torques from close to
the planet. Thus, the tapering is largely motivated by numerics. 

The equation of state is isothermal, so the pressure is  $p=c_s^2
\rho$. Before introducing the planet, $c_s^2=h^2GM_*/R$ is the
locally isothermal sound-speed giving the standard Gaussian
atmosphere. After the planet is introduced, the sound-speed is
modified to $c_s $ given by Eq. \ref{pepeos} which
heats up the disc near the planet relative to the $1/R$ profile. 
This prescription is used to limit mass pile-up near
the planet and reduce numerical errors resulting from this effect. It
also helps to reduce the increase in effective planet mass 
and therefore the effect of self-gravity on
instabilities through the background planetary gap.



\section{Adapting \zeus}\label{adapt_zeus}

\texttt{ZEUS} is a general-purpose astrophysical fluid dynamics code 
based on finite-difference schemes. Several versions
are publicly
available\footnote{\url{http://lca.ucsd.edu/portal/software}},
including  
\texttt{ZEUS-2D} \citep{stone92,stone92b,stone92c} which \fargo 
is based on, and its three-dimensional version \texttt{ZEUS-3D}. The
version adapted here is \zeus, developed and described by
\cite{hayes06}, which places emphasis on parallel computing. 
\zeus is capable of
solving hydrodynamics, self-gravity, magnetohydrodynamics and radiative
transfer, although the latter two modules are disabled in this
study. 


\subsection{Source terms}

The most important new source term is the gravitational forces on disc
fluid due to the planet potential. The acceleration $-\nabla\Phi_p$ is
added directly to to the momentum equations to update each velocity
component ($u_r, u_\theta, u_\phi$) as part of the source term step in
\zeus. Indirect disc and planet potentials are also added, to account
for the non-inertial reference frame.

All simulations presented have zero physical viscosity
(artificial viscosities is employed by \zeus to treat
shocks). 
However, a module for viscous forces has also been added to \zeus. Velocity
components can be updated by the full viscous stress tensor
immediately after the main source step. The magnitude of the viscous
force is characterised by kinematic viscosity $\nu$, which can be
a constant or $\nu = \alpha c_s H$, where $\alpha$ is a constant. 
The effect of viscosity on vortex and edge modes in 3D is not expected
to differ to that in 2D discs, which has already been investigated. 



\subsection{Disc potential}
\zeus was chosen for its ability to solve 3D
self-gravity in spherical polar co-ordinates with
parallelisation. This is an uncommon feature among public codes. 
The discretised Poisson equation is a linear system, $A\bm{x}=\bm{b}$. 
\zeus solves this using the conjugate gradient (CG) method \citep[see][for
details]{hayes06}. It is an iterative scheme that  
minimises the quadratic form  
 $f(\bm{x}) = \frac{1}{2}\bm{x}^TA\bm{x} - \bm{b}^T\bm{x}$. 
At the $k^\mathrm{th}$ iteration, the CG method finds
the scalar $c_{k+1}$ and conjugate direction $\bm{p}_{k+1}$ (such that
$ \bm{p}_i^T A\bm{p}_j=0$ if $i\neq j$) to give the improved solution
as $\bm{x}^{(k+1)} = \bm{x}^{(k)} + c_{k+1}\bm{p}_{k+1}$ \citep[e.g.][]{press92, barrett94}.  


\subsubsection{Boundary potential}
A potential solver based on \cite{boss80}  has been added to \zeus to 
obtain potentials at $r$- and $\theta$-boundaries, which are set as Dirichlet boundary
conditions for the CG solver. The potential is periodic in $\phi$. 
The potential and density are expanded in spherical 
harmonics, so that
\begin{align}\label{sph_trans}
  \Phi = \sum_{l=0}^{\infty}\sum_{m=-l}^{m=l} \Phi_{lm}(r) Y_{lm}(\theta,\phi);\quad
  \rho = \sum_{l=0}^{\infty}\sum_{m=-l}^{m=l} \rho_{lm}(r) Y_{lm}(\theta,\phi).
\end{align}
In practice the sum is truncated at $l=\lmax$ and 
$m=\mmax$. To further reduce computation cost, only $l+m=$even
modes are used, corresponding to symmetry with respect to the
disc midplane.  
$\rho_{lm}$ is found by direct integration: 
\begin{align}\label{sph_trans_rho}
  \rho_{lm} =
  \int_{0}^{2\pi}\int_{\theta_\mathrm{min}}^{\theta_\mathrm{max}} \rho Y_{lm}^*
  \sin{\theta}d\phi d\theta.
\end{align}
This integration is parallelised by domain decomposition along with the
hydrodynamics. Each processor holds a copy of the global $\rho_{lm}(r)$ but
performs the integration over its
local sub-domain of the global disc. The result is written to the 
appropriate sub-portion of the global $\rho_{lm}(r)$. A global
reduction is then performed, summing contributions to $\rho_{lm}(r)$
from all angles that is held on different processors.

After expansion, the Poisson equation becomes
\begin{align}\label{1d_poisson}
  \frac{d^2\Phi_{lm}}{ds^2} + \frac{d\Phi_{lm}}{ds} - l(l+1)\Phi_{lm} =
  4\pi G\rho_{lm}r^2,
\end{align}
where $s\equiv \ln{r}$, and radial  boundary conditions
\begin{align}
  \left.\left[\frac{d}{ds} + (l+1)\right]\Phi_{lm}\right|_{r_o} = 0; \quad
  \left.\left[\frac{d}{ds} - l\right]\Phi_{lm}\right|_{r_i} = 0  
\end{align}
are applied. 
The discretised version of  Eq. \ref{1d_poisson} is a
tridiagonal matrix equation. It can be solved relatively
quickly with the \texttt{LAPACK} software package  
to yield $\Phi_{lm}(r)$. The potential at $r$ and $\theta$ boundaries
can be found from Eq. \ref{sph_trans}. Note that the solution
method outline here assumes no mass outside the computational domain.  

\subsection{Planetary migration}

Planetary migration is mostly neglected in order to focus
on the issue of gap stability. However, a fifth order Runge-Kutta
integrator \citep{press92} for the equation of motion of the planet
has been implemented in \zeus. During one hydrodynamic time step $\Delta t$, 
the planet's position and velocity is integrated from $t\to t+\Delta
t$ with adaptive sub-steps but using the density field at
$t$. If only the upper disc ($z>0$) is simulated, then the vertical
force is set to zero and the planet remains in the midplane. 



\section{Gap stability in 3D}\label{3d_stability}
The substantial cost of a 3D numerical calculation prevent extensive
parameter studies. Instead, described below are specific simulations
demonstrating gap instability. The general numerical setup is as follows. 

The disc is divided into $N_r\times
N_\theta \times N_\phi$ zones in the $(r,\theta,\phi)$ directions,
uniformly spaced in $(\theta,\phi)$
and logarithmically spaced in polar $r$. During $t\leq t_0 \equiv 10P_0$ the
disc is evolved without the planet. 
Between $t_0 \leq t \leq 2t_0$, the planet mass is switched on.
If the disc is self-gravitating then for $0 \leq t \leq t_0$, 
$(\lmax, \mmax)=(48,0)$ is set for the boundary potential and $g_\theta$ is smoothly switched on. 
For $t>t_0$, $(\lmax, \mmax)=(16,10)$ is set for the boundary potential. 
Hydrodynamic boundary conditions are
outflow in radius, reflecting in $\theta$ and periodic in $\phi$.

\subsection{Vortex instability in a non-self-gravitating
  disc}\label{3dvortex} 
In this simulation the disc model is $Q_o=4$. The domain is 
$r\in[1,10]$ and $n_H=2$ scale-heights above and below the
midplane.  The resolution is
$(N_r, N_\theta, N_\phi) = (256,64,696)$, corresponding to $16$ cells
per $H$ in $\theta$, but only $\simeq 5.6$ cells per $H$ in $r$ and
$\phi$. The planet mass is $M_p = 3\times10^{-4} M_*$ and is fixed
in the midplane at radius $r_p=5$ after its introduction. The EOS parameter is
$h_p=0.7$. 

\begin{figure}
    \centering
    \includegraphics[scale=.31,clip=true,trim=0.74cm 0.64cm 2.08cm 
    0cm]{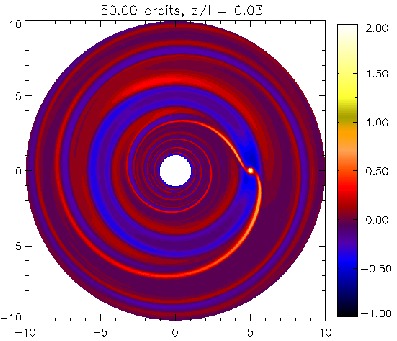}\includegraphics[scale=.31,clip=true,trim=0.74cm
    0.64cm 2.08cm 
    0cm]{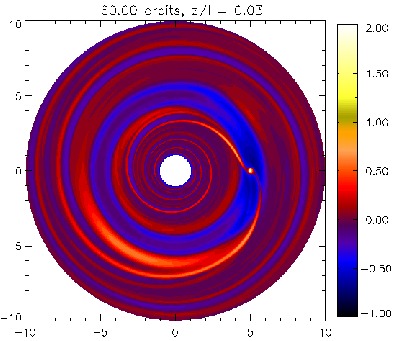}\includegraphics[scale=.31,clip=true,trim=0.74cm   
    0.64cm 2.08cm
    0cm]{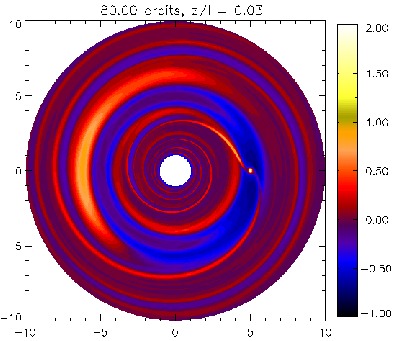}\includegraphics[scale=.31,clip=true,trim=0.74cm  
    0.64cm 0cm 0cm]{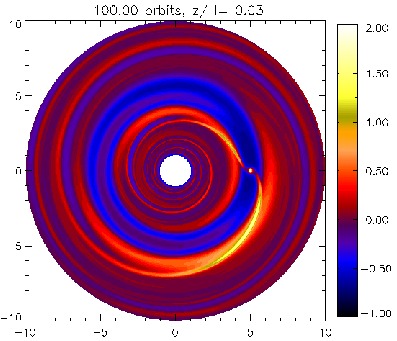}\\
    \includegraphics[scale=.31,clip=true,trim=0.74cm 0.64cm 2.08cm 
    0cm]{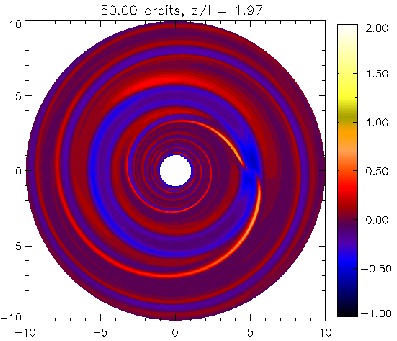}\includegraphics[scale=.31,clip=true,trim=0.74cm
    0.64cm 2.08cm 
    0cm]{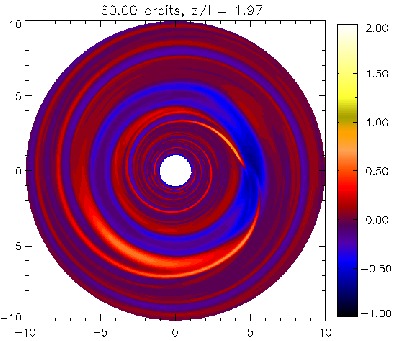}\includegraphics[scale=.31,clip=true,trim=0.74cm   
    0.64cm 2.08cm
    0cm]{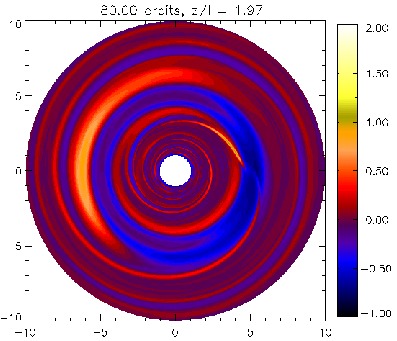}\includegraphics[scale=.31,clip=true,trim=0.74cm  
    0.64cm 0cm 0cm]{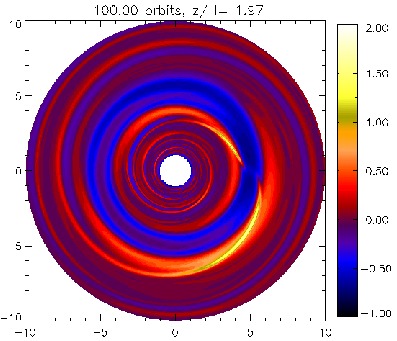}
    \caption{Relative density perturbation near the midplane (top) and 
      near $z=2H$ (bottom).    
    \label{ch5_vortex2}}
\end{figure}

\begin{figure}
    \centering
    \includegraphics[scale=.33]{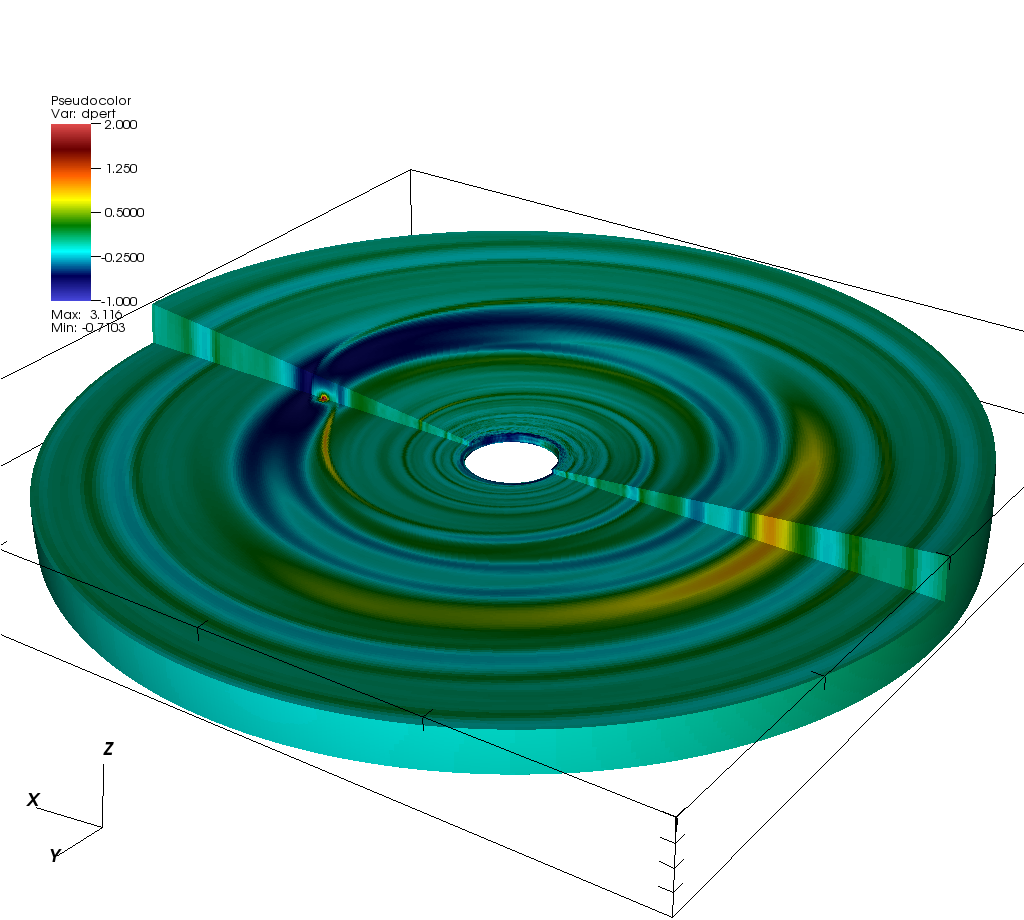}
    \caption{Three-dimensional visualisation of the vortex
      instability at $t=80P_0$. The colour bar is relative density
      perturbation. 
    \label{ch5_vortex2_3d}}
\end{figure}

Fig. \ref{ch5_vortex2} show density perturbations
relative to the azimuthally-averaged density field at $t=10P_0$. Two
cuts in $\theta$ are shown, corresponding to the midplane and at
$z=2H$. The vortex instability clearly persists for planetary gaps in 
a three-dimensional disc. The relative density perturbation has no
vertical dependence (except near the planet since it is
embedded). This  is also seen in the three-dimensional plot in
Fig. \ref{ch5_vortex2_3d}. Since the background density is essentially
Gaussian in $z$, the vortex density field will also be Gaussian
vertically.

The evolution is very similar to 2D discs simulated in Chapter
\ref{paper1}---\ref{paper2}. The only noticeable difference is that in 2D without
self-gravity, 3 vortices develop initially then merge
(Fig. \ref{vortices_sg_nsg}). In the 3D run,  one vortex develops from
the beginning (although a weak, second vortex can be identified
in Fig. \ref{ch5_vortex2} at $t=60P_0$). Reduction in the number of vortices from 2D to 3D
was also noted by \cite{meheut10}. However, in quasi-steady state the vortex in
3D and 2D are essentially identical.  
These results show that the vortex instability is a 2D instability
. Thus, the condition for
instability can be largely determined by vertically integrated
properties of the disc. 



\begin{figure}[!ht]
    \centering
    \includegraphics[scale=.67,clip=true,trim=0.0cm  
    1.0cm 0cm 0cm]{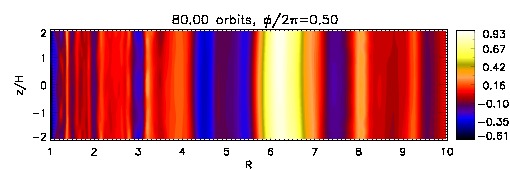}\\
    \includegraphics[scale=.67,clip=true,trim=0.0cm  
    0.0cm 0cm 0.99cm]{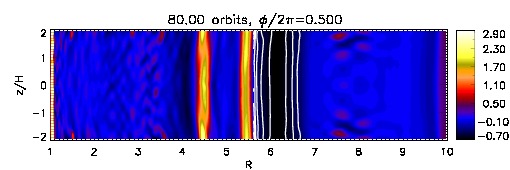} 
    \caption{Relative perturbation of density (top) and
      vertical component of vortensity $\eta_z$ (bottom) 
     in a meridional plane 
      through the vortex at $R\in[5.6,6.7]$ (white contour lines).  
    \label{ch5_vert_cut}}
\end{figure}

The vortex vertical structure is shown in Fig. \ref{ch5_vert_cut}. 
The vortex lies just outside the outer gap edge at $R\simeq 5.6$. Its
associated relative density and vortensity perturbations have very
weak vertical dependence, despite the background being
Gaussian. Stratification has no effect on the instability. 



\begin{figure}[!ht]
  \centering
  \includegraphics[scale=.55,clip=true,trim=0.0cm 1.0cm 1.68cm
  0cm]{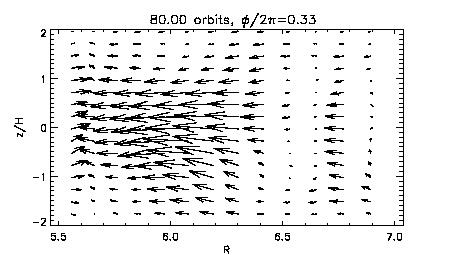}\includegraphics[scale=.55,clip=true,trim=1.7cm 1.0cm 1.68cm 0cm]{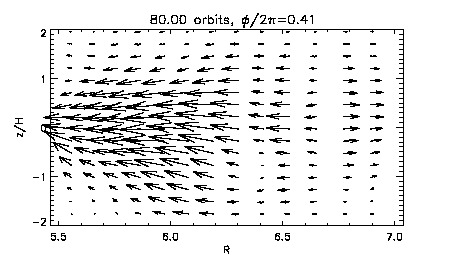}\\
  \includegraphics[scale=.55,clip=true,trim=0.0cm 1.0cm 1.68cm
  0cm]{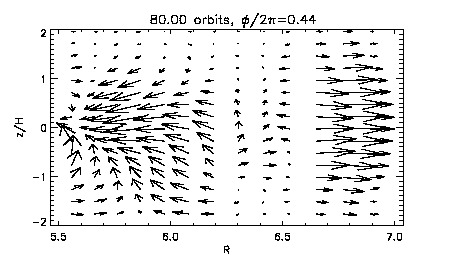}\includegraphics[scale=.55,clip=true,trim=3.6cm
  13.7cm 5.33cm 6.5cm]{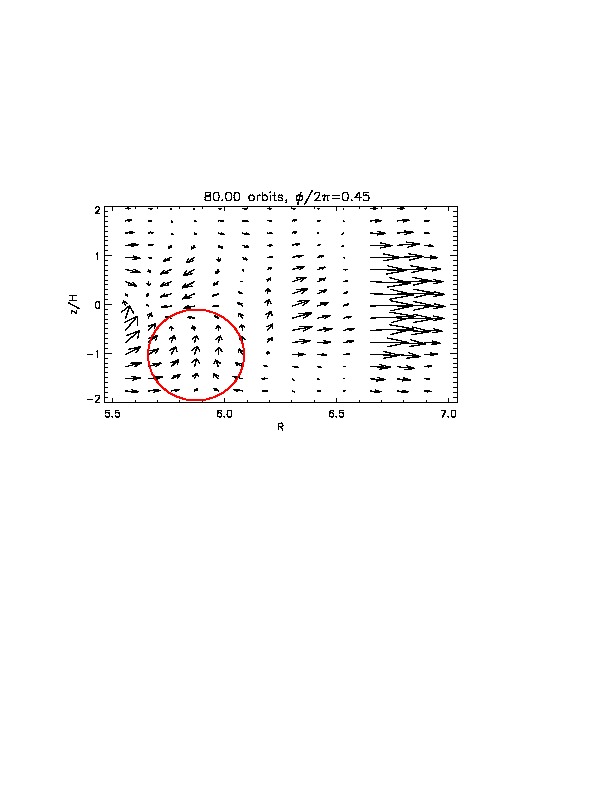}\\
  \includegraphics[scale=.55,clip=true,trim=0.0cm 0.0cm 1.68cm 0cm]{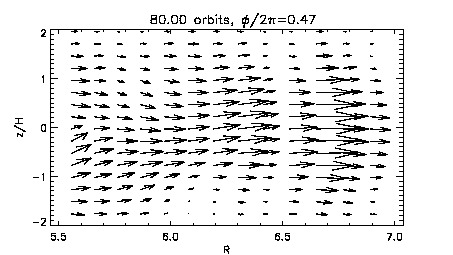}\includegraphics[scale=.55,clip=true,trim=1.7cm 0.0cm 1.68cm 0cm]{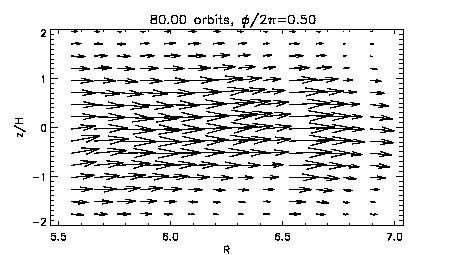}
  \caption{Meridional flow in vertical planes through the
    vortex. Velocity vectors have been scaled by the local
    density.  The vertical velocity has also been
    divided by $H$. The circled region highlights flux towards the
    midplane near the vortex centroid. 
    \label{ch5_stream}}
\end{figure}

The mass flux $\rho\bm{u}$ of the main vortex at
$t=80P_0$ is drawn in Fig. \ref{ch5_stream} for several meridional
planes. 
As expected for an anticyclonic vortex, the radial flow
transits from inwards to outwards as one traverses 
anti-clockwise (increasing $\phi $) through the vortex. The transition
point is the vortex centroid. The
$\phi/2\pi=0.45$ slice shows that the transition involves
vertical flux, but its magnitude is small even near the midplane,
where the mass density is largest. This means the vortex
centroid has very little motion. There is no evidence for strong
meridional circulation found by \cite{meheut10}, but the flow towards the
midplane highlighted in Fig. \ref{ch5_stream} suggest vertical motion
near the vortex centroid, 
as seen in \citeauthor{meheut10}. Note that symmetry about
$z=0$ is broken, so mass flux can overshoot across the
midplane.       


\subsection{Vortex-planet interaction}
  
\begin{figure}[!ht]
    \centering
    \includegraphics[width=\linewidth]{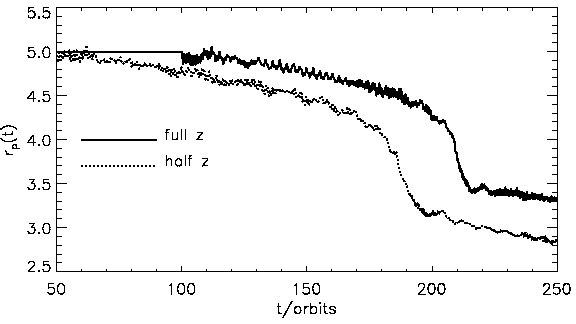}
    \caption{Vortex-induced migration of a Saturn-mass planet in 3D
      discs. The run `full z' simulates the entire disc and
      `half z' only simulates the upper half-space ($z>0$). 
    \label{ch5_orbit3_nsg}}
\end{figure}

Fig. \ref{ch5_orbit3_nsg} shows the planet orbital radius after
it is released at $t=100P_0$ for the simulation in
\S\ref{3d_stability}. Also shown is a second simulation where only
$z>0$ was simulated and the planet released at $t=50P_0$ (before vortices fully
develop). Results are very similar to 2D, the migration 
composing of slow and rapid phases, the latter induced by the vortex
associated with the inner gap edge. Since the interaction is gravitational,
and the perturbed density field associated with the vortex 
has no vertical structure, it is not surprising that the same interaction occurs in
3D as in 2D. 

\vspace{-0.5cm}
%
%

\subsection{Vortex modes with self-gravity}
For 3D self-gravitating simulations, the computational domain is 
$r\in[1,25]$ and $0 \leq z \leq 2H$. For simulations in this
subsection the resolution is $(N_r,
N_\theta, N_\phi)=(256, 32, 512)$.  Jovian planets
are used to open the gap because the steeper
edges imply higher growth rates than gaps opened by a Saturn-mass
planet. This shortens the simulation time. The planet is fixed at
$r_{p}=10$ and simulated up to $t=50P_0$. The EOS parameter is
$h_p=0.5$. 

\begin{figure}[!ht]
  \centering
 \includegraphics[scale=.4,clip=true,trim=.85cm 0.64cm 2.2cm
  0cm]{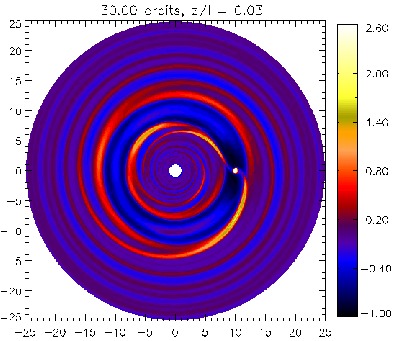}\includegraphics[scale=.4,clip=true,trim=.85cm
  0.64cm 2.2cm
  0cm]{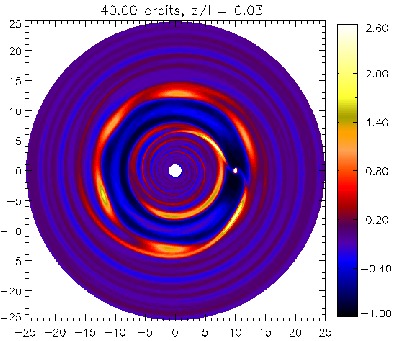}\includegraphics[scale=.4,clip=true,trim=.85cm
  0.64cm 0.0cm
  0cm]{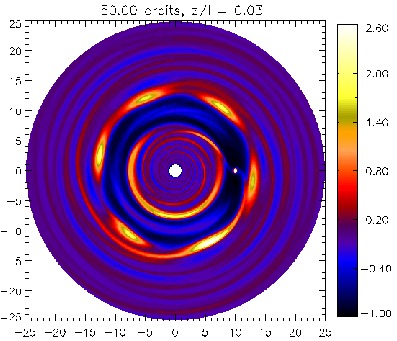}\\
  \includegraphics[scale=.4,clip=true,trim=.85cm 0.64cm 2.2cm
  0.69cm]{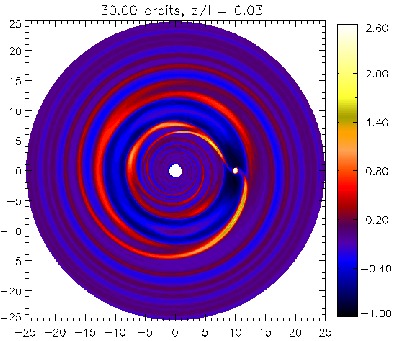}\includegraphics[scale=.4,clip=true,trim=.85cm
  0.64cm 2.2cm
  0.69cm]{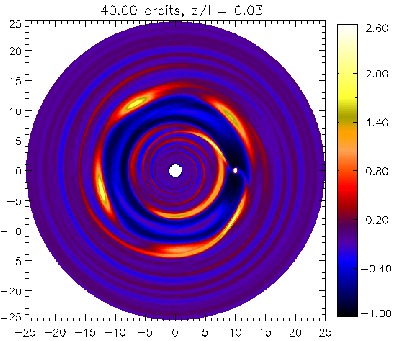}\includegraphics[scale=.4,clip=true,trim=.85cm
  0.64cm 0.0cm
  0.69cm]{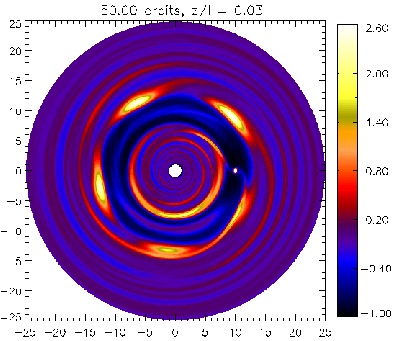}\\
  \includegraphics[scale=.4,clip=true,trim=.85cm 0.64cm 2.2cm
  0.69cm]{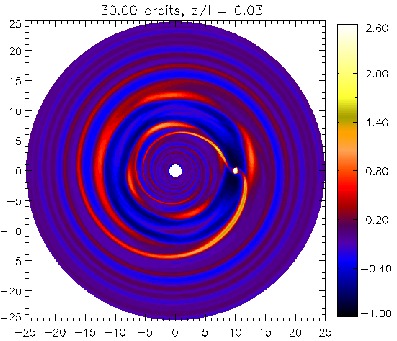}\includegraphics[scale=.4,clip=true,trim=.85cm
  0.64cm 2.2cm
  0.69cm]{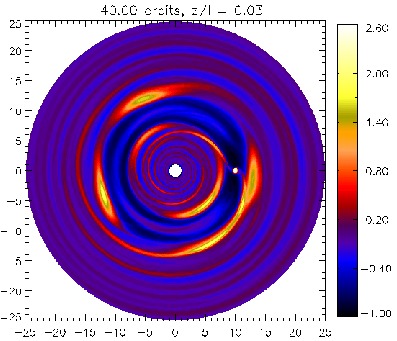}\includegraphics[scale=.4,clip=true,trim=.85cm
  0.64cm 0.00cm
  0.69cm]{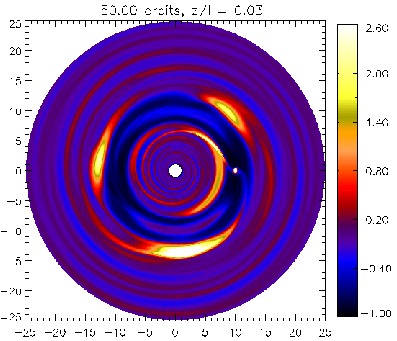}
  \caption{Vortex instability in self-gravitating
    3D disc models $Q_o=3$ (top), $Q_o=4$ (middle) and $Q_o=8$
    (bottom). The relative
    density perturbation near the midplane is shown.
    \label{ch5_vortex5_vortex6}}
\end{figure}

Fig. \ref{ch5_vortex5_vortex6} compares vortex formation in the
$Q_o=3,\,4,\,8$ disc models, with $M_p=10^{-3}M_*$. 
It confirms that the effect of
self-gravity on the vortex instability discussed in Chapter
\ref{paper2}, persists in 3D. Comparing the outer gap edge at
$t=30P_0$ shows self-gravity is stabilising, because in $Q_o=8$
vortices can already be identified but not so readily seen in
$Q_o=4$. At $t=40P_0$, more vortices develop with lowering $Q_o$
because stabilisation is more effective for low $m$, thus higher $m$
vortex modes are favoured, just as in 2D. 
The 6-vortex and 5-vortex configuration remains
in $Q_o=3,\,4$, respectively, at $t=50P_0$ whereas only 3 remain in
$Q_o=8$ (reduced from 4). This is due to resisted merging as vortices
behave like co-orbital planets with increasing self-gravity. Again,
this was seen in 2D.

\begin{figure}[!ht]
    \centering
    \includegraphics[scale=.67,clip=true,trim=0.0cm  
    1.cm 0cm 0.5cm]{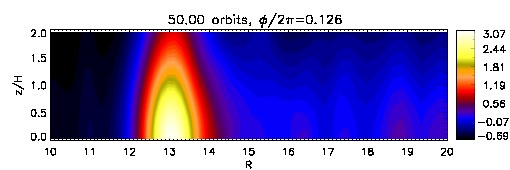}\\
    \includegraphics[scale=.67,clip=true,trim=0.0cm  
    1.cm 0cm 0.99cm]{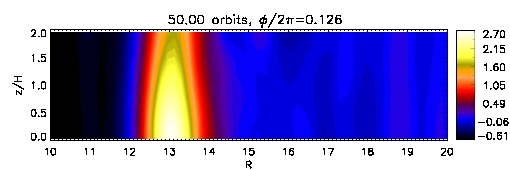} \\
    \includegraphics[scale=.67,clip=true,trim=0.0cm  
    0.0cm 0cm 0.99cm]{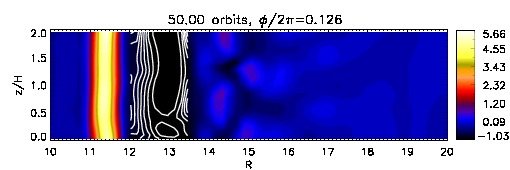}
    \caption{Density perturbation relative to $t=0,\,10P_0$ (top, middle) 
      and perturbation to the vertical component of vortensity
      relative to $t=10P_0$ (bottom). The 
      disc model is $Q_o=4$ with self-gravity. 
    \label{ch5_vortex5_vert_cut}}
\end{figure}

The vertical structure of a vortex in the $Q_o=4$ disc is shown in
Fig. \ref{ch5_vortex5_vert_cut}. Because the initial disc is not in
self-gravitating equilibrium, the density perturbation relative to $t=0$
and to $t=10P_0$ (just after $g_\theta$ is fully switched on) are both
shown. The increasing perturbation towards the midplane is partly due
to vertical self-gravity increasing the background midplane density. 
Between $t=10P_0$ and $t=50P_0$ there may also be additional
stratification due to vertical self-gravity of the vortex, since the
midplane density is largest.
Stratification  is reduced if perturbation is 
taken relative to $t=10P_0$, so the instability indeed tends to involve
the entire column of fluid. 

\begin{figure}[!ht]
  \centering
  \includegraphics[scale=.69,clip=true,trim=2.4cm
  12.8cm 5.33cm 6.71cm]{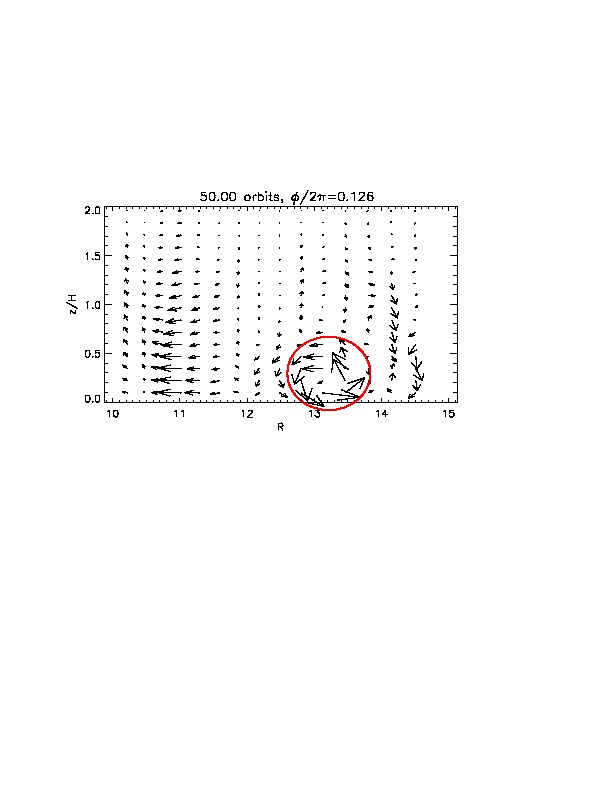}
  \caption{Mass flux corresponding to 
    Fig. \ref{ch5_vortex5_vert_cut}. 
    \label{ch5_vortex5_stream}}
\end{figure}

The vortensity field in Fig. \ref{ch5_vortex5_vert_cut} shows some 
disturbance along the contour lines at the vortex boundary, as well as 
closed contours within the vortex, which is associated with
anti-clockwise motion indicated by mass flux vectors in 
Fig. \ref{ch5_vortex5_stream}. \cite{meheut10} found similar flow to occur in
pairs --- clockwise ($r<r_c$) and anti-clockwise ($r>r_c$), where
$r_c$ denotes the vortex centroid. 
 Such symmetry may not be expected
 here since there is a planet located on one side of the gap edge. 
It is also  likely that the flow here has been
modified by self-gravity of the vortex and/or background.  
However, despite additional vertical structure and flow, the
vortensity perturbation remains largely columnar.    


\clearpage
\subsection{Edge modes} 
The disc model $Q_o=1.5$ is used to demonstrate edge modes. 
The resolution is increased to $(N_r,
N_\theta, N_\phi) = (256,64,516)$, on account of 
reduction in effective scale-height due to vertical self-gravity.  

\begin{figure}[!ht]
  \centering
  \includegraphics[scale=.4,clip=true,trim=.85cm 0.64cm 2.2cm 
  0cm]{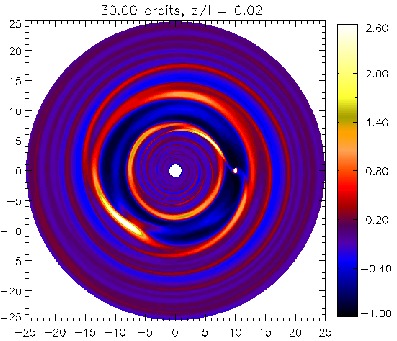}\includegraphics[scale=.4,clip=true,trim=.85cm
  0.64cm 2.2cm 
  0cm]{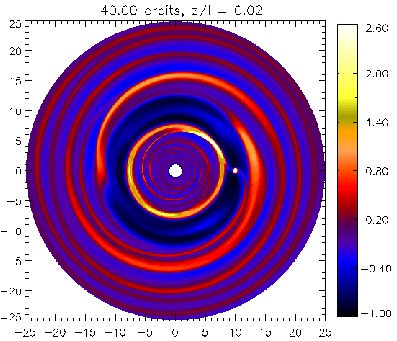}\includegraphics[scale=.4,clip=true,trim=.85cm   
  0.64cm 0.0cm
  0cm]{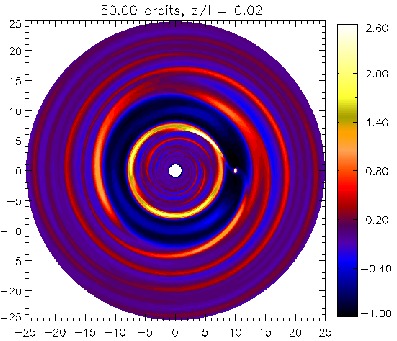}\\
  \includegraphics[scale=.4,clip=true,trim=.85cm 0.64cm 2.2cm 
  0.cm]{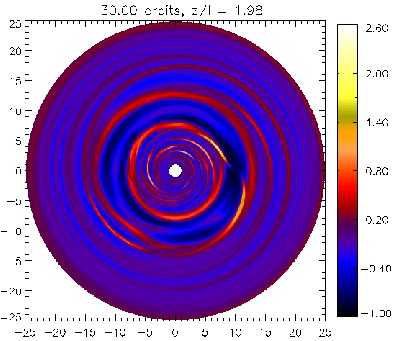}\includegraphics[scale=.4,clip=true,trim=.85cm
  0.64cm 2.2cm 
  0.cm]{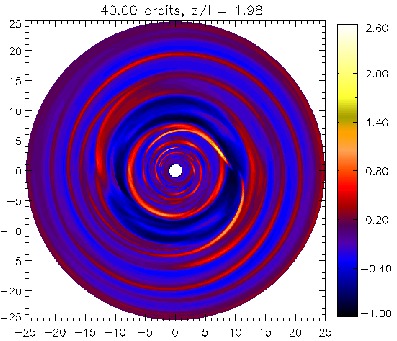}\includegraphics[scale=.4,clip=true,trim=.85cm   
  0.64cm 0.00cm
  0.cm]{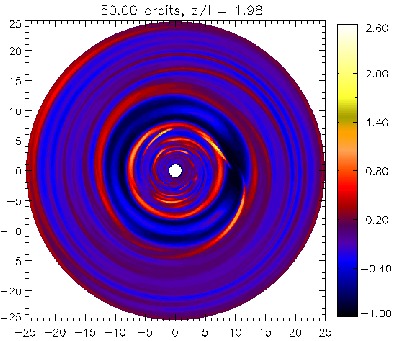}
  \caption{Edge mode in the 3D disc model $Q_o=1.5$. The density
    perturbation relative to $t=10P_0$ is shown near the midplane
    (top) and $z=2H$ (bottom).  
    \label{ch5_edge1HR}}
\end{figure}

\begin{figure}[!ht]
    \centering
    \includegraphics[scale=.33,clip=true,trim=0cm 1.5cm 0cm
    3cm]{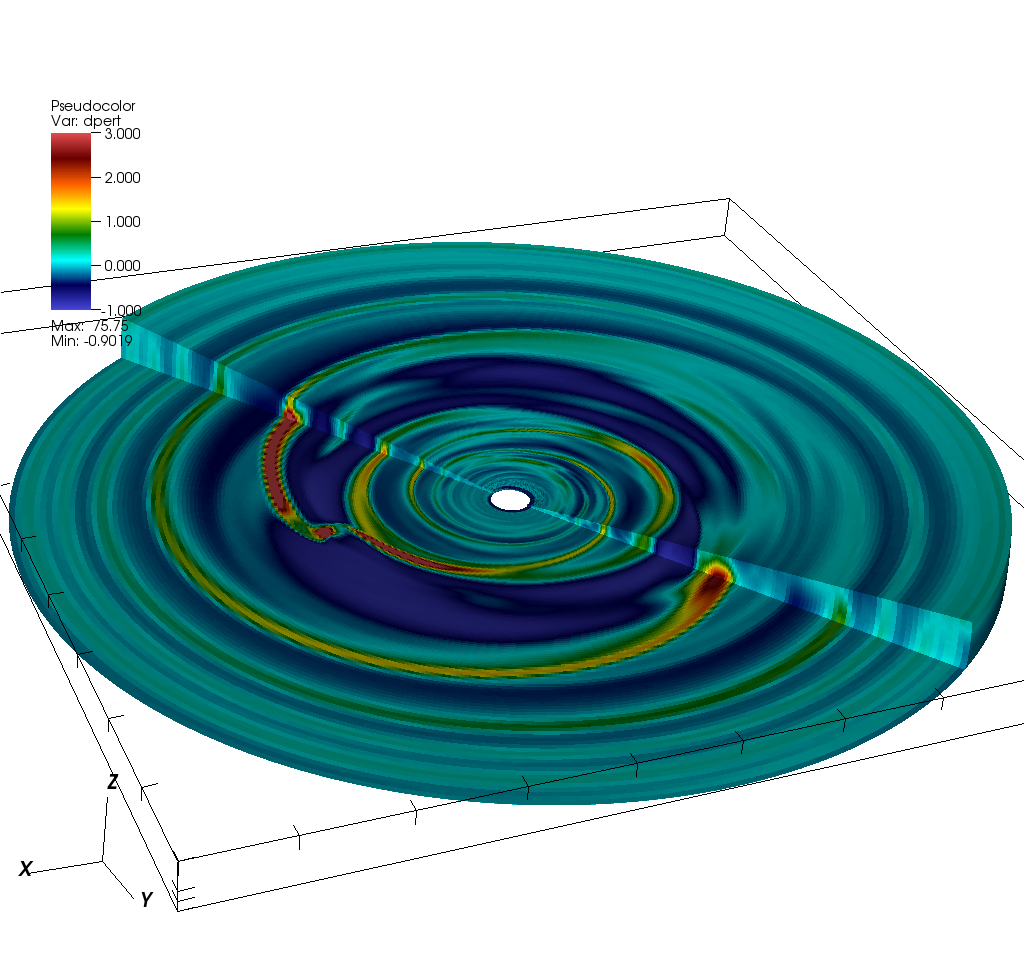} 
    \caption{Three-dimensional visualisation the edge mode instability
      for the simulation in Fig. \ref{ch5_edge1HR}. The snapshot is
      taken at $t=35P_0$. One of the edge mode spiral arms coincide with the
      planetary wake. The colour bar is relative density
      perturbation. 
    \label{ch5_edge1HR_3d}}
\end{figure}

Fig. \ref{ch5_edge1HR} shows the development of an $m=2$ edge mode 
and Fig. \ref{ch5_edge1HR_3d} is a three-dimensional
visualisation. Perturbations are calculated relative to $t=10P_0$.   
The behaviour near the midplane is similar to 2D 
simulations, but the edge mode amplitude decreases noticeably away from the 
midplane. The background density ($t=10P_0$) near $z=2H$ has been
reduced by vertical self-gravity from $t=0$. This means that perturbations high in
the atmosphere appear enhanced, but their actual densities are small compared to the
midplane. 


\begin{figure}[!ht]
    \centering
    \includegraphics[scale=.67,clip=true,trim=0.0cm  
    1.cm 0cm 0.5cm]{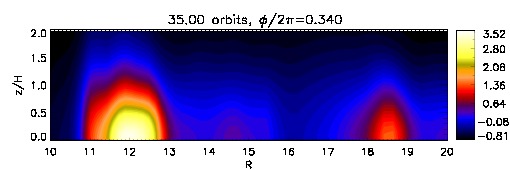}\\
    \includegraphics[scale=.67,clip=true,trim=0.0cm  
    1.cm 0cm 0.99cm]{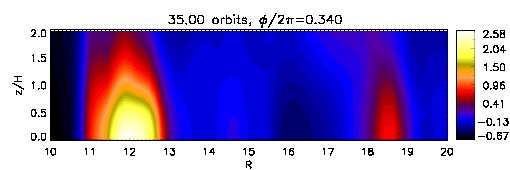} \\
    \includegraphics[scale=.67,clip=true,trim=0.0cm  
    0.0cm 0cm 0.99cm]{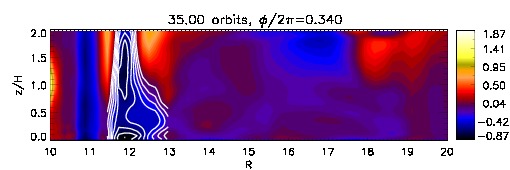}
    \caption{Density perturbation relative to $t=0,\,10P_0$ (top, middle)
     and perturbation to the vertical component of vortensity
      relative to $t=10P_0$ (bottom) in a vertical plane
      through the edge mode in Fig. \ref{ch5_edge1HR_3d}.
    \label{ch5_edge1HR_vert_cut}}
\end{figure}

\begin{figure}[!ht]
  \centering
  \includegraphics[scale=.69]{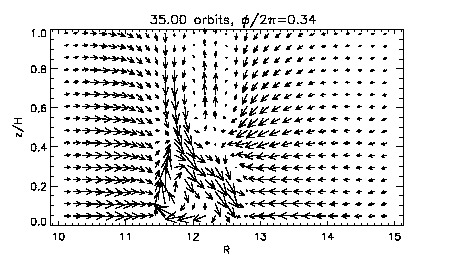}
  \caption{Mass flux corresponding to 
    Fig. \ref{ch5_edge1HR_vert_cut}. Note the vertical axis only covers
    $z<H$. 
    \label{ch5_edge1HR_stream}}
\end{figure}

Fig. \ref{ch5_edge1HR_vert_cut}---\ref{ch5_edge1HR_stream} shows the
vertical structure of the edge mode near the gap edge. 
The density 
perturbation is even more stratified than a self-gravitating vortex 
. Most of the disturbance is contained within $z<H$ and the
corresponding vortensity perturbation is no longer columnar.  
Mass flux goes into the over-density radially from either side, as well
as vertically from above. Since the density perturbation in the midplane is large 
(up to $2$---3 times over-dense than the background), this blob of fluid 
perturbs the surrounding flow gravitationally. The vertical flux is likely attributable 
to this effect. Recall that the vertical boundaries are solid, so flow must reflect 
off it. 


\subsection{Outwards migration induced by edge modes} 
This simulation is the 3D version of the fiducial case in Chapter
\ref{edge_migration}. The disc model is $Q_o=1.7$ and the resolution
is $(N_r,N_\theta,N_\phi) = (256,32,512)$. $N_\theta$ is
reduced to enable longer simulation times. The planet mass is now $M_p =
0.002M_*$ and it is allowed to migrate for $t>20P_0$. 
Since torques originating from close to the planet were 
found to be important in the 2D simulations, the torque cut-off
parameter is reduced to $\epsilon_c=0.5$ (about 3.5 cells). 

\begin{figure}
  \centering
  \includegraphics[scale=.67]{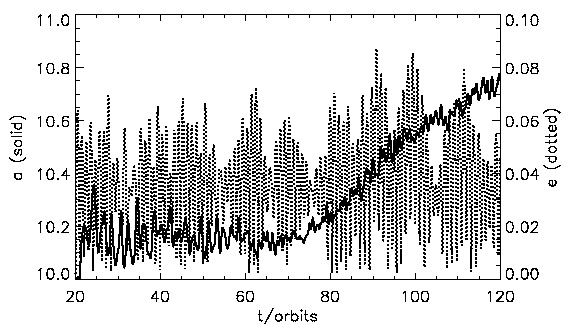}
  \caption{Keplerian semi-major axis $a$ and eccentricity $e$ for
    planet ($M_p=0.002M_*$) in the $Q_o=1.7$ three-dimensional disc model.
    \label{ch5_edge2_migration}}
\end{figure}

Fig. \ref{ch5_edge2_migration} and Fig. \ref{ch5_edge2_migration}  shows 
the migration history and snapshots of
 density perturbation. These figures can be compared to the 2D
case, Fig. \ref{ch4_Q1d7_migration}---\ref{ch4_Q1d7_polarxy}. Whereas
the 2D disc developed $m=2$--3  edge modes, only $m=2$ was seen in
3D. This could be due to the much reduced resolution in 3D (by a
factor of 4 in $r$ and $\phi$ compared to 2D), or reduction in the
strength of self-gravity due to disc thickness.

\begin{figure} 
  \centering
  \includegraphics[scale=.5,clip=true,trim=0cm 0cm 1.7cm
  0cm]{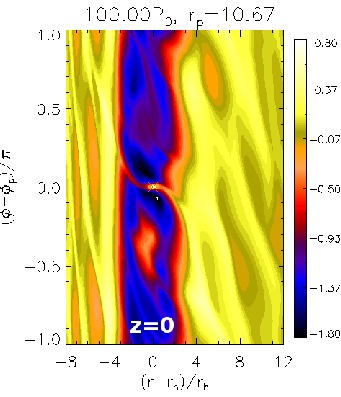}\includegraphics[scale=.5,clip=true,trim=2.21cm 0cm 1.7cm
  0cm]{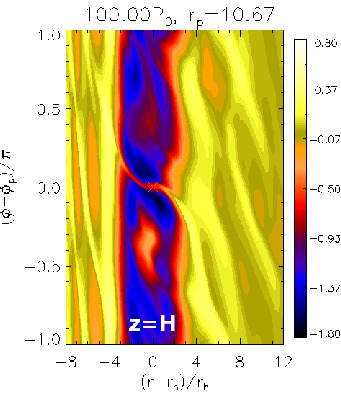}\includegraphics[scale=.5,clip=true,trim=2.21cm 0cm 0.0cm
  0cm]{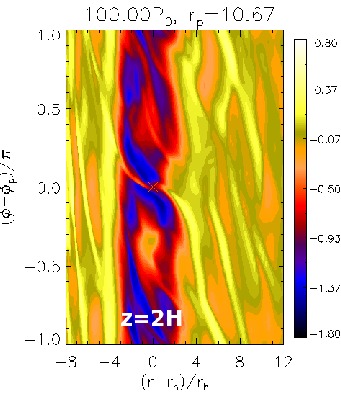}  
  \caption{Logarithmic density perturbation
    $\log{\left[\rho/\rho(t=10P_0)\right]}$ at three heights, during
    outwards  migration of a planet induced by edge modes in a 3D disc. 
  \label{ch5_edge2_polarxy}}
\end{figure}

Despite the low resolution,  this simulation demonstrates that edge
modes can induce outwards migration in 3D discs. Outwards migration
only begins around $50P_0$ after release, whereas it starts immediately in the 2D
run. The migration in 3D is also on average slower than in 2D.  
This is probably because the applied tapering 
(for numerical considerations) artificially reduces the torque magnitudes. 
However, the existence of edge  
modes does not depend on treatment close to the planet. Their 
development will likely bring material into the gap  
for interaction with the planet, because they are associated with the gap edge.   
This is 
shown in Fig. \ref{ch5_edge2_polarxy}, which is qualitatively very
similar to the 2D run. The planet is asymmetrically positioned with
respect to gap edges. Edge mode spiral arms protrude into the gap 
 and supply an over-density within the coorbital region of the planet upstream to it.  
As the spiral pattern pass by from $\phi>\phi_p$ to $\phi < \phi_p$,  
edge mode material in the planet's coorbital region is expected to
execute inward horseshoe turns ahead of the planet. This contributes positively to
the disc-on-planet torque. 


\vspace{-0.25cm}
\section{Summary and discussion}\label{summary}
The first 3D numerical simulations demonstrating
planetary gap instability were presented.  The
instabilities behave similarly in 2D and 3D. 
In retrospect, 2D is a better approximation for self-gravitating discs 
than for non-self-gravitating discs because vertical self-gravity 
compresses the midplane. 


The vertical structure of the vortex instability was
first studied in non-self-gravitating discs by
\cite{meheut10}. They found prominent circulatory 
 flow in the $(R,z)$ plane, but these 
features are not seen in the current simulations.   
 However, \citeauthor{meheut10} used prescribed disc profiles which
lead to strong instability (growth rate $\simeq
0.4\Omega$) whereas profiles here are induced by a
planet, for which 2D linear calculations gave smaller growth rates (a few per
cent of $\Omega$). There could also be influences from the planet. 
Simulations here show vortices and edge mode 
spirals have increasing vertical dependence as the strength of 
self-gravity is increased. This is probably due to additional vertical 
acceleration due to self-gravity by the higher midplane density. 


Computational cost present significant obstacle to 3D
simulations. Vortices and edge modes are global instabilities
associated with localised gradients. Thus entire discs must be
simulated whilst maintaining good resolution locally. 
In the 3D runs here, resolution in $(r,\phi)$ is
only 4---6 cells per $H$, compared to  16 in 2D
simulations. Improved resolution and a convergence study
should be performed in the future. Despite numerical limitations, the
similarity between 3D and 2D results validates the razor-thin disc
approximation employed in previous Chapters. In fact, it may be more
advantageous to use 2D in some situations, e.g. massive discs and planet migration,  because high resolution is 
achievable.  
Migration should only depend on the density field, which
has essentially the same structure in 2D and 3D.



\def\baselinestretch{1}
\chapter{Conclusions and future work}
\graphicspath{{Conclusions}}

\section{Main results} 



\subsection{Unstable gap structure}
Planetary gaps are characterised by thin vortensity rings outlining
its edges. They are generated across shocks induced by the planet. 
Their formation can be anticipated from models of the planet-perturbed
flow and expected shock location. For planetary masses of Saturn or
above, these shocks extend well into the gap and the rings reside in
the planet's co-orbital region. The existence of vortensity extrema
satisfies the necessary condition for instability.  


\subsection{Vortex modes and induced migration}
In a low viscosity, non-self-gravitating disc, corotation radii of
unstable modes are associated with local vortensity minima at gap
edges. Their nonlinear outcome is the formation of a few vortices
that merge on dynamical timescales. The result is one
azimuthally-extended vortex circulating just outside either gap
edge. 

For the standard disc models adopted, differential Lindblad
torques favour inwards migration and lead to gravitational scattering
of the inner vortex outwards. The planet therefore receives an inwards
kick of a few Hill radii. Migration alternates between a slow phase,
when gap and vortex formation takes places, and brief rapid inward
jumps. 

The role of vortices in such non-smooth migration can be understood as a 
modified form of type III migration. 
Vortices provide a coorbital mass deficit when
 it is circulating on the gap edge, and remove it when they flow across
the gap and fill it to some extent. 


\subsection{Effect of self-gravity on vortex modes}
Increasing self-gravity is stabilising for the vortex mode, if
the background disc profile remains fixed. However, in practice 
the effective planetary mass increases with the strength of
self-gravity due to disc material bound to the planet. This results in
steeper gap edges with increasing surface density scales,
and this destabilises the vortex mode. 

The combined effect of self-gravity through the background profile and
through the linear response is to shift the most unstable vortex mode
to higher azimuthal wave-numbers. This leads to more vortices 
with increasing disc mass. These self-gravitating vortices are localised in
azimuth as well as radius, and they can behave like co-orbital
planets. Their mutual gravitational interaction allows them to survive
against vortex merging for a longer period of time than in
non-self-gravitating discs.     


\subsection{Edge modes in massive discs}
The vortex mode is suppressed in sufficiently massive discs. Instead,
unstable modes have co-rotation radii at local vortensity
maximum, which lie closer to the planet than vortensity minimum. 
These edge modes can be interpreted  as a disturbance associated with
the gap edge, which induces activity in the smooth part of the disc
via self-gravity. Thus they are global modes. For the disc models
adopted here, they are seen predominantly as one-sided spiral arms in the disc
beyond the planet's orbital radius. 

Edge modes associated with the outer gap edge make the disc-on-planet
torque on average more positive than that in stable discs.
Outwards migration is thus promoted. A Saturn-mass planet can be
scattered by an edge mode spiral and jump outwards significantly on
orbital timescales. More massive Jovian planets reside in deeper gaps but 
can migrate outwards on timescales of 10's of orbits. 
 These effects originate from
material brought into the gap by the edge mode. 


\subsection{Validity of  razor-thin discs}
The use of 2D discs to study planetary gap stability has been
validated by counterpart 3D numerical simulations. Vortex modes and
edge modes behave similarly in both geometries. Not surprisingly, the
vertical dependence of their structure increases with the strength of
self-gravity. 


\section{Directions for future work}
Attention has mostly been placed on understanding the phenomenon found
in each of the preceding Chapters. The trade-off is the realism of
disc models and the size of parameter space that can be practically
surveyed. Extensions to this work are listed below. 


\subsection{Three-dimensional linear calculations} 
Although the non-self-gravitating 3D simulations did not reproduce
vertical structure found by \cite{meheut10}, it is clear
that vortices involve vertical motion. In order to gain a better
understanding of three-dimensional effects, if any,  it would be ideal
to have 3D linear stability analysis to complement nonlinear
simulations. 

A simple starting point would be three-dimensional, locally
isothermal discs with radial structure, such as a surface density
bump \citep[e.g. profiles used by][]{li00}. The vertical structure is
Gaussian, so if the disc occupies $z\in[-\infty,\infty]$, one can
expand the vertical dependence of perturbations as
$W(R,z) = \sum_{l=0}^{\infty} W_l(R)\he_l(z)$, where $\he_l$ is
the $l^\mathrm{th}$ Hermite polynomial. $W_0$ would represent the 2D
solution and $W_2$ would be the first 3D contribution to an even
vortex mode. This leads to the system of
coupled first order ODE's analysed by \cite{tanaka02}, but the problem
here would be solving for the homogeneous solution, with
eigenfrequencies to be determined. Preliminary results show that 3D
has negligible effect on growth rates and, e.g. $|W_2|\ll
|W_0|$, so perturbations are indeed columnar. Nonetheless, the small
vertical velocities still grow exponentially 
and this should be important in the nonlinear regime.

However, it is more realistic to have finite vertical boundaries,
particularly for the vortex modes. This is because they are easily
suppressed by viscosity, and would only exist in `dead
zones' of protoplanetary discs which have sufficiently low
viscosity. However, dead zones exist close to the disc midplane and
do not occupy the entire vertical extent \citep{gammie96}. In this case one requires
the solution to a 2D partial differential equation eigenvalue
problem. 




\subsection{Planet-edge mode interactions}
Systematic studies of planetary migration in massive discs have only
begun recently \citep{michael11,baruteau11}. These first
simulations consider disc masses of $M_d/M_* \geq 0.14$. Such discs
are gravitationally active even without a planet.      

Lower mass discs that are gravitationally stable without a planet can become 
gravitationally active via edge modes associated 
with planetary gaps. Edge mode properties depend on planetary 
properties, 
but the unstable disc has significant back-reaction on 
planet migration, which could subsequently affect further edge mode
development. An example is edge mode-induced migration moving the planet
to a region even more prone to edge modes. 

These inter-dependencies will be explored with high resolution 2D 
numerical simulations. This is an expansion of Chapter \ref{edge_migration},
which focused on the migration mechanism, to a parameter survey. 
Questions include the relation between planet and disc masses
that allow edge modes, and how the extent of migration depend on planetary and disc 
properties.      



\subsection{Numerical simulations with  \zeus}
Three-dimensional simulations in Chapter \ref{3d} were primarily performed 
to confirm 2D results. 
Future numerical studies with \zeus will focus on properties of vortex 
and edge modes in 3D self-gravitating discs. 

Further development of \zeus is needed. Currently, 
the non-self-gravitating initial conditions are evolved 
towards equilibrium. This inevitably leaves disturbances in the disc
, relative to which perturbations are defined. Although subsequent 
perturbations due to planet, vortices and edge modes are much larger than the noise in the 
background, it would be ideal to have a module to
to generate clean initial discs in equilibrium with self-gravity. 
Development of polytropic disc models is also planned, as a 
preliminary step towards including thermodynamics, which is important for 
gravitational instabilities. 

Many interesting issues arise in 3D. The vertical structure of vortex and edge modes
as a function of gap properties should be investigated. 
Vertical boundary 
conditions (including the treatment of the gravitational potential in terms of $\lmax, \mmax$ ) 
may be important for vortex modes because they tend to involve entire fluid columns, 
i.e. they are global in $z$. The evolution of 3D self-gravitating disc vortices, 
including the possibility of self-gravitational collapse, vortex merging (a phenomenon 
common in 2D) and their migration \citep[so far only simulated in 2D by][]{paardekooper10}
should be studied. These numerical works require improved numerical resolutions.


\addtocontents{toc}{\protect\setcounter{tocdepth}{0}}
\appendix
\cleardoublepage
\addappheadtotoc
\appendixpage
\chapter{Upper limit on the horseshoe width}\label{xslimit}

It is possible to place an upper limit on the horseshoe width $x_s$
that is valid  in the limit  of  either zero  or constant
pressure. Consider a local Cartesian 
frame $(x,y)$ that  has origin at the planet and hence co-rotates with
it with the Keplerian 
angular velocity $\Omega_p =\sqrt{GM_*/r_p^3}.$ The $x$ direction
corresponds to increasing radius and $y$ is azimuthal direction (same
sense as $\Omega_p$). In this frame,  consider fluid elements approaching the
planet from $(x=x_0>0,\,y=\infty)$ with velocity
$(0,-3\Omega_px_0/2).$ Suppose such a fluid element executes a horseshoe
turn crossing the co-orbital radius  at position $(0,y)$ with velocity $(u_x,0)$.
The maximum value of $x_0$, below which such motions occur, is
determined as follows. 
 
The equation of motion governing a fluid element or particle, in the
absence of pressure forces, is
\begin{align}\label{horseshoe}
  \frac{D\bm{u}}{Dt}+2\Omega_p\hat{\bm{z}}\times\bm{u} = 
  -\nabla\Phi_\mathrm{tot}, 
\end{align}
where
\begin{align}
  \Phi_\mathrm{tot}= -\frac{GM_p}{\sqrt{x^2+y^2}}-\frac{3}{2}\Omega_p^2x^2.
\end{align}
$ \Phi_\mathrm{tot}$  contains contributions from the gravitational
potential of the planet of mass $M_p$ (with the softening length
neglected) and  the tidal potential associated with
the central object. From  Eq. \ref{horseshoe} it follows that
the Jacobi invariant   
\begin{align}\label{jacobi}
  J = \frac{1}{2}|\bm{u}|^2 + \Phi_\mathrm{tot}
\end{align}
is constant along a particle path. Equating $J$  evaluated at $(x,y)=(x_0,\infty)$
to $J$ evaluated at  $(0,y)$
gives
\begin{align}\label{eq1}
  -\frac{3}{8}\Omega_p^2x_0^2 = \frac{1}{2}u_x^2 -\frac{GM_p}{y}.
\end{align}
The steady-state Euler equation of motion for $u_y$ evaluated at $(0,y)$ is
\begin{align}\label{eq2}
u_{x}\frac{\p u_y}{\p x} +2\Omega_p u_x = -GM_p/y^2.
\end{align}
In the neighbourhood of $(0,y)$ for the type of  streamline
being considered,  $u_x<0$ and    $ \p_x u_y<0$ because the particle
is going from $x>0$ to $x<0$ by executing a U-shape turn. This implies
\begin{align}\label{vxlimit}
  |u_x|>\frac{GM_p}{2\Omega_p y^2}.
\end{align}
Combining Eq. \ref{eq1} and Eq. \ref{vxlimit} gives
\begin{align}\label{limit}
  -\frac{3}{8}\Omega_p^2x_0^2 > \frac{1}{2}\left(\frac{GM_p}{2\Omega_p y^2}\right)^2
  -\frac{GM_p}{y}.
\end{align}
Writing $x_0$ and $y$ in units of the Hill radius, i.e. setting
$x_0 = \hat{x}_0 r_h$ and $y = \hat{y} r_h $, gives
\begin{align}
  \hat{x}_0 < \sqrt{\frac{8}{\hat{y}} - \frac{3}{\hat{y}^4}} \lesssim 2.3.
\end{align}
Thus because the maximum possible  value of the right hand side of the above is $2.3,$
particles executing a U-turn could not have originated
further than $2.3r_h$ from the planet's orbital radius.  
This is comparable to the value of $2.5r_h$ 
that has been estimated from hydrodynamic simulations
\citep{artymowicz04b,paardekooper09}. 

\section{Application to co-orbital vortices}\label{coorb_vortices}

It is straight forward to apply the equations above to describe
gravitational interaction between two vortices in the 
shearing sheet. The equations have exactly the same form, with a
slight change of notation $x\to X,\,y\to Y$ where $X,Y$ denote the
relative displacement of the two vortices in the shearing sheet 
frame. $M_p\equiv qM_*$ becomes the sum of vortex masses.  
The frame is now centred on one of the vortices, its average
orbital radius from the central star being $r=r_p$.  

Non-dimensionalising by $r_p$, Eq. \ref{limit} becomes
\begin{align}\label{min_sep1}
  \frac{3}{8}\hat{X}_0^2  < \frac{q}{\hat{Y}} - \frac{q^2}{8\hat{Y}^4}  
\end{align}
where $\hat{X}_0 = X_0/r_p$, $\hat{Y} = Y/r_p$, 
and the angular speed of the rotating frame $\Omega_p=\sqrt{GM_*/r_p^3}$ has been used.  
Eq.  \ref{min_sep1} 
relates the initial mutual radial separation of the vortex pair,
$\hat{X}_0$, to their minimum azimuthal separation, $\hat{Y}$. 

Eq. \ref{min_sep1} is useful for the case there vortices are just able
to undergo U turns. For fixed $q$, the function
\begin{align}\label{fy}
  f(\hat{Y};q) = \frac{q}{\hat{Y}} - \frac{q^2}{8\hat{Y}^4}
\end{align}
has a maximum value at $\hat{Y} = (q/2)^{1/3},$
corresponding to the
maximum conceivable  initial separation $X_0=X_s,$ where
\begin{align}\label{xs}
  X_s = 2^{2/3}q^{1/3}r_p.
\end{align}
If $X_0>X_s$ then Eq. \ref{min_sep1}  cannot be satisfied and there
can be no mutual horseshoe turns (as in the test particle and planet case). 
For initial separations  $X_0<X_s$, Eq. \ref{min_sep1} implies that
the minimal inter-vortex distance (in $Y$) must exceed  $q^{1/3}/2$ (so that
$f,\hat{X}_0^2 >0$).  Now for sufficiently large $q,$ $q^{1/3}/2$ will be larger
than the critical separation  for merging (set by hydrodynamics),  so
merging is avoided during the encounter.

It is interesting to compare Eq. \ref{xs}  to the estimate of the
horseshoe half-width $x_s$ of  \cite{paardekooper09}. They found $x_s= 
1.68(q/h)^{1/2}r_p$ based on hydrodynamic simulations for low mass
planets. Equating $x_s$ and $X_s$ with $h=0.05$ gives
$q=8.9\times10^{-5}.$ This should give the minimum $q$ for which
pressure effects could be ignored. Inserting this value in
\cite{murray00}'s model of coorbital satellites gives a minimal
separation of $0.58$ (with other parameter values estimated in \S\ref{vortices_planets}), which is 
close to simulation results (Fig. \ref{kida_expts}). Hence, if the 
vortex-pair interaction in Chapter \ref{paper2} is purely gravitational, a single vortex
behaves in a similar way to $\sim 15$ Earth masses, i.e. a low-mass
protoplanet. 

Considering a vortex size of order $H$, the vortex-to-star mass ratio  
is $q\sim \pi H^2\Sigma/M_*\simeq h^3/Q$. For a self-gravitating vortex where the Toomre 
$Q\sim 1$, $q\simeq h^3=1.25\times10^{-4}$ for $h=0.05$,
slightly exceeding the threshold value above.
Hence the pressureless treatment of self-gravitating
vortex-vortex interactions should be acceptable for the purpose 
of explaining the resisted-merging of self-gravitating vortices.


\chapter{Linear model for flow perturbed by a planet}\label{alt_linear} 
\graphicspath{{Chapter1/figures/}}

When modelling the pre-shock flow in the
presence of a planet (\S\ref{pre_shock_flow}), 
the pressureless momentum equations were
integrated directly. Here, a simple linear model for the pre-shock
velocity field is presented, which was found to give qualitatively
correct results compared to hydrodynamic simulations.    

Low mass planets cannot have shocks close to $r=r_p$ because
flow is purely azimuthal and becomes subsonic close to the planet (in the planet's frame).  
The goal of this calculation 
is to introduce non-zero radial velocity by treating the planet
potential as a 
perturbation. The planet mass $M_p$ is assumed to be 
sufficiently small to allow linearisation, but large enough to ignore
pressure. The zero-pressure,  
steady state momentum equations in the shearing sheet are 

\begin{align}
  &  u_x\frac{\p u_x}{\p x} +u_y\frac{\p u_x}{\p y} -2\Omega_p u_y =
  -\frac{\p\Phi_p}{\p x}+3\Omega^2 x\\ 
  &  u_x\frac{\p u_y}{\p x} +u_y\frac{\p u_y}{\p y} + 2\Omega_p u_x =
  -\frac{\p \Phi_p}{\p y}. 
\end{align}

\noindent The centre of the frame co-rotates with the planet. Linearising
about the Keplerian flow

\begin{align}
  &u_x\to 0+\dd u_x\\
  &u_y\to -\frac{3}{2}\Omega_p x +\dd u_y,
\end{align}

\noindent where $\dd$ denote small Eulerian perturbations, gives

\begin{align}
  &\frac{3}{2}x\Omega_p\frac{\p\dd u_x}{\p y} +2\Omega_p\dd
  u_y=\frac{\p \Phi_p}{\p x}\\ 
  \notag 
  \Rightarrow& \frac{\p\dd u_x}{\p y} +\frac{4}{3x}\dd
  u_y=\frac{2}{3\Omega_p 
    x}\frac{\p\Phi_p}{\p x} 
\end{align}
\noindent from the $x$-momentum equation. The linearised $y$-momentum 
equation is:
\begin{align}\label{vx}
  &\dd u_x\left(-\frac{3}{2}\Omega_p\right)-\frac{3}{2}x\Omega_p\frac{\p \dd
    u_y}{\p y} 
  +2\Omega_p\dd 
  u_x=-\frac{\p \Phi_p}{\p y}\\\notag
  \Rightarrow&\dd u_x = 3x\frac{\p\dd u_y}{\p y}
  -\frac{2}{\Omega_p}\frac{\p \Phi_p}{\p y}.  
\end{align}


\noindent Eliminating $\dd u_x$ gives
\begin{align}\label{vyODE}
  \frac{\p^2\dd u_y}{\p y^2}+k^2\dd u_y =
  \frac{k}{\Omega_p}\frac{\p^2\Phi_p}{\p
    y^2}+\frac{k^2}{2\Omega_p}\frac{\p\Phi_p}{\p x}\equiv F,
\end{align}
where $k\equiv 2/3x$. Eq. \ref{vyODE} is an ODE with $x$ as a
parameter. Multiplying by $k^{-2}$ and considering the limit $x\to0$ ( 
$k^{-1}\to\infty$) demands the balance:
\begin{align}\label{smallx}
  \dd u_y&=\frac{1}{2\Omega_p}\frac{\p\Phi_p}{\p x},
\end{align}
which implies, from Eq. \ref{vx}, that close to $x=0$,
\begin{align}
  \dd u_x&=\frac{3x}{2\Omega_p}\frac{\p^2\Phi_p}{\p x\p
    y}-\frac{2}{\Omega_p}\frac{\p\Phi_p}{\p y}. 
\end{align}
\noindent The expression for $\dd u_y$ is
equivalent to Eq. 28 of \cite{paardekooper09} when pressure is
neglected, and $\dd u_x$ is equivalent to their Eq. 30 if
$\p_x\Phi_p$ is neglected.


\indent Eq. \ref{vyODE} describes a forced oscillator and solutions are
readily obtained by writing 
\begin{align}
  \dd u_y=A(y)\cos{ky}+B(y)\sin{ky},
\end{align}
and imposing
\begin{align}
  A^\prime\cos{ky}+B^\prime\sin(ky)=0,\notag
\end{align}
where the prime denotes $\p_y$. For $x>0$, the coefficients are
written as
\begin{align}
  A&=k^{-1}\int^\infty_yF(y^\prime)\sin{ky^\prime}dy^\prime,\label{osc_int1}\\
  B&=-k^{-1}\int^\infty_yF(y^\prime)\cos{ky^\prime}dy^\prime.
\end{align} 
So that $\dd u_y,\,\dd u_x\to0$ as $y\to\infty$ for fixed $x$. The
$\p_y\Phi_p$ part of the forcing $F$ can be integrated by parts
(twice) to give explicit velocity expressions
\begin{align}\label{velexpressions}
  \dd u_y&=-\frac{k\Phi_p}{\Omega_p}-\frac{k^2GM_p}{\Omega_p}
  \int^\infty_y\frac{\sin{k(y-y^\prime)}dy^\prime}{(x^2+y^{\prime
      2}+\epsilon_p^2)^{1/2}} 
  +\frac{GM_p}{3\Omega_p}\int^\infty_y\frac{\sin{k(y^\prime-y)}dy^\prime}{(x^2+y^{\prime
      2}+\epsilon_p^2)^{3/2}},\\ 
  \dd
  u_x&=-\frac{2k^2GM_p}{\Omega_p}\int^\infty_y\frac{\cos{k(y-y^\prime)}dy^\prime}{(x^2+y^{\prime  
      2}+\epsilon_p^2)^{1/2}}
  -\frac{2GM_p}{3\Omega_p}\int^\infty_y\frac{\cos{k(y^\prime-y)}dy^\prime}{(x^2+y^{\prime
      2}+\epsilon_p^2)^{3/2}}\label{osc_int2} \\
  \phantom{\dd u_x}&=
  2B\cos{ky}-2A\sin{ky}-\frac{2}{\Omega_p}\frac{\p\Phi_p}{\p y}. \notag
\end{align}
The second line for $\dd u_x$ follows from Eq. \ref{vx} once $A,\,B$
are known, and this is the expression used in practice. Since $k\propto
x^{-1}$, linearisation fails for small $x$. 

\begin{figure}[!ht]
\center
\subfigure[$M_p=2.8\times10^{-4}M_*$]{\includegraphics[width=0.49\linewidth]{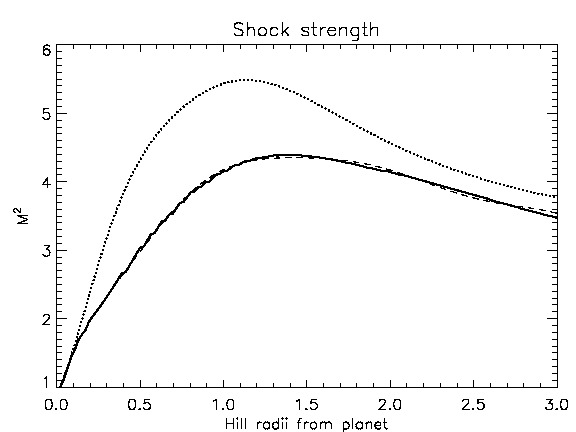}}
\subfigure[$M_p=2\times10^{-4}M_*$\label{jump_machsq_lin_m2}]{\includegraphics[width=0.49\linewidth]{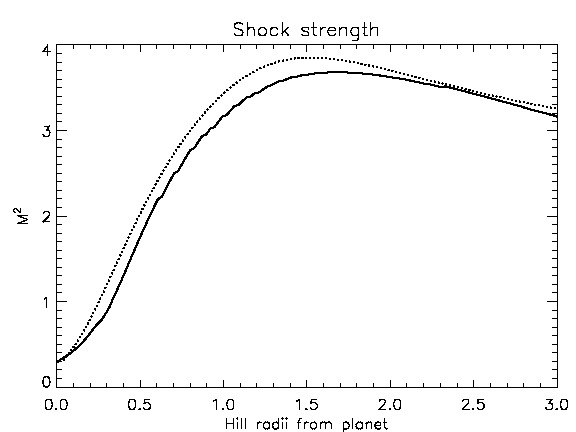}} 
\caption{Perpendicular Mach number squared, $M^2$, along the outer
  spiral shock induced by a giant planet. The location of the shock
  front is taken from simulation. The horizontal axis is the
  $x$ co-ordinate of the shock in the shearing sheet. The solid line
  is $M^2$ measured from simulation, with the dashed curve as a
  polynomial fit, and the dotted line is $M^2$ calculated from the
  velocity field given by the linear model (but with perturbation to
  the $y$-velocity neglected).\label{jump_mach_lin}}
\end{figure}

The planet-perturbed velocity field is now available for use. To first 
approximation, let the velocity field be $(\dd u_x,u_y)$, i.e.
the $y$-velocity perturbation is neglected in comparison to Keplerian
flow. The unperturbed uniform density field,
$\Sigma=\Sigma_0\times10^{-4}$, is assumed when the surface density is needed. 
Taking the shock front from simulation data, Fig. \ref{jump_mach_lin}
compares shock strength predicted by 
the above calculation to that found in disc-planet simulations. The
linear solution over-estimates $M^2$ but its position of maximum, at
$r-r_p = 1.1r_h,\, 1.5r_h$ (for
$M_p=2.8\times10^{-4}M_*,\,2\times10^{-4}M_*$ respectively) is close
to simulation results. In particular, the 
$M_p=2\times10^{-4}M_*$ curve follows closely to simulation. This
simple linear calculation adequately predicts the qualitative
behaviour of $M^2$ along a  given shock. More importantly it confirms
the planet-induced $u_x$ enable shocks near $r=r_p$, and that
vortensity is generated within about a Hill radius from the planet (which is 
the region of increasing $M^2$ with respect to $x$).

\begin{figure}[!ht]
\center
\subfigure[$M_p=2.8\times10^{-4}M_*$]{\includegraphics[width=0.49\linewidth]{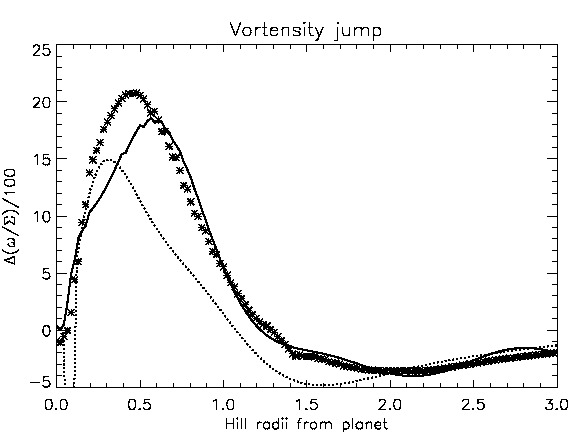}}
\subfigure[$M_p=2\times10^{-4}M_*$]{\includegraphics[width=0.49\linewidth]{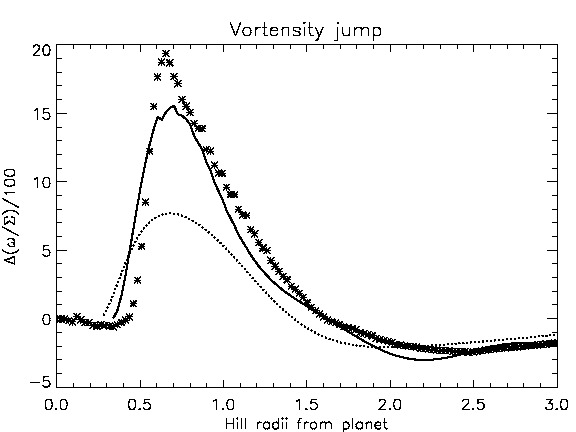}}
\caption{Vortensity jump across the outer spiral shock induced by a giant
  planet. The location of the shock front is taken from
  simulation. Asterisks are jumps measured
  from simulation. The solid line is the predicted jump given pre-shock data
  from simulation  using the analytic jump expression
  (Eq. \ref{vortensity_jump}). The dotted curve is the jump 
  predicted using the linear model for pre-shock velocity and 
using the unperturbed uniform surface density in the pre-shock flow. \label{jump_vortensity_lin}}
\end{figure}

Fig. \ref{jump_vortensity_lin} compare vortensity jumps.
Linear theory under-estimates vortensity generation, probably
because $|[\omega/\Sigma]|\propto
1/\Sigma$ and the unperturbed uniform density was used, which
is larger than the actual pre-shock surface density (this is seen in
Fig. \ref{shock_curve2_} as diverging particle paths approaching
the U-turn). The agreement between the linear model and
simulation is better in $M_p=2\times10^{-4}M_*$, in terms of position of
maximum vortensity jump and shock strength, since linearisation become
better approximation as $M_p$ decreases (despite the pressureless
approximation becoming worse).

\chapter{Shocks induced by a planet and the vorticity jump across it}

\section{Estimating shock fronts}\label{shockfront}

\cite{papaloizou04} presented a method based on characteristic rays 
to estimate the shape of wakes induced by a
planet. It was applied to Keplerian
flow. Their formalism is outlined below and applied to general 
two-dimensional flow. Note that the method applies to local or global
frames. 

Consider a frame which rotates at the planet's 
fixed circular orbital frequency.  
The hydrodynamic variables are assumed to have wave-like spatial and time
dependencies, $\exp{i(\bm{k}\cdot\bm{x} - \sigma t)}$, where
$\bm{k}$ is a wavenumber and $\sigma(\bm{k},\bm{x}, t)$ a frequency.
Seeking a local dispersion relation from the hydrodynamic equations
yields 
\begin{align}\label{loc_disp}
  \sigma = -\bm{k}\cdot\bm{v} + c_s |\bm{k}|.
\end{align} 
The wavefronts or characteristic curves $y(x)$ are related to the
dispersion relation by 
\begin{align}\label{wavefront}
  \frac{dy}{dx} &= \frac{\p\sigma/\p k_y}{\p\sigma/\p k_x}\\
  &=\frac{-\hat{v}_y + k_y/|\bm{k}|}{-\hat{v}_x +
    k_x/|\bm{k}|}\notag\\
  & = \frac{-\hat{v}_y(1+\beta^2)^{1/2} +
    \beta}{-\hat{v}_x(1+\beta^2)^{1/2} + 1},\notag
\end{align}
where $\hat{v}_y=v_y/c_s$ is the non-dimensional velocity and similarly for $v_x$. 
The second line in
Eq. \ref{wavefront} applies the dispersion relation
Eq. \ref{loc_disp}, and the third line follows by defining
$\beta\equiv k_y/k_x$. 

The solution of interest is the characteristic
ray  with $\sigma=0$, corresponding to a
stationary wave in the planet frame. Setting the frequency to zero, 
Eq. \ref{loc_disp} reads:
\begin{align}\label{zero_freq}
  0=-\hat{v}_x - \beta\hat{v}_y + (1+\beta^2)^{1/2},
\end{align}
which is a quadratic equation for $\beta$ with solution
\begin{align}\label{beta_solution}
\beta = \frac{\hat{v}_x\hat{v}_y\pm(\hat{v}^2
  -1)^{1/2}}{1-\hat{v}_y^2},  
\end{align}
where $\hat{v}^2 = \hat{v}_x^2+\hat{v}_y^2$. 
Using Eq. \ref{zero_freq} to replace $(1+\beta^2)^{1/2}$ in
Eq. \ref{wavefront} gives  
\begin{align}
  \frac{dy}{dx} = \frac{\beta(1 -\hat{v}_y^2) - \hat{v}_x\hat{v}_y}{1
    -\hat{v}_x^2 - \beta \hat{v}_x\hat{v}_y }.  
\end{align}
Finally, inserting the expression for $\beta$ from
Eq. \ref{beta_solution} gives
\begin{align}
\frac{dy}{dx}&=\frac{\pm(\hat{v}^2
  -1)^{1/2}(\hat{v}_y^2-1)}{(\hat{v}^2-1)\pm\hat{v}_x\hat{v}_y(\hat{v}^2 -1)^{1/2}} \\   
&=\frac{\hat{v}_y^2-1}{\hat{v}_x\hat{v}_y\pm(\hat{v}^2
  -1)^{1/2}},
\end{align}
which can be solved once a flow field is given. The solution $y(x)$   
correspond to  wakes induced by the planet, with the minus (plus)
sign corresponding to wakes exterior (interior) to the planet.  These
wakes develop into shocks for sufficiently large planetary masses.   


\section{Vorticity jump across a  steady isothermal
  shock}\label{vortjump} 
\begin{figure}[!hb]
  \center
  \includegraphics[width=\linewidth]{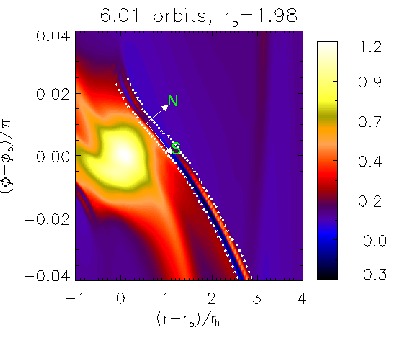}
  \caption{Vorticity generation across the outer spiral shock 
    induced by  planet.  The absolute vorticity is shown on a linear   
    scale. Arrows define directions normal ($N$, equivalent to
    $x_2$) and tangential ($S$, equivalent to $x_1$ ) to the shock. 
    The upper (lower) dotted line indicate pre-shock (post-shock) 
    flow, since in the  planet frame, for $r>r_p$ material flows
    towards the planet from $\phi>\phi_p$.  
\label{vortgen_shock}} 
\end{figure}

\graphicspath{{Chapter1/figures/}}
\indent Consider an isothermal  shock
that is stationary in a frame  rotating with 
 angular velocity  $\Omega_p\hat{\mathbf{z}}$. 
In order to evaluate the vorticity jump across the shock it is convenient
to use a two dimensional orthogonal coordinate system  $(x_1,x_2)$ 
defined in the disc mid-plane such that  one 
of the curves $x_2 = {\rm constant} = x_s$  coincides with the shock.
The curves  $x_1 = {\rm constant}$  will then be normal  to the shock where they
intersect it.  In addition the coordinates  are set up so  that
$(x_1, x_2, z)$ is a right handed system. This co-ordinate system is
depicted in Fig.\ref{vortgen_shock} where distances along the normal
($x_2$) and tangential ($x_1$) directions to the shock are labelled
$N,\,S$ respectively. 

The $\hat{\mathbf{z}}$-component of relative vorticity $\omega_r$  can then be written as
\begin{align}\label{vorticity_def}
\omega_r=\frac{1}{h_1h_2}\left(\frac{\p (u_2 h_2)}{\p x_1}
-\frac{\p (u_1h_1)}{\p x_2}\right),
\end{align}
where $(h_1,h_2)$  are the components of the  coordinate  scale factor.

Note that on $x_2= x_s,$ $u_1$ is the velocity tangential  to the
shock and is unchanged across the shock. The normal component $u_2$
and other state variables  undergo a jump from pre-shock values to
post-shock values on normally traversing  the curve $x_2= x_s.$ 
For an isothermal shock
\begin{align}\label{shockc}
\frac{u_{2,\mathrm{post}}}{u_2}=M^{-2}=\frac{\rho}{\rho _{\mathrm{post}}},
\end{align}
where $M=u_2/c_s$ is the pre-shock perpendicular Mach number. Here
and in similar expressions below connecting pre-shock and post-shock
quantities,  post-shock values are denoted with a subscript
$_\mathrm{post}$ while pre-shock quantities are left without a
corresponding subscript. \noindent Thus the jump in relative vorticity
is 
\begin{align}\label{jump1}
[\omega_r]=\omega_{r,\mathrm{post}} -\omega_r.
\end{align}
Quite generally the  $x_1$ component of the  equation of motion for a steady state flow  is
\begin{align}
\frac{u_1}{h_1}\frac{\p u_1}{\p x_1}+\frac{u_2}{h_2}\frac{\p u_1}{\p x_2} -u_2\left(
\frac{u_2}{h_1 h_2}\frac{\p h_2}{\p x_1} -\frac{u_1 }{h_1 h_2}\frac{\p
  h_1}{\p x_2}\right)
=
-\frac{1}{\rho h_1}\frac{\p p}{\p
  x_1}-\frac{1}{h_1}\frac{\p\Phi_\mathrm{eff}}{\p x_1}+2\Omega_p
u_2,\label{eqm1} 
\end{align}
or equivalently
\begin{align}
\frac{u_1}{h_1}\frac{\p u_1}{\p x_1}+\frac{u_2}{h_1}\frac{\p u_2}{\p x_1} -(2\Omega_p+\omega_r)u_2
=
-\frac{1}{\rho h_1}\frac{\p p}{\p x_1}-\frac{1}{h_1}\frac{\p\Phi_\mathrm{eff}}{\p x_1},\label{eqm2}
\end{align}
where $p$ is the pressure and $\Phi_\mathrm{eff}$ the total potential
(the latter quantity being continuous across the shock). 
From Eq. \ref{eqm2}, the relative vorticity can be written in the form   
\begin{align}\label{vorticity_alt}
\omega_r=\frac{1}{h_1}\frac{\p u_2}{\p x_1}+\frac{u_1}{u_2 h_1}\frac{\p u_1}{\p x_1}
+\frac{1}{\rho u_2  h_1}\frac{\p p}{\p x_1}+\frac{1}{u_2
  h_1}\frac{\p\Phi_\mathrm{eff}}{\p x_1}-2\Omega_p. 
\end{align}
 Applying Eq. \ref{vorticity_alt} to give an expression for the post shock
relative vorticity, and using Eq. \ref{shockc} 
to express  the post-shock normal velocity and density
in terms of the corresponding pre-shock quantities, the vorticity jump
can be written in the form 
\begin{align}\label{vorticity_alt1}
[\omega_r]=\frac{1}{h_1}\frac{\p (M^{-2}u_2)}{\p x_1}+\frac{M^2 u_1}{u_2 h_1}
\frac{\p u_1}{\p x_1}
+\frac{M^2}{u_2 h_1}\frac{\p\Phi_\mathrm{eff}}{\p x_1} +
\frac{1}{\rho u_2  h_1}\frac{\p p_\mathrm{post}}{\p x_1}-(\omega_r+2\Omega_p),
\end{align}
where continuity of $\rho u_2$ across the shock has been used in the
pressure term. Adopting  a locally isothermal equation of state
implies  $p_\mathrm{post}=M^2 p.$ 
Substituting this into Eq. \ref{vorticity_alt1}, while  making
use of Eq. \ref{vorticity_alt} together with  
the relation $u_2=c_s M$ gives 
\begin{align}\label{vorticity_alt2}
[\omega_r]\equiv [\omega] =-\frac{c_s(M^2-1)^2}{M^2h_1}\frac{\p M}{\p
  x_1}+(M^2-1)\omega  
-\frac{(M^4-1)}{Mh_1}\frac{\p c_s}{\p x_1},
\end{align}
where $\omega= 2\Omega_p+\omega_r$ is the absolute vorticity.

Setting $h_1dx_1 =dS$ with $dS$ being the corresponding element of  distance measured parallel
to the shock, this takes the form
\begin{align}\label{vorticity_alt3}
[\omega_r]\equiv [\omega] =-\frac{c_s(M^2-1)^2}{M^2}\frac{\p M}{\p
  S}+(M^2-1)\omega
-\frac{(M^4-1)}{M}\frac{\p c_s}{\p S}.
\end{align}
This gives the vorticity jump across a shock in terms of  pre-shock quantities 
measured in the rotating frame 
in which it appears steady. It is important to note that
Eq. \ref{vorticity_alt3} applies specifically  in the right handed coordinate system
adopted with shock location given by $x_2=x_s.$ 
If instead $x_1=x_s$ was adopted for this location, the signs of the derivative terms
would be reversed as in the expression (2.23) given in \cite{kevlahan97}.
 Note too that the last term on the right hand side
that is proportional to the gradient of $c_s$ along the shock arises from
the assumption of a {\it locally}   isothermal equation of state and is not present
in the treatment given by \cite{kevlahan97}.

\chapter{Approximate equations in linear calculations}\label{lep_qep}

The full governing equation for linear
perturbations in a self-gravitating disc with fixed temperature
profile was summarised in Chapter \ref{paper2} as
\begin{align}\label{approx_gov}
  \mathcal{L}(W) = 0.
\end{align}
An expression for the linear operator $\mathcal{L}$ is given by
Eq. \ref{operator}. The eigenvalue $\sigma$ is contained
in $\mathcal{L}$ as $\bar{\sigma}$ in a complicated manner. 
However, once appearances of $\sbar$ as denominators are removed by multiplying
Eq. \ref{approx_gov} by the appropriate factor, one can neglect various powers of
$\sbar$, to approximate the above equation in quadratic and
linear order in $\sigma$.

\section{Governing equation to $\sigma^2$}
In the quadratic approximation, the governing equation is
\begin{align}
  (\sigma^2 \mathcal{L}_2 + \sigma \mathcal{L}_1 + \mathcal{L}_0)W = 0,
\end{align}
where
\begin{align}
  \mathcal{L}_0 =&  m\Omega\kappa^2\mathcal{D}_1 + m\Omega[\kappa^2/r - \kappa^{2\prime}
  + \kappa^2\Sigma^\prime/\Sigma + 2m^2\Omega\Omega^\prime]\mathcal{D}_2 \notag\\
  &+m\{\kappa^2[2r(\Omega^\prime + \Omega\Sigma^\prime/\Sigma) -
  m^2\Omega] 
  - 2r\Omega[\kappa^{2\prime} + m^2\Omega(\Omega\Sigma^\prime/\Sigma -
  \Omega^\prime)]\}\mathcal{D}_3/r^2 \notag \\
  &+m(2c_s^{2\prime}\kappa^2\Omega/r - 2m^2\Omega^3c_s^{2\prime}/r - \kappa^4\Omega)\mathcal{D}_4  \\
  \mathcal{L}_1 =& \kappa^2\mathcal{D}_1 + (\kappa^2/r - \kappa^{2\prime}
  + \kappa^2\Sigma^\prime/\Sigma + 4m^2\Omega\Omega^\prime)\mathcal{D}_2 -
  m^2(\kappa^2+ 4r\Omega^2\Sigma^\prime/\Sigma)\mathcal{D}_3/r^2  \notag\\
  &-(4m^2\Omega^2c_s^{2\prime}/r + \kappa^4)\mathcal{D}_4\\
  \mathcal{L}_2 =& 2m\Omega^\prime \mathcal{D}_2 - 2m(\Omega^\prime +
  \Omega\Sigma^\prime/\Sigma)\mathcal{D}_3/r - 2m\Omega c_s^{2\prime}\mathcal{D}_4/r.
\end{align}
The operators $\mathcal{D}_i$ are:
\begin{align}
  \mathcal{D}_1 =& c_s^2\frac{d^2}{dr^2} + c_s^{2\prime}\frac{d}{dr} +
  \mathcal{I}_2,\\
  \mathcal{D}_2 =& c_s^2\frac{d}{dr} + \mathcal{I}_1,\\
  \mathcal{D}_3 =& c_s^2 + \mathcal{I}_0,\\
  \mathcal{D}_4 =& 1.
\end{align}
The operators $\mathcal{I}_i$ are defined by  Eq. \ref{iiop}. 

\section{Governing equation to $\sigma$}
The approximation for corotational or vortex modes made in Chapter \ref{paper1}
can also be recast in a standard eigenvalue form as above. Recall that in
this approximation, $\bar{\sigma}$ is only retained when
it appears as $1/\bar{\sigma}$ associated with the vortensity gradient term. In other words, 
$D\simeq \kappa^2$ and $D^\prime\simeq \kappa^{2\prime}$. 
The simplified equation is then
\begin{align}
  \mathcal{A}W = \sigma \mathcal{B}W,
\end{align}
where
\begin{align}
  \mathcal{A} = &m\Omega[\mathcal{D}_1 + (1/r + \Sigma^\prime/\Sigma -
  \kappa^{2\prime}/\kappa^2)\mathcal{D}_2] \notag\\
  &+[2m\Omega(\Sigma^\prime/\Sigma + \Omega^\prime/\Omega -
  \kappa^{2\prime}/\kappa^2)/r - m^3\Omega/r^2]\mathcal{D}_3 \notag\\
  &+ m\Omega(2c_s^{2\prime}/r - \kappa^2)\mathcal{D}_4,\\
  \mathcal{B} = & - \mathcal{D}_1 - (\Sigma^\prime/\Sigma + 1/r -
  \kappa^{2\prime}/\kappa^2)\mathcal{D}_2 + m^2\mathcal{D}_3/r^2 + \kappa^2\mathcal{D}_4. 
\end{align}
Self-gravity was neglected in Chapter \ref{paper1}, in which case the operators $\mathcal{I}_i$
are set to zero. 



\chapter{Artificial vortices in an accretion disc}\label{kidasetup}
The artificial Kida-like vortices used in \S\ref{hydro2} were set up by
introducing velocity perturbations $(\dd u_r,\dd u_\phi)$ that 
correspond to the  elliptical vortices of \cite{kida81} in 
an incompressible shear flow. 

Consider a small patch of the disc, whose centre $(r_p, \phi_p)$
rotates at angular speed $\Omega_p$ about the primary and set up local
Cartesian co-ordinates $x=r_p(\phi-\phi_p),\,y=r_p-r$. In the $(x,y)$
frame, there exists a Kida vortex solution for incompressible flow,
whose velocity field
$(u_x, u_y)$ is
\begin{align}  
   u_x = \frac{3\Omega_p\Theta y}{2(\Theta -1)},\quad
   u_y = -\frac{3\Omega_p x}{2\Theta(\Theta-1)}
\end{align}
inside the vortex core.  The ratio of the vortex
semi-major to semi-minor axis being  $\Theta=a/b$ is a free
parameter.  This velocity field is such that the
 vorticity $\omega$ is constant in the rotating frame. The elliptical boundary of the vortex
is defined such that
 $\omega =-3\Omega_p(1+\Theta^2)/[2\Theta(\Theta-1)] = \omega_v - 3\Omega_p/2 $ inside the
boundary  of the vortex and $\omega = - 3\Omega_p/2$ outside. The quantity
 $\omega_v= -3\Omega_p(1+\Theta)/[2\Theta(\Theta-1)]$ is then
the vorticity of the vortex core relative to the background.
Being negative, this corresponds to an anticyclonic vortex.
In order to introduce perturbations corresponding
to Kida vortices, perturbations $\dd u_r \equiv
-u_y,\,\dd u_\phi \equiv  u_x$ are introduced inside a specified elliptical boundary with an  exponential decay
outside. The boundary is fixed  by specifying  $\Theta = 8$
and semi-major axis $a=H(r_p)$ where the  reference radius is
$r_p=1$.

\chapter{Energy densities in linear theory of edge modes}\label{energy}

In order to understand edge modes, in  \S\ref{linear_paper3} the
thermal-gravitational energy  (TGE) was defined  and the different 
contributions to it were discussed. There, a factor $D^2$ was 
applied to the TGE per unit length and to  each contributing term 
in order to overcome numerical difficulties associated with Lindblad
resonances. For completeness, the behaviour of the  TGE and the
various contributions to it is considered here without the additional
factor  $D^2.$  

\begin{figure}[!ht]
  \centering
  \includegraphics[width=1.0\textwidth]{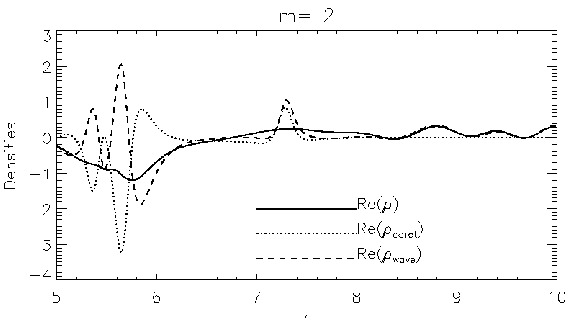}
  \caption{ The thermal and gravitational 
     energy (TGE)  per unit length  (solid) computed from 
    eigensolutions for the fiducial case $Q_o=1.5$ from linear
    theory. The  contributions from the  vortensity term (dotted line) and other
    terms (dashed line), defined by Eq. \ref{totalenerg}, are also shown.  
    The relatively small 
    bumps near $r=7$ are numerical.  However as they are the
    same magnitude for both contributions, incorporating them
      does  not affect conclusions
    concerning the overall energy balance in the system. 
\label{energy_den_original}}
\end{figure}

Fig. \ref{energy_den_original} shows very smilier features to the
corresponding  curves
discussed in \S\ref{energy_balance}. $\mathrm{Re}(\varrho)$ is negative around
corotation, again signifying a self-gravity driven edge mode. Beyond
$r\simeq 6.4$, $\mathrm{Re}(\rho)$ becomes positive due to pressure. The
 most extreme peak at $r\simeq 5.6$ coincides with the 
background vortensity edge (Fig. \ref{linear_basic}). The negative
contribution from the 
corotation term is larger in magnitude than that from the positive wave term,
resulting in $\mathrm{Re}(\rho)<0$. The co-rotation radius is such that $r_c < 5.6$,  
making the shifted frequency $\sbar_R(5.6) = m[\Omega(5.6) -
\Omega(r_c)]<0$. Also at $r=5.6$, $d\eta/dr < 0$, resulting in 
$\mathrm{Re}(\rho_\mathrm{corot}) <0$ at this point. 

$\mathrm{Re}(\rho_\mathrm{wave})$ and $\mathrm{Re}(\rho_\mathrm{corot})$ have 
relatively small spurious bumps around $r=7.2$  that are associated with
the outer Lindblad resonance and are numerical.
The eigenfunctions $W,\, \dd\Phi$ are well-defined
without singularities there, but evaluating
$\rho_\mathrm{corot}$  and $\rho_\mathrm{wave}$ involves division by
$D$, which can amplify numerical errors at Lindblad
resonances where $D\to 0$. Integrating $\mathrm{Re}(\rho)$ over
$r\in[5,10]$ gives 
$ U\equiv\mathrm{Re}\int_5^{10}\rho dr < 0.$ 
The TGE is negative, which means gravitational energy dominates over
the  pressure contribution  for  $r\geq 5$. Integrating the
contributions to the  TGE separately gives
\begin{align*}
  & U_\mathrm{corot}\equiv\mathrm{Re}\int_5^{10} \rho_\mathrm{corot} dr\simeq -0.94|U|,\\
  & U_\mathrm{wave}\equiv\mathrm{Re}\int_5^{10} \rho_\mathrm{wave} dr\simeq -0.077|U|.
\end{align*}
The integration range includes the OLR and thus the
spurious bumps in $\rho_\mathrm{wave}$ and
$\rho_\mathrm{corot}.$ However,
$U$ still approximately equals $ U_\mathrm{corot} + U_\mathrm{wave}$. The correct
energy balance is maintained despite  being subject to
possible numerical error due to the diverging factor $1/D.$  
The above means that, aside from the spurious bumps, 
the energy  density values for $r\in[5,10]$ may still be used to
interpret energy balance. As 
as before, the TGE  is predominantly accounted for by  the vortensity
term, since $|U_\mathrm{corot}/ U|\sim 0.9$.



\addtocontents{toc}{\protect\setcounter{tocdepth}{1}}


\end{document}